\documentclass[a4paper,11pt]{article}
\pdfoutput=1 

\usepackage{jheppub} 

\usepackage[utf8]{inputenc}
\usepackage[T1]{fontenc} 
\usepackage{multirow}
\usepackage{physics}
\usepackage{slashed}
\usepackage{xifthen}
\usepackage{subfig}
\usepackage{amsmath}
\usepackage{caption}
\usepackage{listings}
\usepackage{pdflscape}
\usepackage{graphics}
\usepackage{placeins}
\usepackage{xfrac}
\usepackage[dvipsnames]{xcolor}
\usepackage{standalone}
\usepackage{enumitem}

\usepackage{tikz}
\usepackage[compat=1.1.0]{tikz-feynman}
\tikzfeynmanset{warn luatex=false}
\tikzfeynmanset{/tikzfeynman/warn luatex = false}

\newcommand{\reffig}[1]{fig.~\ref{#1}}

\newcommand*{\halfway}{0.5*\pgfdecoratedpathlength+.5*5pt}
\tikzset{->-/.style={decoration={markings, mark=at position #1 with {\arrow{latex}}},postaction={decorate}},
	->-/.default=\halfway}
	
\makeatletter
\newcommand{\subalign}[1]{%
  \vcenter{%
    \Let@ \restore@math@cr \default@tag
    \baselineskip\fontdimen10 \scriptfont\tw@
    \advance\baselineskip\fontdimen12 \scriptfont\tw@
    \lineskip\thr@@\fontdimen8 \scriptfont\thr@@
    \lineskiplimit\lineskip
    \ialign{\hfil$\m@th\scriptstyle##$&$\m@th\scriptstyle{}##$\hfil\crcr
      #1\crcr
    }%
  }%
}
\makeatother
\allowdisplaybreaks
\DeclareMathOperator{\sgn}{sgn}

\chardef\MyArticleWithColor=\pdfcolorstackinit page direct{0 g}

\definecolor{mmblue}{rgb}{0.37, 0.51, 0.71}
\definecolor{mmyellow}{rgb}{0.88, 0.61, 0.14}
\definecolor{mmgreen}{rgb}{0.56, 0.69, 0.19}
\definecolor{mmred}{rgb}{0.92, 0.39, 0.21}

\title{Numerical Loop-Tree Duality:\\ contour deformation and subtraction}

\author[]{Zeno Capatti,}
\author[]{Valentin Hirschi,}
\author[]{Dario Kermanschah,}
\author[]{Andrea Pelloni}
\author[]{and Ben Ruijl}

\affiliation[]{ETH Z\"urich,\\
R\"amistrasse 101, %
8092 Z\"urich, Switzerland}

\emailAdd{zeno.ca@gmail.com}
\emailAdd{valentin.hirschi@gmail.com}
\emailAdd{d.kermanschah@gmail.com}
\emailAdd{benruyl@gmail.com}
\emailAdd{apelloni90@gmail.com}

\abstract{We introduce a novel construction of a contour deformation within the framework of Loop-Tree Duality for the numerical computation of loop integrals featuring threshold singularities in momentum space.
The functional form of our contour deformation automatically satisfies all constraints without the need for fine-tuning. We demonstrate that our construction is systematic and efficient by applying it to more than 100 examples of finite scalar integrals featuring up to six loops.
We also showcase a first step towards handling non-integrable singularities by applying our work to one-loop infrared divergent scalar integrals and to the one-loop amplitude for the ordered production of two and three photons. This requires the combination of our contour deformation with local counterterms that regulate soft, collinear and ultraviolet divergences.
This work is an important step towards computing higher-order corrections to relevant scattering cross-sections in a fully numerical fashion.
}

\begin{document} 
\maketitle
\flushbottom

\section{Introduction}

The Large Hadron Collider (LHC) is entering its high luminosity data acquisition phase and is thus transitioning from being a discovery experiment to a precision measurement one. For this new goal, accurate theoretical predictions are necessary in order to ensure that theoretical uncertainties remain at or below the level of experimental ones. In particular, this involves the computation of higher-order corrections to the cross-sections of relevant scattering processes, which are built by considering processes with additional unresolved partons (real-emission type of contributions) and additional loop degrees of freedom (virtual type of contributions).
These two classes of contributions are separately divergent but combine into a finite quantity in virtue of the Kinoshita-Lee-Nauenberg theorem~\cite{Kinoshita:1962ur,Lee:1964is}.

Traditionally, the computation of these two components is performed using very different approaches and the deep connection relating their degenerate infrared degrees of freedom is only realised through dimensional regularisation~\cite{tHooft:1972tcz,Bollini:1972ui,Ashmore:1972uj} at the very end of the computation.
Indeed, real-emission contributions are typically computed numerically through the introduction of subtraction counterterms~\cite{Frixione:1995ms,Catani:1996vz,GehrmannDeRidder:2008ug,Currie:2016bfm,Czakon:2010td,Boughezal:2015dra,DelDuca:2016csb,Somogyi:2009ri,DelDuca:2016ily,Caola:2017dug,Herzog:2018ily,Magnea:2018hab} or some form of phase-space slicing~\cite{Catani:2007vq,Grazzini:2017mhc,Cieri:2018oms,Boughezal:2016wmq,Boughezal:2019ggi,Gaunt:2015pea,Cacciari:2015jma,Currie:2018fgr}, whereas the evaluation of their virtual counterparts is mostly carried out purely analytically, thus realising the cancellation of infrared singularities at the integrated level.
A notable exception is the computation of \emph{inclusive} Higgs production at N$^3$LO accuracy~\cite{Anastasiou:2014vaa}, which was performed through \emph{reverse-unitarity}~\cite{Anastasiou:2002yz,Anastasiou:2002wq}.
Even though the same technique was further developed to accommodate the Higgs rapidity distribution in ref.~\cite{Dulat:2018bfe}, it is clear that this approach is not applicable to fully differential high-multiplicity processes.
Furthermore, despite impressive advances in the mathematical aspects of the \emph{reduction} of scattering amplitudes to master integrals~\cite{Chetyrkin:1981qh,Baikov:1996iu,Gehrmann:1999as,Anastasiou:2000mf,Smirnov:2005ky,vonManteuffel:2014ixa,Lee:2008tj,Ruijl:2017cxj,Anastasiou:2004vj,vonManteuffel:2012np,Lee:2013mka,Maierhoefer:2017hyi,Smirnov:2019qkx,Frellesvig:2019uqt,Kosower:2018obg,Ita:2015tya}, and their subsequent computation by means of differential equations~\cite{Gehrmann:1999as,Kotikov:1990kg,Papadopoulos:2014hla,Henn:2013pwa,Lee:2014ioa,Lee:2017oca,Primo:2016ebd}, it is believed that the computation of many relevant higher-order corrections to important processes (e.g. NNLO corrections to $p p \rightarrow t \bar{t} H$ and $p p \rightarrow t \bar{t} b \bar{b}$) will remain intractable with this traditional approach, in part due to the increase in the number of scales relevant to the problem and because of the appearance of new mathematical structures in the form of generalised elliptic polylogarithms~\cite{Adams:2015gva,Broedel:2017kkb,Broedel:2017siw,Passarino:2016zcd,Broedel:2019hyg}.

Numerical alternatives have been developed for the direct evaluation of loop integrals through sector decomposition~\cite{Binoth:2000ps,Anastasiou:2007qb,Lazopoulos:2007ix,Smirnov:2008py,Carter:2010hi,Borowka:2017idc} of their Feynman parametrisation or semi-numerical solutions~\cite{Francesco:2019yqt,Bonciani:2019jyb,Czakon:2008zk} of the system of differential equations relating them. This lead to the flagship computations of the NNLO corrections to the processes $p p \rightarrow H H$~\cite{Borowka:2016ypz,Borowka:2016ehy} and $p p \rightarrow t \bar{t}$~\cite{Czakon:2013goa}, where the exact dependency on all quark masses was kept.
Although these achievements demonstrate the superiority of numerical approaches in selected cases, they still suffer from the scalability issue inherited from their reliance on the analytical reduction of the complete amplitude to master integrals.

In light of the above overview of the research field of precise collider predictions, we choose to pursue an alternative construction which considers a purely numerical integration of the virtual contribution in momentum space. One particular benefit from such an approach lies in the prospect of bypassing the reduction to scalar integrals by considering the numerical integration of complete amplitudes directly (see existing results for one-loop amplitudes in refs.~\cite{Gong:2008ww,Becker:2010ng,Becker:2011vg,Becker:2012aqa} and first steps for applications to higher-loop finite scalar integrals in ref.~\cite{Becker:2012bi}).
Working in momentum space is especially appealing when also performing the loop energy integral(s) analytically using residue theorem. This energy integration yields the \emph{Loop-Tree Duality} (LTD) which provides an alternative representation for the loop integral containing terms with as many on-shell constraints as there are loops, making them effectively trees.
This aligns the measure of phase-space and LTD integrals, thus making LTD ideally suited to pursue the ambitious goal of directly combining real-emission and virtual contributions and compute them numerically at once by realising the local cancellation of their infrared singularities.
As with reverse-unitarity, this \emph{direct-unitarity} treatment explicitly maintains the aforementioned connection between real-emission and virtual contributions which is lost when computing them separately or using Feynman parametrisation.
Pioneering work of ref.~\cite{Kilian:2009wy,Soper:1998ye,Soper:1999xk,Nagy:2003qn,Assadsolimani:2009cz}
demonstrated the potential of carrying out this numerical programme by applying it at one loop.
However, during the last decade, the NLO revolution and the successes of analytical methods for the computation of many NNLO-accurate $2\rightarrow2$ cross-sections mostly overshadowed such purely numerical approaches. That is until recently, when groundbreaking new results from traditional analytical techniques arguably slowed down, thus opening the way for more numerical alternatives.

Since such radically different purely numerical approaches have to be developed from the ground up, they will not immediately catch up with the impressive analytical work performed by the community over the last two decades.
Instead, we proceed incrementally and build progressively towards the complete numerical evaluation of higher-order corrections while making sure at every step that our partial results are robust and make no compromise regarding generality in terms of the perturbative order and process considered.
We started this endeavour with ref.~\cite{Capatti:2019ypt} where we derived a general formulation of LTD by iteratively applying one-dimensional residue theorem. We showed how the duality relation hence obtained can easily be constructed algorithmically for any loop count and topology, and we tested it by applying it to many integrals without threshold singularities. In that regime, we could perform the integration of the LTD integrand directly as it does not require any contour deformation or counterterms. 

The first part of this work concerns the natural follow-up to ref.~\cite{Capatti:2019ypt}: regulating threshold singularities in order numerically integrate loop integrals evaluated with physical kinematics.
We achieve this by constructing a contour deformation in the ($3 n$)-dimensional complex integration space, designed in accordance with the constraints imposed by the causal prescription of Feynman propagators and by the matching conditions stemming from analytic continuation.
Contour deformations for numerical integration have been considered in the past~\cite{Gong:2008ww,Becker:2012bi,Buchta:2015wna}, and we present a novel variant well-suited to our multi-loop LTD expression. In order to ensure that our construction is correct for arbitrary \mbox{(multi-)loop} integrals, we apply it to more than a hundred qualitatively different examples, always finding agreement with the analytical benchmark (when available).
We also demonstrate in this way that the convergence rate of our current numerical implementation already renders it competitive. Finally, we discuss optimisation strategies to explore in future work that can improve results further.

The second part of the paper is dedicated towards applying our numerical programme to the computation of divergent scalar diagrams and of physical amplitudes. We consider divergent scalar box and pentagon topologies and the one-loop correction to the ordered production of two and three photons from a quark line.
This amplitude involves soft and collinear singularities that correspond to pinched threshold singularities where no regulating contour deformation is allowed.
This type of singularities can therefore only be regulated by the introduction of ad-hoc counterterms or through a direct combination with real-emission contributions.
In this work, we consider the former. In the case of scalar integrals we introduce a method to remove all IR divergences from one-loop diagrams. When considering complete amplitudes we combine our contour deformation and LTD integrand with the infrared and ultraviolet counterterms presented in ref.~\cite{Anastasiou:2018rib,Anastasiou:2019xxx}.

The outline of this work is as follows. In sect.~\ref{sec:ltd_intro}, we fix our notation by recalling our general multi-loop LTD expression. We construct a general contour deformation in sect.~\ref{sec:contour_deformation}. In sect.~\ref{sec:amplitudes}, the subtraction procedure for one-loop scalar integrals and amplitudes is discussed. In sect.~\ref{sec:optimisation}, we discuss various optimisations for our numerical integration. In sect.~\ref{sec:numerical_implementation}, we discuss our numerical implementation and we show our results in sect.~\ref{sec:results}. Finally, we present our conclusion in sect.~\ref{sec:conclusion}.

\section{Loop-Tree Duality}
\label{sec:ltd_intro}

In this section, we fix the notation and summarise our findings presented in ref.~\cite{Capatti:2019ypt}. A general $n$-loop integral in four-momentum Minkowskian space can be rewritten as an integral over the Euclidean space of the three-dimensional spatial part of the loop momenta. The integrand in that case is the sum of residues obtained by iteratively integrating out the energy variables one after the other by applying residue theorem. Each residue identified in this manner corresponds to a particular \emph{spanning tree} (i.e. a tree graph that connects all vertices) of the underlying loop graph, or equivalently, to a particular \emph{loop momentum basis} (i.e. the $n$ edges that complete a spanning tree back to the original $n$-loop graph) together with a specific set of signs for the energy solutions of the on-shell conditions fixing the residue location, which we call the \emph{cut structure}. \par
More precisely, we start from the following  $n$-loop integral
\begin{align}
    \label{eq:integraldefinition}
    I = \int \prod_{j=1}^n \frac{\mathrm{d}^4k_j}{(2\pi)^4} \frac{N}{\prod_{i \in \mathbf{e}} D_i},
    \quad D_i = q_i^2 - m_i^2 + \mathrm{i}\delta,
\end{align}
where $\mathbf{e}$ is the set of indices labelling the edges of the connected graph identifying the integral considered and the numerator $N$ is a regular function of the loop momenta.
We assume the Feynman propagators to be pairwise distinct with on-shell energies ${\pm E_i = \pm\sqrt{\vec{q}_i^{\,2} + m_i^2 - \mathrm{i}\delta}}$.
The momentum flow in a graph is uniquely determined by the choice of (consistent) signature vectors $\mathbf{s}_i=(s_{i1},\dots,s_{in})$, $s_{ij} \in \{\pm1,0\}$ for each propagator, such that $q_i^\mu = \sum_{j=1}^n s_{ij} k_j^\mu + p_i^\mu$, where $p_i^\mu$ is a shift that depends on external momenta. \par

We consider the integration of the energies in a fixed arbitrary order, set by $(k_1^0,\dots,k_n^0)$, each along the real line\footnote{As discussed in ref.~\cite{Runkel:2019yrs}, our final expression in eq.~\eqref{eq:LTDmaster} is also correct in the case of complex-valued external momenta, due to the fact that the right-most column of the matrix appearing in eq.~\eqref{eq:imaginaryinterplay} does \emph{not} include the imaginary part $\Im[{p_i^0}]$ of the external momenta. We note however, that the correct interpretation of the absence of this term in eq.~\eqref{eq:imaginaryinterplay} for complex-valued external kinematics is that the energy integrals are no longer performed along the real line but instead along a path including only one out of the two complex energy solutions of each propagator.} and closing on an arc of infinite radius in \emph{either} the upper (with winding number $\Gamma_j = + 1$) \emph{or} the lower ($\Gamma_j = -1$) complex half-plane.
We assume the integrand to vanish for large loop momenta, so that we can consider the integral along this arc to be zero, thus allowing us to relate the original integral to the sum of residues at poles located within the contour. \par

When carrying out this iterative integration of the loop energies and collecting residues, one finds that some residues may lie within or outside the integration contour depending on the spatial part of the loop momenta.
This would be an unfortunate complication, but we conjectured and verified explicitly that only the residues that unconditionally lie within the integration contour contribute to the integral, and moreover with the same prefactor, whereas all other \emph{conditional} residues are subject to exact cancellations~\cite{Capatti:2019ypt}.
We write the \emph{dual integrand} corresponding to one particular residue of the original integrand $f=N/\sum_{i \in \mathbf{e}} D_i$ identified by the loop momentum basis choice $\mathbf{b}=(b_1,\dots,b_n)$, $b_j \in \mathbf{e}$ (corresponding to the list of propagators put on-shell for this residue) as
\begin{align}
    \label{eq:masterresidue}
    \mathrm{Res}_\mathbf{b}[f]
    &=
    \frac{1}{\prod\limits_{i \in \mathbf{b}} 2E_{i}}
    \frac{N}{
    \prod\limits_{i\in \mathbf{e}\setminus\mathbf{b}} D_i}
    \Bigg\rvert_{\{q^0_j = \sigma_j^\mathbf{b} E_j\}_{j\in \mathbf{b}}}
\end{align}
with $\boldsymbol{\sigma}^\mathbf{b}=(\sigma_1^\mathbf{b},\dots,\sigma_n^\mathbf{b})$, $\sigma_j^\mathbf{b} \in \{\pm1\}$. It describes a residue that is within the contour for all loop momentum configurations if 
\begin{align}
    \prod_{r=1}^n \Theta\left(\Gamma_r \Im [k_{\mathbf{b},r}^{\boldsymbol{\sigma}}]\right) = 1,
    \quad
    \forall \vec{k}_j \in \mathbb{R}^{3},
\end{align}
where
\begin{equation}
    \Im [k_{\mathbf{b},r}^{\boldsymbol\sigma}]
    = \frac{\det
    \begin{pmatrix}
         & \sigma_{1} \Im [E_{i_1}] \\
        (s_{b_{j_1}j_2})_{\substack{1\leq j_1\leq r \\1\leq j_2 < r}} &  \vdots \\
        & \sigma_{r} \Im [E_{i_r}] \\
    \end{pmatrix}
    }{\det\left((s_{b_jj})_{1\leq j\leq r}\right)},
\label{eq:imaginaryinterplay}
\end{equation}
which for a choice of integration order, contour closure and momentum routing (determined by $(\vec{k}_1,\dots,\vec{k}_n)$, $\Gamma_j$ and $s_{ij}$ respectively), is satisfied unconditionally for exactly \emph{one} configuration of signs, the cut structure, denoted by $\boldsymbol{\sigma}^\mathbf{b}$. \par

Therefore, the original integral of eq.~\eqref{eq:integraldefinition} is identically equal to the resulting LTD expression
\begin{align}
\label{eq:LTDmaster}
    I = (-\mathrm{i})^n \int \prod_{j=1}^n \frac{\mathrm{d}^3\vec{k}_j}{(2\pi)^3} \sum_{\mathbf{b}\in \mathcal{B}} \mathrm{Res}_{\mathbf{b}}[f],
\end{align}
where $\mathcal{B}$ is the set of all loop momentum bases. \par

We stress again that the functional form of the LTD expression is implicitly dependent on the chosen order for the integration of loop energies, the contour closure choices and the particular momentum routing chosen for the original integral. However, we verified explicitly that one always \emph{numerically} obtains the same result for the sum of residues for given values of the spatial part of the loop momenta (set in a particular basis).
In order to facilitate the understanding of the central result of eq.~\eqref{eq:LTDmaster}, as well as to give some insight on its derivation, we provide an explicit two-loop example in appendix~\ref{sec:ltd_two_loops}. Finally, we provided as ancillary material of ref.~\cite{Capatti:2019ypt} a {\sc\small Python} implementation of the automated derivation of the cut structure for arbitrary loop topologies. Beyond its practical value, this code also demonstrates that \emph{explicitly} unfolding eq.~\eqref{eq:LTDmaster} can be done without any computational overhead.

The dual integrands can become singular on surfaces which may be labelled by the residue corresponding to the particular dual integrand in which they appear (specified through the loop basis $\mathbf{b}$) and the particular propagator of that dual integrand that becomes on-shell (specified through the propagator index $i$). These \emph{singular surfaces} are of the form
\begin{align}\label{decomp_surf}
    \xi_{\mathbf{b},i,\alpha_i} \equiv
    \sum_{j \in \mathbf{b}} \alpha_j E_j + \alpha_i E_i + \tilde{p}_i^{0,\mathbf{b}} = 0,
\end{align}
with $\alpha_i \in \{\pm1\}$ for $i \in \mathbf{e}\setminus\mathbf{b}$ and $\alpha_j = s_{ij}^\mathbf{b} \sigma^\mathbf{b}_j \in \{0,\pm1\}$ for $j \in \mathbf{b}$, where $s_{ij}^\mathbf{b}$ and $\tilde{p}_i^{\mu,\mathbf{b}}$ are implicitly defined through the change of basis $q_i^\mu = \sum_{j \in \mathbf{b}} s_{ij}^\mathbf{b} q_j^\mu + \tilde{p}^{\mu,\mathbf{b}}_i$
induced by the loop momentum basis $\mathbf{b}$ identifying this surface.
The singular surfaces $\xi$ can be separated into two classes: \emph{E}- and \emph{H-surfaces}.
E-surfaces are defined by the property of having all signs $\alpha_k, k \in \mathbf{b}\cup\{i\}$ equal, unless $\alpha_k$ is zero. We call the particular sign that all $\alpha_k$ are equal to (when not being zero) the \emph{surface sign}.
We factor out the surface sign and name the resulting E-surface $\eta^{\mathbf{b},i}$. From this point on, we consider every E-surface to have a positive sign for all energies:
\begin{equation}
    \label{eq:Esurface_positive_energies_def}
    \eta_{\mathbf{b},i} \equiv
    \sum_{j \in \mathbf{b}} E_j + E_i + p^{0}_{\eta} = 0.
\end{equation}
E-surfaces are convex and bounded.
H-surfaces are then defined by having at least one positive and at least one negative $\alpha_k$ and they are labelled $\gamma_{\mathbf{b},i,\alpha_i}$.

A particularly elegant feature of LTD is that the sum of dual integrands forming eq.~\eqref{eq:LTDmaster} only becomes singular on E-surfaces, as the singularities from H-surfaces cancel pairwise thanks to a mechanism referred to as \emph{dual cancellations}~\cite{LTDRodrigoOrigin2008,Aguilera-Verdugo:2019kbz}.
For $\delta=0$, an E-surface has a non-empty set of real solutions in $\vec{k}=(\vec{k}_1,\dots,\vec{k}_n) \in \mathbb{R}^{3n}$ if it satisfies
\begin{align}
\label{eq:exisitence_condition}
    (p^{0,\mathbf{b}}_i)^2 - (\vec{p}^{\,\mathbf{b}}_i)^2
    \geq 
    \left(
        \sum_{j \in \mathbf{b}}\alpha_j m_j + \alpha_i m_i
    \right)^2
    \quad
    \text{and}
    \quad
    p^{0,\mathbf{b}}_i < 0.
\end{align}
When both sides of this inequality are exactly zero, the E-surface has no interior since its minor axis is zero, and the E-surface corresponds to the location on an infrared collinear and/or soft singularities of the integral. We refer to them as \emph{pinched} E-surface, with the important property that singularities they correspond to cannot be regularised via a contour deformation of the loop momenta integration phase-space.

For $\delta > 0$ an E-surface $\eta$ is uniquely regulated by the imaginary prescription
\begin{align}
\label{eq:imag-prescription}
    \sgn \Im[\eta] = -1.
\end{align}
We do not find it particularly useful to work out the imaginary part of the the squared propagators appearing in eq.~\eqref{eq:LTDmaster} (referred to as \emph{dual propagator} in ref.~\cite{LTDRodrigoOrigin2008}). Instead, we prefer to stress that the relevant imaginary part of the E-surface equations induced by the causal prescription has a simple definite sign. As it will be made clear later, this observation is indeed the only relevant one in regard to the construction of a contour deformation that satisfies physical requirements and regulates threshold singularities. 

\section{Contour Deformation}
\label{sec:contour_deformation}

Numerical integration of Feynman diagrams and physical amplitudes in momentum space originated with the early attempts by Davison E. Soper in \cite{LTD_Soper_1} and \cite{LTD_Soper_2}, in which the LTD formalism was applied to virtual diagrams at one loop in order to then integrate the cross-section directly. Interestingly, the author also explicitly mentions and utilises the mechanism of \emph{local} real-virtual cancellations to render the integrand finite at the location of the non-integrable soft and collinear singularities. In order to avoid so-called scattering singularities, referred to in our work as one-loop E-surfaces, the author devised a contour deformation capable of satisfying the relevant constraints.

Several methods have since been developed for integrating diagrams and amplitudes directly in four-dimensional loop momentum space. A first success was the computation of one-loop photon amplitudes in ref.~\cite{Gong_Nagy_Soper}, followed by refs.~\cite{Becker:2010ng, BeckerMultiLoop2012, BeckerMasses2012, BeckerEfficiency2012} which generalised the formalism beyond one loop and applied it to more challenging integrals. The especially inspiring feature of this series of publication is the focus on constructing a provably exact deformation, through the concept of anti-selection and dynamic scaling of the deformation.

Around the same time when these techniques were developed, a different line of work expanded on LTD and, specifically, on its aspects relevant for the (3$n$)-dimensional numerical integration of integrals, amplitudes and cross sections \cite{Hernandez-Pinto:2015ysa, LTDRodrigoNumerical2017, Rodrigo_last}.
The contour deformation presented in these works is based on a linear combination of vectors normal to the existing E-surfaces, weighted by adjustable parameters and dampened by exponential functions with unspecified width; the deformation proves to be correct for simple threshold structures and in the limit of arbitrarily small dampening widths. Results obtained in this way however highlighted for the first time the potential of numerical integration over the spatial degrees of freedom resulting from the LTD identity.

In this section we will construct a reliable and exact deformation that is valid for an arbitrary number of loops and legs. We will give specific examples in order to illustrate how to implement the deformation constraints for complicated singular structures, especially on intersections of multiple E-surfaces.

As long as an integral only features non-pinched threshold singularities, it is possible to engineer a contour deformation yielding a finite result for the integral. The absorptive part of the integral is correct provided that the contour deformation considered satisfies requirements imposed by physical conditions, in particular \emph{causality}. In relativistic quantum mechanics, causality is originally realised in Feynman propagators via the $i\delta$-prescription or, equivalently, by the request that the theory is in the range of validity of Gell-Mann and Low's theorem~\cite{GellMann:1951rw}. In the LTD formalism, an imaginary prescription on propagators remains and, although its formal expression is more complicated than $i\delta$, it still holds that on E-surfaces this prescription sign is fixed (i.e. it does not depend on either external nor loop kinematics, see eq.~\eqref{eq:imag-prescription}).

Contour integration of threshold singularities requires to analytically continue the LTD integrand by replacing its dependence on the chosen basis of loop momenta $\vec{k}$, by the complex variable $\vec{k}-\mathrm{i}\vec{\kappa}\in(\mathbb{C}^{3})^n$, where $\vec{k}=(\vec{k}_1,\ldots,\vec{k}_n) \in (\mathbb{R}^{3})^n$ and ${\vec{\kappa}=(\vec{\kappa}_1,\ldots,\vec{\kappa}_n)\in (\mathbb{R}^{3})^n}$.
The spatial momenta associated with each propagator are a linear combination of the vectors in the chosen loop momentum basis plus an affine term:
\begin{equation}
    \vec{q}_j(\vec{k})=\sum_{i=1}^n s_{ji}\vec{k}_i+\vec{p}_j=\vec{Q}_j(\vec{k})+\vec{p}_j.
\end{equation}
Once analytically continued, these spatial momenta then also acquire an imaginary part:
\begin{equation}
    \vec{q}_j(\vec{k}-\mathrm{i}\vec{\kappa})=\vec{q}_j(\vec{k})-\mathrm{i}\vec{Q}_j(\vec{\kappa}).
\end{equation}
Each surface $\eta$ has an associated energy shift $p^0_\eta$, defined in eq.~\eqref{eq:Esurface_positive_energies_def} as a specific linear combination of the energies of external particles.

An approximation of the imaginary part of the E-surface $\eta$ can be obtained from the first order term of its Taylor expansion in $\lVert \vec{\kappa} \rVert$:
\begin{equation}
    \label{eq:deformed_esurface_imag_part}
    \mathrm{Im}[\eta(\vec{k}-\mathrm{i}\vec{\kappa})]=-  \vec{\nabla}_{\!\!\vec{k}} \eta(\vec{k}) \cdot \vec{\kappa} + \mathcal{O}(\lVert \vec{\kappa} \rVert^2),
\end{equation}
The quantity $\vec{\nabla}_{\!\!\vec{k}} \eta(\vec{k})$, henceforth denoted as $\vec{\nabla}\eta$, is the outward pointing normal vector to the surface $\eta(\vec{k})=0$.
The contour deformation is defined in the ($3n$)-dimensional complex space and we parametrise it as $\vec{k}-\mathrm{i}\vec{\kappa}(\vec{k})$.
It must satisfy constraints affecting two of its key characteristics, the direction and magnitude of the vector field $\vec{\kappa}(\vec{k})$:
\begin{description}
\item[Direction:] The deformation vector $\vec{\kappa}(\vec{k})$ must induce a sign of the imaginary part of the E-surface equation that matches the sign enforced by the causal prescription whenever $\vec{k}$ lies on a singular E-surfaces. This imposes conditions on the direction of the vector field $\vec{\kappa}(\vec{k})$. We derive these conditions by comparing the sign of the LTD prescription on E-surfaces (eq.~\eqref{eq:imag-prescription}) with the sign of the imaginary part of E-surfaces that results from the deformation (eq.~\eqref{eq:deformed_esurface_imag_part}). We obtain:
\begin{equation}
    \sgn [\vec{\nabla} \eta \cdot \vec{\kappa}]=+1, \text{ when } \eta(\vec{k})=0. 
\end{equation}
\item[Magnitude:] The norm of the deformation vector is limited by three constraints:
\begin{description}
 \item[Integrand continuity:] The LTD expression can be seen as a function of the on-shell energies of the internal particles $E_i=\sqrt{\vec{q}_i^{\,2}+m_i^2-\mathrm{i}\delta}$. These square roots have to be evaluated on a well-defined Riemann sheet. Thus the contour must not cross the branch cuts of any of the involved square roots.
%
\item[Complex pole constraint:] By extending the domain of the LTD integrand from $\mathbb{R}^{3n}$ to $\mathbb{C}^{3n}$ through the replacement of its functional dependency on $\vec{k}$ with $(\vec{k},\vec{\kappa})$, we find that in addition to real-valued poles (corresponding to the existing E-surfaces), the integrand also features complex-valued poles located at $(\vec{k},\vec{\kappa})$, with $\vec{\kappa}\neq\vec{0}$.
We stress that these \emph{complex poles} exist for \emph{all} E-surface equations: those (pinched or not) already having solutions for real loop momenta $(\vec{k},\vec{0})$ as well as those that do not and which are referred to as \emph{non-existing} E-surfaces (in regard to the fact that their existence condition of eq.~\eqref{eq:exisitence_condition} is not fulfilled).

\noindent According to Cauchy's theorem, the result of the contour-deformed integral will only be identical to that of the original defining integral over the spatial part of the loop momenta in the real hyper-plane, if and only if the volume defined by this real hyper-plane and the deformed contour does not contain any of such complex poles.
The magnitude of the contour deformation must therefore be constrained to be small enough so as to exclude these complex poles. 
%
\item[Expansion validity:] The causal constraint on the direction of the contour deformation as well as the complex pole constraint are derived from the Taylor expansion of each energy function $E_i$. We must therefore impose that the norm of the contour deformation vector field is such that the complex argument of each square root defining an energy remains within the range of validity of its expansion.
\end{description}
\end{description}
The next section~\ref{sec:one_loop_construction} presents the one-loop contour deformation direction constraints and our approach for solving them.
We will refer explicitly to illustrative examples that introduce key concepts of our work.
The precise and complete description of our construction of a contour deformation valid for an arbitrary number of loops and legs is presented in sect.~\ref{sec:general_solution}.

\subsection{Pedagogical construction at one loop}
\label{sec:one_loop_construction}

Consider a one-loop scalar box diagram in the LTD representation after having explicitly solved the on-shell constraint:
\begin{align}
    I
    &=
    -\mathrm{i} \int \frac{\mathrm{d}^3 \vec{k}}{(2\pi)^3}
    \sum_{b=1}^4\frac{1}{2E_b}
    \prod_{\substack{i=1 \\ i\neq b}}^4
    \left.\frac{1}{D_i}\right\vert_{q_b^0=E_b} \\
    &=
    -\mathrm{i}\int \frac{\mathrm{d}^3 \vec{k}}{(2\pi)^3} 
    \sum_{b=1}^4\frac{1}{2E_b}
    \prod_{\substack{i=1 \\ i\neq b}}^4 \frac{1}{\eta_{bi}\gamma_{bi}}.
\end{align}
where we used that at one loop the dual propagator factorises into the product of an E- and an H-surface, as ${D_i(k)\vert_{q_b^0=E_b} = \eta_{bi}(\vec{k})\gamma_{bi}}(\vec{k})$.
At one loop, one can also simplify the loop basis identifier $\mathbf{b}$ and write it as the index $b\in\mathbf{e}=\{1,2,3,4\}$ corresponding to the single LTD cut considered.
Thanks to the mechanism of dual cancellations, the sum of all dual integrands is only singular on E-surfaces which, at one loop, are two-dimensional rotational ellipsoids in spatial loop momentum space.
All of the potential singular E-surfaces of this scalar box appear as zeros of the functions
\begin{equation}
    \eta_{bi}(\vec{k})
    \equiv
    \sqrt{(\vec{k}+\vec{p}_{b})^2+m_b^2}+\sqrt{(\vec{k}+\vec{p}_{i})^2+m_i^2}-p_{b}^0+p_{i}^0,
    \quad
    p_{i}\equiv\sum_{j=1}^{i} p^{\,\text{ext}}_j,
\end{equation}
with $i,b \in \mathbf{e}$, $i\neq b$, and for given four-momenta of the four external legs $p^{\,\text{ext}}_j$, ${j \in \{1,2,3,4\}}$.
The number of E-surfaces that have solutions for real loop momenta has an upper bound based on the topology and the number of legs $N$. For one-loop topologies, an upper bound on the total number of \emph{existing} E-surfaces is $N(N-1)/2$, since we require $b \neq i$ and using the fact that if $\eta_{bi}$ exists, $\eta_{ib}$ cannot exist.

\begin{figure}
    \centering
    \includegraphics[width=0.6\textwidth]{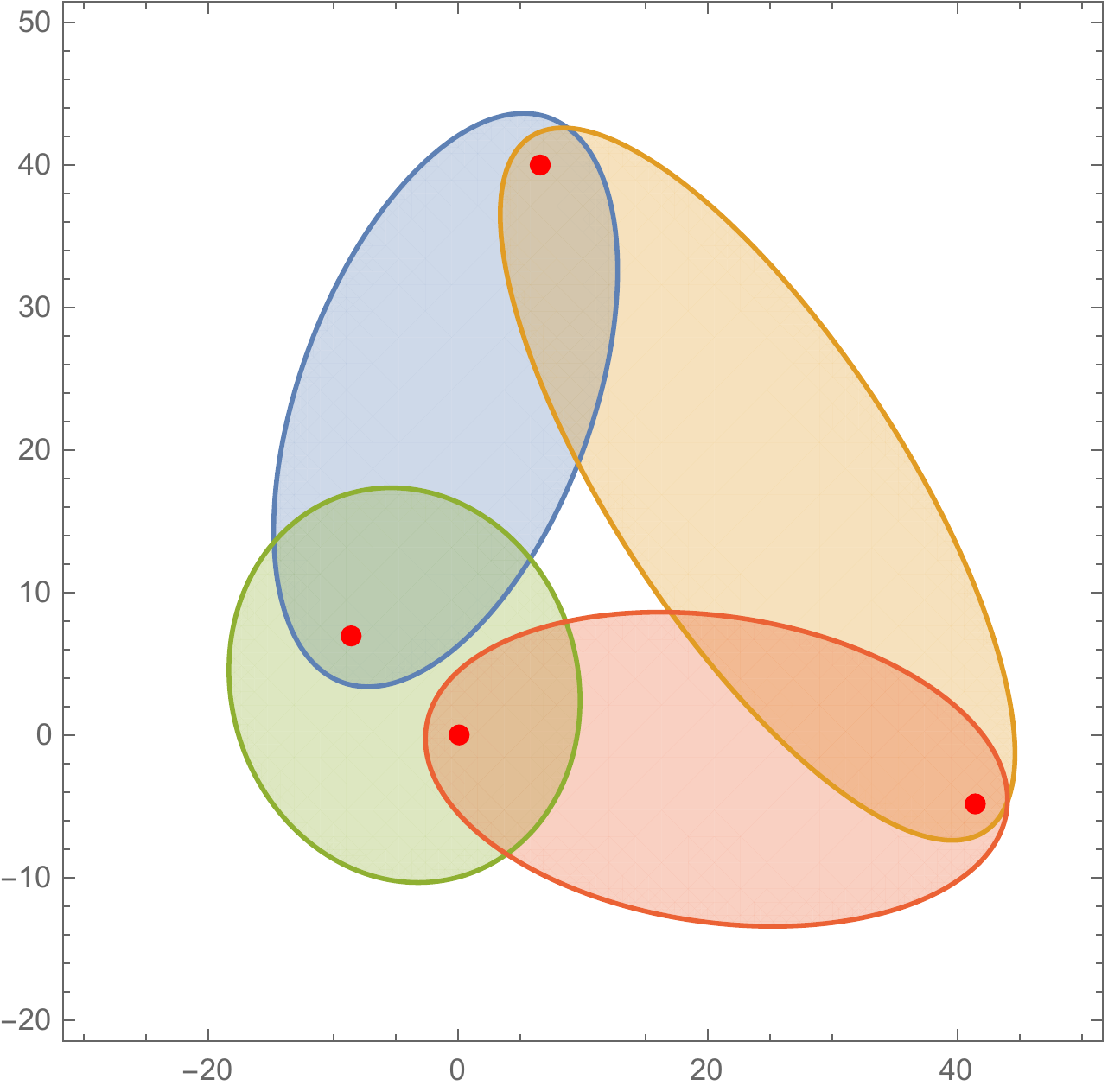}
    \caption{A $k_z=0$ section of the singular structure of the example configuration \texttt{Box4E}. It has four singular E-surfaces with four (partially shared) focal points coloured in red.}
    \label{fig:box_4e}
\end{figure}

The singularity structure of the LTD expression can be studied by focusing on particular singular E-surfaces and their intersections. In order to do this, we define the boundary and interior operators as
\begin{align}
    \partial\eta_{bi}&=\{\vec{k}\in\mathbb{R}^3 \ | \ \eta_{bi}(\vec{k})=0\}, \\
    \partial^-\eta_{bi}&=\{\vec{k}\in\mathbb{R}^3 \ | \ \eta_{bi}(\vec{k})<0\}.
\end{align}
The E-surface $\eta_{bi}$ exists, that is $\partial \eta_{bi}\neq\emptyset$, if $(p_{b}^0-p_{i}^0)^2-(\vec{p}_{b}-\vec{p}_{i})^2\geq(m_i+m_b)^2$ and $p_{b}^0-p_{i}^0\geq0$. If two ellipsoids $\eta, \ \eta'$ exist and intersect, then $\partial \eta \cap \partial \eta'\neq \emptyset$. Furthermore, if they intersect without being tangent, they also \emph{overlap}: $\partial^- \eta \cap \partial^- \eta'\neq \emptyset$. 
As an illustrative example, we now set particular values for the external box kinematics, which we refer to as \texttt{Box4E},
\begin{align}
\label{eq:box4e_kin}
\begin{split}
p_1^\text{\,ext} &=
(\phantom{-}14.0,-\phantom{0}6.6, -40.0,  \phantom{-}0), \\
p_2^\text{\,ext} &=
(-43.0, \phantom{-}15.2, \phantom{-}33.0,  \phantom{-}0), \\
p_3^\text{\,ext} &=
(-17.9, -50.0, \phantom{-} 11.8,  \phantom{-}0), \\
p_4^\text{\,ext} &= -p_1^\text{\,ext}-p_2^\text{\,ext}-p_3^\text{\,ext}
\end{split}
\end{align}
and list the resulting four members of the set of existing E-surfaces
 $\mathcal{E}=\{\textcolor{mmblue}{\boldsymbol{\eta_{12}}}, \textcolor{mmyellow}{\boldsymbol{\eta_{13}}}, \textcolor{mmgreen}{\boldsymbol{\eta_{42}}}, \textcolor{mmred}{\boldsymbol{\eta_{43}}}\}$,
\begin{align}\label{esurfaces_box4E}
\begin{split}
\textcolor{mmblue}{\boldsymbol{\eta_{12}}} &=
  \sqrt{(-6.6 + k_x)^2 + (-40 + k_y)^2 + k_z^2} + 
  \sqrt{(8.6 + k_x)^2 + (-7 + k_y)^2 + k_z^2}
  - 43,\\
\textcolor{mmyellow}{\boldsymbol{\eta_{13}}} &=
  \sqrt{(-6.6 + k_x)^2 + (-40 + k_y)^2 + k_z^2} + 
  \sqrt{(-41.4 + k_x)^2 + (4.8 + k_y)^2 + k_z^2}
  - 60.9,\\
\textcolor{mmgreen}{\boldsymbol{\eta_{42}}} &=
  \sqrt{k_x^2 + k_y^2 + k_z^2} +
  \sqrt{(8.6 + k_x)^2 + (-7 + k_y)^2 + k_z^2}
  - 29,\\
\textcolor{mmred}{\boldsymbol{\eta_{43}}} &=
  \sqrt{k_x^2 + k_y^2 + k_z^2} +
  \sqrt{(-41.4 + k_x)^2 + (4.8 + k_y)^2 + k_z^2}
  -46.9.
\end{split}
\end{align}
The four E-surfaces in eq.~\eqref{esurfaces_box4E} are coloured according to the colour scheme used in fig.~\ref{fig:box_4e}. 
A focal point is the loop momentum $(k_x,k_y,k_z)$ that sets the argument of an energy square root to zero. Each ellipsoid has two focal points, indicated with red dots in the figure.
The energy shift $p_{i}^0-p_{b}^0$ is the length of the major axis. The particular external kinematic configuration chosen in eq.~\eqref{eq:box4e_kin} has no component along the $k_z$-axis and therefore the particular section $k_z=0$ corresponds to the plane where the four E-surfaces have a maximal extent.

According to eq.~\eqref{eq:imag-prescription} we require the imaginary part on any E-surface $\eta$ to always be negative: $\sgn(\Im[\eta])=-1$.
By replacing $\vec{k}\rightarrow \vec{k}-\mathrm{i}\vec{\kappa}(\vec{k})$ and expanding the E-surface equations to first order in $||\vec{\kappa}||$, we find that the prescription reads
\begin{equation}
    \vec{\kappa}\cdot\vec{\nabla}\eta_{bi}
    =
    \vec{\kappa}\cdot\Bigg(\frac{\vec{k}+\vec{p}_b}{E_b}+\frac{\vec{k}+\vec{p}_i}{E_i} \Bigg)>0, \ \ \forall \vec{k}\in \partial\eta_{bi}, \ \forall \eta_{bi} \in \mathcal{E} \,,
    \label{eq:deformation_direction}
\end{equation}
which imposes that on any point on the E-surface, $\vec{\kappa}(\vec{k})$ should point outwards of the E-surface. On the intersection of many E-surfaces, the combined prescriptions impose that $\vec{\kappa}(\vec{k})$ must simultaneously point outwards of all of the intersecting E-surfaces.

\begin{figure}[h]
    \centering
    \includegraphics[width=0.6\textwidth]{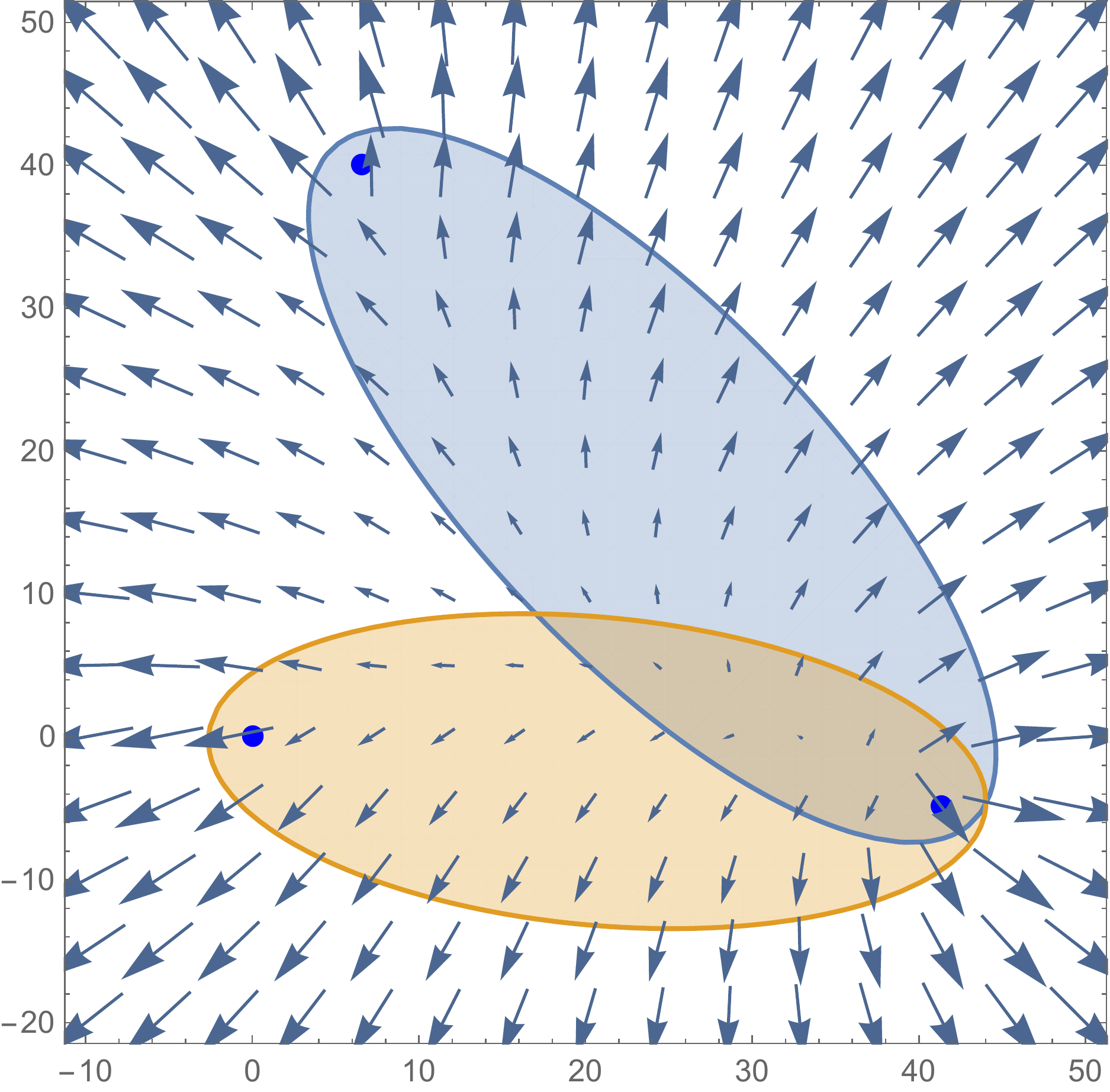}
    \caption{A correct deformation for two E-surfaces represented in the $k_z=0$ plane constructed by summing the normal vector fields of each of the two E-surfaces.}
    \label{fig:deformation_with_normals}
\end{figure}

One choice that always satisfies the condition of eq.~\eqref{eq:deformation_direction} for one single E-surface as well as for two intersecting E-surfaces is the sum of their respective normal vector fields, as shown in fig.~\ref{fig:deformation_with_normals}.
A similar deformation was proposed in ref.~\cite{LTDRodrigoNumerical2017}, where the deformation field $\vec{\kappa}(\vec{k})$ is written as a linear combination of the normal fields weighted by an exponential dampening factor that ensures that each normal field vanishes away from its defining E-surface. 
This particular choice of deformation vector is unsatisfactory when more than two E-surfaces exist, since 
\begin{itemize}
\item there could be triple intersections where the sum of the normal vectors is not guaranteed to be correct, unless the coefficients of the decomposition on normal vector fields is fine-tuned (and made dynamical functions of the real part of the loop momenta) so as to induce a vector with a valid direction and
\item contributions from various E-surfaces may spoil the validity of the deformation direction on another surface. Again this must be avoided by fine-tuning the strength of the dampening factors affecting each normal field.
\end{itemize}

\begin{figure}
    \centering
    \begin{minipage}{.49\textwidth}
    \centering
    \includegraphics[width=\textwidth]{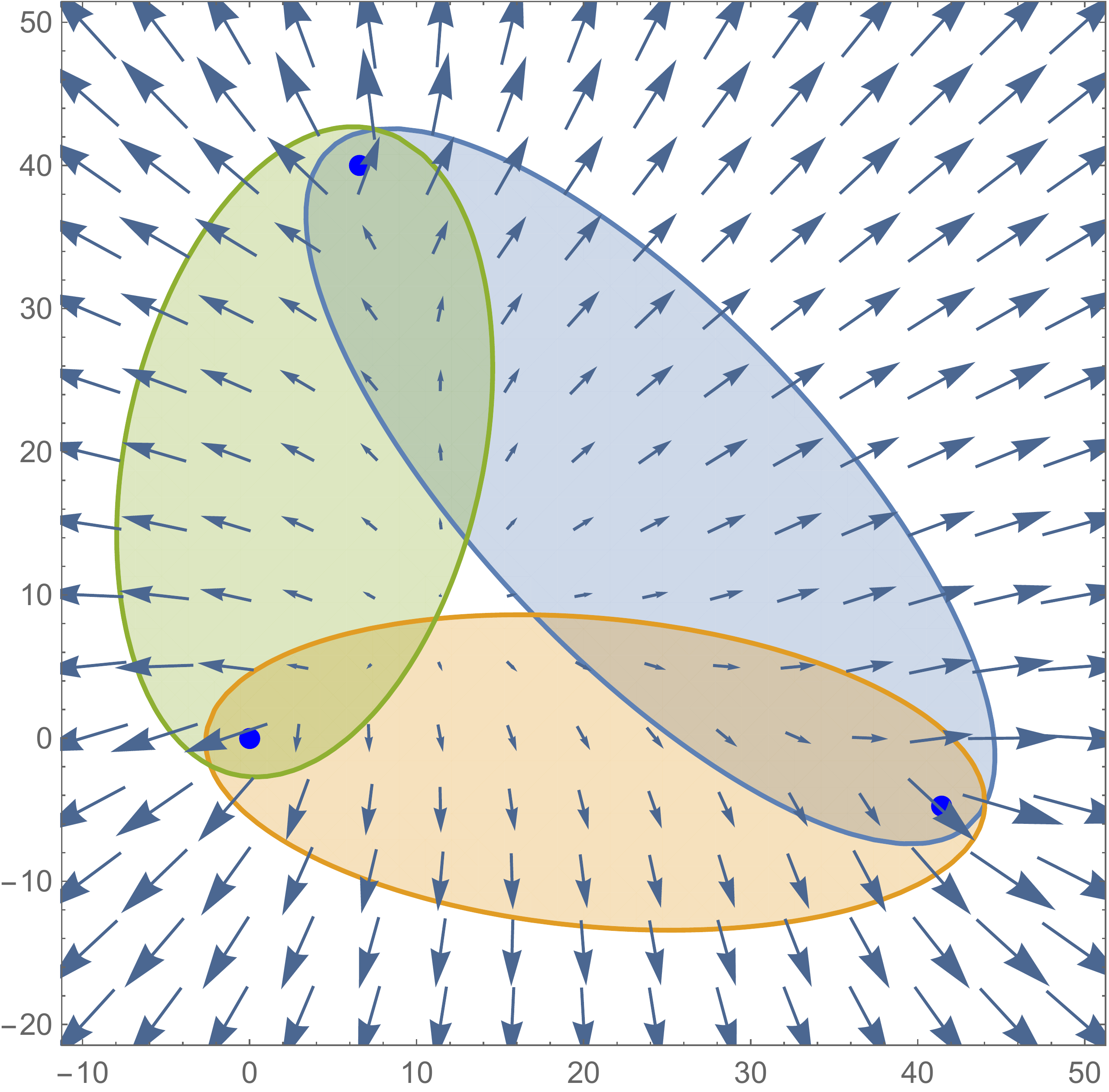}
    \end{minipage}
    \begin{minipage}{.49\textwidth}
    \centering
    \includegraphics[width=\textwidth]{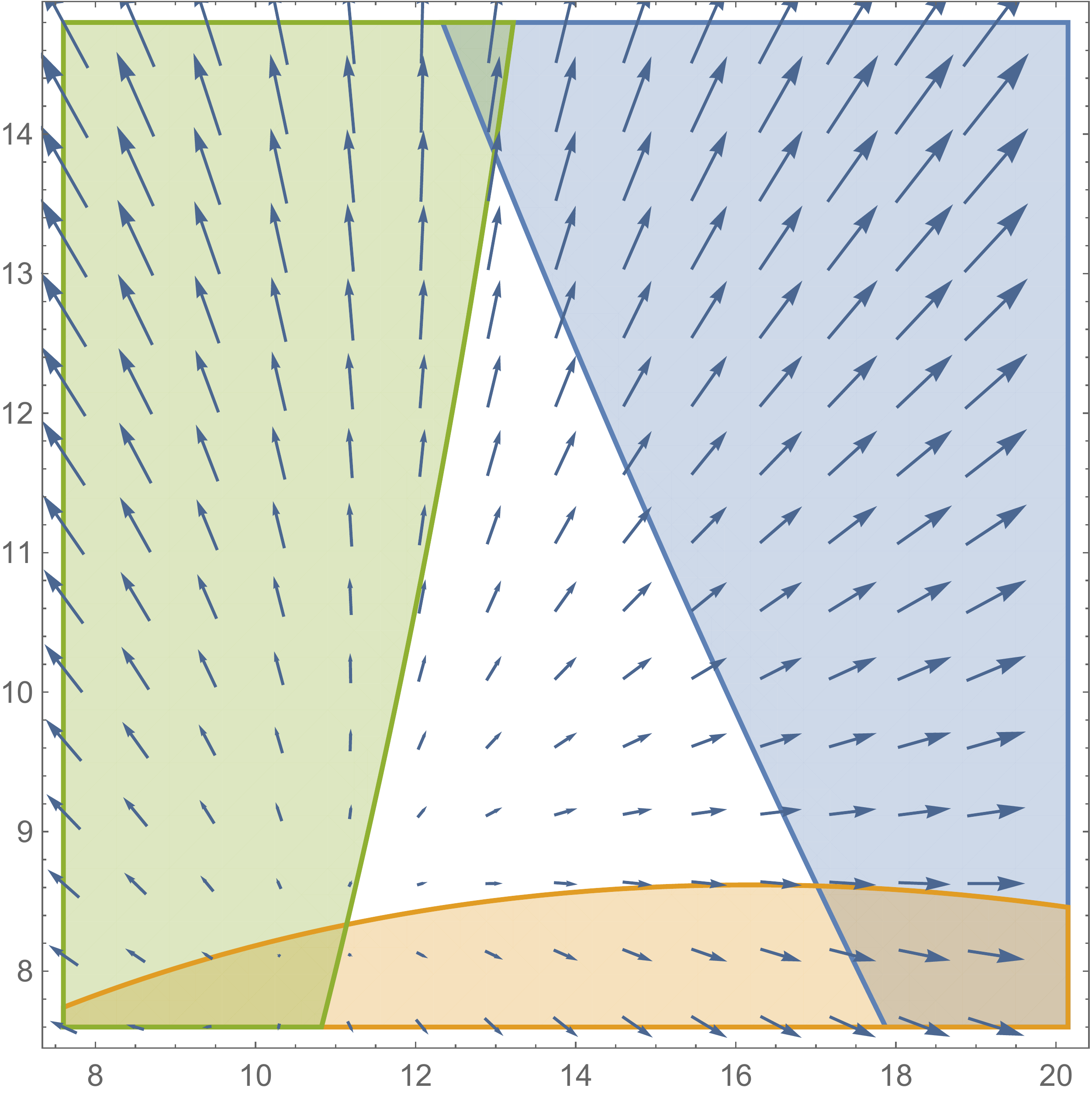}
    \end{minipage}
    \caption{An example of an incorrect deformation vector field constructed by adding the normal vector fields of three E-surfaces. This particular case requires fine-tuning of the normalisation of each of the three fields added in order to obtain a valid deformation.}
    \label{fig:normal_fail}
\end{figure}

In fig.~\ref{fig:normal_fail} we give an example with three E-surfaces, where a naive unweighted sum of normal vectors does not yield a valid deformation. 
By using fine-tuned dampening of the normal vector fields from each E-surface, such cases may be avoided but this does require an ad-hoc treatment and can lead to poor numerical convergence.

The next subsection introduces the concept of \emph{deformation sources} which we will use to build a deformation that avoids the shortcomings discussed in this section when considering normal fields.

\subsubsection{Deformation sources}

Since E-surfaces are convex surfaces, given a point $\vec{s}$ within the interior of an E-surface $\partial^-\eta$, the radial field $\vec{v}_{\vec{s}}(\vec{k}) \equiv \vec{k}-\vec{s}$, centered at $\vec{s}$, satisfies the causal prescription $\Im[\eta]|_{\vec{k}-i\vec{v}_{\vec{s}}} <0$ on any point on the surface, where $\eta(\vec{k})=0$.
We note that the interior of the \emph{intersection} of a set $F\subseteq\mathcal{E}$ of E-surfaces again defines a convex volume and therefore we analogously have that, for any given point $\vec{s}$ in this volume, that is $\vec{s}\in\bigcap_{\eta\in F}\partial^-\eta$, the corresponding radial field $\vec{v}_{\vec{s}}$ simultaneously satisfies the causal prescription of \emph{all} on the E-surfaces in $F$ and, especially, on their intersections.
We call such a point $\vec{s}$ a \emph{deformation source} for the overlapping set $F$. 
For a case in which there exists a single point $\vec{s}$ simultaneously in the interior of \emph{all} of the existing E-surfaces, then the radial deformation field $\vec{\kappa}(\vec{k}) \propto (\vec{k}-\vec{s})$ satisfies the causal prescription on all the threshold singularities.

\begin{figure}
    \centering
    \begin{minipage}{.49\textwidth}
    \centering
    \includegraphics[width=\textwidth]{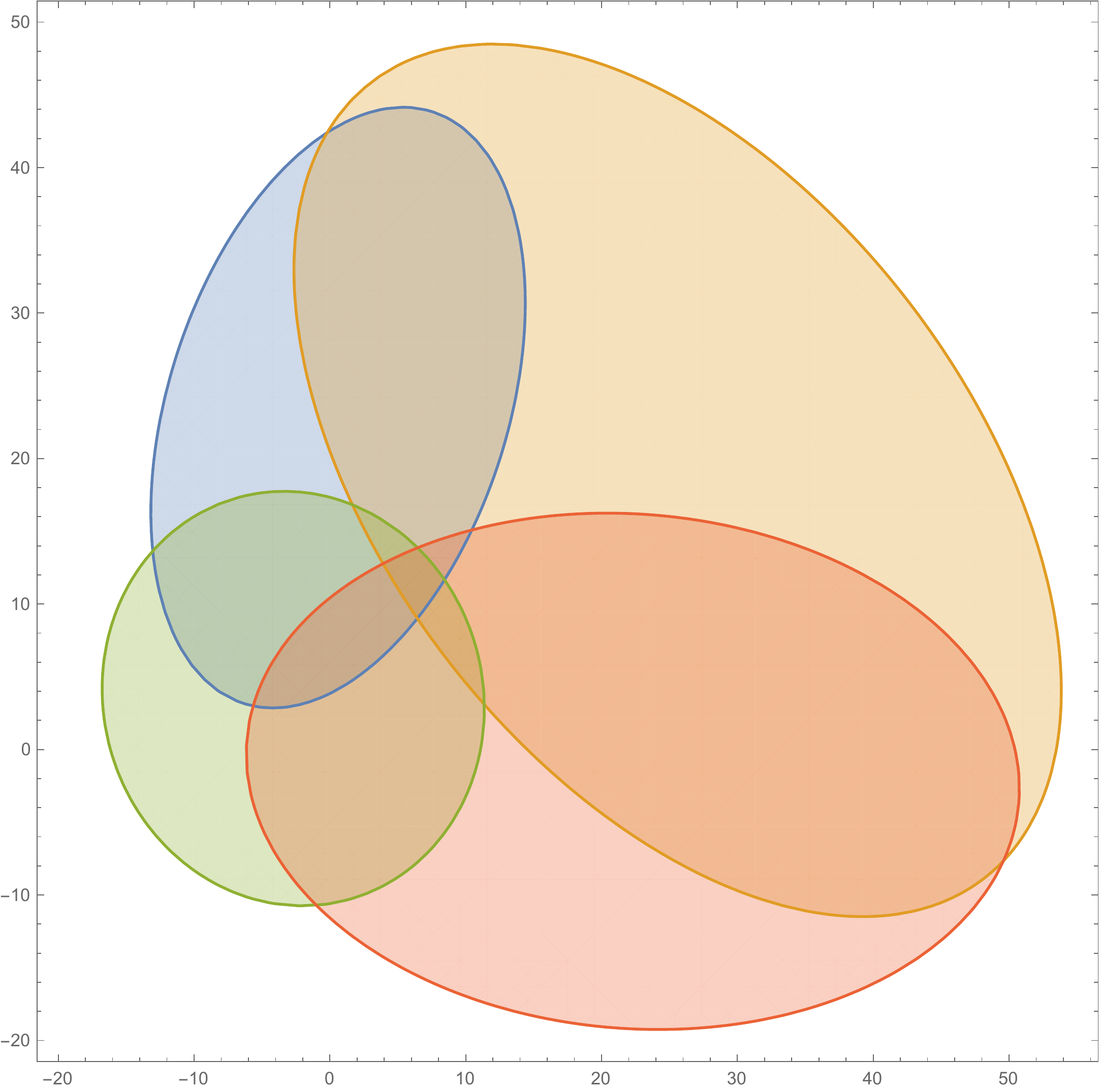}
    \end{minipage}
    \begin{minipage}{.49\textwidth}
    \centering
    \includegraphics[width=\textwidth]{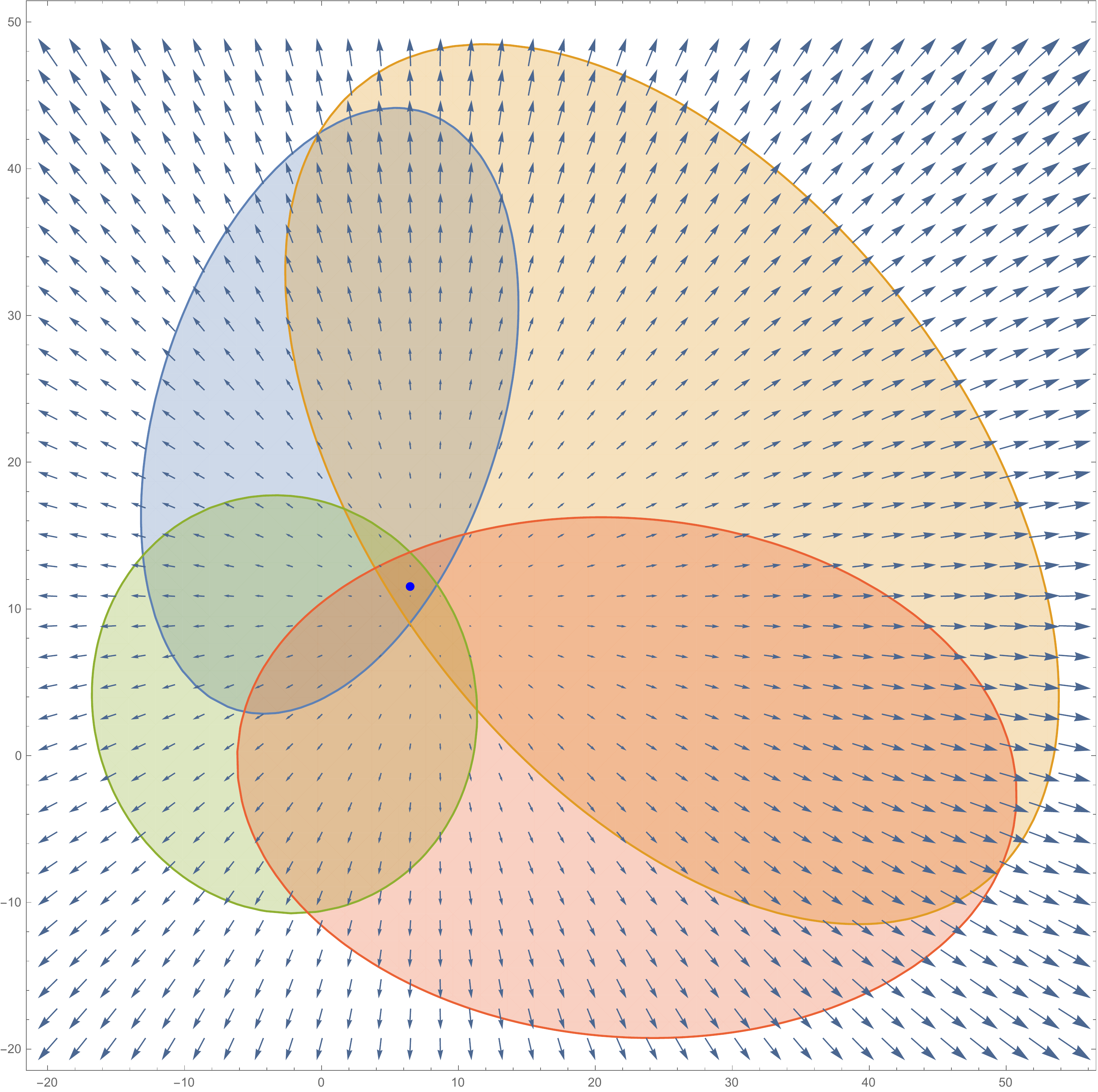}
    \end{minipage}
    \caption{A correct deformation using the radial field $\vec{k}-\vec{s}$ generated by a single source $\vec{s}$ contained in the interior of all four E-surfaces.}
    \label{fig:one_overlap_structure}
\end{figure}

When there is no single point simultaneously in the interior of all E-surfaces, one can construct a deformation vector written as the sum of radial fields centered at different locations, and adequately multiplied by an anti-selector function disabling the effect of the radial field on all the E-surfaces in which the point is not contained. 
The anti-selection is constructed such that the individual terms building the deformation vector fields are always ``additive'' in their ability to satisify the causality requirements. Indeed, a crucial aspect of our design of the deformation is the adoption of a model in which contributions that may spoil the direction on a particular threshold singularity are excluded (i.e. ``anti-selected''), as opposed to a model that enables (i.e. ``selects'') the correct contributions on the particular thresholds they are designed for. 

\begin{figure}
    \centering
    \begin{minipage}{.49\textwidth}
    \centering
    \includegraphics[width=\textwidth]{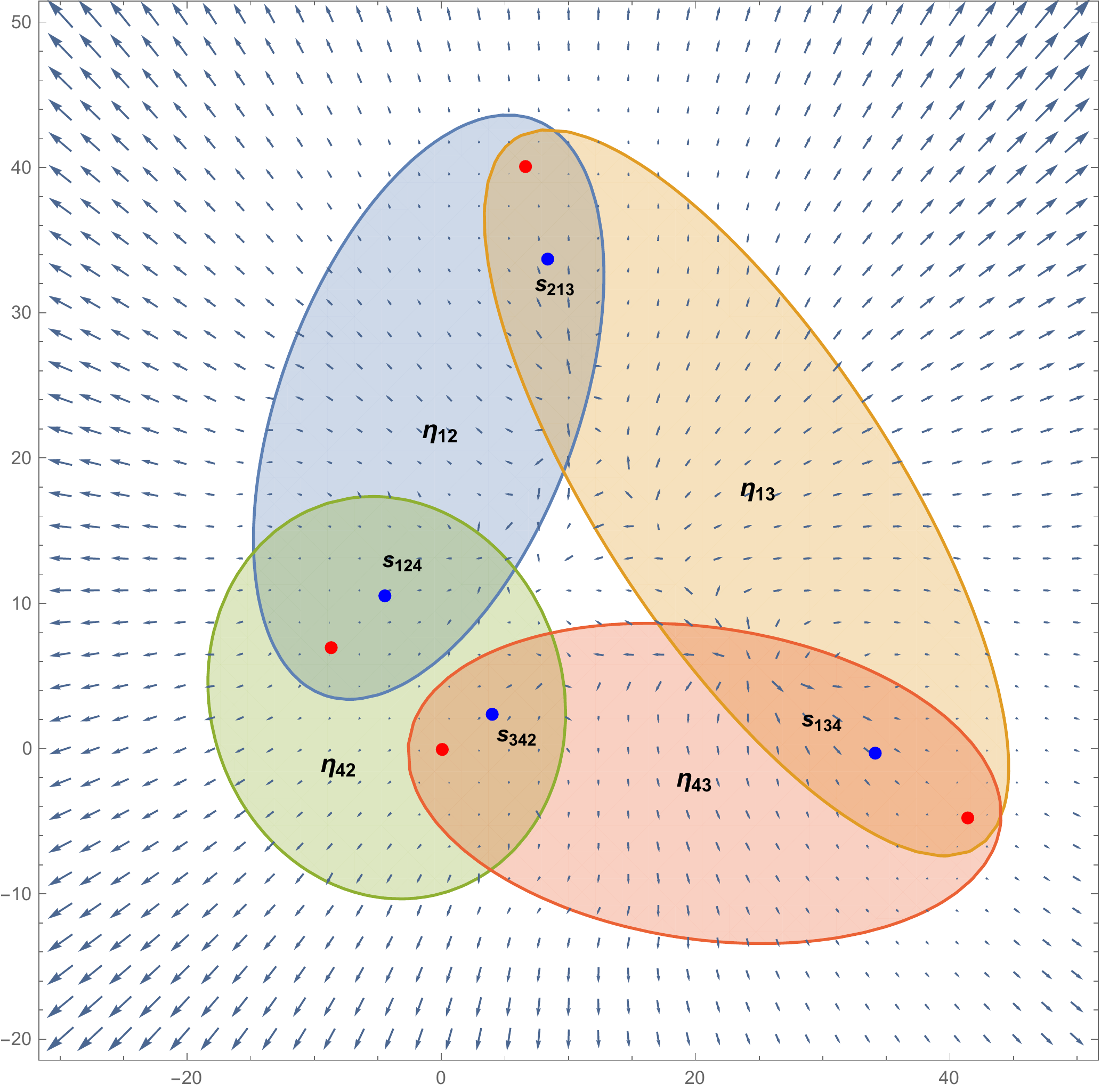}
    \end{minipage}
    \begin{minipage}{.49\textwidth}
    \centering
    \includegraphics[width=\textwidth]{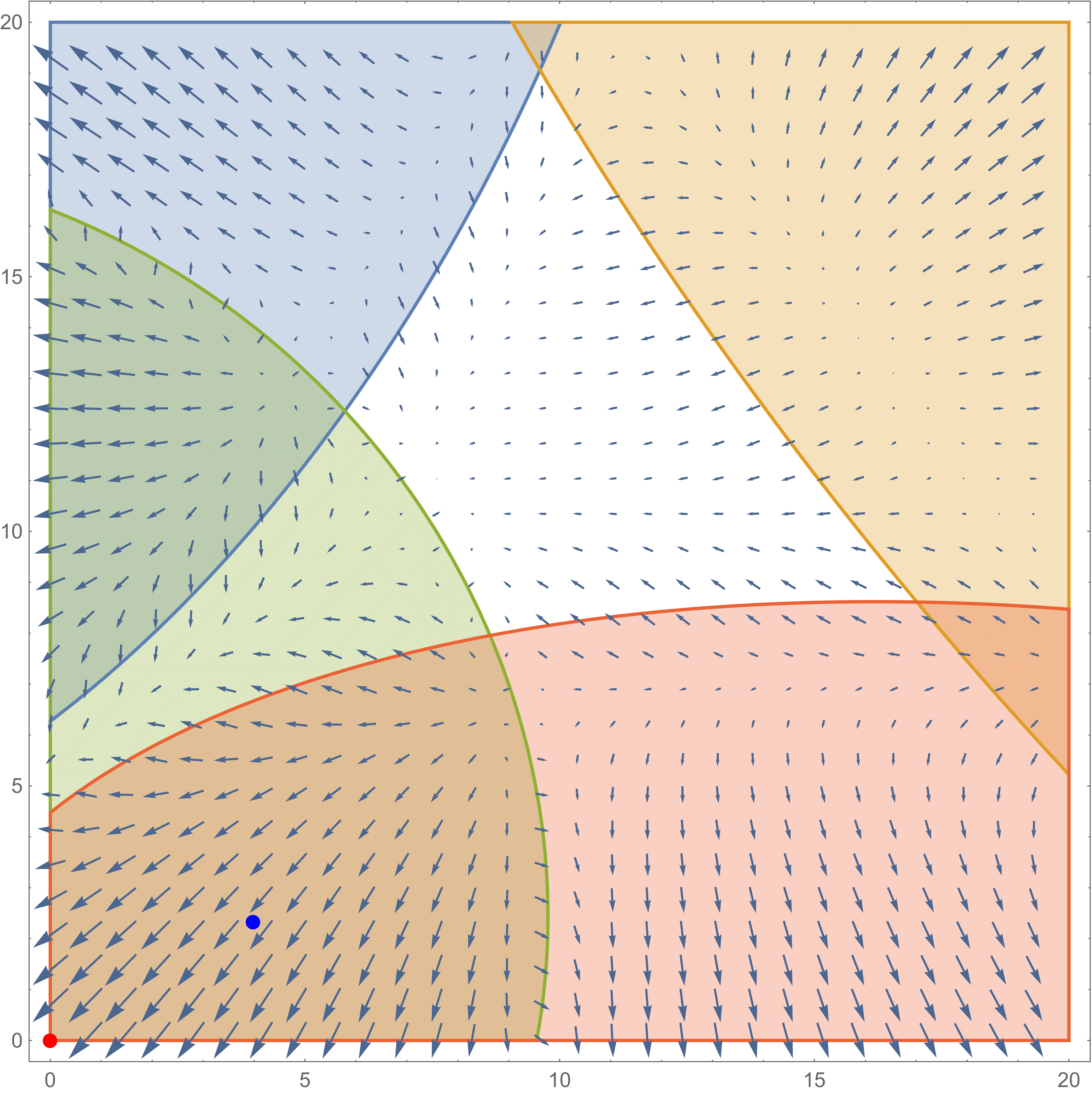}
    \end{minipage}
    \caption{A correct deformation direction with functional form described by eq.~\eqref{deffo} for \texttt{Box4E} using four sources which are excluded on those E-surfaces whose interior does not contain the source. The right plot is a zoom-in on the central region.}
    \label{fig:multi_overlap_structure}
\end{figure}

We illustrate more specifically how an \emph{anti-selection} model is preferable to a \emph{selection} one by highlighting the shortcomings of the latter when applied to the previously introduced \texttt{Box4E} configuration whose four E-surfaces are shown in fig.~\ref{fig:multi_overlap_structure} in the $k_z=0$ plane.
The ``selection'' model would in this case amount to combine all four radial fields as follows (the discussion of the analogous construction of ref.~\cite{LTDRodrigoNumerical2017} that involves normal fields would be similar):
\begin{eqnarray}
\label{eq:selection_model_deformation}
    \vec{\kappa}_{\rm selection\_model} &=& \vec{v}_{\vec{s}_{124}} \left(\bar{T}(\eta_{12})+\bar{T}(\eta_{42})\right) \nonumber\\
    &+& \vec{v}_{\vec{s}_{213}} \left(\bar{T}(\eta_{12})+\bar{T}(\eta_{13})\right) \nonumber\\
    &+& \vec{v}_{\vec{s}_{134}} \left(\bar{T}(\eta_{13})+\bar{T}(\eta_{43})\right) \nonumber\\
    &+& \vec{v}_{\vec{s}_{342}} \left(\bar{T}(\eta_{43})+\bar{T}(\eta_{42})\right),
\end{eqnarray}
where the selection function\footnote{The selection function chosen in ref.~\cite{LTDRodrigoNumerical2017} is an exponential Gaussian of adjustable width $A_{bi}$: $\exp(-\frac{ (\eta_{bi}(\vec{k}) \gamma_{bi}(\vec{k}))^2 }{A_{bi}})$.} simply is one minus the anti-selection function $T(\eta_{bi})$ defined as follows:
\begin{eqnarray}
\bar{T}(\eta_{bi})&=&1-T(\eta_{bi})\label{eq:selector}\\
T(\eta_{bi})&=&\frac{\eta_{bi}(\vec{k})^2}{\eta_{bi}(\vec{k})^2 + M^2 \left(p_{i}^0-p_{b}^0\right)^2 },\label{eq:antiselector}
\end{eqnarray}
where $M$ is an adjustable free parameter, and $p_{i}^0-p_{b}^0$ is the length of the major axis of the E-surfaces $\eta_{bi}$, which provides a measure for the size of the E-surfaces. Another possible choice is to substitute the normalisation $-p_{b}^0+p_{i}^0$ with $\sqrt{(p_{i}^0-p_{b}^0)^2-(\vec{p}_{i}-\vec{p}_{b})^2-(m_i+m_b)^2}$, which is the minor axis length of the E-surface. The choice of $M$ provides an estimate of how rapidly $T(\eta)$ saturates to one when $\vec{k}$ is further away from the surface $\eta_{bi}$.

The deformation of eq.~\eqref{eq:selection_model_deformation} stemming from the selection model is problematic for mainly two reasons:
\begin{itemize}
\item On the threshold E-surface $\eta_{12}$, the deformation receives contributions mostly from $\vec{v}_{\vec{s}_{124}}$ and $\vec{v}_{\vec{s}_{213}}$ (which do satisfy the causal prescription) but also from $\vec{v}_{\vec{s}_{134}}$ and $\vec{v}_{\vec{s}_{342}}$ (which may not satisfy the causal prescription) since the suppression factor induced by their respective selection function is \emph{small} on this surface, but \emph{not zero}. This implies the necessity of fine-tuning the suppression parameters which may be a difficult task when E-surfaces with very different causal constraints lie close to each other.
\item On the intersection of two E-surfaces, for example $\partial\eta_{12}\cap\partial\eta_{13}$, three of the four radial deformation fields $\vec{v}_{\vec{s}_{124}}$,$\vec{v}_{\vec{s}_{213}}$ and $\vec{v}_{\vec{s}_{134}}$ are active \emph{without any} suppression, even though only $\vec{v}_{\vec{s}_{213}}$ is guaranteed to be correct on this particular intersection.
\end{itemize}
One may think of alleviating the intersection problem by simply removing such intersections from the selector function applied to the deformation sources that are invalid:
\begin{eqnarray}
\label{eq:selection_model_deformation2}
    \vec{\kappa}_{\rm selection\_model\_improved} &=& \vec{v}_{\vec{s}_{124}} \left(\bar{T}(\eta_{12})T(\eta_{13})+\bar{T}(\eta_{42})T(\eta_{43}) \right) \nonumber\\
    &+& \vec{v}_{\vec{s}_{213}} \left(\bar{T}(\eta_{12})T(\eta_{42})+\bar{T}(\eta_{13})T(\eta_{43})\right) \nonumber\\
    &+& \vec{v}_{\vec{s}_{134}} \left(\bar{T}(\eta_{13})T(\eta_{12})+\bar{T}(\eta_{43})T(\eta_{42})\right) \nonumber\\
    &+& \vec{v}_{\vec{s}_{342}} \left(\bar{T}(\eta_{43})T(\eta_{13})+\bar{T}(\eta_{42})T(\eta_{12})\right).
\end{eqnarray}
However, this solution is again not exact since even though $\bar{T}(\eta_{42})T(\eta_{43})$ and $\bar{T}(\eta_{43})T(\eta_{42})$ are small quantities on $\partial\eta_{12}\cap\partial\eta_{13}$, they are not identically zero.
In fact, it is impossible to build a \emph{continuous} selection function that identically vanishes on a particular intersection of E-surfaces while at the same time being identically unity when evaluated anywhere on one of the intersecting E-surfaces but outside of the intersection.

The above shows that if the the contour deformation is required to be correct (i.e. independently of its parameters), the radial deformation fields must be combined using an \emph{anti}-selection paradigm that also avoids referring directly to intersections of E-surfaces, since one cannot continuously (anti-)select them.
In the example of \texttt{Box4E}, we achieve this by constructing the final deformation vector $\vec{\kappa}$ as follows:
\begin{eqnarray}\label{deffo}
\vec{\kappa}&=& \vec{v}_{\vec{s}_{124}} T(\eta_{13}) T(\eta_{43}) \nonumber\\
    &+&  \vec{v}_{\vec{s}_{213}} T(\eta_{42}) T(\eta_{43}) \nonumber\\
    &+&  \vec{v}_{\vec{s}_{134}} T(\eta_{42}) T(\eta_{12}) \nonumber\\
    &+&  \vec{v}_{\vec{s}_{342}} T(\eta_{12}) T(\eta_{13})
\end{eqnarray}
which \emph{exactly} satisfies the causal requirements for $\vec{k}$ on $\partial\eta_{12}$ and $\partial\eta_{12}\cap\partial\eta_{13}$:
\begin{eqnarray}
\vec{\kappa} (\vec{k}) |_{\vec{k} \in \partial \eta_{12}} &\overset{!}{=}& \vec{v}_{\vec{s}_{124}} T(\eta_{13}) T(\eta_{43})
 +  \vec{v}_{\vec{s}_{213}} T(\eta_{42}) T(\eta_{43}) \nonumber\\
\vec{\kappa} (\vec{k}) |_{\vec{k} \in \partial \eta_{12}\cap \partial \eta_{13}} &\overset{!}{=}& \vec{v}_{\vec{s}_{213}} T(\eta_{42}) T(\eta_{43}).
\end{eqnarray}


In general, the minimal set of sources required for constructing a valid deformation with this anti-selection model is obtained by determining the \emph{maximal overlap structure} of the E-surfaces, which we will formally define in sect.~\ref{sec:general_solution}. For \texttt{Box4E}, this structure is $\{\{\eta_{01}, \eta_{13}\},\{\eta_{13},\eta_{43}\},\{\eta_{42},\eta_{43}\},\{\eta_{12},\eta_{42}\}\}$.
After the maximal overlap structure has been determined, one has to construct source points in the interior of each overlap listed in the maximal overlap structure. Details about our strategy for choosing these particular points are given in sect.~\ref{sec:source_determination}.

Now that we have introduced and illustrated the key concepts underlying our construction of a valid deformation direction, we formalise it for an arbitrary number of loops and legs.

\subsection{General solution to constraints on direction}
\label{sec:general_solution}
In the absence of UV and IR non-integrable divergent behaviours, E-surfaces are the only singularities in the space of loop momenta that  need to be regulated by a contour deformation. In sect.~\ref{sec:one_loop_construction}, we have shown that we have to construct a vector field pointing outwards on every E-surface. In this section we study this constraint in more detail. We remind the reader of the simplified notation identifying $(\vec{k}_1,\ldots,\vec{k}_n)$ with $\vec{k}$ that combines all coordinates of the $n$-loop integration space.

E-surfaces are the boundary of convex, bounded volumes. We write the E-surface manifold as $\partial\eta$ and its convex interior as $\partial^-\eta$, that is:
\begin{align}
\partial\eta & \equiv \{\vec{k}=(\vec{k}_1,\ldots,\vec{k}_n) \ | \ \eta(\vec{k})=0\},\\
\partial^-\eta & \equiv \{\vec{k}=(\vec{k}_1,\ldots,\vec{k}_n) \ | \ \eta(\vec{k}) < 0\}.
\end{align}
The radial field $\vec{k}-\vec{s}$ centred at point $\vec{s}$ has a strictly positive projection on any normal to the surface if and only if it is inside the surface itself:
\begin{equation}
(\vec{k}-\vec{s})\cdot \nabla \eta (\vec{k}) > 0 \ \ \forall \vec{k}\in \partial \eta \hspace{0.4cm}\text{       iff      }\hspace{0.4cm} \vec{s}\in \partial^{\tiny -} \eta .
\end{equation}
In general, given a set of E-surfaces $F$ and a point in their interior:
\begin{equation}
\vec{s}_F \in \bigcap\limits_{\eta \in F} \partial^-(\eta) \,,
\end{equation}
then $\vec{k}-\vec{s}_F$ will have positive projection on all normal vectors of E-surfaces in $F$ and thus satisfies the causal prescription for all E-surfaces in $F$. We call $\vec{s}_F$ the \emph{source} of the set $F$.

The aforementioned construction of the deformation field $\vec{k}-\vec{s}_F$ provides a systematic solution to the hard problem of constructing a deformation vector on the intersection of all E-surfaces in $F$, where many causal constraints need to be satisfied simultaneously.

In order to extend the applicability of the construction, we need to generalise it to more than one set of overlapping E-surfaces. Given the set of all existing E-surfaces $\mathcal{E}$, we define the \emph{overlap structure}
\begin{equation}
\mathcal{O}=\Big\{F\subseteq\mathcal{E} \ | \  \bigcap_{\eta\in F}\partial^- \eta\neq \emptyset \Big\}.
\end{equation}
Thus $\mathcal{O}$ contains all possible sets of overlapping E-surfaces. One can immediately conclude that, if a set $F$ is in $\mathcal{O}$, then any subset $F'\subseteq F$ is in $\mathcal{O}$.

%
Since a deformation vector $\vec{k}-\vec{s}_F$ is not guaranteed to satisfy the causal prescription on any point on an E-surface in $\mathcal{E}\setminus F$, one has to identify the sets of overlaps $F_1,\ldots,F_N$ such that, among the radial fields $\vec{k}-{\vec{s}_{F_1}},\ldots,\vec{k}-{\vec{s}_{F_N}}$ generated by such overlaps, there is at least one satisfying the correct causal direction on any point on an E-surface and, especially, on any intersection of them. Such a set with the least amount of elements is referred to as the \emph{maximal overlap structure} $\mathcal{O}^{(\rm max)}$ and does not contain any set of E-surfaces that is a subset of another set in $\mathcal{O}$:
\begin{equation}\label{eq:omax}
\mathcal{O}^{(\rm max)}=\{F\in \mathcal{O} \ | \  \nexists F'\in\mathcal{O} \text{ with } F\subset F'\} \,.
\end{equation}
The set $\mathcal{O}^{(\rm max)}$ is the minimal set that ensures that one can build the final deformation without requiring special treatment for the intersections of E-surfaces (i.e. (anti-)selection thereof). 
Determining the maximal overlap structure is a challenging problem and is discussed in sect.~\ref{sec:source_determination}.

In order to construct the deformation field for $\mathcal{E}$, each element $F\in\mathcal{O}^{(\rm max)}$ is associated to a source $\vec{s}_F$ whose corresponding radial deformation field $\vec{k}-{\vec{s}_F}$ is imposed to vanish on any E-surface \emph{not} contained in $F$. This task is performed by a positive, bounded and smooth anti-selector function $g_F$ satisfying the following constraints
\begin{equation}
g_F(\vec{k})=\begin{cases}
\begin{aligned}
&0& &\text{if} \hspace{0.5cm} \vec{k}\in \partial \eta, \ \forall \eta\in\mathcal{E}\setminus F \\
&a(\vec{k})>0 & &\text{if} \hspace{0.5cm} \vec{k}\in \Big(\bigcup_{\eta\in F}\partial\eta\Big)\setminus\Big(\bigcup_{\eta'\in \mathcal{E}\setminus F}\partial \eta'\Big)
\end{aligned}
\end{cases}.
\end{equation}
In practice, we build $g_F(\vec{k})$ from the same E-surface anti-selector building block $T(\eta_{bi})$, already introduced in eq.~\eqref{eq:antiselector}:
\begin{equation}
T(\eta)=\frac{\eta(\vec{k})^2}{\eta(\vec{k})^2 + M^2 {p_{\eta}^0}^2 },
\end{equation}
which can be combined as follows to build $g_F(\vec{k})$:
\begin{equation}
g_F(\vec{k}) = \prod_{ \eta \in \mathcal{E} \setminus F } T(\eta).
\end{equation}
Equipped with this anti-selection, we can now define a deformation field $\vec{\kappa}_F$ valid for all E-surfaces in $F$ (and their intersections) which does not contribute (i.e. it is \emph{exactly} zero) to the deformation applied on any E-surface in $\mathcal{E}\setminus F$:
\begin{equation}
\vec{\kappa}_F=\alpha_F(\vec{k})(\vec{k}-\vec{s}_F)g_F(\vec{k}), \hspace{0.3cm} \alpha_F(\vec{k})\in[0,\infty),
\end{equation}
where the overlap function $\alpha_F(\vec{k})$ is, for now, any positive function which is non-zero on any E-surface contained in $F$.
The construction of the final deformation can now be completed by adding together all vectors $\vec{\kappa}_F$ where $F$ ranges through at least all the elements of the maximal overlap set.
We are now ready to write down a complete deformation field which satisfies the causal constraints stemming from \emph{all} E-surface, independently of any deformation hyperparameter:
\begin{equation}\label{eq:minimal_causal_deformation}
\vec{\kappa}_\emptyset (\vec{k})=\sum_{F\in\mathcal{O}^{(m)}} \alpha_F(\vec{k}) ( \vec{k}-\vec{s}_F )g_F(\vec{k}), \hspace{0.5cm} \ \alpha_F(\vec{k})\in(0,\infty).
\end{equation}
The above \emph{minimal} deformation field is what we used at one loop throughout this paper, including for producing the results presented in sect.~\ref{sec:results}. As we shall see in sect.~\ref{sec:continuity_constraint}, beyond one-loop it becomes necessary to consider additional deformation fields to accommodate particular continuity constraints of the integrand.

We stress that supplementing the minimal deformation with additional causal fields can be performed without spoiling the causal properties of the individual terms because of the nature of the anti-selector functions.
In fact, the sum $\vec{\kappa}_{F}+\vec{\kappa}_{F'}$ of two individually valid deformation vector fields $\vec{\kappa}_{F}$ and $\vec{\kappa}_{F'}$ is also causally correct. More precisely, thanks to the anti-selection functions contained in $\vec{\kappa}_{F}$ and $\vec{\kappa}_{F'}$, we have that their sum is:
\begin{itemize}
    \item correct for $\vec{k}$ lying on an E-surface $\eta$ in $F$ \emph{or} an E-surface $\eta'$ in $F'$, but \emph{not} on any intersection of $\eta$ and $\eta'$, that is on all points
    \begin{equation}
    \vec{k}\in\bigcup_{\eta\in F\cup F'}\partial \eta \setminus \bigcup_{\substack{\eta\in F \\ \eta\in F' }}\partial\eta\cap \partial\eta'
    \end{equation}
    \item exactly zero on the above-mentioned intersections as well as on any surface $\eta$ \emph{not} in $F$ \emph{nor} in $F'$, that is on any point 
    \begin{equation}
    \vec{k}\in\Big(\bigcup_{\eta\in \mathcal{E}\setminus(F\cup F')}\partial \eta\Big) \cup \Big(\bigcup_{\substack{\eta\in F \\ \eta\in F' }}\partial\eta\cap \partial\eta'\Big),
    \end{equation}
\end{itemize}
thus ensuring that $\vec{\kappa}_{F}+\vec{\kappa}_{F'}$ also satisfies all causal prescriptions if the deformation fields $\vec{\kappa}_{F}$ and $\vec{\kappa}_{F'}$ already do.
Another example of a deformation field that can be added is the sum of all appropriately anti-selected normal vectors of each E-surface. Thanks to this additive property of anti-selected deformation fields, one particular generalisation of eq.~\eqref{eq:minimal_causal_deformation} is obtained by adding additional \emph{support sources} from a set $O$ of overlaps taken from the set $\mathcal{O}$:
\begin{equation}\label{causal-def}
\vec{\kappa}_O(\vec{k})=\sum_{F\in\mathcal{O}^{(\text{max})}\cup O} \alpha_F(\vec{k}) ( \vec{k}-\vec{s}_F )g_F(\vec{k}), \hspace{0.5cm} O\subseteq \mathcal{O}, \ \alpha_F(\vec{k})\in(0,\infty).
\end{equation}
The dependence of $\vec{\kappa}_O(\vec{k})$ on $O$ underlines the aforementioned fact that adding to the minimal deformation vector --- that is, the one constructed from $\mathcal{O}^{(\text{max})}$ --- any deformation vector constructed from an extra overlap $F\in\mathcal{O}$ cannot spoil the causal constraints already satisfied by $\vec{\kappa}_\emptyset$. More generally, it is also possible to add multiple radial fields generated by several sources from the same overlap $F$, although this is equivalent to adding a single radial field stemming from a different source in the same overlap.
Adding support sources may improve numerical convergence and we intend to explore this possibility more systematically in future work.

The particular strategy for selecting a near-optimal source point $\vec{s}_F$ within a given overlap $F$ is an implementation detail that we will discuss in sect.~\ref{sec:numerical_implementation}.
The next section turns to the problem of assigning the correct normalisation to the deformation field constructed in this section. In particular, we will derive a necessary expression for the prefactors $\alpha_F(\vec{k})$.

\subsection{General solution to constraints on magnitude}
\label{sec:magnitude_constraints}
Once a procedure is established for constructing the correct deformation direction for a generic multi-loop integral, it remains to investigate conditions on the magnitude of this deformation. When writing the deformation vector field as $\lambda \vec{\kappa}(\vec{k})$, determining the normalisation of the deformation amounts to setting the value of $\lambda$. Constraints on the magnitude can be formulated locally for every $\vec{k}$ and can thus be satisfied by scaling parameters that are a continuous function of loop momenta $\lambda=\lambda(\vec{k})$. For numerical stability it is typically advantageous to set the scaling parameter and the overlap function as large as possible while still satisfying the constraints.

The magnitude of the deformation is bounded by three conditions in the LTD framework:
\begin{itemize}
\item the continuity constraint (sect.~\ref{sec:continuity_constraint}), 
\item the expansion validity constraint (sect.~\ref{sec:expansion_constraint}), 
\item the complex pole constraint (sect.~\ref{sec:threshold_displacement}).
\end{itemize}

Scaling parameters satisfying each of these constraints individually are denoted by $\lambda_{\mathrm{cc}}(\vec{k})$, $\lambda_{\mathrm{e}}(\vec{k})$ and $\lambda_{\mathrm{p}}(\vec{k})$ respectively.
An overall scaling function $\lambda(\vec{k})$ satisfying all three constraints can then be constructed as
\begin{equation}
    \lambda(\vec{k})=\text{min}\{\lambda_{\mathrm{cc}}(\vec{k}),\lambda_{\mathrm{e}}(\vec{k}),\lambda_{\mathrm{p}}(\vec{k}),\lambda_{\mathrm{max}}\},
\end{equation}
where $\lambda_{\mathrm{max}}\in (0,\infty)$ is the maximum allowed value of the magnitude of the deformation. Although $\lambda_{\mathrm{max}}$ is effectively a hyperparameter and thus subject to optimisation, the correctness of the deformation is independent of it. All the results presented in this work have been obtained by setting $\lambda_{\mathrm{max}}=10$.

We will see that the continuity constraint also imposes conditions on the overlap function $\alpha_F(\vec{k})$ and the choice of overlap set $O$ for eq.~\eqref{causal-def}, thus arriving at the final expression for $\vec{\kappa}(\vec{k})$ that we will give in eq.~\eqref{multi-loop-def}. Our final expression of the contour deformation is then:
\begin{equation}
    \label{eq:final_deformation_expression}
    \vec{k} \rightarrow \vec{k} - \mathrm{i} \lambda(\vec{k}) \vec{\kappa}(\vec{k})
\end{equation}

\subsubsection{Continuity constraint}
\label{sec:continuity_constraint}

The request that the integrand is continuous on the contour adds constraints to the deformation vectors that have to be satisfied for all values of $\vec{k}$, and specifically require that the argument of any square root appearing as energies of any on-shell particle never crosses the negative real axis, consistently with the choice of the principal square root branch. The energy can be written as a function of $\vec{k}-\mathrm{i}\vec{\kappa}$:
\begin{equation}
    E_j(\vec{k}-\mathrm{i}\vec{\kappa})=\sqrt{\vec{q}_j(\vec{k})^2+m_j^2-2\mathrm{i}\vec{q}_j(\vec{k})\cdot \vec{Q}_j(\vec{\kappa})-\vec{Q}_j(\vec{\kappa})^2}.
\end{equation}
and thus the requirement of integrand continuity imposes that for any value of $\vec{k}$ and $\vec{\kappa}$:
\begin{equation}\label{continuity-constraint}
    \vec{q}_j(\vec{k})\cdot \vec{Q}_j(\vec{\kappa})\neq 0 \text{ if }\vec{q}_j(\vec{k})^2+m_j^2-\vec{Q}_j(\vec{\kappa})^2< 0 \ . 
\end{equation}

Consider now a small ball centred at ${\vec{k}}^*$ with $\vec{q}_j(\vec{k}^*)=0$: then $\vec{\kappa}(\vec{k})$ has a constant direction throughout the infinitesimal volume of the  ball (unless $\vec{\kappa}(\vec{k}) \propto \vec{k}-{\vec{k}}^*$).
Since $\vec{q}_j(\vec{k})$ spans all possible directions in this neighbourhood, it implies that there is always a continuous set of points containing $\vec{k}^*$ and such that $\vec{q}_j(\vec{k})\cdot \vec{Q}_j(\vec{\kappa})=0$.
If $\vec{q}_j(\vec{k})^2+m_j^2$ is smaller than $\vec{Q}_j(\vec{\kappa})^2$ on such points, then eq.~\eqref{continuity-constraint} is violated. 
One concludes that on all the points where $ \vec{q}_j(\vec{k})\cdot \vec{Q}_j(\vec{\kappa})=0$, including at $\vec{k}^*$, one must have $\vec{Q}_j(\vec{\kappa})^2 \leq \vec{q}_j(k)^2+m_j^2$.
Instead of imposing this constraint on this continuous set of points only, we instead impose it everywhere, resulting in the following stronger (and simpler) version:
\begin{equation}\label{eq:branchcut_scaling}
\vec{q}_j(\vec{k})^2+m_j^2-\vec{Q}_j(\vec{\kappa})^2\ge 0 \hspace{0.3cm} \forall j\in \mathbf{e} \, ,
\end{equation}
which restricts the argument of the square root to lie in either the first or fourth complex quadrant.
At one loop, given that $\vec{q}_j(\vec{k})=\vec{k}+\vec{p}_j$, this constraint can be satisfied by just using for the deformation from eq.~\eqref{eq:minimal_causal_deformation} a scaling which imposes the deformation to always be lower in magnitude than $E_j(\vec{k}), \ \forall j$, that is 
\begin{equation}\label{causal-def-2}
\lambda_\mathrm{cc}(\vec{k})= \text{min}_{j\in \bf{e}}\Bigg\{\frac{\epsilon_{\mathrm{cc}} E_j(\vec{k})}{\lVert \vec{\kappa}_\emptyset(\vec{k})\rVert}\Bigg\} \, ,
\end{equation}
where $\epsilon_{\mathrm{cc}}$ is a parameter that we set to 0.95.

The only problematic points are when a focal point of a massless internal propagator $j$, i.e. a solution of the equation $\vec{q}_j(\vec{k}^*)=0$, $j\in \bf{e}$, coincides with a point on another E-surface. According to eq.~\eqref{causal-def-2} this implies that $E_j(\vec{k}^*)=0$ and thus $\lambda_\mathrm{cc}(\vec{k}^*)=0$, although the point is also located on an E-surface and thus requires a non-zero deformation. However, these points can be shown to be specific to the frame of reference initially chosen for the calculation and can be easily removed with a Lorentz boost (see section~\ref{sec:lorentz_invariance}).

For multi-loop integrals satisfying the continuity constraint is not straightforward; indeed, consider an existing two loop surface equation for a massless diagram
\begin{equation}
    \lVert \vec{k_1}+\vec{p_1}\rVert+\lVert \vec{k_1}+\vec{k_2}+\vec{p_2}\rVert+\lVert \vec{k_2}+\vec{p_3}\rVert-p_1^0-p_3^0+p_2^0=0.
\end{equation}
It admits as a solution the point $(\vec{k_1},\vec{k_2})=(-\vec{p_1},\vec{k_2}^*)$, where $\vec{k_2}^*$ is a solution of the lower dimensional E-surface equation $\lVert \vec{k_2}+\vec{p_2}-\vec{p_1}\rVert+\lVert \vec{k_2}+\vec{p_3}\rVert-p_1^0-p_3^0+p_2^0=0$. Since $\vec{k_1}+\vec{p_1}=0$, a continuity constraint as in eq.~\eqref{causal-def-2} scales the deformation to zero, although the point itself is on a singular surface, and thus requires deformation.

Strictly speaking, this dilemma is absent for diagrams with only massive internal propagators, as the masses act as regulators (i.e., $\text{Re}[E_j^2]>m^2-\vec{Q}_j(\vec{\kappa})^2$) and forbid the deformation to be scaled to zero. However, in such cases a small mass imposes an unnecessarily strict constraint on the deformation in the neighbourhood of the corresponding focal point.

In order to remedy this problem, we observe that, given any proper subset of $\mathbf{c}\subset \mathbf{b}$ of a loop momentum basis $\mathbf{b}$, there is a proper subspace of the space of loop variables such that $\vec{Q}_j(\vec{\kappa})=0 \hspace{0.3cm} \forall j\in \mathbf{c}$, since the system is not full rank. This can be used to construct deformation vectors satisfying all causal constraints and branch cut constraints simultaneously on the portion of E-surfaces which lie on the subspaces $\vec{q}_j(\vec{k})=0 \hspace{0.3cm} \forall j\in \mathbf{c}$.
Indeed, let $\vec{\kappa}=\vec{k}-\vec{s}$, then
\begin{equation}\label{source_subspace}
\vec{Q}_j(\vec{k}-\vec{s})\big|_{\vec{q}_j(\vec{k})=0}=-\vec{q}_j(\vec{s})=0 \hspace{0.3cm} \forall j\in \mathbf{c} \,
\end{equation}
imposes conditions on $\vec{s}$ which make the radial field $\vec{k}-\vec{s}$ automatically satisfy the continuity constraints in the neighbourhood of the subspace.
The source determined this way is now partially constrained by the request that it satisfies the continuity condition without the use of a function directly suppressing the radial field on the subspace $\vec{q}_j(\vec{k})=0$, $\forall j\in \mathbf{c}$. One can now try to construct a deformation vector from sources satisfying eq.~\eqref{source_subspace}, by additionally imposing it has a causal direction on any E-surface when restricted to the subspace itself. More specifically, given the restriction of the E-surface to the subspace identified by $\mathbf{c}$,
\begin{equation}
    \eta_\mathbf{c}(\vec{k})=\eta(\vec{k})|_{\{\vec{q}_j(\vec{k}
    )=0\}_{j\in \mathbf{c}}},
\end{equation}
the overlap structure is restricted to this subspace as well and can be defined as
\begin{equation}
\mathcal{O}_\mathbf{c}=\Big\{F\subseteq\mathcal{E}\ | \  \bigcap_{\eta\in F}\partial^- \eta_\mathbf{c}\neq \emptyset \Big\} \,,
\end{equation}
which is contained in the original overlap structure, that is $\mathcal{O}_\mathbf{c}\subseteq\mathcal{O}$. 
Given any element $F\in\mathcal{O}_\mathbf{c}$, one can thus obtain a source $\vec{s}_{F}^{\,\mathbf{c}}$ that satisfies the following convex constraints:
\begin{equation}
\vec{s}_{F}^{\,\mathbf{c}}\in \bigcap_{\eta\in F}\partial^- \eta_\mathbf{c}, \hspace{0.3cm}\vec{q}_j(\vec{s}_{F}^{\,\mathbf{c}})=0 \ \ \forall j\in \mathbf{c}, \ F\in\mathcal{O}_\mathbf{c} \,.
\end{equation}
Therefore, one can define a radial field $\vec{k}-\vec{s}_F^{\,\mathbf{c}}$ which will be non-zero on the subspace identified by $\mathbf{c}$ while still satisfying the continuity constraint and providing a causal direction on the portion of the E-surfaces in $F$ and their intersections contained in the subspace identified by $\mathbf{c}$. In order to not spoil causality outside the subset overlapping E-surfaces as contained in the subspace we will use a properly anti-selected deformation vector 
\begin{equation}
  \vec{\kappa}^{\,\mathbf{c}}_F=(\vec{k}-\vec{s}_{F}^{\,\mathbf{c}})g_F(\vec{k}) \,.
\end{equation}
As before, $\vec{\kappa}^{\,\mathbf{c}}_F$ will not violate causality constraints outside of the subspace, since the anti-selector function $g_F$ will take care of setting the deformation to zero on E-surfaces corresponding to different overlaps in the subspaces characterised by $\mathbf{c}$ and all the E-surfaces not appearing in the subspace.
Analogously to sect.~\ref{sec:general_solution}, one can define the maximal overlap set in the subspace $\mathbf{c}$ 
\begin{equation}
\mathcal{O}_\mathbf{c}^{(\rm max)}=\Big\{F\subseteq\mathcal{O}_\mathbf{c}\ | \  \nexists F'\in\mathcal{O}_\mathbf{c} \text{ with } F\subset F' \Big\} \,,
\end{equation}
and thus construct a causal deformation vector when restricting integration to the subspace $\mathbf{c}$,  
\begin{equation}
  \vec{\kappa}^{\,\mathbf{c}}_\emptyset(\vec{k})=\lambda^\mathbf{c}(\vec{k})\sum_{F\in\mathcal{O}_\mathbf{c}^{(\mathrm{max})}}(\vec{k}-\vec{s}_{F}^{\,\mathbf{c}})g_F(\vec{k}) \,.
\end{equation}
$\vec{\kappa}^{\,\mathbf{c}}_\emptyset(\vec{k})$ is exactly the deformation constructed in eq.~\eqref{causal-def} from the overlap structure obtained in the subspace identified by $\mathbf{c}$, with all the overlap functions $\alpha^\mathbf{\mathbf{c}}_F(\vec{k})$ chosen equal to a single function $\lambda^\mathbf{\mathbf{c}}(\vec{k})$, which ensures that  $\vec{\kappa}^{\,\mathbf{c}}_\emptyset(\vec{k})$ satisfies the continuity constraint on any subspace different than $\mathbf{c}$. That is:
\begin{equation}
\lambda^\mathbf{c}(\vec{k})=\text{min}_{j\in E\setminus \mathbf{c}}\Bigg\{\frac{\epsilon_{\mathrm{cc}} E_j(\vec{k})}{\lVert \vec{Q}_j(\vec{\kappa}^{\,\mathbf{c}}_\emptyset(\vec{k}))\rVert}, 1 \Bigg\} \,.
\end{equation}
In order to construct the final multi-loop deformation vector field, it is necessary to associate a deformation vector to each strict subspace $\mathbf{c}\in\mathcal{P}=\bigcup_{\mathbf{b}\in\mathcal{B}}\big(\mathcal{P}(\mathbf{b}) \setminus \{\mathbf{b}\}\big)$, where 
 $\mathcal{P}(\mathbf{b})$ is the power set of the loop momentum basis $\mathbf{b}$.
We finally obtain
%
\begin{equation}\label{multi-loop-def}
\vec{\kappa}(\vec{k})=\frac{1}{|\mathcal{P}|}\sum_{\substack{\mathbf{c}\in \mathcal{P}
}}\frac{\lambda^\mathbf{c}(\vec{k})}{|\mathcal{O}_\mathbf{c}^{(\mathrm{max})}|}\sum_{F\in\mathcal{O}_\mathbf{c}^{(\mathrm{max})}}(\vec{k}-\vec{s}_{F}^{\,\mathbf{c}})g_F(\vec{k}),  
\end{equation}
where $g_F(\vec{k})$ is the previously defined anti-selector function. Observe that eq.~\eqref{multi-loop-def} is equal to eq.~\eqref{eq:minimal_causal_deformation} at one loop since $\mathcal{P}=\{ \emptyset \}$. Furthermore, since eq.~\eqref{multi-loop-def} can be constructed from eq.~\eqref{causal-def} by setting 
\begin{equation}
    O=\bigcup_{\substack{\mathbf{c}\in \mathcal{P}
    }}\mathcal{O}_\mathbf{c}^{(\mathrm{max})}, \hspace{0.3cm} \alpha_F(\vec{k})=\frac{\lambda^\mathbf{c}(\vec{k})}{|\mathcal{P}||\mathcal{O}_\mathbf{c}^{\mathrm{max}}|}, \hspace{0.3cm} s_{F}=s_F^\mathbf{c} \hspace{0.3cm}\ \forall F\in\mathcal{O}^{(\mathrm{max})}_\mathbf{c},
\end{equation} 
it immediately follow that $\vec{\kappa}(\vec{k})$ is a causal deformation vector. One can observe that in the limit $\vec{q}_j(\vec{k})\rightarrow 0$ the deformation satisfies the continuity constraint $\vec{Q}_j(\vec{\kappa})^2<\vec{q}_j(\vec{k})^2$ without necessarily being identically zero. We stress that, although the continuity constraint is satisfied on all subspaces and neighbouring points, there is no insurance that it is still the case away from it. Thus, as already mentioned, the final deformation vector must be given an overall scaling factor:
\begin{equation}
    \lambda_\mathrm{cc}(\vec{k})= \text{min}_{j\in \mathbf{e}}\Bigg\{\frac{\epsilon_{\mathrm{cc}} E_j(\vec{k})}{\lVert \vec{Q}_j(\vec{\kappa}(\vec{k}))\rVert}\Bigg\} \,,
\end{equation}
which is now not suppressing the deformation to zero on subspaces.

This concludes the construction of a general contour deformation which works both in the case of massive or massless propagators, satisfying all causal constraints.

\subsubsection{Complex pole constraint}
\label{sec:threshold_displacement}

The analytically continued LTD integrand is singular at complex locations other than the real location of thresholds. These complex poles must not be included in the region of space between the deformed contour and the real hyperplane for the final result to be correct. This is consistent with the request that the integral on the contour matches the original one defined on $\mathbb{R}^{3n}$. 

\begin{figure}[h]
    \centering
    \includegraphics[width=\textwidth]{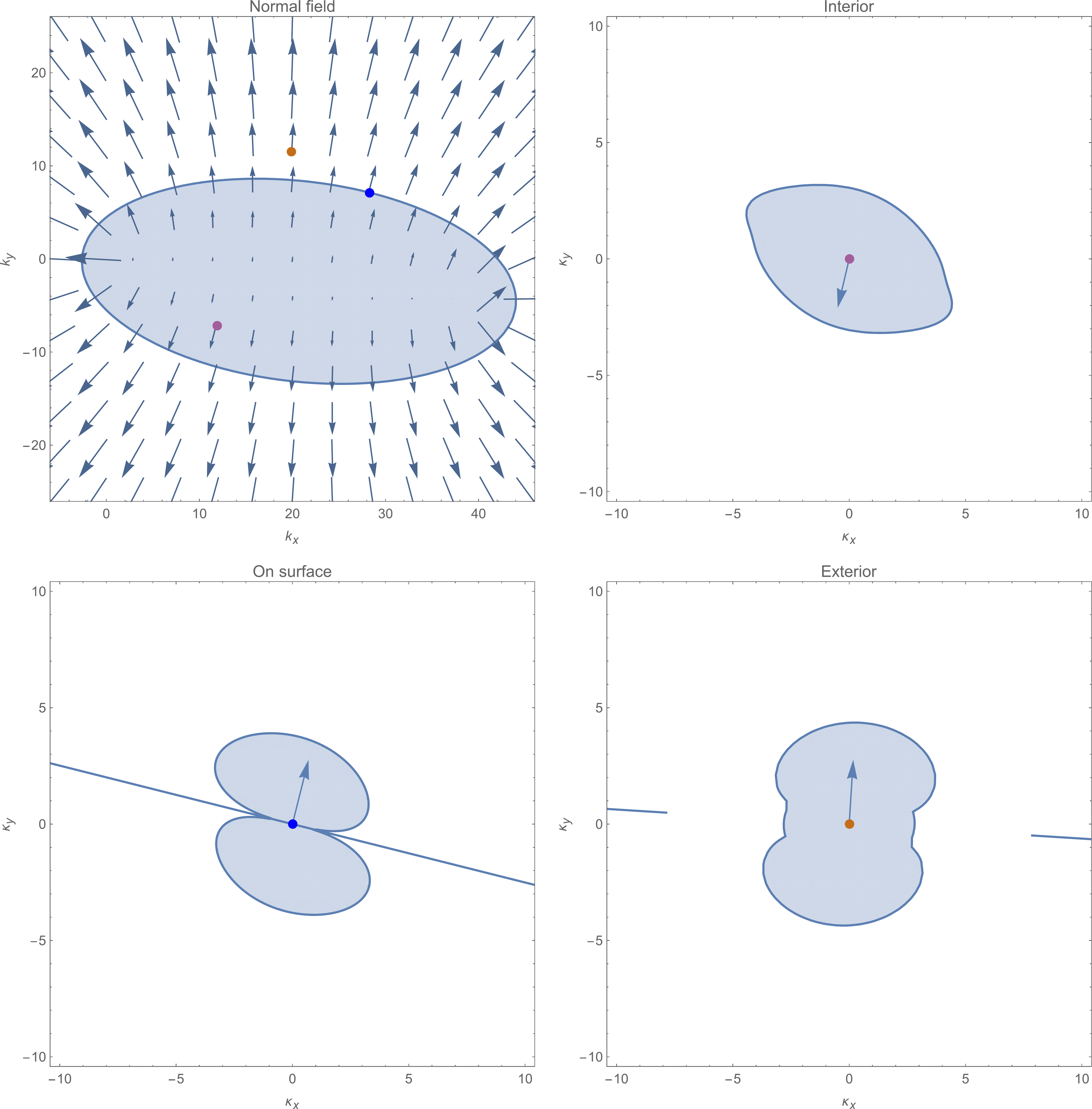}
    \caption{On the top left, an E-surface with its own normal field in $(\vec{k}_x,\vec{k}_y)$ space. Three points, one in the interior of the E-surface (purple), one on the E-surface surface (blue) and one on the exterior of the E-surface (orange) are highlighted. In the other three pictures, one can find, for each of the highlighted points, the $(\vec{\kappa}_x, \vec{\kappa}_y)$ space showing the forbidden line stemming from eq.~\eqref{pole-condition2} as well as the region allowed by the scaling of eq.~\eqref{eq:pole_check_lambda_determination} which guarantees that the deformation does not cross complex poles.}
    \label{fig:pole_locations}
\end{figure}

The approximate complex pole location can easily be found when the square roots of E-surfaces are expanded up to second order in $\lVert \vec{\kappa}\rVert$ and the truncated expressions for the real part and imaginary part are set to zero: 
\begin{equation}\label{pole-condition}
     \eta(\vec{k})-\sum_{i} \sqrt{a_i}\frac{a_i c_i-b_i^2}{2 a_i^2}=0 \ \ \text{ and } \sum_{i} \frac{b_i}{\sqrt{a_i}}=0,
\end{equation}
where the sum runs over all square roots expressing the energies appearing in the surface $\eta$ (see eq.~\eqref{eq:Esurface_positive_energies_def}), with the following coefficients
\begin{align}\label{abc_definition}
\begin{split}
&   a_i=\vec{q}_i(\vec{k})^2+m_i^2, \\
&   b_i=\vec{q}_i(\vec{k})\cdot \vec{Q}_i(\vec{\kappa}), \\
&   c_i=\vec{Q}_i(\vec{\kappa})^2 \,.
\end{split}
\end{align}
Eq.~\eqref{pole-condition} can be solved in the variable $\vec{\kappa}\in\mathbb{R}^{3n}$, for given $\vec{k}$, which provides a parametrisation of the singular surface for the analytically continued integrand. Any point satisfying $\eta(\vec{k})<0$ will admit no solution since the triangle inequality ensures that $a_i c_i-b_i^2>0$, whereas points satisfying $\eta(\vec{k})=0$ will have $\vec{\kappa}=0$ as a unique solution: the latter poles are the original E-surface boundary around which there is initially an intent to deform. Writing $\vec{\kappa}=\lVert \vec{\kappa} \rVert \hat n_{\vec{\kappa}}$, we find that for $\eta(\vec{k})>0$ there is a $(3n-2)$-dimensional set of solutions which entirely lies on the hyperplane $\hat n_{\vec{\kappa}}\cdot\vec{\nabla}\eta=0$ and which is radially symmetric with respect to the origin. This is illustrated for a two-dimensional example in fig.~\ref{fig:pole_locations}.

Whether a pole is included within the contour can be established according to the following guiding principle: given a parametrised deformation vector $\vec{\kappa}(\vec{k})$, the deformation contour will flatten out to become the original real space as the magnitude of the deformation $\lVert\vec{\kappa}(\vec{k})\rVert$ is sent to zero. Thus, if a pole is contained in the region between the contour and the real hyperplane for a given $\vec{\kappa}(\vec{k})$, $\vec{\kappa}$ can be scaled down such that the pole is exactly on the surface. 

The request that the contour does not include any pole thus translates into a set of allowed values of $\kappa$ for the deformation contour: $\vec{\kappa}$ is an allowed value if rescaling it so that $\vec{\kappa}\rightarrow \lambda\vec{\kappa}$, there exists no value of $\lambda\in(0,1]$ such that a solution of eq.~\eqref{pole-condition} is exactly on the contour. This immediately allows to state that, for given $\vec{k}$, any value of $\vec{\kappa}$ satisfying
\begin{equation}\label{pole-condition2}
\eta(\vec{k})-\sum_{i} \sqrt{a_i}\frac{a_i c_i-b_i^2}{a_i^2}\le 0 \ \ \text{ and } \sum_{i} \frac{b_i}{\sqrt{a_i}}= 0 \text{ and } \eta(\vec{k})> 0
\end{equation}
is not allowed. 
Once the contour is explicitly parametrised as $\vec{k}-\lambda\mathrm{i}\vec{\kappa}(\vec{k})$, the constraint on the allowed values of the deformation can be dynamically satisfied by using the treatment of ref.~\cite{Gong_Nagy_Soper}, which can be applied to any quadratic equation in the scaling parameter characterising the location of complex poles.
Specifically, this treatment allows $\lambda$ to take a large value whenever the imaginary part of the complex-valued surface is reasonably high in absolute value, as in these cases the deformation $\vec{\kappa}$ is far from the hypersurface orthogonal to the normal, which contains all the poles and forbidden areas. When $\vec{\kappa}$ approaches the surface orthogonal to the normal field, its value is constrained to yield a positive value for the real part of the surface $\eta(\vec{k})$. In this way, the forbidden region eq.~\eqref{pole-condition} is never reached. 
More specifically, given 
\begin{equation}
    \eta(\vec{k}-\mathrm{i}\vec{\kappa})+o(\lVert\vec{\kappa} \rVert)=A+2\mathrm{i}\lambda B-\lambda^2 C, 
\end{equation}
with
\begin{align}\label{abc_definition2}
    A=\eta(\vec{k}), \quad
    B=\frac{1}{2}\sum_{i} \frac{b_i}{\sqrt{a_i}}, \quad
    C=\sum_{i} \sqrt{a_i}\frac{a_i c_i-b_i^2}{a_i^2},
\end{align}
one has that there is no value of $\vec{\kappa}$ such that eq.~\eqref{pole-condition2} is satisfied if 
\begin{equation}
    \lambda_\eta^2=\begin{cases}
    \begin{aligned}
        &\frac{A}{4C}  & &\text{ if } 2B <  A \\
        &\frac{B}{C} - \frac{A}{4C}  & &\text{ if } 0< A <2 B \\
         &\frac{B}{C} - \frac{A}{2C}  & &\text{ if } A <0 \,.
    \end{aligned}
    \end{cases}
\end{equation}
Finally, one can calculate and collect a scaling parameter $\lambda_\eta$ for each existing or non-existing, pinched or non-pinched E-surface, and write
\begin{equation}
    \label{eq:pole_check_lambda_determination}
    \lambda_{\mathrm{p}}=\text{min}_{\eta}\{\lambda_\eta\}.
\end{equation}
It is important to include non-existing E-surfaces, as they may still have complex solutions. 

It is particularly illuminating, in order to understand the relevance and location of the complex poles, to observe how the zeros of the original E-surface equation morph into the zeros of the real part of the complex valued E-surface equation. The location of the ``displaced'' threshold is implicitly determined through the equation
\begin{equation}\label{init}
0=\mathrm{Re}\big[\sum_i \sqrt{a_i+2i\lambda b_i-\lambda^2 c_i}+p_\eta^0\big]=\sum_i \sqrt{a_i}+p_\eta^0-\sum_i \lambda^2\sqrt{a_i}\frac{c_ia_i-b_i^2}{2a_i^2}+\mathcal{O}(\lambda^3).
\end{equation}

This implicit equation defines a surface which is in general very different from the original E-surface, although it is clear that in the limit $\lambda\rightarrow 0$, the two surface equations will be the same (see sect.~\ref{sec:visualisation} for visualisations). In the second order truncation in $\lambda$, it is also clear that the interior region of the displaced surface will necessarily contain the interior region of the original E-surface, since $c_i a_i-b_i^2>0$, $\forall \vec{k}$.
A rough bound on the volume of its interior region can be obtained by truncating the expansion of the square root to the next-to-leading order in the real part and requiring the correction to be smaller than $\epsilon_{\mathrm{th}}$ (see sect.~\ref{sec:expansion_constraint}):
\begin{equation}
0=\mathrm{Re}\big[\sum_i \sqrt{a_i+2i\lambda b_i-\lambda^2 c_i}+p_\eta^0\big]\ge\sum_i \sqrt{a_i}(1-\epsilon_{\text{th}}^2)+p_\eta^0, \ \ \lambda\ll 1.
\end{equation}
This equation can thus be used to provide an upper bound for the volume of the displaced threshold, in the form of another E-surface with the same focal points and larger constant term.

It is interesting to note that the real part of the complex-valued E-surface equation is negative in the interior region of the displaced threshold, and positive outside.
It means that no forbidden values of the deformation can be crossed in the region outside the displaced threshold.
However, inside the original E-surface, no pole is allowed.
Thus, the region of loop momentum integration space which may lead to forbidden values of the deformation (when there is no appropriate dynamic scaling) is all contained between the original E-surface and the displaced threshold. An example of this behaviour is shown in fig.~\ref{fig:pole_check_effect}.

\subsubsection{Expansion validity}
\label{sec:expansion_constraint}

The causal constraint on the direction and the complex pole constraint are formulated in the limit of a small deformation vector norm $\lVert\vec{\kappa}\rVert$. In this limit, the imaginary part of $\eta$ takes an especially simple form, as it prescribes that the projection of the deformation vector on the normal of $\eta$ must always be positive. Likewise, the complex pole constraint admits an especially simple and elegant solution when $\eta$ is expanded up to second order. This constraint also concerns the magnitude of the vector. Consider the energy 
\begin{equation}
    E_j(\vec{k}-\mathrm{i}\vec{\kappa})=\sqrt{a_j+2\mathrm{i}\lambda_j b_j-\lambda^2 c_j},
\end{equation}
where $a_j, \ b_j, \ c_j$ are defined as in eq.~\eqref{abc_definition}. Observe that the chosen stronger version of the continuity constraint eq.~\eqref{eq:branchcut_scaling} already imposes that $b_j<a_j$ and $c_j<a_j$. Thus a way to ensure the feasibility of the expansion is through the same mechanism which ensures that no branch cut is crossed. A more systematic approach to the constraints on the expansion, however, is to ensure that the argument of the square root is small in norm
\begin{equation}\label{full_lambda_dependency}
    \Bigg|2\mathrm{i}\lambda_j \frac{b_j}{a_j}-\lambda_j^2 \frac{c_j}{a_j}\Bigg|<\epsilon_{\mathrm{th}}\,,
\end{equation}
which leads to the condition
\begin{equation}\label{full_lambda}
    \lambda_j^2\le -2\frac{b_j^2}{c_j^2}+\sqrt{4\frac{b_j^4}{c_j^4}+\epsilon_{\mathrm{th}}^2\frac{a_j^2}{c_j^2}} \,.
\end{equation}
This is effectively equivalent to requiring that the square root is expanded when its argument is contained within a disc of radius $\epsilon_{\mathrm{th}}$. The overall expansion validity constraint can be satisfied by setting $\lambda_{\mathrm{e}}$ equal to the minimal $\lambda_j$ for all energies $E_j$:
\begin{equation}
    \lambda_{\mathrm{e}}^2=\text{min}_{j\in\mathbf{e}}\Bigg\{-2\frac{b_j^2}{c_j^2}+\sqrt{4\frac{b_j^4}{c_j^4}+\epsilon_{\mathrm{th}}^2\frac{a_j^2}{c_j^2}} \Bigg\}
\end{equation}

Another approach is to directly compare higher-order corrections to the leading order terms in the expansion. The odd orders are imaginary, whereas the even ones are real. The expansion to third order reads
\begin{equation}
 \sqrt{1+2\mathrm{i}\lambda\frac{b_j}{a_j}-\lambda^2\frac{c_j}{a_j}}=
   1+
   \mathrm{i}\lambda\frac{b_j}{a_j}
   -\frac{\lambda^2}{2}\Bigg(\frac{c_j}{a_j}-\frac{b_j^2}{a_j^2}\Bigg)
   +\frac{\mathrm{i}\lambda^3}{2}\Bigg(\frac{b_j c_j}{a_j^2}-\frac{b_j^3}{a_j^3}\Bigg)+\mathcal{O}(\lambda^4) \,,
\end{equation}
which, when compared with the expression
\begin{equation}
\sqrt{1+2\mathrm{i}\lambda\frac{b_j}{a_j}-\lambda^2\frac{c_j}{a_j}}=z_0+\mathrm{i}\lambda z_{1}-\lambda^2 z_2+\mathrm{i}\lambda^3 z_3 + \mathcal{O}(\lambda^4) \,,
\end{equation}
yields the relation
\begin{equation}
    \frac{z_2}{z_0}=\frac{z_3}{z_1} \,,
\end{equation}
whose significance relies on the fact that suppressing the importance of the next-to-leading order with respect to the leading order of the expansion of the imaginary part also achieves the same for the real part. Suppression of this ratio can be obtained by imposing
\begin{equation}\label{ratio}
  \lambda^2\frac{z_2}{z_0}=\frac{\lambda^2}{2}\frac{a_j c_j-b_j^2}{a_j^2}<\epsilon_{\mathrm{th}}, \ \ \epsilon_{\mathrm{th}} \in(0,1), \ \ \forall j\in E \,.
\end{equation}
This shows that the choice $\lambda^2<\epsilon\frac{a_j}{2b_j}$ makes the next to leading order contribution to the imaginary part dominate over the leading order when $b_j$ is small with respect to $\epsilon a_j$. As a consequence, the choice of the scaling of the deformation is constrained by the condition that
\begin{equation}\label{expansion_threshold}
    \lambda_{\mathrm{e}}^2=\epsilon_{\text{th}}^2 \ \text{min}_{j\in \mathbf{e}}\Bigg\{\frac{2a_j^2}{a_j c_j-b_j^2}\Bigg\}, \hspace{0.3cm} \epsilon_{\text{th}}\in(0,1). 
\end{equation}

The practical advantage of eq.~\eqref{full_lambda_dependency} is that it is true to any order in the expansion, while its downside resides in a non-obvious interpretation of the expansion parameter $\epsilon$. On the other hand, while eq.~\eqref{ratio} only considers terms up to third order and does not account for the relevance of higher orders, it constrains the corrections to the imaginary and real parts simultaneously and consistently with only one expansion parameter. This parameter signifies the relative size of the higher-order correction with respect to the leading one. 

The most conservative approach is to impose both constraints, but in practice we found good results by imposing eq.~\eqref{expansion_threshold} only, which is what we used for producing the results presented in this work.

\subsubsection{Visualisation of the contour deformation and its effects}
\label{sec:visualisation}

In sect.~\ref{sec:one_loop_construction} we constructed and visualised the deformation vector field for a one-loop configuration with four pairwise overlapping E-surfaces, called \texttt{Box4E}. In this section we will study the interplay between the contour deformation and the integrand in more detail.

\begin{figure}
    \centering
    \subfloat[\label{fig:kappa_ev_a}]{\includegraphics[width=0.24\textwidth]{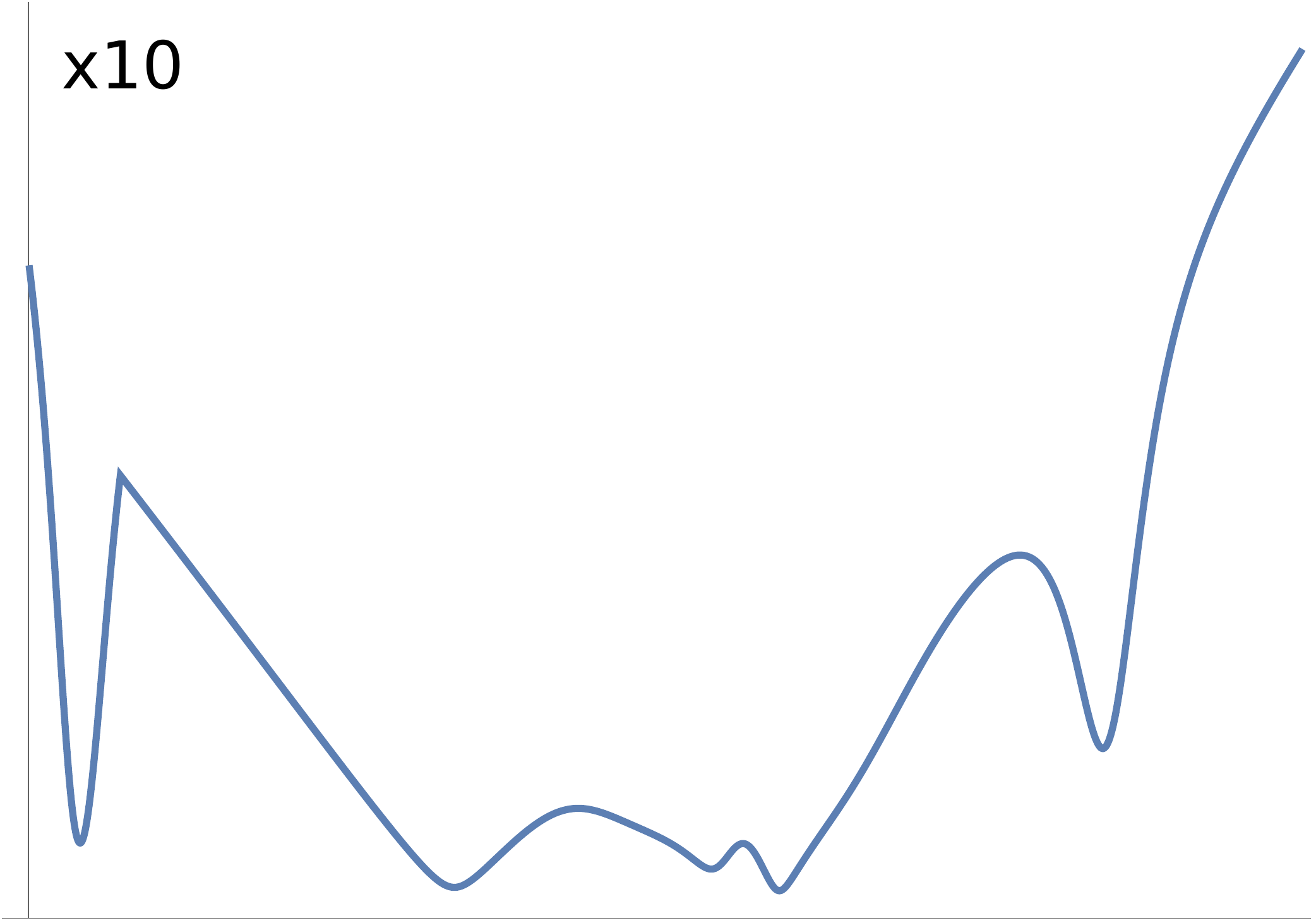}}
    \subfloat[\label{fig:kappa_ev_b}]{\includegraphics[width=0.24\textwidth]{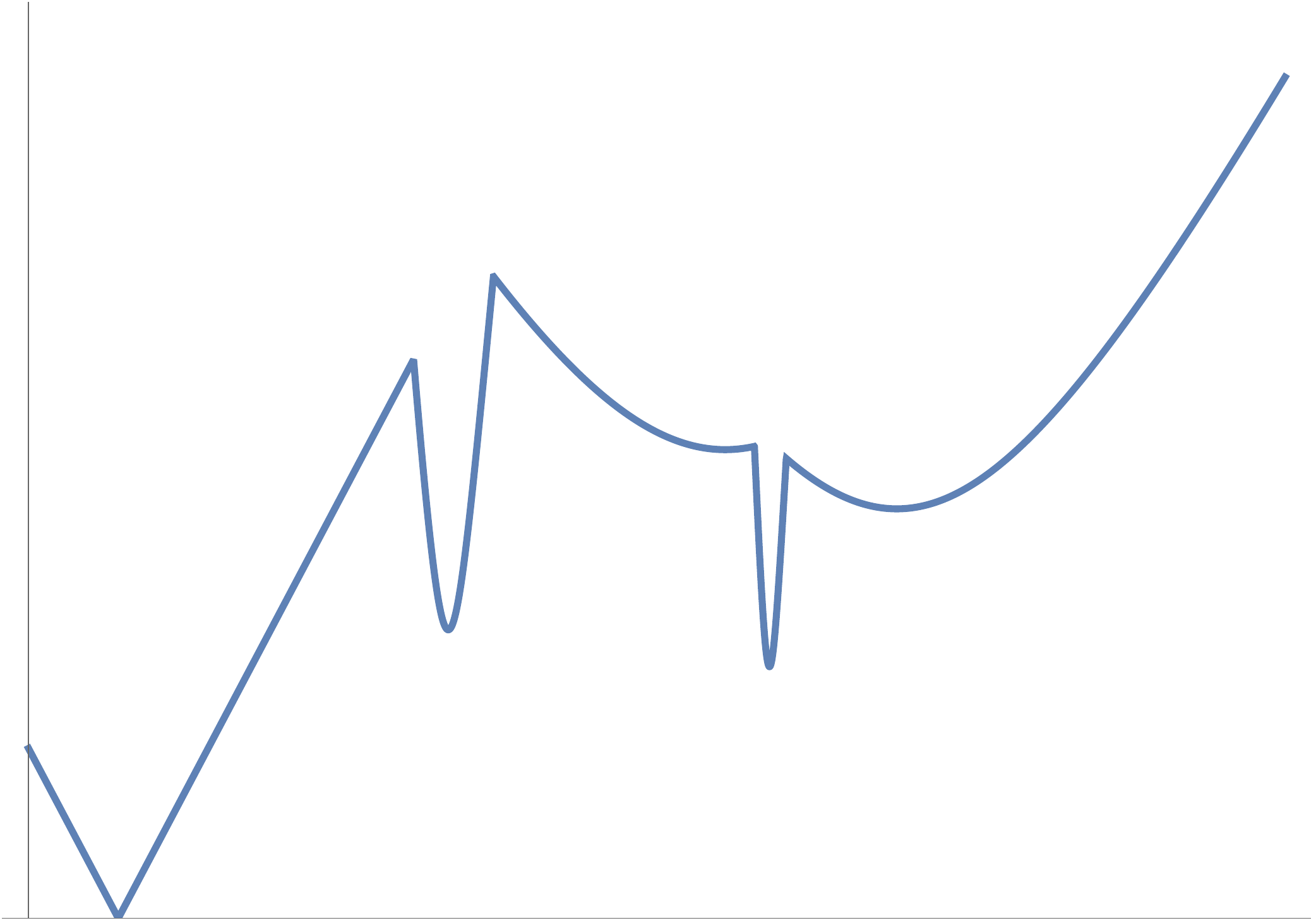}}
    \subfloat[\label{fig:kappa_ev_c}]{\includegraphics[width=0.24\textwidth]{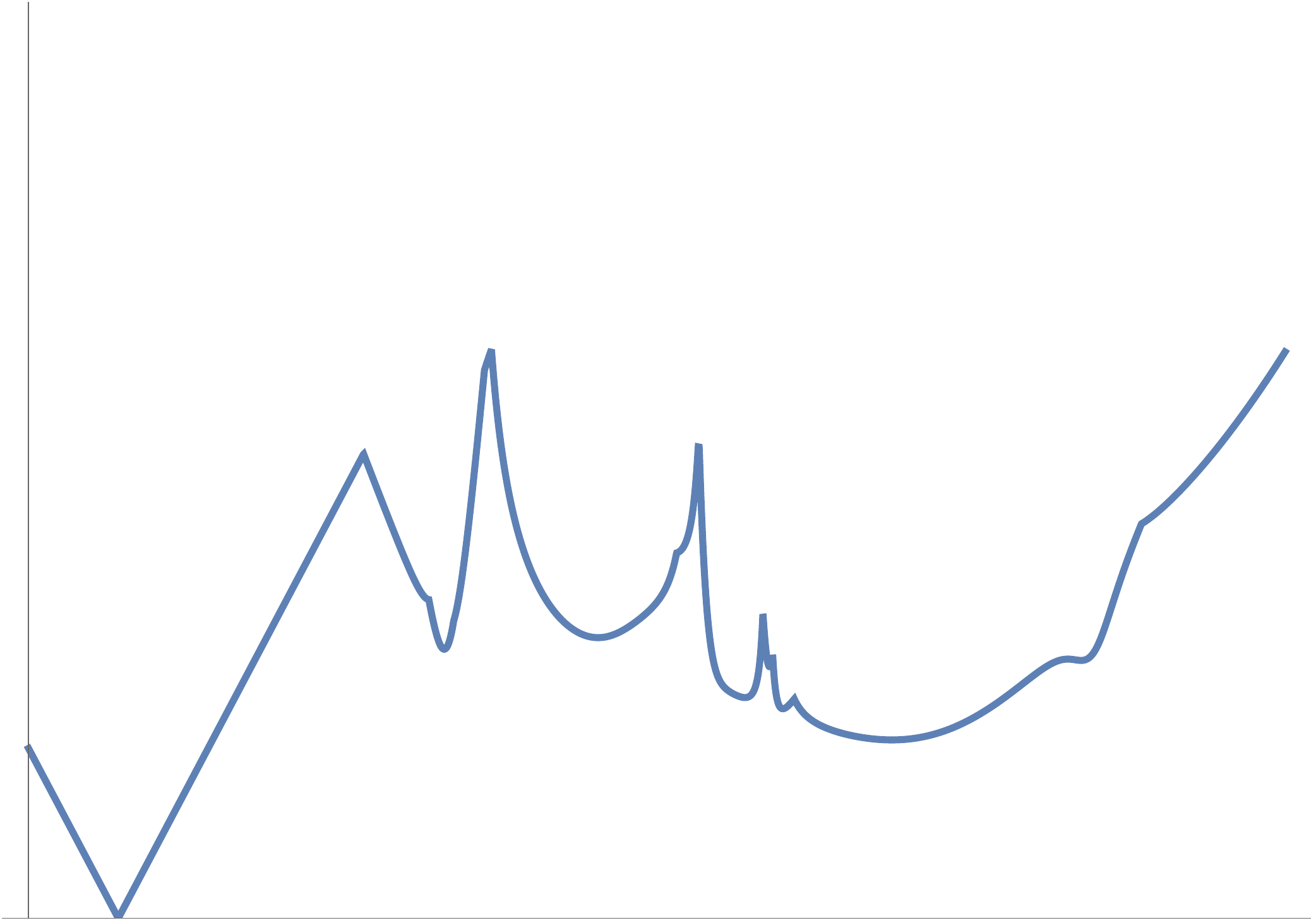}}
    \subfloat[\label{fig:kappa_ev_d}]{\includegraphics[width=0.24\textwidth]{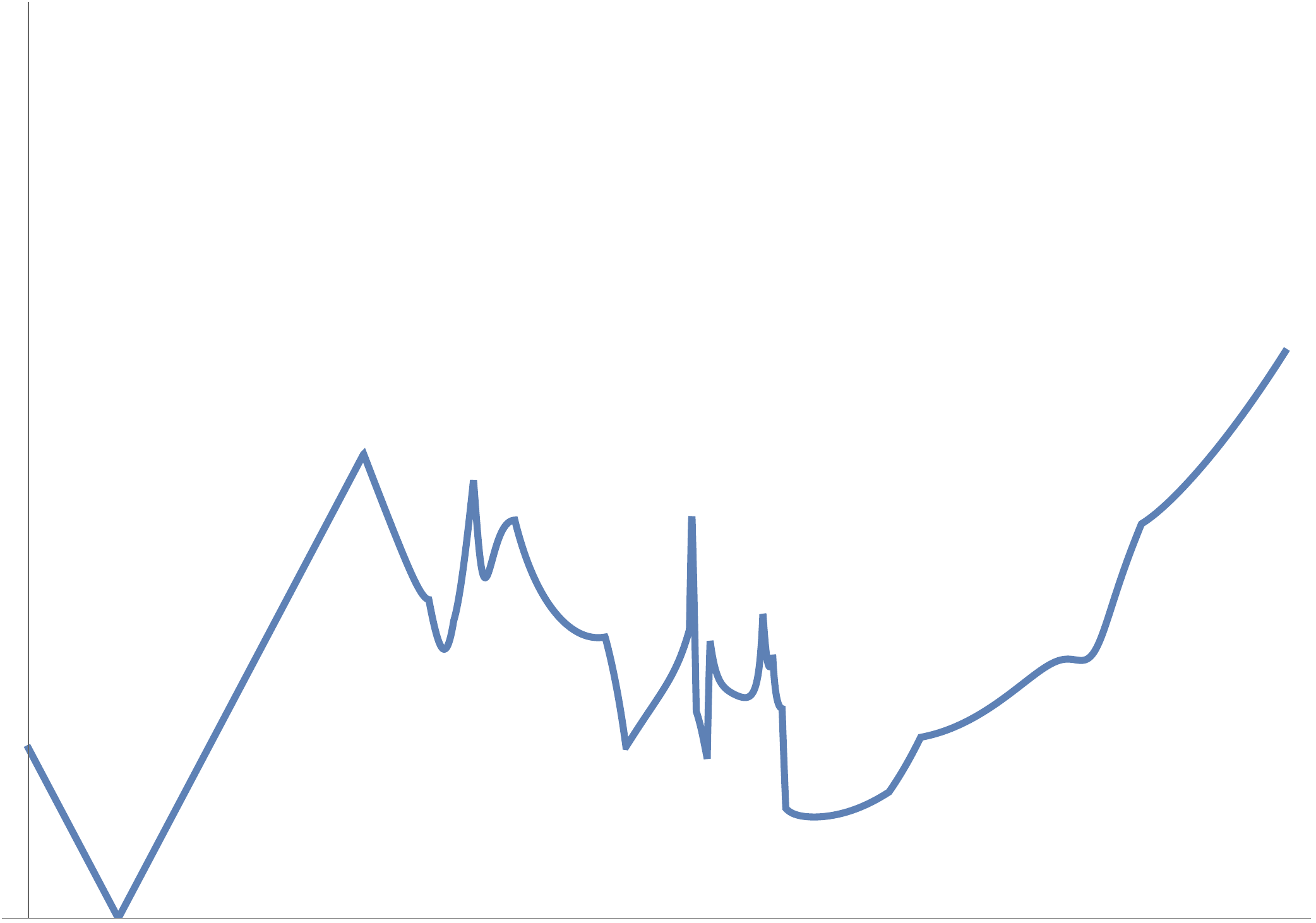}}
    \caption{The magnitude of the deformation vector $\lVert\vec{\kappa}\rVert$ along a line segment in loop momentum space (see fig.~\ref{fig:line_annotated}) is subject to constraints in sect.~\ref{sec:magnitude_constraints}. From left to right: a) no magnitude constraints, b) only continuity constraint enabled, which forces $\lVert\vec{\kappa}\rVert$ to 0 on focal points c) continuity and expansion constraint enabled, d) all constraints enabled, i.e. continuity, expansion and complex pole constraint. The scale of a) is ten times larger than the scale of the other plots, since we use a non-restrictive $\lambda_{\text{max}}=10$. The image is scaled down for comparison.}
    \label{fig:kappa_evolution}
\end{figure}

First, we investigate the properties of the contour deformation $\vec{k}-\mathrm{i}\vec{\kappa}$.
Various aspects of the direction of the deformation vector $\vec{\kappa}$ were already discussed in sect.~\ref{sec:one_loop_construction}.
In this section, we highlight details about the deformation magnitude $\lVert\vec{\kappa}\rVert$, specifically, the impact of the three conditions it is subject to, as laid out in sect.~\ref{sec:magnitude_constraints}.
The magnitude $\lVert\vec{\kappa}\rVert$ can be studied at various stages in the construction of a deformation that will eventually satisfy all physical constraints.
In fig.~\ref{fig:kappa_evolution} we break down the construction of $\vec{\kappa}$ into four stages:
\begin{enumerate}[label=(\alph*)]
    \item The deformation vector is subject to none of the constraints described in sect.~\ref{sec:magnitude_constraints} and the deformation magnitude is therefore determined alone by the superposition of all radial source fields.
    \item We impose the continuity constraint, introduced in sect.~\ref{sec:continuity_constraint}. It guarantees continuity of the integrand, since branch cuts of the square roots involved cannot be crossed thanks to this constraint.
    \item The conditions on the direction of $\vec{\kappa}$, described in sect.~\ref{sec:general_solution}, as well as the complex pole constraint in sect.~\ref{sec:threshold_displacement}, rely on expansions in $\lVert\vec{\kappa}\rVert$. We limit the magnitude $\lVert\vec{\kappa}\rVert$ with the expansion constraint given in sect.~\ref{sec:expansion_constraint} in order to remain in the range of validity of said expansion.
    \item The volume enclosed between the real hyper-plane and the contour deformation must not include any of the pole located at complex-values of the loop momenta. In order to guarantee this, we impose the complex pole constraint discussed in sect.~\ref{sec:threshold_displacement}. It again limits the magnitude of $\vec{\kappa}$.
\end{enumerate} \par



After these four steps, the deformation vector field $\vec{\kappa}$ is such that the integral is well-defined and yields the physically correct result.
In fact, an E-surface $\eta$ that has real solutions $\vec{k}\in (\mathbb{R}^3)^n$ of the equation $\eta=0$ when the deformation is inactive ($\vec{\kappa}=\vec{0}$), has no more real solutions when the deformation is active.
We therefore visualise the effect of the deformation on the E-surfaces.
The deformed E-surface $\eta$ defines two regions of interest: the zeros of its real part $\Re\eta$, and the zeros of its imaginary part $\Im\eta$.
\begin{figure}[h]
    \centering
    \includegraphics[width=\textwidth]{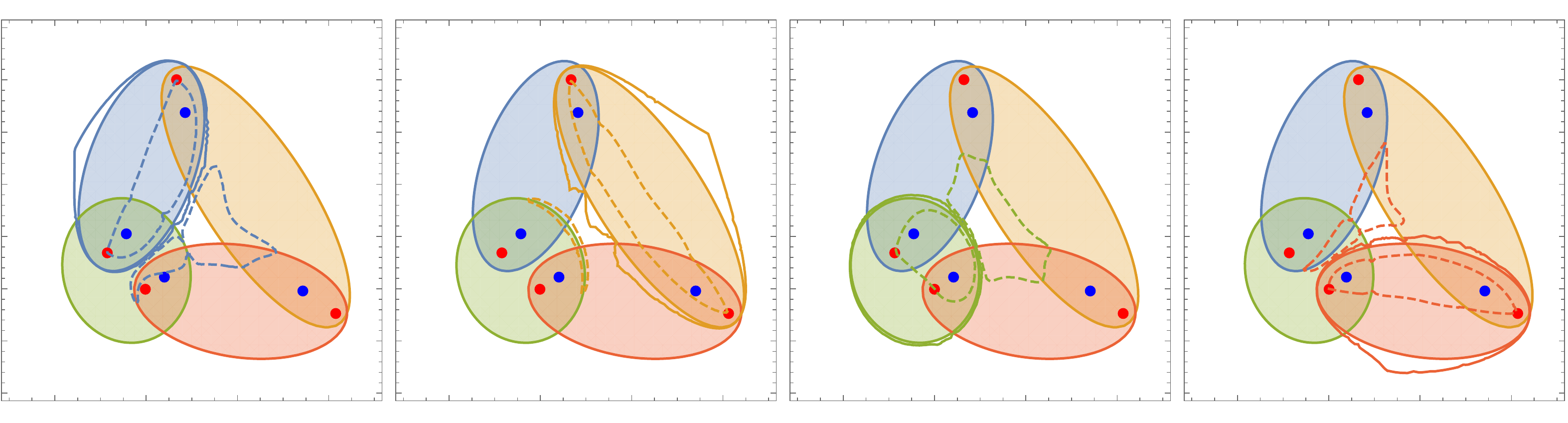}
    \caption{Surface structure in integration momentum space before and after deformation:
    For each existing E-surface $\eta$ of the \texttt{Box4E}, we show the solutions to $\eta=0$ without contour deformation, i.e. $\vec{\kappa}=\vec{0}$, as ellipses with shaded interiors.
    With a contour deformation enabled, the zeros of the real part of the complex E-surface $\Re \eta = 0$ are shown as solid lines, where the zeros of the imaginary part $\Im \eta = 0$ are dashed.
    Note that on the integration contour, each E-surface develops one surface where $\Re \eta = 0$ and two separated surfaces where $\Im \eta = 0$.
    Additionally, on each focal point (red) it holds that $\Im \eta = 0$, which is not apparent in this plot.}
    \label{fig:deformed_surfaces2}
\end{figure}
In fig.~\ref{fig:deformed_surfaces2} we display the two regions of interest one-by-one for each of the four E-surfaces.
With respect to the smooth elliptic surface described by $\eta=0$, when the deformation is switched off, the regions $\Re\eta(\vec{k}-\mathrm{i}\vec{\kappa})=0$ and $\Im\eta(\vec{k}-\mathrm{i}\vec{\kappa})=0$ can be seen as a displacement of $\eta$ into complex space.
It is crucial here that these two regions do not intersect.
If they did, i.e. the real and imaginary part of the E-surface equations were simultaneously zero, there exists a solution to the deformed E-surface equation $\eta=0$, which cannot be allowed by our contour deformation, since $\lVert\vec{\vec{\kappa}}\rVert$ satisfies the complex pole constraint.
To showcase this exact scenario, we refer to the side-by-side comparison in fig.~\ref{fig:pole_check_effect}, where we used two deformation vector fields $\vec{\kappa}$, a correct one and one that is not subject to the complex pole constraint.
Its effect is subtle in this case, as it moves the real and imaginary solutions only marginally but essentially, as it renders the integral divergent without it.
\begin{figure}[p]
    \centering
    \includegraphics[width=0.4\textwidth]{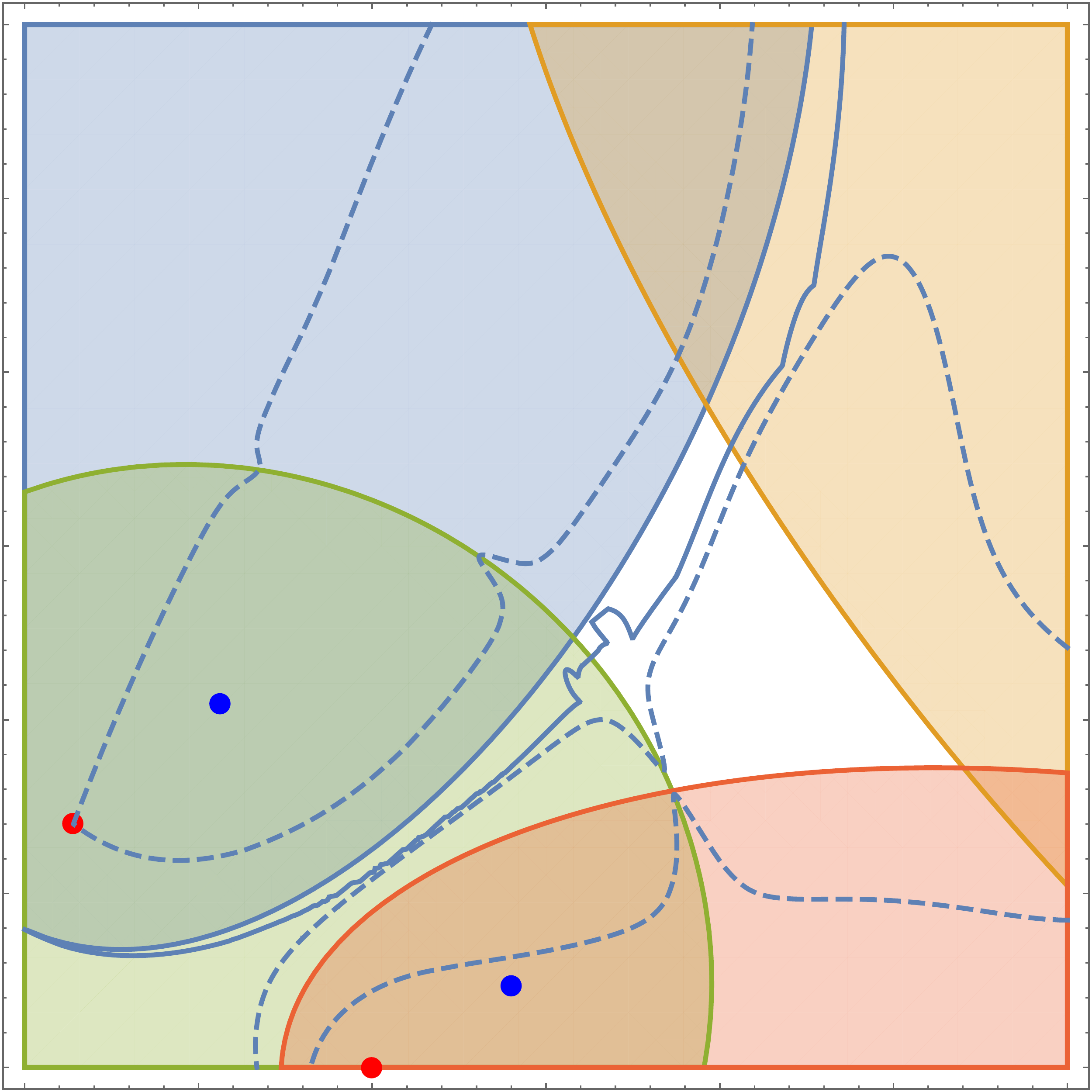}
    \includegraphics[width=0.4\textwidth]{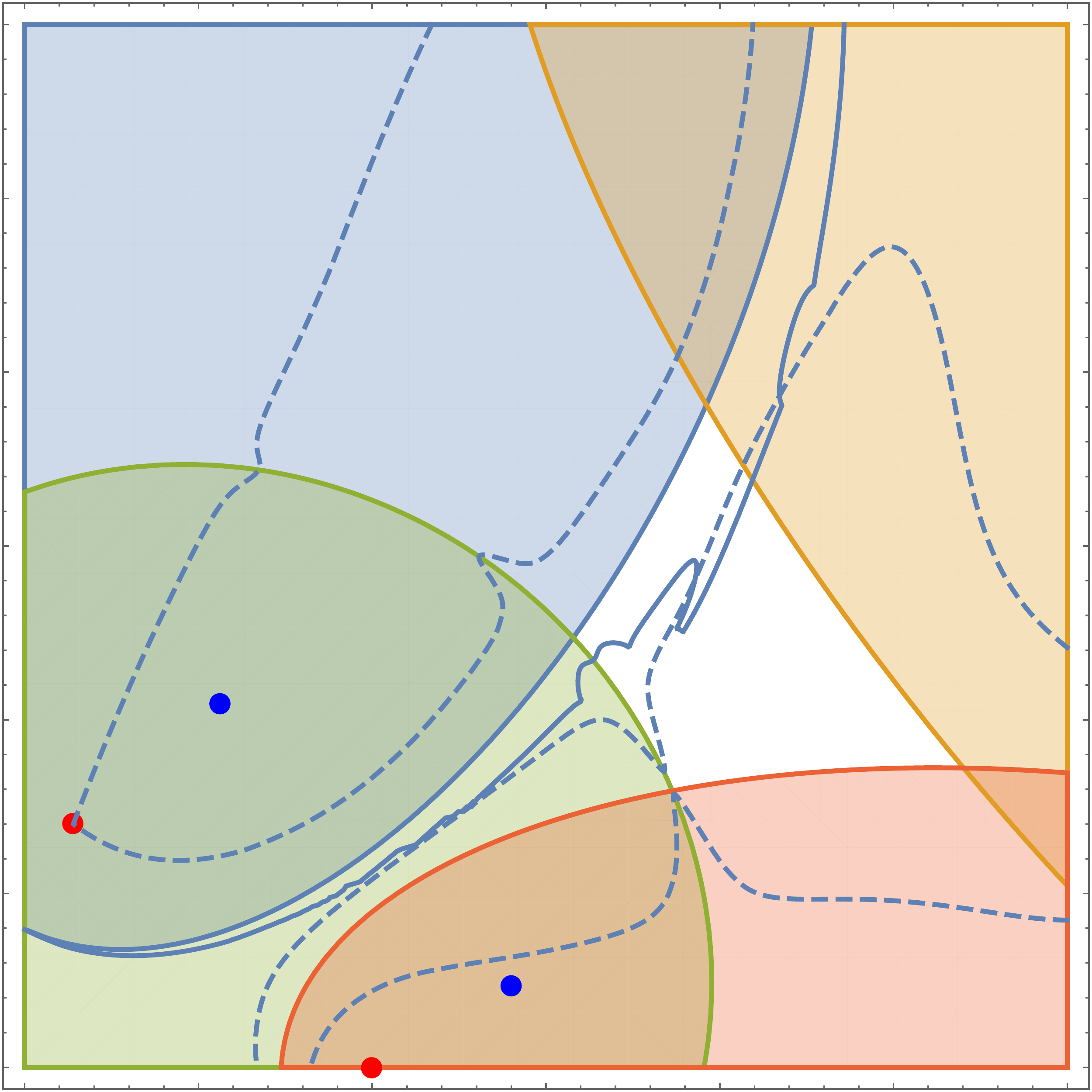}
    \caption{A deformation that satisfies the complex pole constraint, described in sect.~\ref{sec:threshold_displacement}, prevents that an E-surface $\eta$ has zeros in the integration space.
    Left: a deformation rescaled by the complex pole constraint. Right: a deformation that violates the complex pole constraint. The integrand has poles where the dashes line meets the solid line.}
    \label{fig:pole_check_effect}
\end{figure}
We take a more detailed look at the region between the four E-surfaces of the \texttt{Box4E}, as displayed in fig.~\ref{fig:center_region}.
It contains the full deformation vector field $\vec{\kappa}$ and the regions of vanishing real or imaginary part of the deformed E-surface. \par
\begin{figure}[p]
    \centering
    \includegraphics[width=0.6\textwidth]{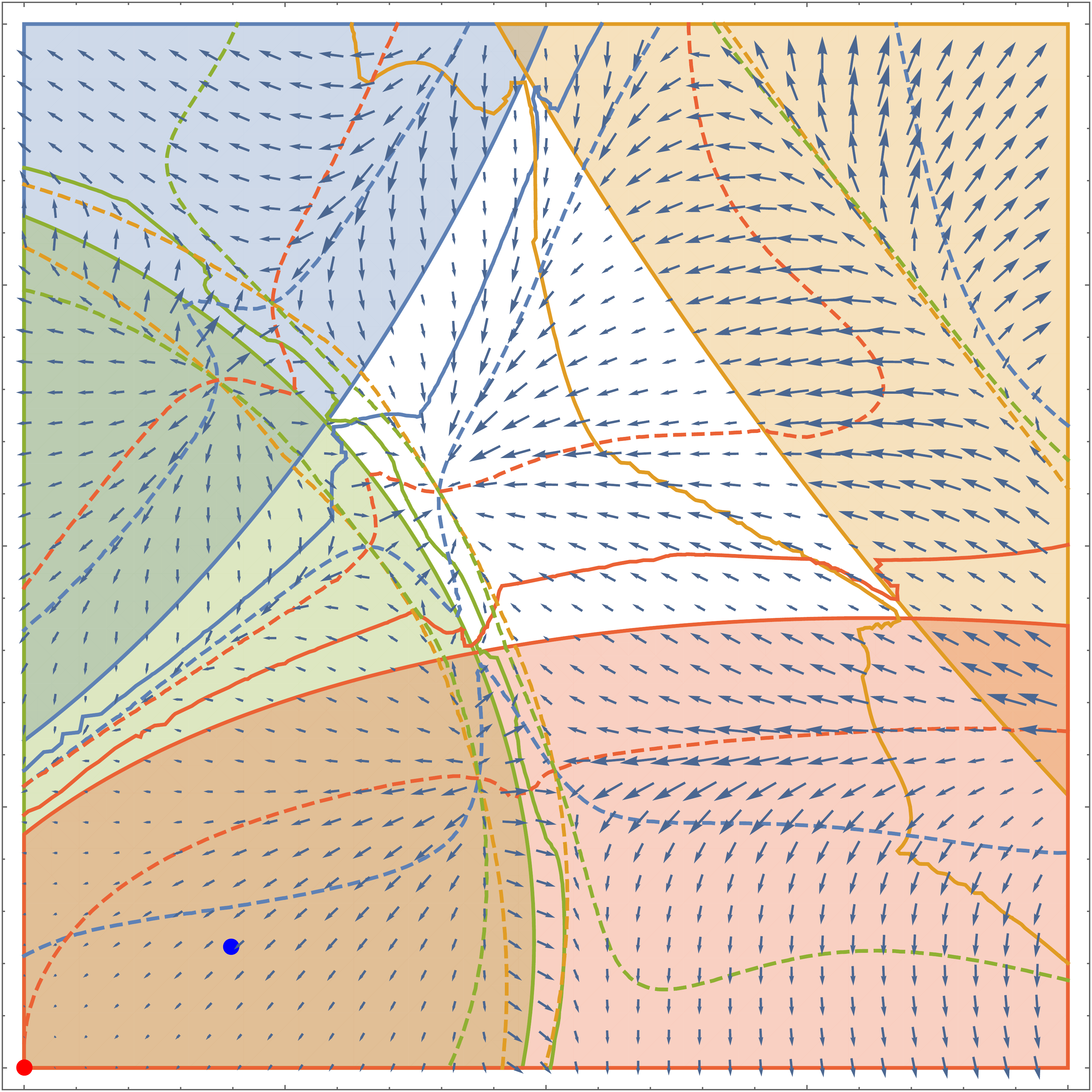}
    \caption{A zoom-in on the centre region between the E-surfaces $\eta$ (shaded) of \texttt{Box4E}: The deformation vector field $\vec{\kappa}$ (blue arrows) as well as the solutions to $\Re\eta=0$ (solid) and $\Im\eta=0$ (dashed) show the complicated interplay between direction, magnitude and displacement of the surfaces in complex space.
    The deformation vector $\vec{\kappa}$ vanishes on the focal point (red).
    The dashed lines meet when the deformation vanishes, i.e. $\vec{\kappa}=\vec{0}$.}
    \label{fig:center_region}
\end{figure}
\begin{figure}
    \centering
    \includegraphics[width=0.9\textwidth]{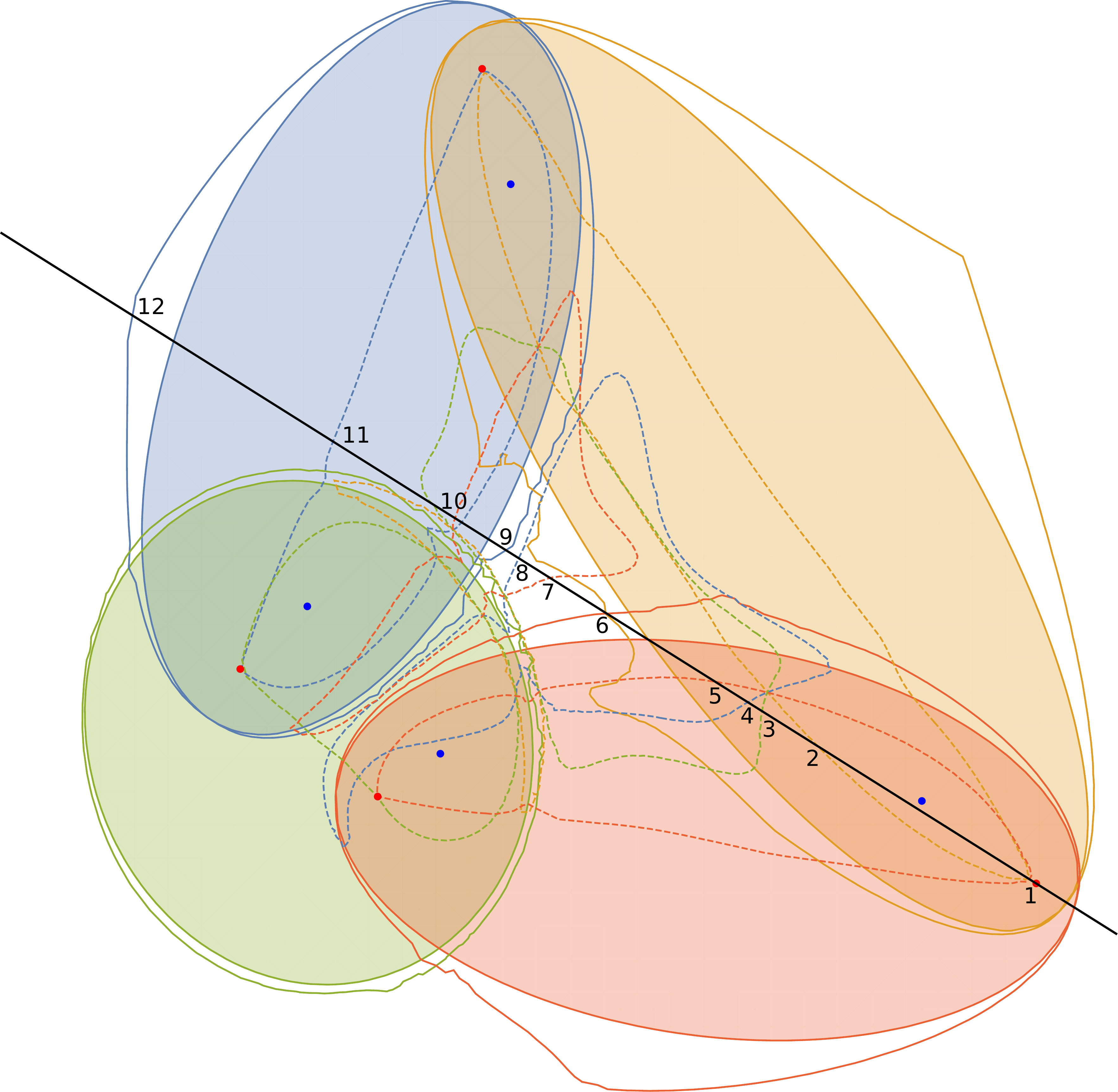}
    \caption{A line segment in integration momentum space: At annotated point 1, the line crosses a focal point. Features 2 to 12 are crossings with the line and the points, where either the real (solid) or the imaginary part (dashed) of the deformed E-surface equations vanishes. See fig.~\ref{fig:kappa_annotated} and fig.~\ref{fig:evaluation_annotated} for more details on the deformation magnitude and the integrand along this line.}
    \label{fig:line_annotated}
\end{figure}
\begin{figure}[p]
    \centering
    \includegraphics[width=0.95\textwidth]{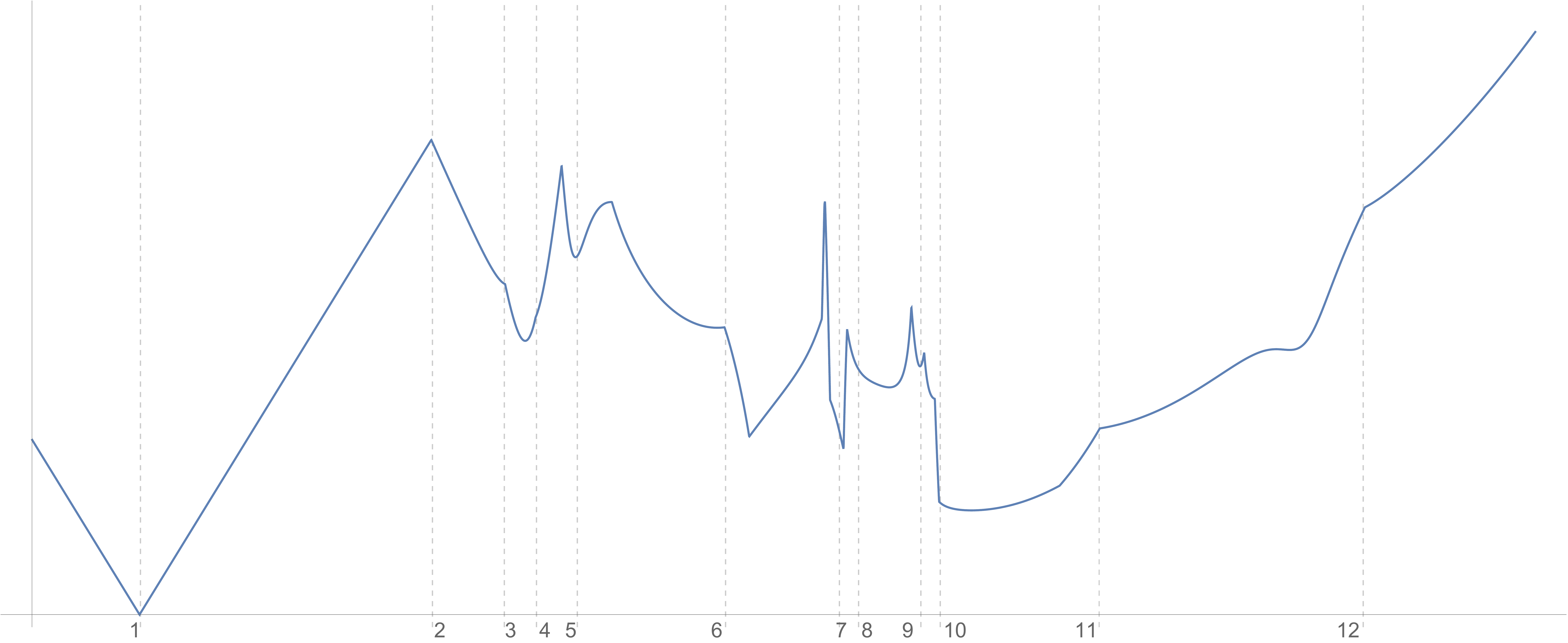}
    \caption{The magnitude of the deformation vector $\vec{\kappa}$ along a line segment (see fig.~\ref{fig:line_annotated}): At annotated point 1, the line crosses a focal point, which forces the deformation to zero. Features 2 to 12 are crossings with the line and the points, where either the real or the imaginary part of the deformed E-surface equations vanishes. These intersections cause a non-smooth behaviour of the deformation vector field $\vec{\kappa}$ due to the constraints on the magnitude.}
    \label{fig:kappa_annotated}
\end{figure}
\begin{figure}[p]
    \centering
    \includegraphics[width=0.95\textwidth]{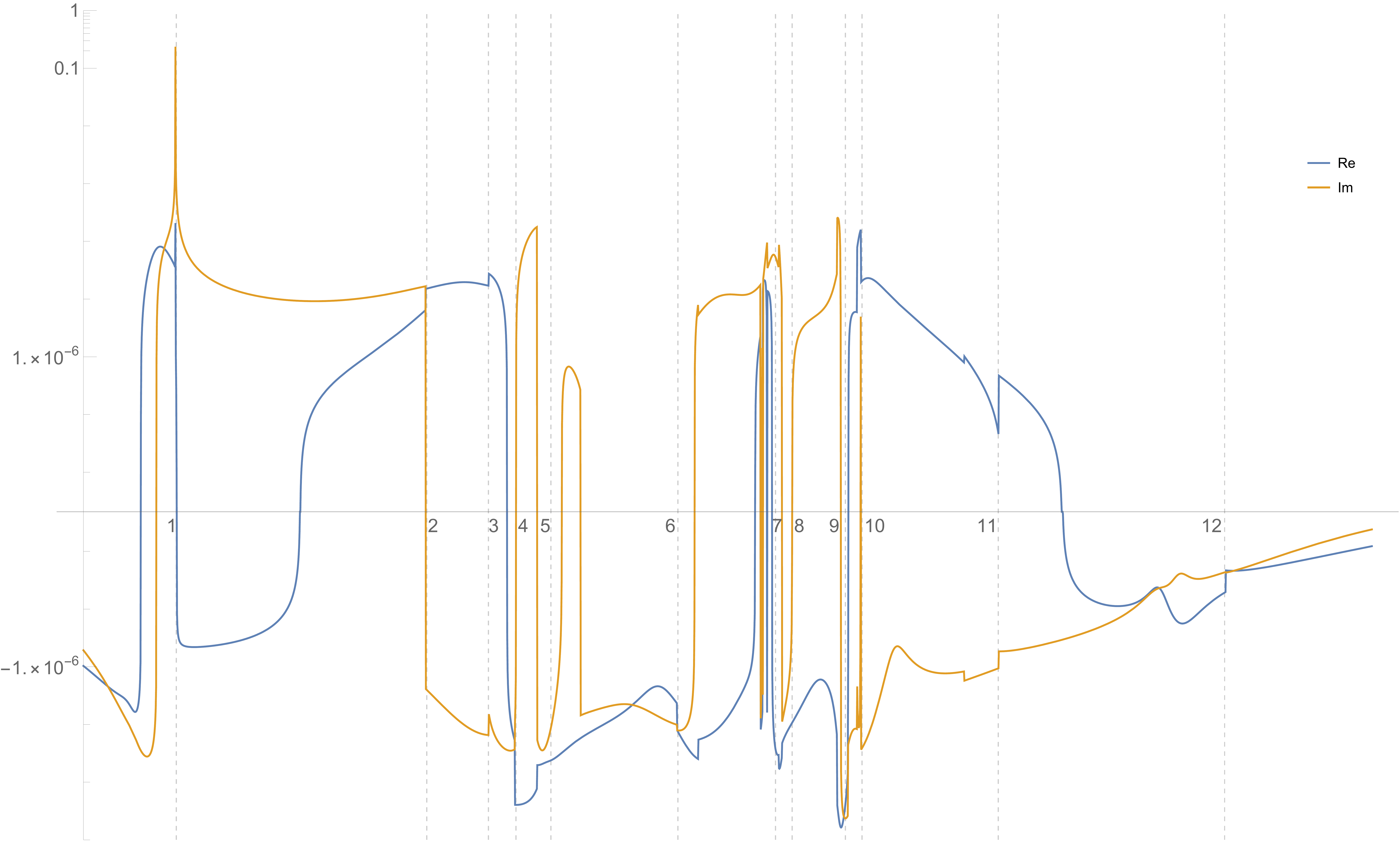}
    \caption{The real (blue) and imaginary part (yellow) of the integrand multiplied by the Jacobian of the contour deformation along a line segment (see fig.~\ref{fig:line_annotated}) on a symmetric log y-axis:
    At annotated point 1, the line crosses a focal point.
    There, the integrand has an integrable singularity.
    Features 2 to 12 are crossings with the line and the points, where either the real or the imaginary part of the deformed E-surface vanishes. On these intersections the deformation vector field $\vec{\kappa}$ is non-smooth (see fig.~\ref{fig:kappa_annotated}), which induces discontinuities in the Jacobian of the contour deformation.}
    \label{fig:evaluation_annotated}
\end{figure}

As a third aspect, we discuss how the deformation magnitude $\lVert\vec{\kappa}\rVert$ affects the integrand.
The connection between magnitude and integrand becomes apparent when studying these quantities on a line segment in integration space.
This line segment is displayed in fig.~\ref{fig:line_annotated}.
We annotated 12 features, where one of them is a focal point and the remaining ones are zeros of either $\Im\eta=0$ or $\Re\eta=0$ of the deformed E-surface $\eta$.
In fig.~\ref{fig:kappa_annotated} we report the deformation magnitude $\lVert\vec{\kappa}\rVert$ along this line.
We see that on the focal point the continuity constraint sets the deformation to zero (feature 1).
At the other features the magnitude constraints lead to a non-smooth behaviour in the deformation vector field.
In fig.~\ref{fig:evaluation_annotated} we study the integrand along the same line.
We observe that on the focal point (feature 1) the integrand is singular.
This is an integrable singularity and can be removed by using multi-channelling in the cut energies (see section~\ref{sec:multichanneling}).

\begin{landscape}
\begin{figure}
    \centering
    \includegraphics[width=\linewidth]{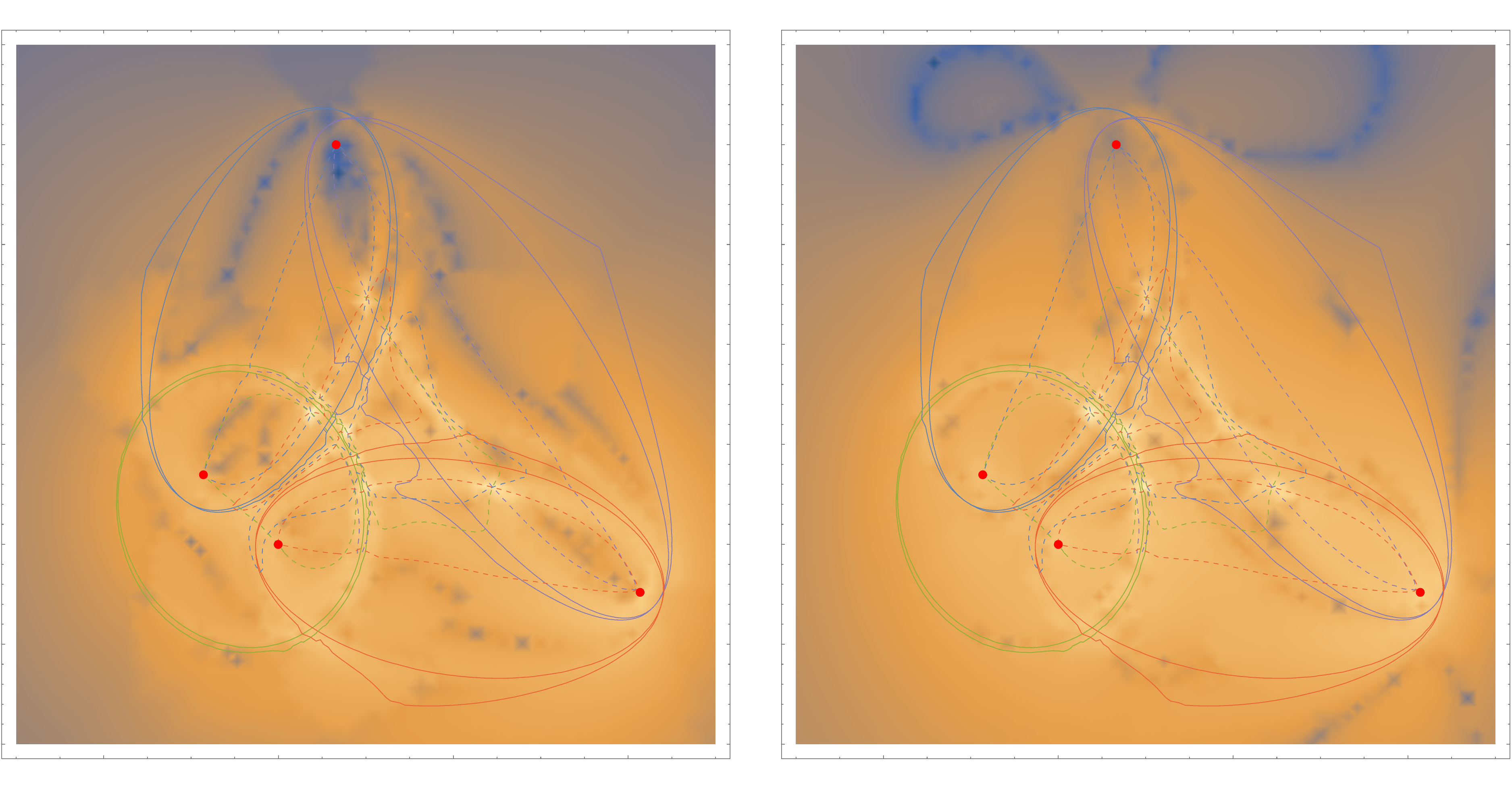}
    \caption{A density plot of the real (left) and imaginary (right) values of the integrand overlaying the undeformed E-surfaces $\eta$ (ellipses) and their solutions to $\Re\eta=0$ (solid) and $\Im\eta=0$ (dashed) with deformation. The real and imaginary part of the integrand has pronounced enhancements (white), where the imaginary parts of multiple E-surface equations are zero. The features for vanishing real and imaginary part of the integrand (blue), have no special significance.}
    \label{fig:heatmap}
\end{figure}
\end{landscape}

Finally, in fig.~\ref{fig:heatmap} we show a density plot of the real and imaginary parts of the integrand, as well as the regions, where the real or imaginary parts of the deformed E-surfaces vanish.
The enhancements in the real or imaginary part of the integrand are directly related to the zeros of the imaginary part of the deformed E-surfaces.
These enhancements are expected when the deformation vanishes close to an E-surface.

\newpage

\newpage
\newcommand{\ifequals}[3]{\ifthenelse{\equal{#1}{#2}}{#3}{}}
\newcommand{\case}[2]{#1 #2} 
\newenvironment{switch}[1]{\renewcommand{\case}{\ifequals{#1}}}{}

\newcommand*{\momsl}[1]{%
  \begin{switch}{#1}
      \case{1}{(-\slashed{k})}
      \case{2}{(-\slashed{k}-\slashed{p}_{2})}
      \case{3}{(-\slashed{k}-\slashed{p}_{23})}
      \case{4}{(-\slashed{k}+\slashed{p}_{15})}
      \case{5}{(-\slashed{k}+\slashed{p}_{1})}
      \case{uv}{(-\slashed{k})}
      \case{s23}{(-\slashed{p}_{23})}
      \case{s15}{(\slashed{p}_{15})}
	\end{switch}
}
\newcommand*{\prop}[1]{%
  \begin{switch}{#1}
      \case{1}{k^2}
      \case{2}{(k+p_{2})^2}
      \case{3}{(k+p_{23})^2}
      \case{4}{(k-p_{15})^2}
      \case{5}{(k-p_{1})^2}
      \case{uv}{k^2-\mu_{uv}^2}
	\end{switch}
}
\newcommand{\pol}[1]{\hat{\slashed\varepsilon}_{#1}}

\clearpage 
\section{Subtraction}
\label{sec:amplitudes}
In the discussion so far, we considered integrals that do not have singularities for loop momenta of large magnitude (ultraviolet (UV) singularities) or soft and/or collinear to external legs (infrared (IR) singularities). For practical applications, such as computing amplitudes of physical processes, this will not be the case, as individual diagrams can contain both UV and IR divergences.

After transforming the integrand using LTD, non-integrable singularities manifest themselves as pinched (squeezed) E-surfaces. For the case of Feynman diagrams with massless internal propagators, this will happen when one or more of the massless external legs become on-shell.
It is however still possible to numerically integrate such integrals, provided that the non-integrable singularities are regulated first.
In general this is achieved by subtracting from the integrand an expression that contains the same pinched E-surface(s) and that approximates the original integral in the limit where the singular surface is approached.
If these subtraction terms (also known as counterterms) are significantly simpler than the original integral, one can integrate them analytically in dimensional regularisation and add them back to the final expression in order to recover the original integral, including all its poles in the dimensional regulator.
In this section we start by presenting a novel method to regulate divergent scalar integrals at one-loop without the introduction of propagators linear in the loop momentum featured in ref.~\cite{Anastasiou:2018rib}. We then discuss the introduction of counterterms for physical amplitudes~\cite{Anastasiou:2019xxx} where only one term is introduced to remove all IR divergences.
This regulated expression can then be integrated using LTD and the contour deformation discussed in sect.~\ref{sec:contour_deformation}.

Note that in this section we refer to the external momenta as $p_i$ for ease of reading.
\subsection{Divergent scalar integrals}
We start by investigating scalar integrals subject to IR divergences at one-loop.
In general, it is convenient to express counterterms in terms of the same building blocks as the original integrand, namely quadratic propagators. 
This allows to use the LTD formalism that has been introduced for the case of finite scalar integrals.
At one-loop, we will show that we can always achieve such subtraction using a linear combination of triangles built by a subset of the original propagators and with coefficients expressed in terms of the kinematic invariants $s_{ij}$.
Since the counterterms involve only propagators already present in the original diagram, they do not introduce any new E-surfaces.

\subsubsection{General one-loop massless scalar integral}
Let us consider an $n$-point function with all the internal propagators massless and with external momenta $p_j$ with $p_j^2=m_j^2$.
We first consider the case where only one leg $i$ is massless ($m_i=0$).
As a consequence, the corresponding scalar integrand will develop a collinear singularity when the loop four-momentum $k$ becomes collinear to the corresponding momentum $p_i$:
\begin{align}\label{eq:collinear_factorization}
\begin{gathered}
\includegraphics[scale=.8,page=1]{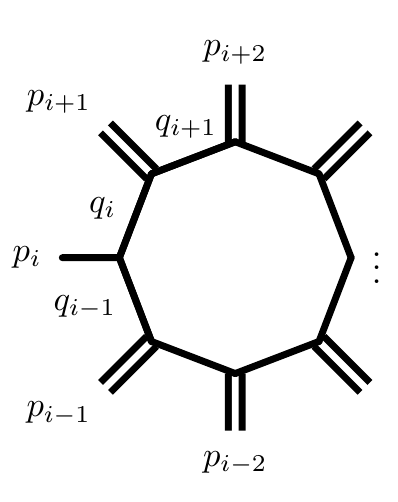}
\end{gathered}
&\quad \xrightarrow{q_i= x\, p_i}\quad
\left(
\begin{gathered}
\includegraphics[scale=.9,page=2]{img/scalar_diags.pdf}
\end{gathered}
\right)
\;\times\;
\underbrace{
\left(
\begin{gathered}
\includegraphics[scale=.8,page=3]{img/scalar_diags.pdf}
\end{gathered}
\right)}_{=:\,c_i(x)}.
\end{align}
In the expression above (where we consider the loop momentum to flow clockwise) we can see how the integrand factorises in the collinear limit.
The integration of this counterterm can be performed as shown in ref.~\cite{Anastasiou:2018rib}.
The variable $x$ is a function of the loop momentum and is defined as follows:
\begin{equation}\label{sudakov}
q_i = x p_i + y v +q_{i,\perp}, \quad \text{where}\quad
\begin{array}{l}
v^2=0,\\v\cdot p_i\neq 0.
\end{array}
\end{equation}
The expression on the l.h.s of eq.~\eqref{eq:collinear_factorization} can be written in an integral form as follows:
\begin{equation}
\mathcal{I}_n := \int \\d^4 k\, I_n(k,\{p_i\}),\qquad\text{where}\qquad
\left\{
\begin{array}{rcl}
I_n(k,\{p_i\})&=&\dfrac{1}{\prod_{i}^n (q_i)^2},\\[10pt]
q_i&=&k + \sum_{j=1}^i p_i, \\
\end{array}
\right.,
\end{equation}
\begin{equation}\label{eq:collinear_factorization_expession}
\mathcal{I}_n\quad \xrightarrow{q_i=x\, p_i}\quad  \int \\d^4 k\ \frac{c_i(x)}{(q_{i-1})^2(q_i)^2} \,.
\end{equation}
The coefficient $c_i(x)$ that multiplies the bubble propagators corresponds to the remaining hard propagators with the loop momentum evaluated in the collinear limit:
\begin{equation}
c_i(x)=\frac{1}{\prod_{j=i+1}^{n+i-2}(x\,p_i+q_{ji})^2},\qquad
q_{ji}:= q_j-q_i.
\end{equation}
The limit shown on the right-hand side of eq.~\eqref{eq:collinear_factorization_expession} could be used to build an IR finite expression by subtracting it from $\mathcal{I}_n$, however such a counterterm introduces propagators that are linear in the loop momentum. Linear propagators yield singular surfaces that are not akin to E-surfaces, implying that the general construction of the contour deformation presented in sect.~\ref{sec:contour_deformation} cannot directly control the properties of the imaginary part of the loop momentum on them. We leave the investigation of solutions for accommodating linear propagators to future work and for now aim at casting the subtraction terms $c_i(x)$ in terms of propagators already present in the original divergent one-loop integral.

We start by considering all possible triangles that factorise the same divergent bubble in the collinear limit. This condition fixes two of the three propagators of the triangle to be the ones that become singular in a specific collinear limit, whereas the third propagator can be chosen to be any of the other ones appearing in the original  $n$-point integral. All such triangles are:
\begin{equation}\label{Tij}
T(i,j) := 
\begin{gathered}
\includegraphics[scale=1,page=4]{img/scalar_diags.pdf}
\end{gathered},\qquad
j\in \mathcal{J}_i:=\{i+1,i+2,\dots,i+n-2\}.
\end{equation}
with periodic conditions on the loop momenta labels.
In the collinear limit, each element $T(i,j)$ factorises one hard propagator $t_{ij}$ whose expression reads:
\begin{align*}
t_{ij}(x)&= \frac{1}{(x p_i + q_{ji})^2} \,.
\end{align*}
Note that each squared momentum in the denominator of our coefficient functions is linear in $x$ because $p_i$ is on-shell, resulting in only one simple pole in the variables $x$.
\\
In order to cancel the divergences of the $n$-point function we need to find a linear combination of $T(i,j)$ with coefficients $a_{ij}(x)$ that satisfies:
\begin{equation*}
\sum_{j\in\mathcal{J}_i}t_{ij}(x)a_{ij}(x)=c_i(x).
\end{equation*}
We can multiply both sides of this expression by the denominator of $c_i(x)$ which is equal to the product of all the possible $t_{ij}$ with $i \neq j$.
We then obtain a polynomial of degree $(n-3)$ in $x$:
\begin{align*}
\sum_{j\in \mathcal{J}_i}\left(\prod_{\substack{r\in \mathcal{J}_i\\ r\neq j}}t^{-1}_{ir}(x)\right)a_{ij}(x)
=1 \,.
\end{align*}
Since we have $(n-2)$ degrees of freedom and we insist that coefficients $a_{ij}(x)$ are free of poles in $x$, one needs to involve all terms $T(i,j)$ in order to solve the equation above (assuming all the poles $t_{ij}(x)$ are distinct).
In particular, an explicit solution can be found by using the roots of the inverse coefficients $t_{ij}^{-1}$:
\begin{equation}\label{oneoff_aj}
a_{ij}=\prod_{\substack{r\in \mathcal{J}_i\\ r\neq j}}t_{ir}\left(-\frac{(q_{ji})^2}{2 p_i \cdot q_{ji}}\right),
\end{equation}
resulting in coefficients that depend only on the external kinematics.

This procedure does not work in the case of degenerate (raised) propagators. 
This can be resolved by considering a subset $\bar{\mathcal{J}}_i\subset \mathcal{J}_i$ which contains only one member of each degenerate subset of propagators with multiplicity $\nu_j$ for $j\in\bar{\mathcal{J}}_i$.
Moreover, we need to generalise eq.~\eqref{oneoff_aj} in order to support the degeneracy of the involved propagators.
In the collinear limit, the linear combination of the elements of this set gives the same singularities as the original integral, provided that:
\begin{align*}
\sum_{j\in  \bar{\mathcal{J}}_i}\left(\prod_{\substack{r\in \bar{\mathcal{J}}_i\\ r\neq j}}t^{-\nu_r}_{ir}(x)\right)a_{ij}(x)
&=1.
\end{align*}
In this case we have $| \bar{\mathcal{J}}_i|$ parameters $a_{ij}$ to constrain a polynomial of degree $n$ with $(|\bar{\mathcal{J}}_i|-1)$ distinct roots. It is then clear that the coefficients $a_{ij}$ take the same values as those given in equation~\eqref{oneoff_aj}. From this point onward, we will only consider one-loop scalar integrals with non-degenerate propagators.

We are now equipped with a method that removes single collinear singularities from integrals with one off-shell external momentum by writing a linear combination of the triangular elements $T(i,j)$.
When more than one external leg has a vanishing mass, we can apply the same procedure for each of them. In this case, we have to be careful when one of the triangles appears in more than one regularisation. For example, when two adjacent momenta are on-shell at the same time, one has $T(i,i+1) = T(i+1,i-1)$.
In this kinematic configuration the corresponding coefficients will be same:
\begin{align*}
a_{i,i+1} = a_{i+1,i-1}, \qquad\text{when} \qquad p_i^2=p_j^2=0 \,.
\end{align*}
Thus, one has to be careful when summing the regulator corresponding to each of the massless external legs in order to avoid double-counting.

We can write one general subtraction term, referred to as $\text{CT}_n$, that can be used for any combination of on/off-shell external momenta of a scalar one-loop $n$-point integral:
\begin{equation*}
\mathcal{I}_{n|\textrm{subtracted}}=\mathcal{I}_n-\text{CT}_n,
\end{equation*}
\begin{align}\label{general_scalar_ct}
\text{CT}_n	&= \sum_{i=1}^n \left(\beta_i T(i,i+1) 
	+\sum_{j=i+2}^{n+(i-3)} a_{ij} T(i,j)\right),
\end{align}
where we introduced the coefficients $\beta_{i}$ used to avoid double counting. Their expression is
\begin{equation}
	\beta_i=
	\begin{cases}
		a_{i,i+1} &:\,p_i^2=0\\
		a_{i+1,i-1} &:\, \text{otherwise}
	\end{cases},
\end{equation}
where we make explicit use of the fact that whenever $p_i$ and $p_{i+1}$ are on-shell at the same time the two coefficients $a_{ij}$ coincide.

Because the constructed collinear counterterms do not depend on the parameter $x$, they completely remove the singularities from pinched E-surfaces, implying that they regulate \emph{both} collinear and soft divergences.
As a consequence, we have that the integral $\mathcal{I}_n-\text{CT}_n$ is finite for all loop momentum configurations.
The original expression $\mathcal{I}_n$ can be recovered by adding back the integrated counterterms. The integrated counterterm consists of $n(n-3)$ distinct one-loop scalar triangles that are straightforward to compute analytically for general external kinematics using dimensional regularisation.
We leave to future work the investigation of the possible multi-loop generalisation of this construction of counterterms that do not involve any propagators that are linear in the loop momenta.

\subsubsection{Explicit example of subtraction for a divergent one-loop scalar box}
For the four-point box topology with massless propagators, there are four counterterms since the sum in eq.~\eqref{general_scalar_ct} over the coefficients $a_{ij}$ is empty. Only the $\beta_i$ are present and take the following expression:
\begin{align}\label{boxCTcoeff}
\begin{split}
\beta_i
=
\begin{cases}
\dfrac{s_{i,i+1} - p_{i+1}^2}{s_{i,i+1}s_{i-1,i}-p_{i+1}^2p_{i-1}^2 }&:\,p_i^2=0\\[10pt]
\dfrac{s_{i,i+1} - p_{i}^2}{s_{i,i+1}s_{i+1,i+2} - p_{i}^2p_{i+2}^2} &:\, \text{otherwise}
\end{cases}
\end{split}\,,
\end{align}where $s_{ij}=(p_i+p_j)^2$. In the particular case where all external momenta are massless and on-shell (i.e. $p_i^2=0$), the final expression of the counterterms reads:
\begin{align}
\begin{split}
\text{CT}_n
	&= \sum_{i=1}^4 \beta_i T(i,i+1) \\
	&=\frac{T(1,2)}{s_{23}} 
	+\frac{T(2,3)}{s_{12}}
	+\frac{T(3,4)}{s_{23}}
	+\frac{T(4,1)}{s_{12}}\,,
\end{split}
\end{align}
which coincides with the results presented in ref.~\cite{Anastasiou:2018rib}, in which this same expression corresponds to the counterterm built for the subtraction of soft singularities (and the authors also concluded that the counterterm cancels all IR divergences in that particular case). In other cases however, and especially beyond one-loop, the counterterms from ref.~\cite{Anastasiou:2018rib} introduce linear propagators of the form of eq.~\eqref{eq:collinear_factorization}.

\subsection{One-loop amplitudes}
The first physical amplitude we study pertains to the production of photons from the scattering of a quark and an anti-quark. For brevity, the order of the photons is kept fixed during this discussion, as performing the integration over all permutations of the final states does not add any complications.

The tree-level contribution for $q \bar q \rightarrow (N-2)\, V$ is defined as
\begin{align}
\label{eq:born}
i\mathcal{A}_0 
	=\hspace{-2em}\begin{gathered}\includegraphics[page=12,scale=.8]{img/ddAAA_diags.pdf}\end{gathered}
  \;=\; C_0\left(\prod_{i=3}^{N-1}\frac{1}{(\sum_{j=2}^i p_j)^2}\right)\,\bar v_2 T_0 u_1 \,,
\end{align}
where all the fermions are assumed to be massless and the coefficients $C_0, \ T_0$ depend on the vector boson considered as a final state. 
If only photons are considered as final states such coefficients are given by:
\begin{equation}
    T_0=\slashed \varepsilon_3 (-\slashed p_{23})\slashed \varepsilon_4\cdots (\slashed p_{15})\slashed \varepsilon_N
    ,\qquad
    C_0=g^3q^3 \,.
\end{equation}
These formulas can easily be extended to the electroweak bosons $W^\pm$ and $Z$ by substituting the photon polarisation vectors with generic ones ${\slashed{\varepsilon}_i\rightarrow \hat{\slashed{\varepsilon}}}$ which also encode the information about the axial and vectorial part of the corresponding boson:
\begin{equation}
    \hat{\slashed{\varepsilon}_i}:=\slashed{\varepsilon}_iP_i,\qquad P_i=c_V-c_A\gamma_5,
\end{equation} 
with projectors defined as
\begin{equation}
P_L:=\frac{1-\gamma^5}{2},\qquad P_Z:=\frac{c_V^d-c_A^d \gamma^5}{2} \,.
\end{equation}

In order to obtain a more general expression we will use this new definition for the polarisation vectors. In the case of photons, all the $P_i$s are proportional to the identity matrix.

In order to compute the one-loop QCD correction to eq.~\eqref{eq:born} one needs to consider all possible insertions of a gluon along the fermionic line.
The IR structure of the relevant diagrams features one or two pinched collinear singularities if the gluon is attached to one or both the external fermion lines, respectively. In the latter case, the diagram also features a soft singularity.

\subsubsection{Counterterms}
\label{sec:counterterms}
If the photons are physically polarised, the only pinched divergences contributing to the IR sector involve a gluon connecting one of the propagators of the tree-level diagram with the external quarks. There are no singularities originating from two internal quarks and an external photon meeting at a vertex and becoming collinear, since the numerator vanishes:
\begin{equation}
[\dots](\slashed k -\slashed p_i)\pol{i}\slashed k[\dots]
\quad	\xrightarrow{k\,=\,x\,p_i} \quad
\bar{x} x[\dots] \,\slashed p_i\pol{i}\slashed p_i[\dots] = 0 \,.
\end{equation}
Since the pinched singularities originate uniquely from insertions of gluons connecting an external fermion to an internal fermion, the Ward identity can be used to regulate all the collinear and soft divergences with a general counterterm.
However, it is necessary to fix a consistent choice of routing for the loop momentum in order for cancelling divergences to be localised in the same region in momentum space, even though they belong to different diagrams.
The general counterterm $I_{\text{IR}}$ reads:
\begin{align}\label{IRCT}
\begin{gathered}\includegraphics[page=11,scale=.8]{img/ddAAA_diags.pdf}\end{gathered}
&= \frac{C_1\,\mu^{2\epsilon} (4\pi)^2}{\prod_{i=3}^{N-1}(\sum_{j=2}^i p_j)^2}\int \frac{\dd^d k}{(2\pi)^d}
	\frac{\bar v_2  \gamma^\mu (\slashed k - \slashed p_2)T_0 (\slashed k + \slashed p_1) \gamma_\mu u_1}{k^2(k+p_1)^2(k-p_2)^2 } \,,
\end{align}
where
\begin{equation}
C_1=\mathrm{i} C_0\frac{C_F\alpha_s}{4\pi}.
\end{equation}
This integration can be performed analytically using Feynman parametrisation, and we obtain:
\begin{align}\label{IRCT_integrated}
I_{\text{IR}}=\mathrm{i}\frac{C_1}{\prod_{i=3}^{N-1}(\sum_{j=2}^i p_j)^2}\left(\frac{4\pi \mu }{-s_{12}}\right)^\epsilon\frac{C_\Gamma}{2-2\epsilon}\left(
       	-2 M_0\frac{2-2\epsilon+\epsilon^2}{\epsilon^2}	
		+ \frac{ M_1}{2\epsilon}	
		+\frac{ M_2}{-s}\right) \,,
\end{align}
where
\begin{align}\label{spin_chains}
\begin{split}
M_0&=\left[\bar v_2 T_0 u_1\right],\\
M_1&=\left[\bar v_2 \gamma^\mu \gamma^\nu T_0 \gamma_\nu \gamma_\mu u_1\right],\\
M_2&=\left[\bar v_2\gamma^\mu\slashed p_1 T_0 \slashed p_2\gamma_\mu  u_1\right],\\
C_\Gamma &= \frac{\Gamma(1-\epsilon)^2\Gamma(1+\epsilon)}{\Gamma(2-2\epsilon)}.
\end{split}
\end{align}

Although subtracting eq.~\eqref{IRCT} from the original integrand allows to completely regulate IR singularities, the subtracted integrand is still divergent in the UV sector. 
This divergence can manifest itself locally, in spite of the integral itself being finite, either due to symmetries of the integrated expression or because the IR and UV poles cancel for integrals that are scaleless in dimensional regularisation.
The behaviour for large momenta is inferred by the scaling of the integrand in these regions, and as a result all log-divergent triangles (one gluon, two fermions) and linearly divergent bubbles (one gluon, one fermion) that appear in the amplitude have to be regulated.
The construction of the counterterm is done by taking the UV limit of each diagram by replacing
\begin{equation}\label{UV_limit}
\frac{\slashed{k}+\slashed{p}}{(k+p)^2}
\quad\rightarrow\quad
\frac{\slashed{k}}{k^2-\mu_{UV}^2},
\end{equation}
where the only relevant momentum is now the loop momentum carried by the exchanged gluon.
The bubble diagram has a leading UV divergence that is linear in the loop momentum. In the context of an analytic integration such contribution integrates to zero because of radial symmetry, although the integrand is locally divergent.
It is therefore necessary to also regulate this leading UV divergence together with the sub-leading one obtained by computing the second order in the Taylor expansion around the UV approximation given by eq.~\eqref{UV_limit}.
An explicit example of this subtraction can be found in appendix~\ref{sec:diagram_expressions}, where eq.~\eqref{A:triangle_UVCT} represents the UV counterterm of a triangle and eq.~\eqref{A:bubble_UVCT} represents the counterterm of a bubble.
The IR counterterm that we introduced is UV divergent and requires regulation as well. Its divergence can be expressed as as a triangle integral and can be subtracted by means of eq.~\eqref{UV_limit}.

The combination of counterterms can be used to build a finite amplitude expression that can be integrated using LTD:
\begin{equation}\label{eq:finite_amplitude}
\mathcal{A}_{\text{finite}}= \mathcal{A} - I_{\text{CT}}\,,\qquad\text{where}\qquad
I_{\text{CT}}
    =\sum_{\substack{\text{UV}\\\text{div.}}}I_{\text{UV}}
    +I_{\text{IR}}
    -I_{\text{UV}_\text{IR}}.
\end{equation}

The counterterm can be integrated analytically with the use of dimensional regularisation.
In the UV contribution to the integrated counterterm we notice that the bubble and the triangle lead to the same value in norm and opposite in sign if constructed according to the substitution rule \eqref{UV_limit}. Thus, the only remaining contribution is
\begin{align}\label{UV_diag_integrated_ct}
\sum_{\substack{\text{UV}\\\text{div.}}}I_{\text{UV}} &= -\mathrm{i}\frac{C_1}{\prod_{i=3}^{N-1}(\sum_{j=2}^i p_j)^2}\left(\frac{4\pi\, \mu^2}{\mu_{\text{UV}}^2}\right)^\epsilon\Gamma(1+\epsilon)\frac{(1-\epsilon)^2}{\epsilon} M_0 \,.
\end{align}

Finally, regulate the IR counterterm with the same technique. The corresponding analytically integrated counterpart reads:
\begin{align}\label{UV_IR_integrated_ct}
I_{\text{UV}_{\text{IR}}} &=-\mathrm{i}\frac{C_1}{\prod_{i=3}^{N-1}(\sum_{j=2}^i p_j)^2}\left(\frac{4\pi\, \mu^2}{\mu_{\text{UV}}^2}\right)^\epsilon\Gamma(1+\epsilon)\frac{1}{4\epsilon} M_1 \;.
\end{align}
The complete expression $I_{\text{CT}}$ can then be expanded in $\epsilon$ up to finite terms and be used to recover the original amplitude once combined with the value coming from numerical integration. 
The integrated counterterm for $q\bar q$ to photons takes the simple form:
\begin{equation}\label{eq:integrated_ct}
I_{\text{CT}}= \dfrac{C_1}{ \prod_{i=3}^{N-1}(\sum_{j=2}^i p_j)^2}\,M_0\dfrac{(4\pi)^\epsilon}{\Gamma(1-\epsilon)} \left[
	\dfrac{1}{\epsilon^2} 
	+ \dfrac{1}{\epsilon} \left(\dfrac{1}{2}+\ln_\mu \right)
	+ \left(4+\dfrac{1}{2} \left(3+\ln_\mu \right)\ln_\mu \right)
	\right]\,,
\end{equation}
where $\ln_\mu = \log(\frac{\mu^2}{-s_{12}})$. Any dependence on $\mu_{\text{UV}}$ has dropped from this final expression. As a consequence, the integration of the finite amplitude will also not depend on the choice of $\mu_{\text{UV}}$. This condition can be used as a further check for the proper cancellation of the divergences.

\subsubsection{Ultraviolet behaviour}
\label{sec:uv_behaviour}
When integrating the LTD expression, one has to take into account that the superficial degree of UV divergence of each dual integrand is higher than that of the sum of its cuts. This is because once the LTD on-shell cuts of the residues are applied, every quadratic propagator scales as $1/|\vec{k}|$ in the UV instead of $1/k^2$.
As a consequence, contrary to the Minkowskian case, the addition of more fermion propagators to the diagram is not suppressing the scaling of the deformation in the UV sector:
\begin{align}
    \int \dd^4 k \,\delta_+(q_j^2)\frac{1}{k^2}\prod_i^N\frac{\slashed q_i}{q_i^2}&\sim k^{2}\, \forall N\,,
\end{align}
compared to the original scaling of the 4D integrand being
\begin{align}
\int \dd^4 k \,\frac{1}{k^2}\prod_i^N\frac{\slashed q_i}{q_i^2}&\sim k^{2-N} \,.
\end{align}
Summing over all the different cuts will however recover the original scaling of $k^{2-N}$. 

If the dual integrand scales faster than $1/|\vec{k}|$ in the UV, the numerical cancellation of large numbers becomes prone to numerical instabilities.
One way avoid such numerical instabilities in the UV region is to approximate the integrand with a better behaved function in the corresponding sector, obtained by taking a UV approximation of the integrand.
The most convenient choice is to replace all the propagators with a common UV one:
\begin{equation}
q^2 \quad\rightarrow\quad (k+p_{\text{UV}})^2 \,.
\end{equation}
This ensures that the approximating function only features a single dual integrand, which directly scales as the $4$-dimensional integrand. 
The numerator can be left unchanged for this approximation. In section~\ref{sec:one_loop_amp} we discuss the effects of this UV approximation.


The UV counterterms can be constructed as shown in sect.~\ref{sec:counterterms} for most integrals, but in the case of a bubble integral, the subleading logarithmic divergence must also be regulated.
The relevant part of the approximation is shown below:
\begin{align}
\begin{split}
\frac{\gamma^\mu\slashed q\gamma_\mu}{q^2}
&\approx
\frac{\gamma^\mu\slashed q\gamma_\mu}{(k+p_{\text{UV}})^2}
	-2k\cdot (q-k-p_{\text{UV}})\frac{\gamma^\mu\slashed q\gamma_\mu}{(k+p_{\text{UV}})^3}
\\&=\frac{\gamma^\mu\slashed q\gamma_\mu}{(k+p_{\text{UV}})^2}
	-\frac{\gamma^\mu\{\slashed k,(\slashed q-\slashed k-\slashed p_{\text{UV}})\}\slashed q\gamma_\mu}{(k+p_{\text{UV}})^3} \,.
\end{split}
\end{align}

Since the UV counterterms have higher-order poles, the LTD formula shown in sect.~\ref{sec:ltd_intro} cannot be applied directly. We discuss how to apply LTD to integrals featuring raised propagators in appendix~\ref{sec:raised_propagators}.

\subsubsection{One-loop amplitude for $q\bar q\rightarrow \gamma_1 \gamma_2 \gamma_3$}
\label{sec:one_loop_amp}
We now study the specific case of the one-loop $d\bar d\rightarrow \gamma_1 \gamma_2 \gamma_3$ amplitude. 
The tree-level diagram of this amplitude is
\begin{align*}
i\mathcal{A}_0 = \begin{gathered}\includegraphics[page=1,scale=.8]{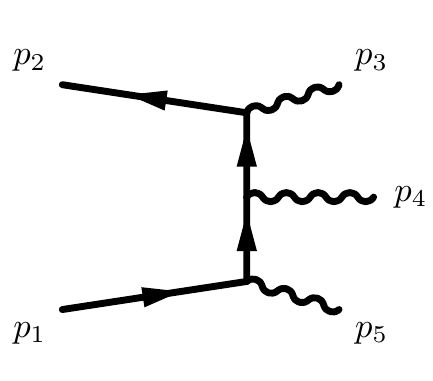}\end{gathered}
    &= C_0\frac{M_0}{\prod_{i=3}^{4}(\sum_{j=2}^i p_j)^2},
\end{align*}
where the coefficients are given by
\begin{equation}
    M_0=\bar v_2 \slashed \varepsilon_1 (-\slashed p_{23})\slashed \varepsilon_2 (\slashed p_{15})\slashed \varepsilon_3 u_1,
    \qquad
    C_0=g^3q^3.
\end{equation}
\begin{figure}[ht]%
 \centering
 \renewcommand*\thesubfigure{D\arabic{subfigure}} 
 \subfloat[]{\includegraphics[page=2,scale=.8]{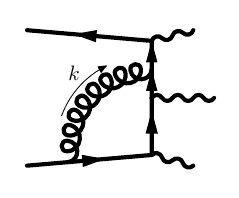}\label{diag:1}}%
 \subfloat[]{\includegraphics[page=6,scale=.8]{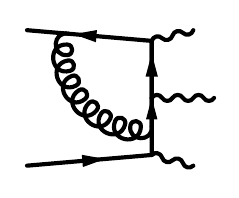}\label{diag:2}}
 \subfloat[]{\includegraphics[page=9,scale=.8]{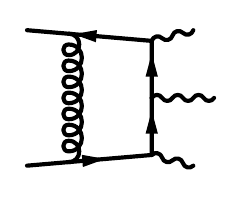}\label{diag:3}}
 \subfloat[]{\includegraphics[page=4,scale=.8]{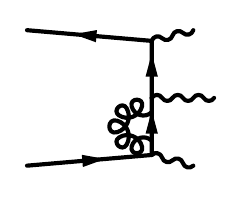}\label{diag:4}}\\
 \subfloat[]{\includegraphics[page=5,scale=.8]{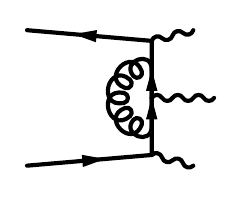}\label{diag:5}}
 \subfloat[]{\includegraphics[page=8,scale=.8]{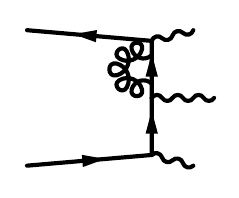}\label{diag:6}}
 \subfloat[]{\includegraphics[page=3,scale=.8]{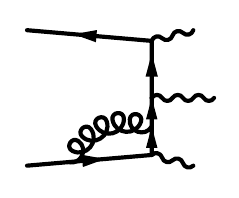}\label{diag:7}}%
 \subfloat[]{\includegraphics[page=7,scale=.8]{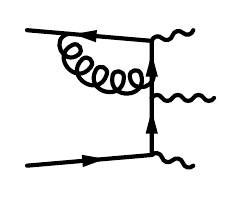}\label{diag:8}}
 \caption{Diagrams contributing to one-loop QCD correction to $q\bar q\rightarrow 3\,\gamma$ amplitude.}%
 \label{fig:1loop_ddAAA}%
\end{figure}
\\
Fig.~\ref{fig:1loop_ddAAA} shows all the diagrams involved in the one-loop QCD correction. 

Diagrams D1 -- D3 and D7 -- D8 are IR divergent: D1 and D7 are divergent when $k$ is collinear to $p_1$ and D2 and D8 are divergent $k$ is collinear to $p_2$, whereas the diagram $D3$ is divergent in both cases and also has a soft divergence.

Despite the fact that the integrated amplitude is UV finite, the local behaviour of the integrand in the UV region needs to be regulated.
This can be done by writing the corresponding counterterms for all UV divergent integrals, specifically D4 -- D8.

In order to ensure that the cancellation occurring across diagrams at the integrated level are also reflected at the local integrand level for the whole amplitude, one must carefully choose the the loop momentum routing of each diagram so as to localise cancelling divergences in the same region of momentum space.
The case at hand is quite easy in that regard, as one can choose the gluon line to have momentum $k$ with momentum flow against the fermionic line for all the diagrams.

Fig.~\ref{fig:1loop_ddAAA} shows the different behaviours when approaching the soft, collinear, and UV limits. 
The different limits are approached by rescaling the loop momentum $k$ by a  factor $\delta$ for the soft and UV limit, while for the collinear limit we use the Sudakov parametrisation of eq.~\eqref{sudakov} with $y$ and $k_\perp$ rescaled by $\delta$ and $\sqrt{\delta}$ respectively.
The different asymptotic scaling $\delta^1$, $\delta^\frac{1}{2}$ and $\delta^{-1}$, prove that the divergences are properly subtracted.

\begin{figure}[p]%
 \centering
 \subfloat[]{\includegraphics[scale=.45]{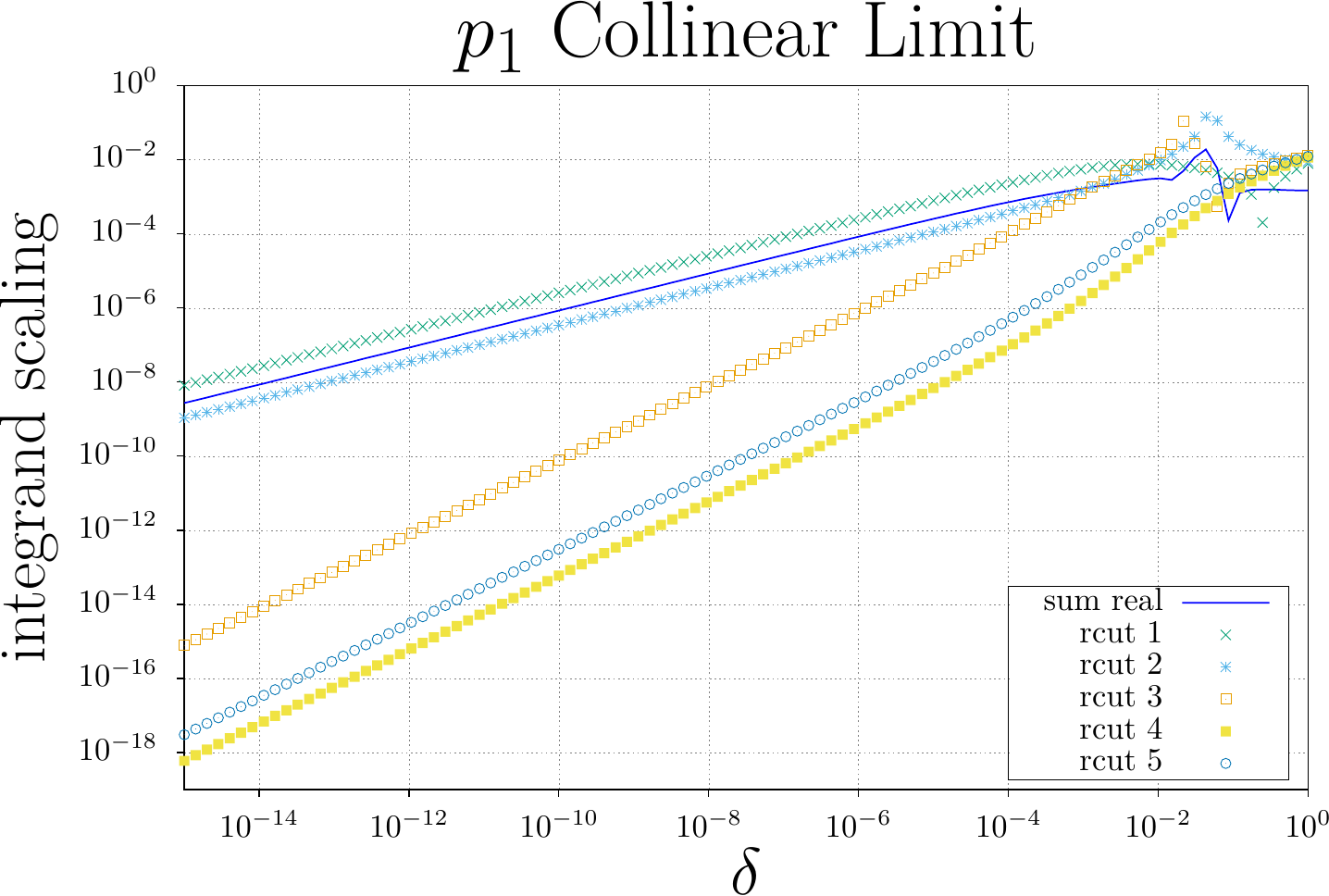}\label{plot:dd3a_p1}}%
 \subfloat[]{\includegraphics[scale=.45]{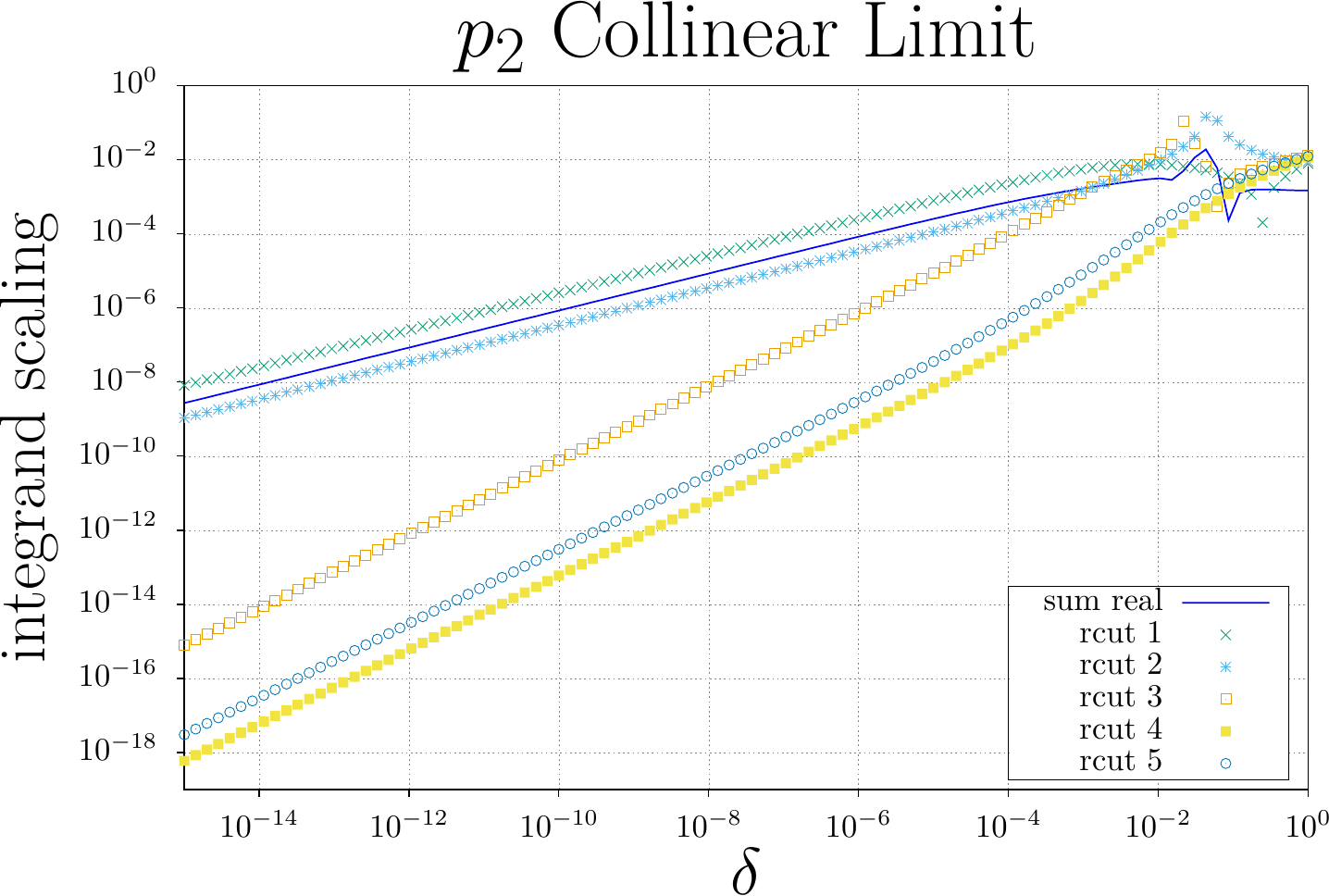}\label{plot:dd3a_p2}}\\
 \subfloat[]{\includegraphics[scale=.45]{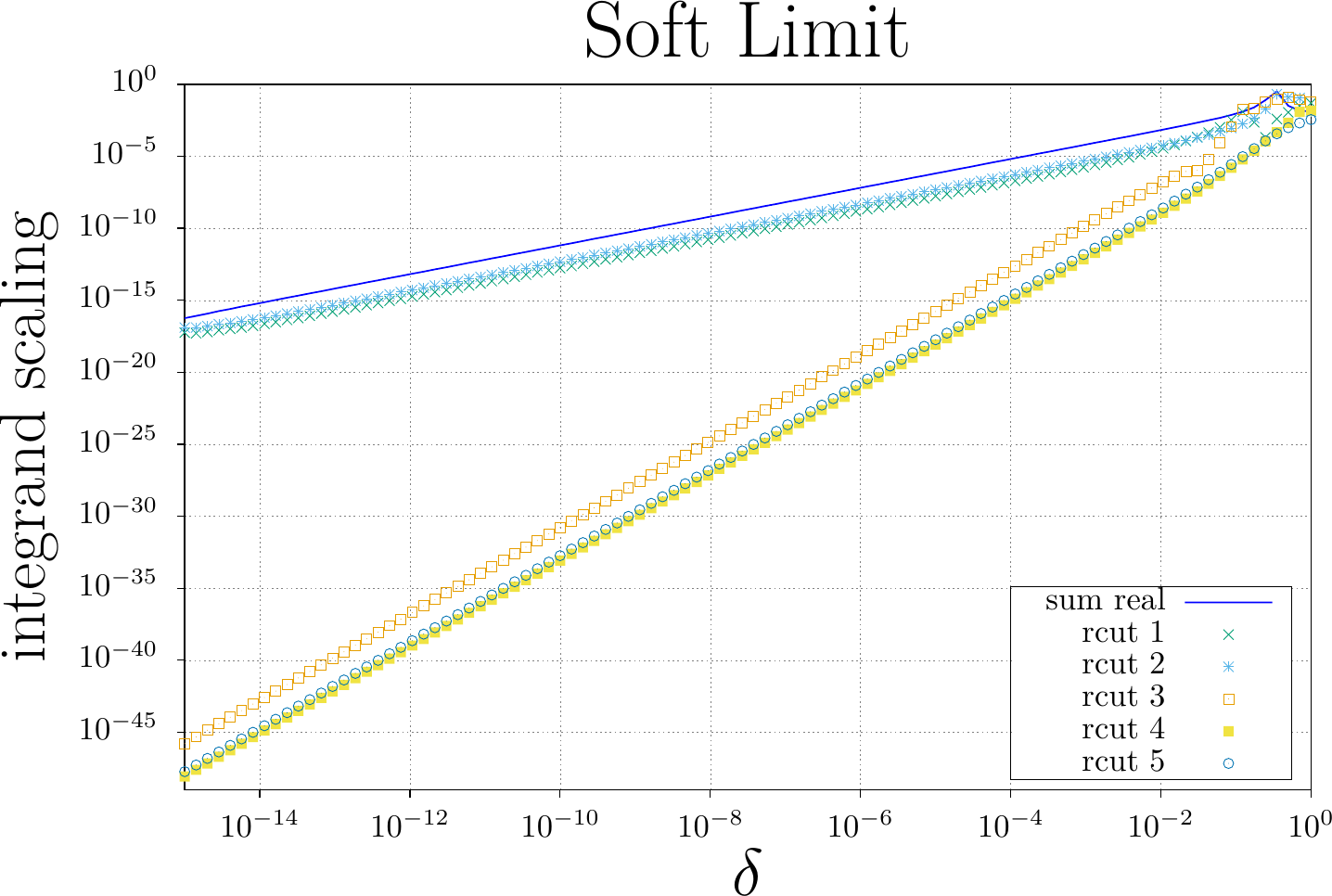}\label{plot:dd3a_soft}}
 \subfloat[]{\includegraphics[scale=.45]{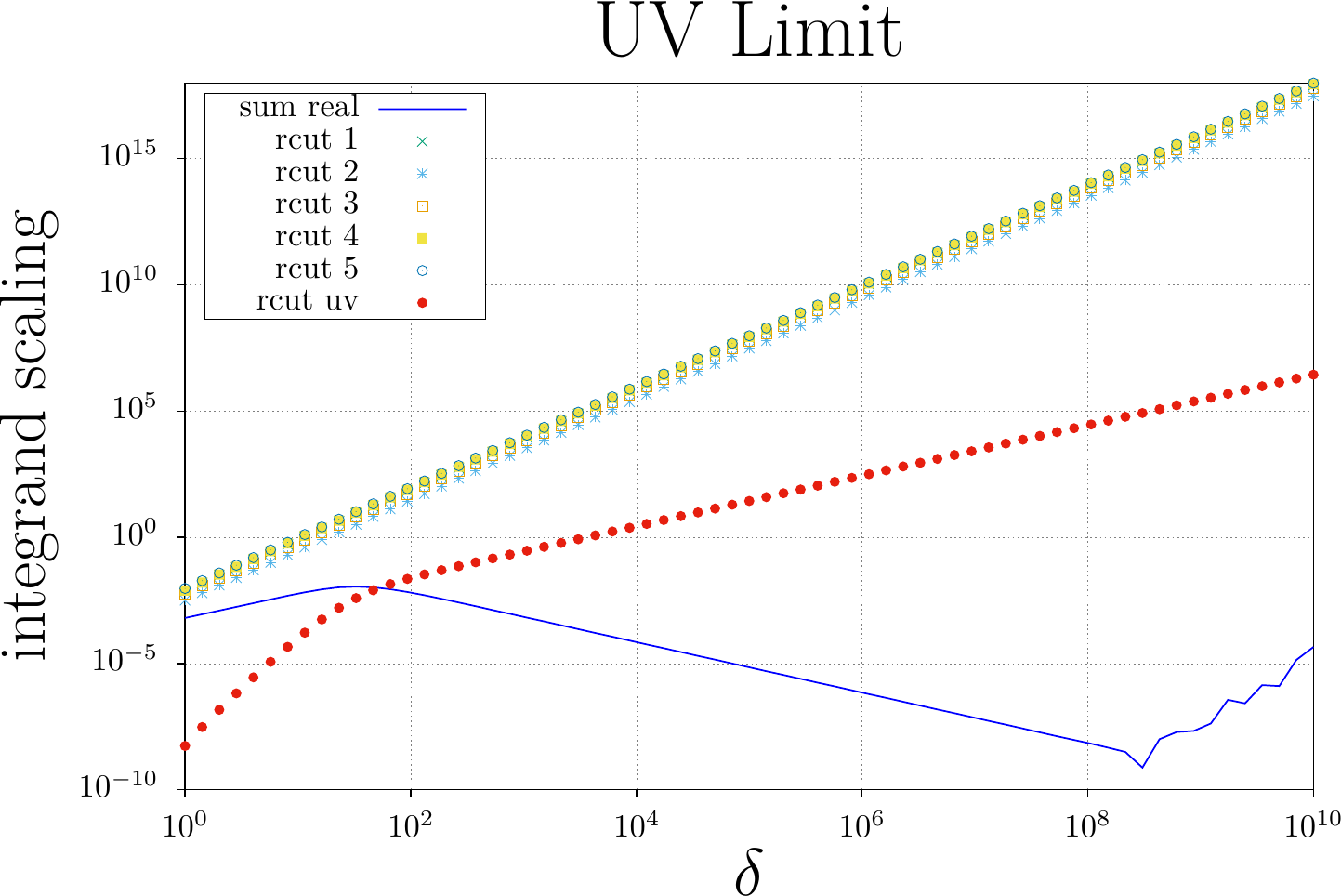}\label{plot:dd3a_uv}}
 \caption{Behaviour of the $q\bar{q}\rightarrow 3 \gamma$ in the different regulated limits. When the various limits are approached linearly in $\delta$ the plots (a--b) show a scaling as $\delta^{\frac{1}{2}}$ whereas (c) goes like $\delta^{1}$ and (d) as $\delta^{-1}$.}%
 \label{fig:1loop_ddAAAlimits}%
\end{figure}

\begin{figure}[p]
\centering
\includegraphics[scale=.75]{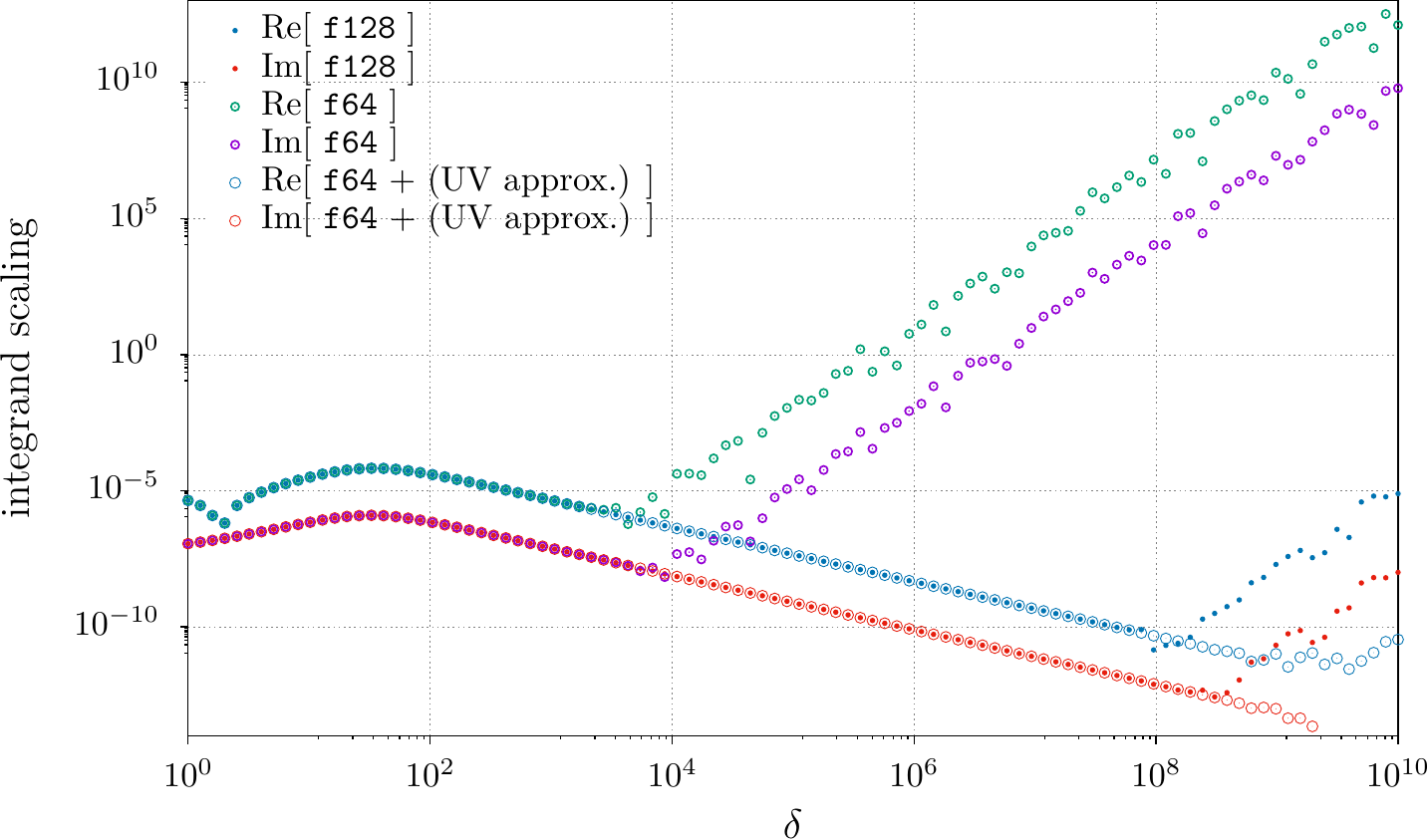}
\caption{Behaviour of numerical instability in the UV due to imprecise cancellations between large numbers from each each dual integrand. The loop momentum $k$ is rescaled by a factor $\delta$ and the real and imaginary part of the amplitude are presented with different precision (double and quadruple) and by expanding the expression around the UV limit as an approximation (see sect.~\ref{sec:uv_behaviour}).}
\label{fig:uv_behavior}
\end{figure}

Despite the use of quadruple precision (\texttt{f128}) to rescue some unstable evaluation of the UV region, we see that the cancellations between dual integrands are broken around $\delta>10^8$ due to numerical instabilities.
In fig.~\ref{fig:uv_behavior} we show how these instabilities spoil the final result in the case of double precision (\texttt{f64}) with and without the use of the approximating function discussed in sect.~\ref{sec:uv_behaviour}.
In the latter case it is possible to push the instability in the far UV and reproduce the behaviour of the quadruple precision evaluation. Where the transition between the approximated function and the all-order amplitude expression occurs, one has has to ensure that the deformation goes to zero, since this region is not analytic.
In both fig.~\ref{fig:1loop_ddAAAlimits} and fig.~\ref{fig:uv_behavior} the rescaled loop momentum is taken to be real and of the same order as $s_{12}$.

\clearpage

\section{Optimisation}
\label{sec:optimisation}

In this section we present various optimisations that we have developed to improve the convergence of our numerical framework.

\subsection{Lorentz invariance}
\label{sec:lorentz_invariance}

The following two subsections are aimed at showcasing the wide range of simplifications made possible by leveraging Lorentz symmetry. Specifically, Lorentz symmetry can be used to both drastically simplify the E-surface overlap structure and eliminate fictitious accidental pinched configurations that may appear for specific external kinematics as a result of competing constraints on the deformation. 

Contrary to symmetry under the (spatial) $SO(3)$ subgroup of the Lorentz group, invariance under boosts is not manifest in the LTD framework. Indeed, Lorentz boosts cause significant changes in the singular structure of the integrand and result in E-surfaces being rescaled and shifted relative to each other: the major axis length of an E-surface, being a linear combination of the energies of the external particles, is not a Lorentz invariant, nor is the distance between any pair of focal points, being a linear combination of the three momentum of the external particles. Conversely, some quantities are Lorentz invariant in the LTD framework: the number of E-surfaces, their existence condition, and some specific features of the overlap structure including, for example, the property of two E-surfaces sharing a focal point.

\subsubsection{Simplified deformation contour for 2-point multi-loop integrals}

A first use-case of the implict realisation of Lorentz invariance in LTD is found in the construction of a surprisingly simple integration contour applicable to any two-point function.
Since the original integral is Lorentz invariant, the single independent external momentum of a two point function can always be boosted in its rest frame. It follows that the spatial momentum shifts in all propagator momenta read
\begin{equation}
\label{eq:all_p_zeros}
    \vec{p}_j = (0, 0, 0), \ \forall j \in \mathbf{e}\,,
\end{equation}
where we recall that $\mathbf{e}$ identifies the list of edges of the loop graph.
Equivalently, we can write $\vec{q}_i(\vec{k})=\vec{Q}_i(\vec{k})$.
A Lorentz boost thus allows to decouple components of $\vec{k}$ from the spatial part of the external momentum. \par

This feature allows for a simpler deformation, characterised by the parameter $\lambda\in (0,1)$, as
\begin{equation}
\vec{k}_j \rightarrow \vec{k}_j (1-{\rm{i}}\lambda).
\end{equation}
This deformation casts squared energies in a particularly simple form, 
\begin{equation}\label{eq:energy_rotated}
    \vec{q}_i(\vec{k}-\mathrm{i}\lambda\vec{k})^2+m_i^2=\vec{q}_i(\vec{k})^2(1-{\rm{i}}\lambda)^2 + m_i^2
\end{equation}
from which follows that because $\lambda<1$, the stronger continuity constraint eq.~\eqref{eq:branchcut_scaling} is always satisfied, since the real part of eq.~\eqref{eq:energy_rotated} is positive and that all focal points coincide with the origin thanks to eq.~\eqref{eq:all_p_zeros}. 
And because $\lambda>0$, the imaginary part of eq.~\eqref{eq:energy_rotated} is positive as well.
It follows that the causal constraints, imposed by LTD, are satisfied everywhere (except at the origin where the deformation scales to zero), since the deformation is guaranteed to never reach the forbidden areas presented in eq.~\eqref{pole-condition2}.
Therefore, the simple deformation vector field $\vec{\kappa}=\mathrm{i}\lambda \vec{k}$ with $\lambda\in (0,1)$, is correct for any two-point function, independently of the number of loops and internal masses.\par

We tested this deformation on a six-loop two-point ladder integral with two sets of kinematic configurations given by $p^2=1$ and masses $m_j^2=0, \ \forall j\in\mathbf{e}$, called \texttt{K}, and $p^2=1$, $m_j^2=0.1 \ \forall j \in \mathbf{e}$ called \texttt{K}$^\star$. We compared the $m^2=0$ numerical result against its analytical counterpart and verified that the procedure is correct. The results are reported in the following table, together with information about the number of dual integrands \texttt{N}$_\mathrm{C}$, the number of existing E-surfaces \texttt{N}$_\mathrm{E}$ and number of {\sc\small VEGAS} Monte-Carlo samples considered \texttt{N}$_\mathrm{p}$:

\vspace{0.5cm}
{
\centering
\resizebox{\columnwidth}{!}{%
\texttt{%
\begin{tabular}{@{}llrrrcllc@{}}
\hline
Topology & Kin. & $\mathtt{N_{\text{C}}}$ & $\mathtt{N_{\text{E}}}$ & $\mathtt{N_{\text{p}} \ [10^9]}$  & Phase & Exp. & \multicolumn{1}{c}{Reference} & Numerical LTD  \\
\hline
\multirow{4}{*}{%
\begin{tabular}{@{}c@{}}
        \begin{tikzpicture}
            \begin{feynman}

            \tikzfeynmanset{every vertex={dot,minimum size=1mm}}

            \tikzfeynmanset{every vertex={empty dot,minimum size=0mm}}
             \vertex (a1);
             
            \vertex[right=0.16666cm of a1] (a2);
            \vertex[right=0.16666cm of a2] (a3);
            \vertex[right=0.16666cm of a3] (a4);
            \vertex[right=0.16666cm of a4] (a5);
            \vertex[right=0.16666cm of a5] (a51);            
            \vertex[right=0.16666cm of a51] (a6);
            
            \vertex[below=1cm of a6] (a7);
            
            \vertex[left=0.16666cm of a7] (a81);	    
            \vertex[left=0.16666cm of a81] (a8);
	    \vertex[left=0.16666cm of a8] (a9);
	    \vertex[left=0.16666cm of a9] (a10);
	    \vertex[left=0.16666cm of a10] (a11);

	    \vertex[left=0.16666cm of a11] (a12);
	    
	    \vertex[right=0.15cm of a6] (d1);
	    \vertex[right=0.15cm of a7] (d2);
	    \vertex[left=0.15cm of a1] (e3);
	    \vertex[left=0.15cm of a12] (e4);

            \tikzfeynmanset{every vertex={dot,minimum size=0.8mm}}

            \vertex[below=0.5cm of a1] (e1);
            
            \vertex[above=0.5cm of a7] (f1);
      
             \tikzfeynmanset{every vertex={empty dot,minimum size=0mm}}           
            
            \vertex[left=0.15cm of e1] (ee1);
            \vertex[right=0.15cm of f1] (ff1);
            
                \diagram*[large]{	
                (e1)--(a2)--(a51)--(f1)--(a81)--(a11)--(e1),
                (a2)--(a11),
                (a3)--(a10),
                (a4)--(a9),
                (a5)--(a8),
                (a81)--(a51),
                
                (ee1)--(e1),
                (ff1)--(f1),
                
                }; 
            \end{feynman}
    \end{tikzpicture} \\
6L2P\end{tabular}}
& \multirow{2}{*}{K}& \multirow{2}{*}{1560}& \multirow{2}{*}{36}& \multirow{2}{*}{0.1\phantom{.0}}& Re& \multirow{2}{*}{-11}& \texttt{-5.9616733} \multirow{2}{*}{\cite{Usyukina:1992jd}} & \texttt{-5.945
~+/-~0.029}\\
& & & & & Im& & \texttt{ 0}& \texttt{-0.009  ~+/-~0.031}\\
\cline{2-9}
& \multirow{2}{*}{K$^\star$}& \multirow{2}{*}{1560}& \multirow{2}{*}{16}& \multirow{2}{*}{0.1\phantom{.0}} & Re& \multirow{2}{*}{-10}& \multicolumn{1}{c}{n/a} & \texttt{-2.9924~+/-~0.0011} \ \\
& & & & & Im& & \multicolumn{1}{c}{n/a} & \texttt{3.9424~+/-~0.0011}\\
\hline
\end{tabular}%
}%
}%
}

\vspace{0.5cm}
The same technique of adding a small imaginary part to the components of the loop momenta corresponding to zero components of all the external momenta can also be considered for the three-(four-)point function. However, in these cases there are only two(one) component(s) that can be set to zero through a boost. The possibility of integrating easily along loop momentum dimensions by adding a small imaginary part to a subset of the components of the loop momenta is the manifestation of a property of two, three and four-point functions already noted in ref.~\cite{Srednyak_2013}.

\subsubsection{Example of overlap structure simplification for a 3-point 2-loop integral}

In general, Lorentz boosts can be used to greatly simplify the overlap structure. For example, we find that the $1 \rightarrow 2$ kinematics of a two-loop ladder diagram with massless propagators (considered here for simplicity), can be written in the following form when boosted in the rest frame of the $p_2+p_3$ system:
\begin{align}
\begin{split}
    p_1 &= \left(m_1^2, 0, 0, 0\right), \qquad  \\
    p_2 &= \left(\sqrt{\omega^2+m_+^2}, 0, 0, \omega \right), \qquad  \\
    p_3 &= \left(\sqrt{\omega^2+m_-^2}, 0, 0, -\omega \right) \,,
\end{split}
\end{align}
with momentum conservation conditions yet to be applied to the energy components. Since in this case any E-surface features at most one focal point with a non-vanishing affine term $\vec{p}_j$, the origin $\vec{k}_i=\vec{0}$ lies within all E-surfaces. Indeed, all E-surfaces of this particular loop integral considered are
\begin{align}
\begin{split}
&    \eta_1(\vec{k})=2\lVert \vec{k}_1\rVert-m_1^2\\
&    \eta_2(\vec{k})=\lVert \vec{k}_1\rVert+\lVert \vec{k}_1+\vec{k}_2\rVert+\lVert \vec{k}_2\rVert-m_1^2\\
&    \eta^{\pm}_3(\vec{k})=\lVert \vec{k}_1\rVert+\lVert \vec{k}_1+\vec{k}_2\rVert+\lVert
    \vec{k}_2+\omega\hat e_z\rVert-\sqrt{\omega^2+m_\pm^2} \\
&     \eta^{\pm}_4(\vec{k})=\lVert \vec{k}_2\rVert+\lVert \vec{k}_2+\omega\hat e_z\rVert-\sqrt{\omega^2+m_\pm^2}\\
&    \eta_5(\vec{k})=2\lVert \vec{k}_2\rVert-m_1^2 \,,
\end{split}
\end{align}
which are all negative when evaluated at $\vec{k}_i=\vec{0}$, indicating that the origin is indeed in the interior of all exisitng E-surfaces. Similar arguments can be used to show that in a physical $2 \rightarrow 2$ process featuring $n$ existing E-surfaces, at least $n-1$ of them must allow for a point in the interior of all of them. 

The boost parameters can themselves be viewed as hyperparameters subject to optimisation and although it is beneficial to boost $2\rightarrow N$ kinematics in the rest frame of the collision, a systematic procedure that maximally optimises the choice of Lorentz frame is still missing.

\subsubsection{Pseudo-pinches}

Pseudo-pinches are singular surfaces at which competing causal or continuity constraints impose the deformation to be zero, although these configurations are non-existent in another frame of reference.
They can be classified as follows:
\begin{enumerate}
    \item Singular subspaces
    \begin{equation}
        \label{eq:singular_suspaces}
        \vec{q}_j(\vec{k})=0, \ \forall j\in \mathbf{c}, \ \mathbf{c} \subseteq \mathbf{b}
    \end{equation}
    with $|\mathbf{c}|$ fixed loop variables and $n-|\mathbf{c}|$ unconstrained loop variables. When  \emph{all} loop momenta configurations $\vec{k}^{(\mathbf{c})}$ satisfying the subspace constraints of eq.~\eqref{eq:singular_suspaces} happen to also lie on \emph{one} particular E-surface $\eta$ (so $\eta(\vec{k}^{(\mathbf{c})})=0$), then no deformation will be allowed on that surface because of the continuity constraint of eq.~\eqref{sec:continuity_constraint}. This situation is accidental as it only happens for particular kinematic configurations and, more importantly, for a particular choice of Lorentz frame. At one loop, this situation corresponds to a focal point being located exactly on an E-surface. 
    \item Intersections of two or more E-surfaces $\eta_1,\ldots,\eta_n$ at a point $\vec{k}$ such that $\exists \eta_i$ with $\vec{\nabla}\eta_i=-\sum_{j\neq i}\alpha_j\vec{\nabla} \eta_j$ and $\alpha_j\ge 0$. This typically happens when two E-surfaces are tangent. We stress here again that, in general, the normal $\vec{\nabla}\eta_i$ to an E-surface $\eta_i$ is a $(3n)$-dimensional vector.
\end{enumerate}
We now illustrate these two different types of accidental pseudo-pinches at one loop.
\paragraph{Case 1}
Let a focus be located exactly on an E-surface. Imposing that the contour does not cross branch cuts of on-shell energies of massless internal particles (using our stronger version of the continuity constraint),
\begin{equation}
\vec{q}_i(\vec{k})^2-\vec{\kappa}^2\ge 0,
\end{equation}
at the point $\vec{q}_i(\vec{k}^*)=\vec{k}^*+\vec{p}_j=0$ implies that $\vec{\kappa}^*=0$. However, since the point is located on a singular E-surface, $\vec{k}^*\in \partial \eta$, a non-zero deformation is required. In this case, the continuity constraint conflicts with the causal constraint. It can be argued that our continuity constraint is stronger than what is minimally required, but even weaker implementations must impose that $\vec{\kappa}^*=0$ in some region containing the focal point.

\paragraph{Case 2} Now let two E-surfaces be tangent. Then, two causal constraints conflict at a point: the normal vectors to the two E-surfaces at the tangent points are opposite in direction, and thus no vector exists having strictly positive projection on both of them.
\vspace{0.3cm}

Both cases are problematic from a conceptual point of view, because they can correspond to kinematic configurations where the deformation breaks down. However, as mentioned earlier, the existence of these cases is accidental and specific to the chosen reference frame for the external kinematics. In both cases, there is an infinite number of infinitesimal Lorentz boosts such that in the boosted kinematics no focal point coincides with any E-surface and no two E-surface are tangent.

This is especially clear in the case of \emph{causally connected} focal points. In order to understand this notion, one can turn to the one-loop example of an E-surface $\eta$ on which lies a focal point $f$ (necessarily, the focal point $f$ cannot coincide with one of the focal points of $\eta$). Now let $\vec{q}_i(\vec{k}_f)=\vec{k}_f+\vec{p}_i=0$ be the equation defining the focal point and let $f'$ be a focal point of the E-surface satisfying the equation $\vec{q}_j(\vec{k}_{f'})=\vec{k}_{f'}+\vec{p}_j=0$. Now consider a boost sending the four-momentum $p_i-p_j$ in its rest frame so that its only non-zero component is the time component. Obviously, this can only be done if $p_i-p_j$ is timelike in which case the two focal points correspond to four-dimensional spacetime coordinates that are causally connected. In this frame of reference, the focal points $f$ and $f'$ overlap and thus $f$ can no longer be located on the surface of the ellipsoid, thereby avoiding the accidental pseudo pinch situation.

Similarly, consider two tangent E-surfaces, and choose one focal point for each E-surface, denoted by $f$ and $f'$, such that their distance in four-dimensional spacetime is timelike. It is now always possible to choose a frame of reference in which the distance between the focal points is zero. In this frame the two E-surfaces share a focal point and thus cannot be tangent.

\subsection{Multi-channelling}
\label{sec:multichanneling}
Improving the numerical efficiency of the numerical integration amounts to finding techniques for reducing the variance of the integrand.
Sharp local enhancements of the integrand, and especially integrable singularities, induce a large variance and can significantly deteriorate the numerical integration. At best, such peaks make the Monte~Carlo (MC) integration converge slowly and at worst they yield an unstable central value, as well as an unreliable estimate of the MC error.

In general, adaptive importance sampling can adjust well to integrands with large variances, provided that their enhancement structure aligns with the integration variables.
However, when the Monte Carlo integrator underestimates the variance of the integrand in some regions of the integration space during the first iterations, it can incorrectly neglect these regions in further iterations.
In such cases, the estimate of the integral will be unreliable, even though the error suggests otherwise.
Even though increasing the number of sampling points in the first iterations can help mitigate this problem, it slows down the integration and reduces the predictive power of the numerical integration.
It is therefore best to first pre-process the integrand so as to remove its sharp enhancements, which is possible when their location and approximate functional form is known.
In this section, we show how this improvement can be systematically implemented for the LTD expression, using a technique known as \emph{multi-channeling} which is commonly used for improving numerical integration in various contexts.

We can write the integrand stemming from the $n$-loop LTD expression as
\begin{align}
    \mathcal{I}
    \equiv
    \left(\frac{-\mathrm{i}}{(2\pi)^3}\right)^n
    \sum_{\mathbf{b}\in\mathcal{B}}\Res_{\mathbf{b}}[f],
\end{align}
where each dual integrand $\Res_{\mathbf{b}}[f]$ features sharp peaks resulting from each propagator put on-shell. Each of these peaks is an integrable singularity when the corresponding propagator is massless. These enhancements for each residue have the following functional form:
\begin{align}
    \Res_{\mathbf{b}}[f]
    \propto
    \prod_{i\in\mathbf{b}} E^{-1}_i,
\end{align}
where $E_i=\sqrt{\vec{q}_i^{\,2}+m_i^2}$, with local extrema at $\vec{q}_i=0$ for $i \in \mathbf{e}$.
In order to take advantage of dual cancellations, i.e. the local cancellations of singularities on H-surfaces among summands of the LTD expression, the dual integrands have to be integrated \emph{together} using a \emph{unique} parameterisation.
We must therefore consider the complete integrand which features the following peak structure
\begin{align}
     \mathcal{I}
     \propto
     \sum_{\mathbf{b}\in\mathcal{B}} \prod_{i\in\mathbf{b}} E^{-1}_i.
\end{align}
In a multi-channeling approach, we seek to flatten these enhancements by first inserting the following expression of unity in the integrand:
\begin{align}
    1 = \frac{
        \sum_{\mathfrak{b}\in\mathcal{B}}\prod_{j\in\mathfrak{b}}E_j^{-1}}{
        \sum_{\mathbf{b}\in\mathcal{B}}\prod_{i\in\mathbf{b}}E_i^{-1}}
\end{align}
and then splitting up the sum in the numerator into $|\mathcal{B}|$ channels, thereby defining an integrand for each channel identified by a basis (or equivalently spanning tree) $\mathfrak{b}\in\mathcal{B}$, whose expression reads:
\begin{align}
    \mathcal{C}_\mathfrak{b}
    \equiv
    \frac{\prod_{j \in \mathfrak{b}} E_j^{-1}}{\sum_{\mathbf{b}\in \mathcal{B}}\prod_{i \in \mathbf{b}} E_i^{-1}}
    \mathcal{I}
    =
    \prod_{j \in \mathfrak{b}} E_j^{-1} \underbrace{\frac{\mathcal{I}}{\sum_{\mathbf{b}\in \mathcal{B}}\prod_{i \in \mathbf{b}} E_i^{-1}}}_{\textrm{no strong enhancement}}
    \propto
    \prod_{i\in\mathfrak{b}} E^{-1}_i \,.
\end{align}
We observe that each channel still features peaks, but only those specitic to $\mathfrak{b}$. This opens the possibility of choosing a \emph{different} parametrisation for each channel, selected so that its Jacobian flattens its enhancement $\prod_{i\in\mathfrak{b}} E^{-1}_i$. We note that a similar multi-channeling approach was used in refs.~\cite{Soper:1998ye, Becker:2012aqa}.
Thanks to the continuity constraint discussed in sect.~\ref{sec:continuity_constraint}, the denominator of the multi-channelling factor does not introduce new integrable singularities when computed with our choice of contour deformation.
More specifically, the integration measure from the spherical parametrisation of the loop momenta in the basis $\mathfrak{b}$ reads\footnote{The change of loop momentum basis always yields a Jacobian of one when keeping boundaries fixed.}:
\begin{align}
\label{eq:spherical_multichannel_parametrisation}
    \left[\mathrm{d}^{3}\vec{k}\right]
    =
    \left[||\vec{q}||^2\mathrm{d}||\vec{q}||\mathrm{d}^2\Omega\right]_\mathfrak{b},
\end{align}
where we introduced the shorthand notation
\begin{align}
    \left[\mathrm{d}^{3}\vec{k}\right]
    \equiv \prod_{j=1}^n \mathrm{d}^3\vec{k}_j,
    \quad
    \left[\mathrm{d}|\vec{q}|\mathrm{d}^2\Omega\right]_\mathfrak{b}
    \equiv \prod_{j\in\mathfrak{b}} ||\vec{q}_j||^2 \mathrm{d}||\vec{q}_j|| \mathrm{d}^2\Omega_j.
\end{align}
We can now choose to integrate each channel $\mathcal{C}_\mathfrak{b}$ separately\footnote{In practice, one can also opt to evaluate each channel successively for each sampling point considered by the integrator. This has the advantage of retaining potential local cancellation across channels but also complicated the overall structure of the integrand that the integrator must adapt to.} and use for each the specific parametrisation of eq.~\eqref{eq:spherical_multichannel_parametrisation}. At one loop, these different parametrisations only differ by a shift of the origin whereas beyond one loop, they also amount to a change of basis in which the loop momenta are expressed.
The resulting integral for each channel then reads:
\begin{align}
\int \left[\mathrm{d}^{3}\vec{k}\right] \mathcal{C}_\mathfrak{b} =
\int \left(\frac{ 
\left[||\vec{q}||^2\mathrm{d}||\vec{q}||\mathrm{d}^2\Omega\right]_\mathfrak{b} 
}{
\prod_{j\in\mathfrak{b}}E_j
}\right)
\left(\frac{\mathcal{I}}{\sum_{\mathbf{b}\in \mathcal{B}}\prod_{i \in \mathbf{b}} E_i^{-1}}\right) \,,
\end{align}
where each of the two factors building the integrand is now free from integrable singularities (or strong enhancement in the case of massive propagators) coming for the cut propagator.
The original integral is then computed as the sum of $|\mathcal{B}|$ channels
\begin{align}
    I = 
    \int
    \left[\mathrm{d}^{3}\vec{k}\right]
    \mathcal{I}
    =
    \sum_{\mathfrak{b}\in\mathcal{B}}
    \int
    \left[||\vec{q}||^2\mathrm{d}||\vec{q}||\mathrm{d}^2\Omega\right]_\mathfrak{b}
    \mathcal{C}_\mathfrak{b} \,.
\end{align}

\begin{figure}[h]
    \centering
    \includegraphics[width=0.8\textwidth]{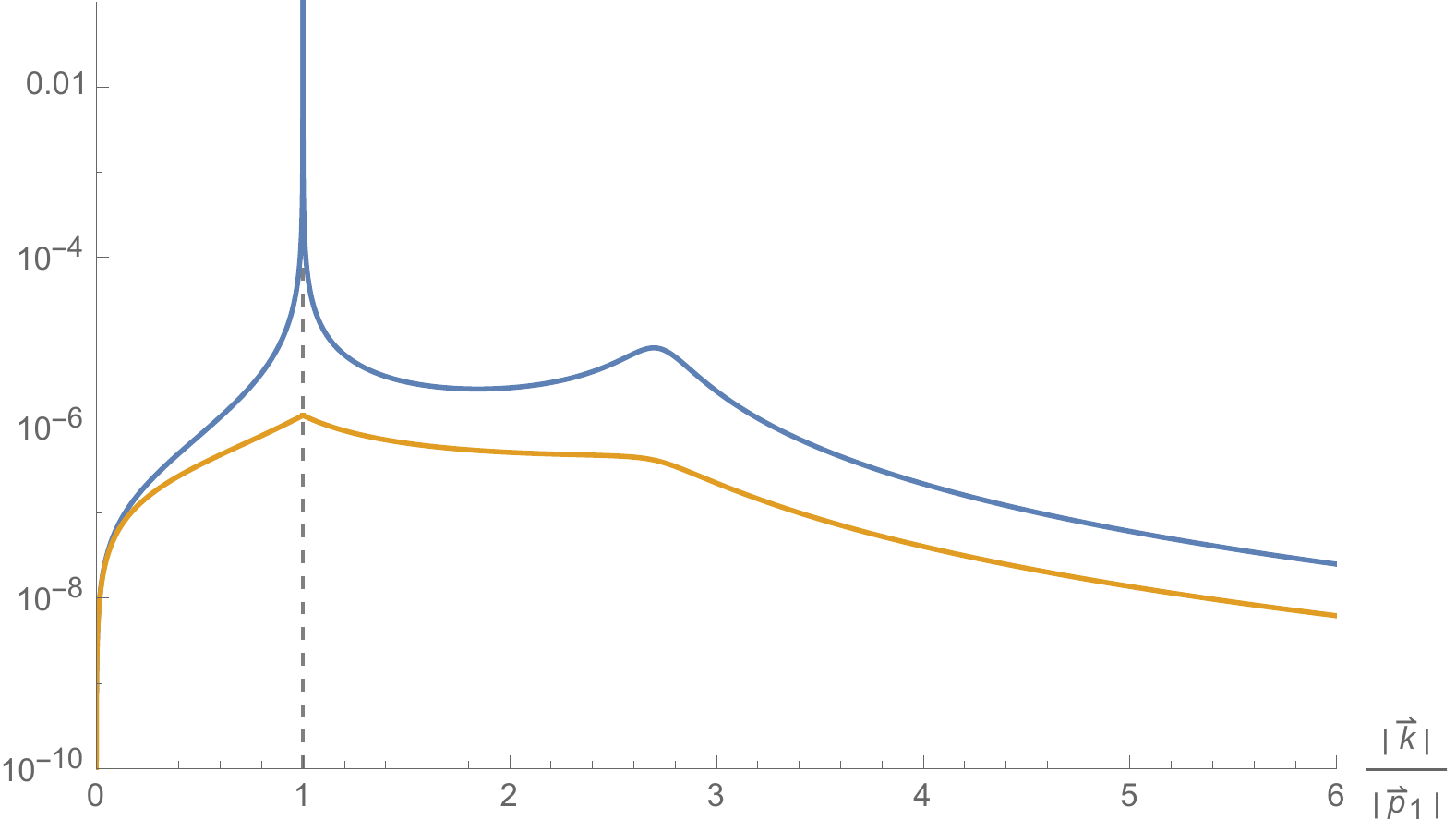}
    \caption{Multi-channelling for a triangle integral with massless propagators with momenta $q_i = k + p_i$, $i \in \mathbf{e}=\{1,2,3\}$, such that $\vec{p}_3=\vec{0}$:
    The LTD integrand $|\vec{k}|^2 \mathcal{I}$ (blue) and the channel $|\vec{k}|^2 \mathcal{C}_{\{3\}}$ (orange) in spherical coordinates along the direction $-\vec{p}_1$.
    Since the triangle has three dual integrands, the LTD integrand $\mathcal{I}$ has three integrable singularites, one for each energy $E_i=0$, $i \in \mathbf{e}$.
    For both integrands, the singularity at $E_3=0$, i.e. when $\vec{k}=\vec{0}$, vanishes when parameterised in spherical coordinates centered at $\vec{k}=\vec{0}$ because of the integration measure.
    The line along $-\vec{p}_1$ goes directly through the singularity at $E_1=0$, i.e. when $\vec{k}=-\vec{p}_1$ and past the one at $E_2=0$ (small bump only since the direction used for this plot is not $\vec{p}_2$ but $\vec{p}_1$) of the LTD integrand.
    In the channel $\mathcal{C}_{\{3\}}$ these two enhancements are flattened and become non-vanishing constants thanks to the multi-channel factor.
    We observe that at $\vec{k}=-\vec{p}_1$ the channel is not differentiable (as well as at $\vec{k}=-\vec{p}_2$).
    }
    \label{fig:evaluation_multi_channel_annotated}
\end{figure}

The effects of multi-channeling are shown in fig.~\ref{fig:evaluation_multi_channel_annotated}, where the peak due to the crossing a focal point is removed.

We note that this multi-channeling approach can be further developed by considering additional channels related to other enhancements coming from E-surfaces and/or infrared limits for example. We leave this investigation to future work.

\section{Numerical implementation}
\label{sec:numerical_implementation}

In this section we discuss various details of our numerical implementation, such as the most challenging aspects associated to the construction of the deformation contour, the evaluation of the Jacobian and consistency checks that are essential for verifying the correctness of the integration contour and guaranteeing the stability of the evaluation of the integrand.

\subsection{Source determination}
\label{sec:source_determination}
Determining the maximal overlap structure requires testing whether there is a point in the interior of a given set of E-surfaces. This problem is convex and, in particular, can be written as a second-order cone program (SOCP). We have used the convex constraint problem rewriter {\sc\small cvxpy}~\cite{cvxpy_rewriting} with the {\sc\small ecos} solver~\cite{bib:Domahidi2013ecos} as a backend to construct a program that ascertains whether a given set of E-surfaces overlap.

Given the aforementioned program, determining the maximal overlap structure $\mathcal{O}^{({\rm max})}$ of eq.~\eqref{eq:omax} is still an NP-hard problem, as the set of possible overlap configurations is exponential in the number of E-surfaces and any algorithm devoted to the determination of $\mathcal{O}^{({\rm max})}$ will have a worse-case complexity that renders it prohibitively slow. In practice however, the class of problems of interest generally features a limited amount of overlapping regions which are shared by many E-surfaces. Indeed, many E-surfaces share one or more focal points, and thus naturally have the focus as a shared interior point. As a consequence of these facts, the algorithm should be constructed so as to take advantage of this heuristic my exploring solution in a top-down order; that is starting with the assumption that all E-surfaces overlap. If all E-surfaces are not in one overlapping set, one E-surface is removed in all possible ways and the test is performed again. Once an overlap is found involving $N$ particular E-surfaces, then the $2^N-1$ subsets of this set never need to be tested again. In order to prevent a combinatorial blow-up, a list of all possible pair-wise intersecting E-surfaces is constructed and used to filter many options when constructing viable subsets. This additional improvement to the heuristic was key in rendering our implementation fast enough for problems with more than 30 E-surfaces, as generating all $2^{30}$ options is too slow.
In practice, the refined algorithm takes only a few seconds to find the solution in the majority of cases. It therefore yields negligible overhead in comparison to time spent in the numerical integration. We note however that for cases involving or more that 40 E-surfaces, it may happen that when our heuristics are not well satisfied, our algorithm cannot determine the maximal overlap structure within \emph{any} reasonable amount of time, as it happened in the case of the loop integral~\ref{tab:Kseries_2loopB}.{\tt 2L8P.K1$^*$} for which we could then not show results.

Once the maximal overlap structure is determined, one must find a point inside each overlap with the extra property to be \emph{optimal} from a numerical convergence point of view. This optimality condition can loosely be approximated by requiring the point to be as far as as possible from all the E-surface defining and enclosing the overlapping volume. The resulting set of point constructed in this manner will serve as the set of deformation sources. The furthest away a source $\vec{s}$ is from all surfaces in the overlap set, the less tangential the deformation $\vec{k}-\vec{s}$ will be when evaluated on the surfaces themselves. For higher-loop cases, the source location is possibly subject to extra requirements due to the continuity constraints within a particular subspace given in eq.~\eqref{source_subspace}.

To \emph{approximate} the optimal centre of the overlap region, which is related to the Chebyschev centre of a convex region, one can solve the convex constrained optimisation problem of maximising the radius $r$ under the constraints that the points $\vec{s} \pm r \hat e_i^{(j)}$ all lie inside all 
E-surfaces $\eta \in F$ for every Cartesian direction in $3n$ dimensions, $\{\{\hat e^{(j)}_i\}_{i=1}^3\}_{j=1}^n$, that is:
\begin{equation*}\label{opt_prob_overlap}
\begin{aligned}
& \text{maximize}
& & r \\
& \text{subject to}
& & \eta\left(\vec{k} \pm r \hat{e}^{(j)}_i\right) \leq 0, \ \forall i=1,2,3 \ \forall j=1,\dots,n \ \forall \eta\in F\\
\end{aligned}
\end{equation*}

Imposing the extra subspace constraints of eq.~\eqref{source_subspace} is most conveniently done by performing a basis change. For example, for given linear constraints $\vec{k}_1=\vec{p}_1$ and $\vec{k}_1+\vec{k}_2=\vec{p}_2$ on vectors $(\vec{k}_1, \vec{k}_2, \vec{k}_3)$, the following system of equations allows to identify the subspace satisfying the constraints and its orthogonal complement
\begin{align}
    \begin{pmatrix}
    1 & 0 & 0\\
    1 & 1 & 0\\
    \multicolumn{3}{c}{\ker(C)}
    \end{pmatrix}
     \begin{pmatrix}
    \vec{k}_1\\
    \vec{k}_2 \\
    \vec{k}_3
    \end{pmatrix}= 
    \begin{pmatrix}
    -\vec{p}_1\\
    -\vec{p}_2\\
    \vec{s}_1
    \end{pmatrix} \,,
\end{align}
where $\ker(C)$ is the kernel of the constraints $C$, $(0, 0, 1)$ in this example. The inverse of the system presented above allows to rewrite the E-surfaces in terms of fixed momenta $\vec{p}_1$, $\vec{p}_2$ and the source variable $\vec{s}_1$.
In this particular subspace example, there remains only three degrees of freedom for setting the source, so that only \emph{three} canonical directions $e_i^{(j)}$ need to be considered when building the constraints on $\vec{s} \pm r \hat e_i^{(j)}$, whereas the original centre finding problem cast without change of basis would require all nine ($3n$).

\subsection{Parameterisation}
The numerical integrator \textsc{\small Cuba}~\cite{Hahn:2004fe} that we use to produce our results generates points in the unit hypercube $[0,1]^{3n}$. These points have to be transformed to $\mathbb{R}^{3n}$ where they then correspond to a particular real-valued sample configuration for the spatial part of the the loop momenta.
Our code provides options for Cartesian maps and spherical maps with hyperbolic and logarithmic scaling for the conformal mapping from $[0,1]$ to $(-\infty,\infty)$. For the results in this paper we used the following spherical and hyperbolic transformation that map each triplet of input variables $(u_1,u_2,u_3) \in [0,1]^3$ to a configuration of the spatial part of one loop momentum $\vec{k}$:
\begin{align}
\begin{split}
r &= b E_{\text{cm}} \frac{ u_1}{1-u_1}\\
\phi &= 2 \pi u_2\\
\theta &= \acos{(-1 + 2 u_3)}\\
\end{split}
&
\begin{split}
k_x &= r \sin{\theta} \cos{\phi}\\
k_y &= r \sin{\theta} \sin{\phi}\\
k_z &= r \cos{\theta}\\
\end{split}
\end{align}
\begin{align}
J = 4 \pi E_{\text{cm}} b\; r^2 \left(1 + \frac{r}{ E_{\text{cm}}b }\right)^2
\end{align}
where $E_{\text{cm}}$ is the centre-of-mass energy of the decay or scattering kinemtics, and
$b$ is a scaling parameter that regulates how much the integrator probes the ultraviolet region.  Our default value for $b$ is $1$.

\subsection{Deformation Jacobian}

The contour deformation $\vec{k} \rightarrow \vec{k} - i \lambda(\vec{k}) \vec{\kappa}$ is effectively parametrised by the real part of the loop-momenta. Determining the resuling Jacobian of this parametrisation analytically is difficult due to off-diagonal contributions in the Jacobian matrix from the generally complicated analytical expression of the deformation magnitude $\lambda(\vec{k})$. In order to bypass this inconvenience, the \emph{exact} Jacobian is calculated numerically using automatic differentiation. This technique is commonly used in machine learning algorithms, such as neural networks. 
Performing the computation with \emph{dual numbers}
\begin{align}
    \begin{pmatrix}
    k_{i_x}\\k_{i_y}\\k_{i_z}
    \end{pmatrix}\rightarrow
    \begin{pmatrix}
    k_{i_x} + \epsilon_{k_{i_x}} \\k_{i_y} + \epsilon_{k_{i_y}} \\k_{i_z} + \epsilon_{k_{i_z}}
    \end{pmatrix} \,,
\end{align}
where the dual components $\epsilon_i$ are subject to the truncation rule $\epsilon_i \epsilon_j = 0$, yields the partial derivatives $\frac{\partial k'}{\partial k_{j_o}}$ as the coefficient of $\epsilon_{k_{j_o}}$.

In our Rust implementation, all routines are generic over floating-point-like types (such as a double-precision floating point number). Since a dual number behaves like a floating point number, the promotion of the arithmetics to dual number can be done transparently from the perspective of our core routines implementing the LTD logic.

\subsection{Consistency checks}
In order to assess the numerical stability of each evaluation, each Monte Carlo sample point is evaluated on numerically different but analytically equivalent integrands, taking advantage of the manifest invariance of the integrand under rotation of the spatial part of every momentum involved (for example, the external momenta, the loop momenta and the sources). If the evaluation of the LTD integrand of a spatially rotated configuration significantly differs (in terms of a sensible adimensional threshold) from the original one, the point is deemed unstable, and we attempt to rescue it by repeating the same exact procedure in quadruple precision. If an unstable point is then considered stable in quadruple precision by performing the same test, then the quadruple-precision result is returned to the integrator. Instead, if the point is still deemed numerically unstable, we set its weight to zero. In practice, even for the more challenging integrals, less than one sample point in a million is numerically unstable in quadruple precision. Furthermore, these exceptional unstable points are often deep in the ultraviolet region and evaluate to values far below the result of the integration and they can therefore safely be set to zero.
We note however that the implementation of a quadruple precision rescuing system was necessary for obtaining many of the results presented in this publication, especially for the computation of amplitude where the ultraviolet behaviour is more relevant (see sect.~\ref{sec:uv_behaviour}).

The correctness of the complex contour deformation is verified by sampling random points on E-surfaces and ensuring that the causality constraint is satisfied. Since finding a parametrisation for E-surfaces is difficult at higher loops, it is more effective to use a bisection strategy to sample points on the E-surfaces. The bisection strategy must be seeded by one point inside the E-surface and one outside. As E-surfaces are bounded, finding a point in the exterior of them is trivial and the most straightforward choice of point in the interior is any of the two focal points of the E-surface. The convexity of E-surfaces then ensures that a unique (correct) solution will be found by the bisection algorithm and that \emph{all} points of a given E-surface can be reached by our approach simply by varying the choice of exterior point. 

To verify the validity of the LTD expression, the occurrence of dual cancellations is explicitly verified. A similar bisection strategy is used to find a point on an H-surface. Then along the bisection line, the LTD integrand is evaluated on points iteratively closer to the H-surface. If the slope of the interpolation between these points is below a chosen adimensional threshold, the dual cancellation is considered successful.
The same setup is also used to verify if the local counterterms used to subtract IR-divergences have the correct scaling behaviour (see section~\ref{sec:amplitudes}).

\section{Results}
\label{sec:results}

The aim of our work is to provide a numerical loop integration technique based on Loop Tree duality which is both robust and generically applicable. It is therefore crucial to accompany the formal derivation of a valid deformation carried out in sect.~\ref{sec:contour_deformation} with illustrative applications that can demonstrate the correctness of the numerical method as well as its practical efficiency. This will be explored in sect.~\ref{sec:scalar_topologies_results}.
We present our numerical results obtained when applying our LTD formulation together with local subtraction counterterms to compute one-loop scalar topologies in sect.~\ref{sec:scalar_results}  and to compute amplitudes for the ordered production of photons from a fermion line in sect.~\ref{sec:amplitude_results} .

\subsection{Multi-loop finite integrals}
\label{sec:scalar_topologies_results}

To demonstrate the practical efficiency and correctness of the deformation, we explore in tables~\ref{tab:explore_1loop}-\ref{tab:Kseries_HigherLoop} a variety of kinematic configurations and many different scalar integral topologies featuring up to four loops (and up to six for cases not necessitating a contour deformation)\footnote{
The exhaustive details (incl. kinematics) necessary for reproducing the results of each integral presented in this section is given in the ancillary material. The integral normalisation matches that of eq.~\eqref{eq:integraldefinition}.}, yielding different combinations of number $N_E$ of unique singular threshold E-surfaces and number $N_S$ of necessary deformation sources. We also indicate the number of dual integrands in the LTD expression of eq.~\eqref{eq:LTDmaster} in the column labelled $N_c$; it corresponds to the number of spanning trees of the topology and also to the number of integration channel it would feature when adopting the multi-channeling procedure discussed in sect.~\ref{sec:multichanneling} (which we do not use in this section, unless otherwise stated).

We also report a shortened representation of the maximal overlap structure $\mathcal{O}^{(\rm max)}$ as a list $L_{\rm max}$ where each entry corresponds to the number of E-surfaces contributing to each maximally overlapping set $F$ contained in $\mathcal{O}^{(\rm max)}$.
We report the discrepancy of our numerical LTD result w.r.t the reference value, relative to each other ($\Delta [\%]$) and relative to the Monte-Carlo error ($\Delta [\sigma]$) reported by the implementation in {\sc\small Cuba}~\cite{Hahn:2004fe} of the {\sc\small Vegas}~\cite{Lepage:1980dq} integrator\footnote{
Similarly to the findings of ref.~\cite{Buchta:2015wna}, we also find significantly more accurate and precise results using the {\sc\small Cuhre} integrator at one-loop.
The results with this integrator are however significantly worse beyond one-loop. For the sake of simplifying the comparison of our results across loop counts, we only report results obtained with the {\sc\small Vegas} integrator.}.
Unless otherwise stated, we consider different fixed statistics of $3\cdot10^9$,  $1\cdot10^9$ and $0.5\cdot10^9$ Monte-Carlo sample points for each of the one-, two-, three- and higher-loop integrals computed\footnote{With typically {\tt n\_start} $\sim$ 1\% of {\tt n\_max} of and {\tt n\_increase} $\sim$ $0.1$ \% of {\tt n\_max} in {\tt Vegas}.}.
For some of the one-loop results (e.g.~\ref{tab:explore_1loop}.{\tt 1L5P.V} and~\ref{tab:explore_1loop}.{\tt 1L6P.IX}), the real part is accidentally small compared to the imaginary part and since the variance of the LTD integrand is of the same order for both phases, we find it relevant to also indicate in the last column of the results table the relative discrepancy of our LTD numerical result on the \emph{modulus} of the complex-valued benchmark result ($\Delta [\%] |\cdot |$).
The timing per PS point ${\rm t/p }$ is reported in microseconds, as measured on a single core of an Intel Xeon CPU E5-2650 v4 @ 2.20GHz CPU.
Throughout this section and unless otherwise mentioned, we keep the deformation hyperparameters fixed to their default values of $\epsilon_{\rm{th}}=0.3$ and $M=0.07$.
These defaults are typically different from what would be the values optimised for each kinematic configuration and/or topology tested, but in this exploratory work we refrained from systematically fine-tuning hyperparameters so as to prevent any bias in our results and be able to fairly showcase the robustness of our approach. However, we will later show two examples where the results from specific integrals could be significantly improved by adjusting the value of the hyperparameter $M$.
Finally, the reference result for all one-loop integrals presented in this section, as well as for the one-loop amplitude computed in sect.~\ref{sec:amplitude_results}, is obtained from the One-Loop Provider {\sc\small MadLoop}~\cite{Hirschi:2011pa,Alwall:2014hca}. {\sc\small MadLoop} uses the OPP~\cite{Ossola:2006us} or Laurent-series expansion~\cite{Mastrolia:2012bu} integrand-level reduction technique as implemented in {\sc\small CutTools}~\cite{Ossola:2007ax} and {\sc\small Ninja}~\cite{Peraro:2014cba,Hirschi:2016mdz}, together with {\sc\small OneLOop}~\cite{vanHameren:2010cp} for the evaluation of one-loop scalar master integrals (containing up to four external legs).

\begin{figure}[ht!]
    \centering
    \includegraphics[width=0.405\textwidth]{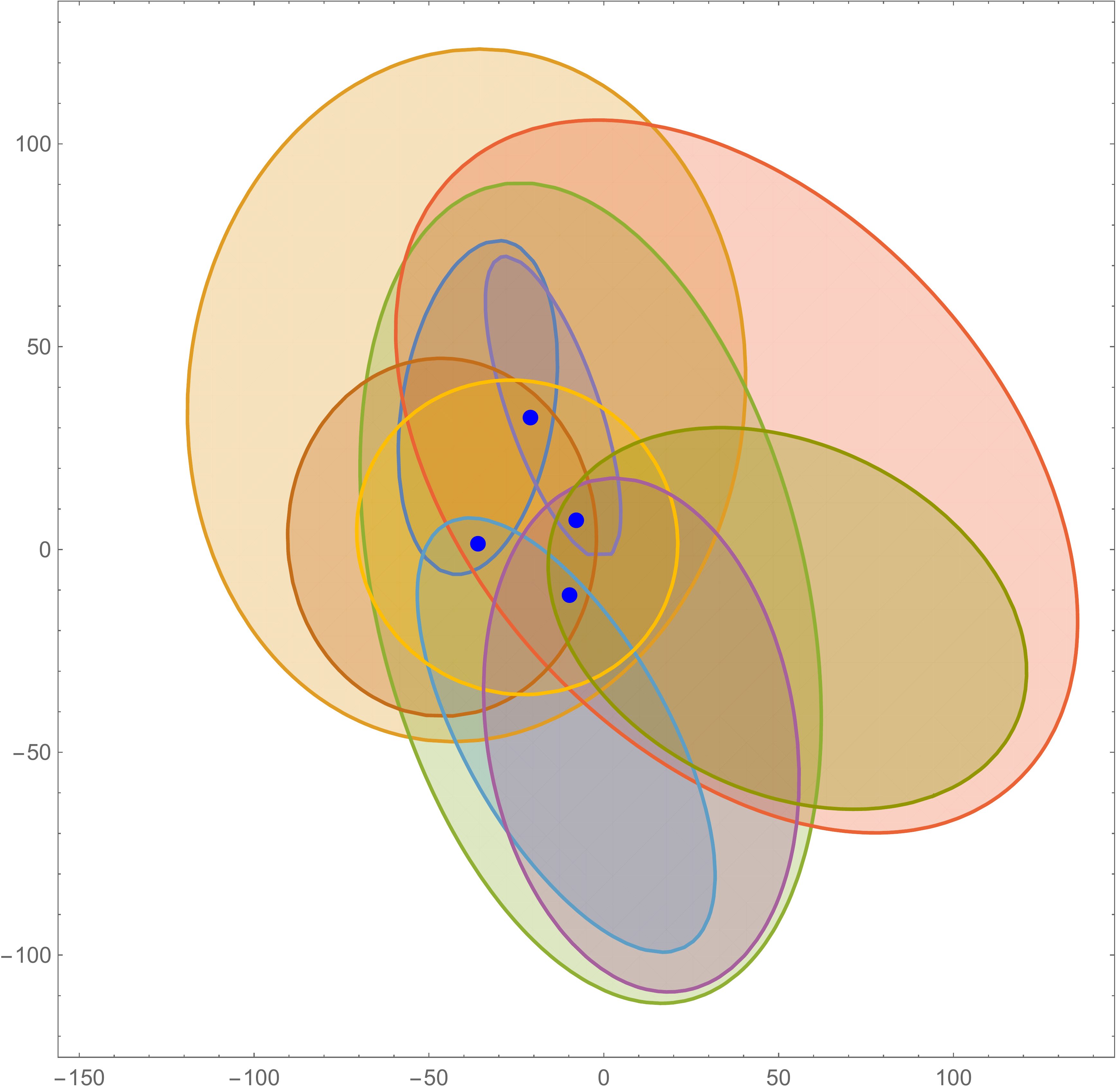}
    \includegraphics[width=0.4\textwidth]{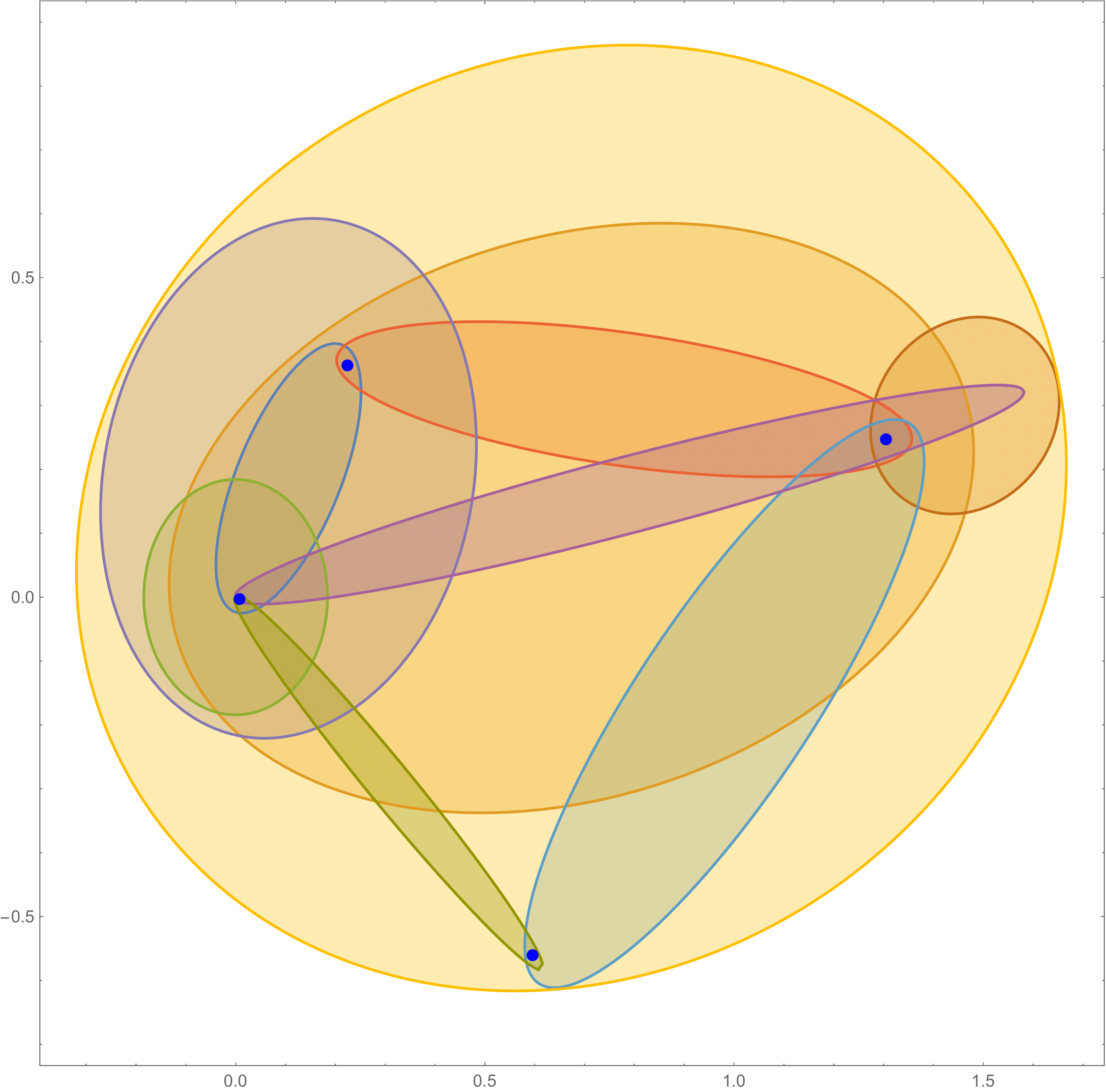}
    \caption{The singular E-surfaces from the two six-point one-loop integrals~\ref{tab:explore_1loop}.{\tt 1L6P.VII} (left) and~\ref{tab:explore_1loop}.{\tt 1L6P.VIII} (right) with different kinematics yielding drastically different maximally overlapping regions. In both cases our construction of the deformation is generated from the combination of four radial fields with sources indicated by blue dots. Additional support sources can potential improve on the worse convergence of integral~\ref{tab:explore_1loop}.{\tt 1L6P.VIII}.}
    \label{fig:hexagons}
\end{figure}

In table~\ref{tab:explore_1loop}, we present results for one-loop five- and six-point scalar integrals for hand-crafted kinematic configurations that correspond to many qualitatively different maximal overlapping situations.
We also include the result for the four-point one-loop integral~\ref{tab:explore_1loop}.{\texttt{Box4E}} which we used as an example throughout this work. The relatively good sub per-mil accuracy obtained for this integral may be surprising in regard to the complexity of the corresponding LTD integrand, depicted in figs.~\ref{fig:evaluation_annotated} and~\ref{fig:heatmap}.
Comparing the Monte-Carlo accuracy and precision obtained for all integrals of table~\ref{tab:explore_1loop}, we observe the general trend that the convergence mildly degrades with an increase in the number of deformation sources and the number of unique threshold E-surfaces.
However, the dominant factor appears to be the shape of the threshold surfaces, which become more elongated as the masses of the external momenta decreases or, more in general, when the hierarchy between the relevant scales in the scattering considered becomes more pronounced.
The integrals~\ref{tab:explore_1loop}.{\tt 1L6P.VII} and~\ref{tab:explore_1loop}.{\tt 1L6P.VIII} are a prime example of this observation as the Monte-Carlo accuracy of the latter integral is much worse despite featuring the same number of unique E-surfaces and deformation sources as the former.
Indeed, the external kinematics of integral~\ref{tab:explore_1loop}.{\tt 1L6P.VIII} yield E-surfaces of very elongated shapes, as hinted by the corresponding maximal overlap structure $L_{\rm max}=[3,5,6,7]$ where one deformation source involves only three out of the total of ten unique threshold E-surfaces.
Fig.~\ref{fig:hexagons} shows a rendering of the E-surfaces from both integrals~\ref{tab:explore_1loop}.{\tt 1L6P.VII} and~\ref{tab:explore_1loop}.{\tt 1L6P.VIII}, which clearly highlights their differences in shape and maximally overlapping regions.

Table~\ref{tab:explore_higherloopA} and table~\ref{tab:explore_higherloopB} show our reproduction of some benchmark multi-loop results from the literature.
The number of sources $N_s$ indicated in this multi-loop case refers to the \emph{total} number of sources, \emph{including} the ones obtained from applying the focal point constraints of eq.~\eqref{source_subspace} that yield the subspace sources corresponding to each set part of the subspace maximal overlap $\mathcal{O}^{(\rm{max})}_{\mathbf{b}}$. On the other hand, the column {\tt $L_{\rm max}$} in the multi-loop case still refers to the cardinality of the sets in $\mathcal{O}^{(\rm{max})}$ (that is, the maximal overlap structure obtained \emph{in the absence} of any focal point constraints).
Furthermore, beyond on loop, the number of channels (i.e. number of dual LTD integrands) $N_c$ is no longer equal to the number of propagators, but instead corresponds to the number of spanning trees which is a quantity specific to each integral topology.

Integrals~\ref{tab:explore_higherloopA}.{\tt 2L6P.a.I} to~\ref{tab:explore_higherloopA}.{\tt 2L6P.f.I} reproduce results from ref.~\cite{BeckerMultiLoop2012}, in which the authors perform a direct integration in four-dimensional Minkowski momentum space. We investigate the exact same \emph{decay} kinematic configurations as the ones considered in that work, which are numerically well-behaved and yield results that are pure phases. We also obtained independent reference results for these two-loop six-point integrals using an alternative numerical computation using {\sc\small pySecDec}~\cite{Borowka:2017idc} and we find only small tensions between all three results.

The multi-loop ladder four-point integrals (\ref{tab:explore_higherloopA}.{\tt 2L4P.c.I}, \ref{tab:explore_higherloopB}.{\tt 3L4P.I}, \ref{tab:explore_higherloopB}.{\tt 4L4P.b.I}, \ref{tab:explore_higherloopB}.{\tt 5L4P.I} and \ref{tab:explore_higherloopB}.{\tt 6L4P.a.I}) are known analytically for massless internal lines~\cite{Usyukina:1992jd}, and a generalisation to $M$x$N$ fishnet topologies (of which integrals~\ref{tab:explore_higherloopB}.{\tt 4L4P.a.I} and~\ref{tab:explore_higherloopB}.{\tt 6L4P.b.I} are two examples) was recently carried out in ref.~\cite{Basso:2017jwq}.
We stress that the five- and six-loop integrals~\ref{tab:explore_higherloopB}.{\tt 5L4P.I},~\ref{tab:explore_higherloopB}.{\tt 6L4P.a.I} and~\ref{tab:explore_higherloopB}.{\tt 6L4P.b.I} are computed for external kinematics yielding no threshold singularities such that the integration can be be performed without any contour deformation.
Furthermore, for these integrals, we used the multi-channeling treatment discussed in sect.~\ref{sec:multichanneling} as we found it to be necessary in order to tame the unbounded integrable singular surfaces that are of large dimensionality at these high loop counts\footnote{When disabling multi-channeling at these higher loop counts, we still found similar convergence pace but often obtained wrong biased central values.}.
The good agreement found for integral~\ref{tab:explore_higherloopB}.{\tt 6L4P.b.I} is the first numerical confirmation of the analytical expression obtained in ref.~\cite{Basso:2017jwq}.

Finally, the two entries {\tt 2L4P.a.I} and {\tt 2L4P.b.I} of table~\ref{tab:explore_higherloopA} present challenging integrals recently considered in ref.~\cite{Frellesvig:2019byn} (in which it appears as topology number B$_{72}$) in the context of the computation of the amplitude for Higgs production in association with a hard jet.
In that work, the exact dependency on the internal quark mass is retained thanks to an original semi-numerical method for solving the system of differential equations relating master integrals. In the case of an internal top quark (\ref{tab:explore_higherloopA}.{\tt 2L4P.a.I}), the authors could validate most of their results against the fully numerical ones obtained from sector decomposition techniques, however the case of the much lighter bottom quark (\ref{tab:explore_higherloopA}.{\tt 2L4P.b.I}) proved to be more challenging for these approaches. The result from numerical LTD agrees with ref.~\cite{Frellesvig:2019byn} and has a numerical integration error only marginally impacted by the different values selected for the internal quark mass.

\begin{figure}[ht]%
\includegraphics[trim={0.8cm 8cm 0 7.5cm},clip, scale=.8]{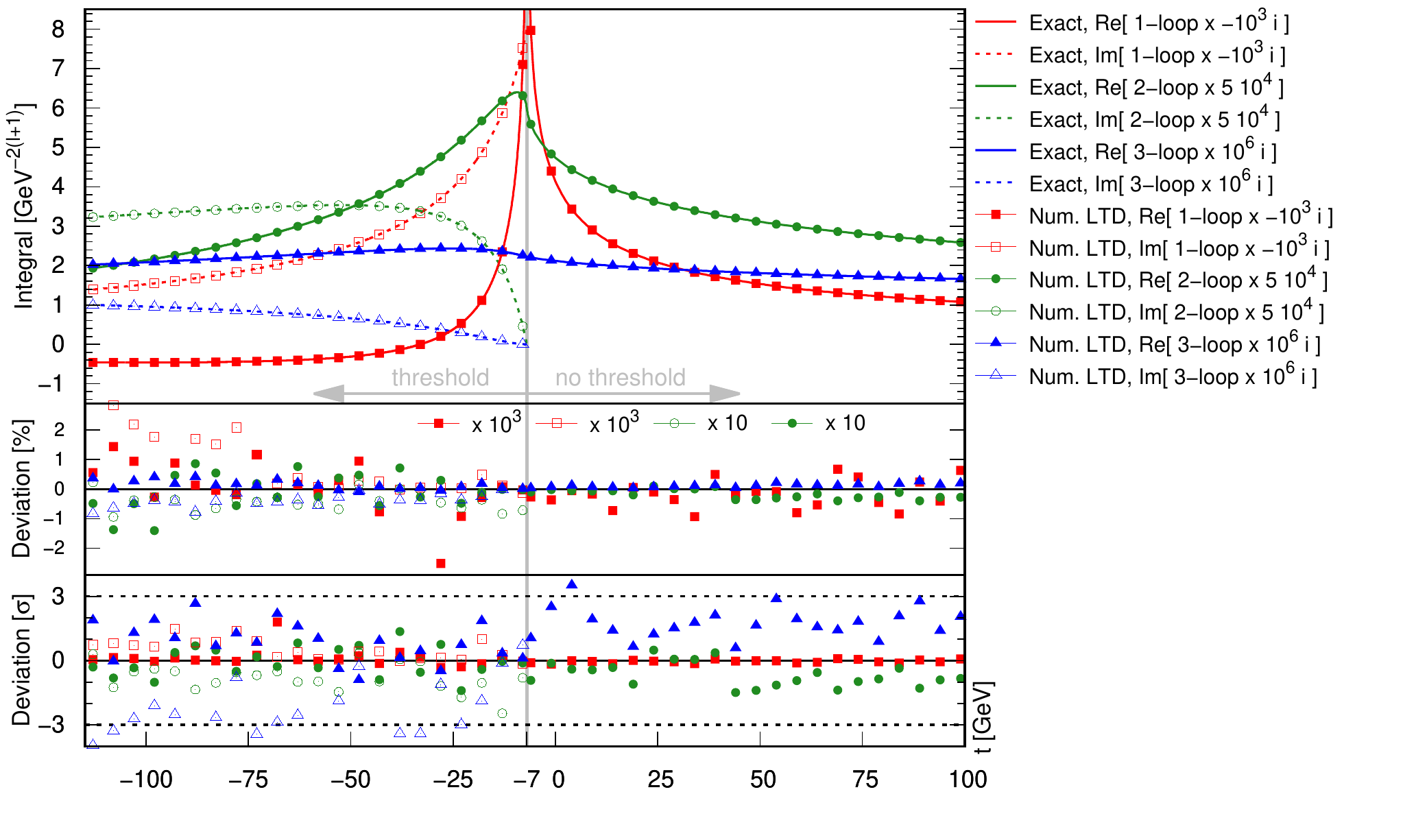}\\
\includegraphics[trim={0.8cm 8cm 0 7.5cm},clip, scale=.8]{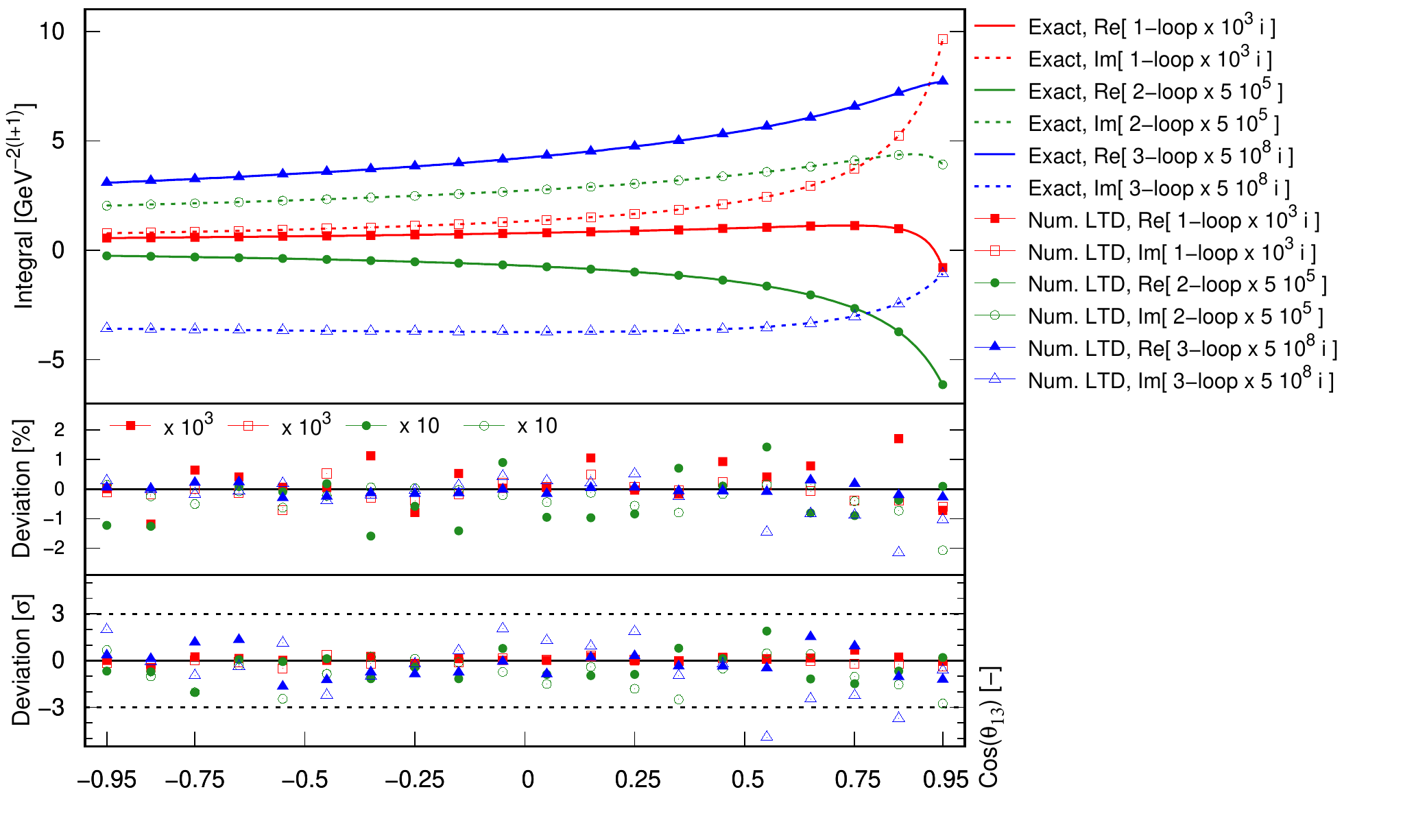}
  \caption{
  \label{plot:boxes_scan}
  Comparison of the exact analytic result from ref.~\cite{Usyukina:1992jd} with our numerical LTD computation for ladder 1-, 2- and 3-loop integrals. The kinematics considered for the top fig. is $p_1^2=-5$ and $p_2^2=p_3^2=p_4^2=(p_1+p_2)^2=-1$ and we scan over different values of the Mandelstamm invariant $t=(p_1+p_3)^2$.
The kinematics of the bottom fig. correspond to a physical $2\rightarrow 2$ scattering with $p_{\{1,2,3,4\}}^2=1$, $s=(p_1+p_2)^2=4.4$ and we scan over different values of the scattering angle $\theta_{13}=\angle (\vec{p_1},\vec{p_3})$.}
\end{figure}

In fig.~\ref{plot:boxes_scan}, we explore the stability of our numerical integration for two different classes of four-point kinematic configurations on one-, two- and three-loop ladder scalar integrals.
The first class of kinematics is unphysical, with $p_1^2=-5$ and $p_2^2=p_3^2=p_4^2=(p_1+p_2)^2=-1$. It is such that the region $(p_1+p_3)^2=t>-7$ can be addressed without any contour deformation, and for which we already showed results in fig.~1 of ref.~\cite{Capatti:2019ypt}.
In the complement region $t<-7$, a threshold singularity develops that corresponds to a \emph{single} E-surface in this particular parametrisation of the kinematics and at any loop count. Our construction of the contour deformation involves 1, 4 and 11 sources ($N_S$) for the 1-, 2- and 3-loop integral respectively. The multi-loop results shown in this upper plot of fig.~\ref{plot:boxes_scan} were obtained with $1$B integration sample points and our default values for the contour deformation hyperparameters.

The second class of kinematics concerns the physical 2-body scattering configuration with $p^2_{1,2,3,4}=1$, $s=(p_1+p_2)^2=4.4$ and a variable scattering angle $\theta_{13}=\angle(\vec{p_1},\vec{p_3})$. This case is far more challenging as it involves 5, 12 and 21 unique existing E-surfaces and necessitates a total of 1, 8 and 49 deformation sources ($N_S$) at 1-, 2- and 3-loop respectively. We note however that the set of maximal overlaps $\mathcal{O}^{(\rm{max})}$ always contains a \emph{single} set $F$ that involves all E-surfaces existing in the particular subspace considered, so that only a single source is necessary for generating a valid deformation in each subspace. The results found and presented in the lower panel of fig.~\ref{plot:boxes_scan} are obtained using modified hyperparameter values $\epsilon_{\rm{th}}=0.5$ and $M=0.05$, together with the multi-channeling treatment described in sect.~\ref{sec:multichanneling} and with a Monte-Carlo statistic of $100$M points for each channel integrated separately.

Fig.~\ref{plot:boxes_scan} demonstrated that numerical LTD is stable for different angular configurations, even when close to the crossing of thresholds in the external kinematics. We have however already observed in the one-loop results of table~\ref{tab:explore_1loop} that the convergence mostly depends on the shape and overlaps of the threshold singularity surfaces, which can become increasingly more complicated for boosted external momenta (that is $|\vec{p_i}|^2 \gg (p_i^0)^2$).
In tables~\ref{tab:Kseries_1loopA} to~\ref{tab:Kseries_HigherLoop}, we therefore seek to more systematically explore the performance of numerical LTD for external \emph{scattering}\footnote{
We find \emph{scattering} type of kinematic configurations to be numerically significantly more challenging than the \emph{decay} kinematics previously considered in the literature and shown in table~\ref{tab:explore_higherloopA}.
} kinematic configurations $p_1\;p_2\; \rightarrow p_3\;\dots\;p_N$ of progressively stronger hierarchies in the scales $m_j^2:=p_j^2$ and $s:=(p_1+p_2)^2$.

We provide our explicit choice of kinematics in the ancillary material and we limit ourselves here to reporting their relevant scales:
\begin{itemize}
    \item {\tt K1} | {\tt K1}$^\star$ : $m_j=1.0+0.1\;(j-1)$, $\sqrt{s}=1.1\;\sum_{j=1}^N m_j$ and $m_{\rm internal}=0.\;|\;0.4$,
    \item {\tt K2} | {\tt K2}$^\star$ : $m_j=1.0$, $\sqrt{s}=3.0\;\sum_{j=1}^N m_j$ and $m_{\rm internal}=0.\;|\;0.25$,
    \item {\tt K3} | {\tt K3}$^\star$ : $m_j=1.0+1.0\;(j-1)$, $\sqrt{s}=3.0\;\sum_{i=j}^N m_j$ and $m_{\rm internal}=0.\;|\;1.001$,
\end{itemize}
where the two different values for the masses of \emph{all} internal propagators correspond to the massive (resp. massless) case labelled with (resp. without) a $^\star$ in the tables. We note that the series of kinematics {\tt K3} features internal propagators with masses set very slightly above that of one of the external momenta.
This specific choice of internal mass is such that the existence condition of some E-surfaces are very close to being fulfilled, thus placing this challenging kinematic very close to crossing a threshold.
Similarly to what can be observed in the scan shown in Fig.~\ref{plot:boxes_scan}, we find numerical LTD to be in general stable even when approaching thresholds.

At one loop (tables~\ref{tab:Kseries_1loopA} and~\ref{tab:Kseries_1loopB}), we observe that the convergence mostly depends on the multiplicity of the external momenta, with a central value in agreement with {\sc\small MadLoop}'s reference beyond the percent level.
At two loops (tables~\ref{tab:Kseries_2loopA} and~\ref{tab:Kseries_2loopB}) and for integrals with more than four external legs, we find the scattering type of kinematics considered to be significantly more challenging than their decay counterpart featured in table~\ref{tab:explore_higherloopA} and we could not obtain a benchmark result from {\sc\small pySecDec}. In those cases, the columns $\Delta [\%]$ and $\Delta [\%] |\cdot|$ refer to the Monte-Carlo precision (and not the discrepancy w.r.t to the benchmark result) relative to the central value, and $\Delta [\sigma]$ is not applicable.

While numerical LTD generally performs well for kinematics featuring weaker hierarchies among its invariants, such as kinematics class {\tt K1}, we found integrals where the convergence for the kinematics {\tt K2} and {\tt K3} was not good enough with our default deformation hyperparameters for the results to be reported in the tables. We note however that adjusting the two contour deformation hyperparameters $\epsilon_{\rm{th}}$ (which governs the strength of the expansion constraint), and $M$ (which governs the strength of the anti-selection) can significantly improve the results. We illustrate this by optimising these two parameters for a particular six-point two-loop integral ({\tt 2L6P.a}) and for the {\tt K2} kinematics.
Using a low-statistics ($50$M points) exploratory scan, we find the optimal value of $(\epsilon_{\rm{th}}$,$M$) to be close to ($0.7$,$0.01$) for this configuration (most of the sensitivity lies in $M$). We then report in the table below the improvement of the convergence (especially strong in the case of massive internal propagators) found w.r.t to our default values ($\epsilon_{\rm{th}}=0.3$,$M=0.07$):

\begin{center}
\texttt{%
\begin{tabular}{@{}llrrrclc@{}}
\hline
Topology & Kin. & $\epsilon_\mathrm{th}$ & $M \ $ & $\mathtt{N_{\text{p}} \ [10^9]}$  & Phase & Exp. &  Numerical LTD  \\
\hline
\multirow{8}{*}{%
\begin{tabular}{@{}c@{}}
    \begin{tikzpicture}
    \begin{feynman}

    \tikzfeynmanset{every vertex={empty dot,minimum size=0mm}}
    \vertex (a1);
    
    \vertex[right=1cm of a1] (a3);
    \vertex[right=0.5cm of a1] (a2);
    \vertex[above=0.5cm of a2] (a4);
    \vertex[below=0.5cm of a2] (a5);
    
    \vertex[left=0.433cm of a2] (b1);
    \vertex[right=0.433cm of a2] (b2);
    
    \vertex[above=0.cm of a1] (c1);
    \vertex[above=0.25cm of b2] (c2);
    \vertex[below=0.25cm of b1] (d1);
    \vertex[below=0.cm of a3] (d2);
    \vertex[below=0.16666cm of a4] (l1);
    \vertex[below=0.33333cm of a4] (l2);
    \vertex[below=0.5cm of a4] (l3);
    \vertex[below=0.66666cm of a4] (l4);
    \vertex[below=0.83333cm of a4] (l5);
    
    \vertex[left=0.15cm of c1] (ec1);

    \vertex[right=0.15cm of d2] (ed2);
    \vertex[right=0.15cm of l1] (el1);
    \vertex[right=0.15cm of l2] (el2);
    \vertex[right=0.15cm of l3] (el3);
    \vertex[right=0.15cm of l4] (el4);
    \vertex[right=0.15cm of l5] (el5);

    \tikzfeynmanset{every vertex={dot,minimum size=0.8mm}}

    \vertex[below=0.cm of a3] (d2);
    
    \vertex[below=0.cm of l1] (l1);
    \vertex[below=0.cm of l2] (l2);
    \vertex[below=0.cm of l3] (l3);
    \vertex[below=0.cm of l4] (l4);
    \vertex[below=0.cm of l5] (l5);
    
        \diagram*[large]{	
        (a1)--[quarter left](a4) -- [quarter left](a3),
        (a5) --[quarter right](a3), 
        (a5)--[quarter left](a1),
        (a4) -- (a5),

        (d2) -- (ed2),
        (l1) -- (el1),
        (l1) -- (el1),
        (l2) -- (el2),
        (l3) -- (el3),
        (l4) -- (el4),
        (l5) -- (el5),
        }; 
    \end{feynman}
    \end{tikzpicture}
    \\
2L6P.a\end{tabular}}
& \multirow{4}{*}{K2}& \multirow{2}{*}{0.3}& \multirow{2}{*}{0.07}& \multirow{2}{*}{3\phantom{.0}}& Re& \multirow{2}{*}{-12}& \texttt{5.12~+/-~0.23}\\
& & & & & Im& & \texttt{-0.56~+/-~0.24} \ \\
\cline{3-8}
& & \multirow{2}{*}{0.7}& \multirow{2}{*}{0.01}& \multirow{2}{*}{3\phantom{.0}} & Re& \multirow{2}{*}{-12} & \texttt{5.13~+/-~0.11}\\
& & & & & Im& & \texttt{-0.26~+/-~0.11} \ \\
\cline{2-8}
& \multirow{4}{*}{K2$^\star$}& \multirow{2}{*}{0.3}& \multirow{2}{*}{0.07}& \multirow{2}{*}{3\phantom{.0}} & Re& \multirow{2}{*}{-11} & \texttt{0.6~+/-~1.1}\\
& & & & & Im&  & \texttt{-3.7~+/-~0.7} \ \\
\cline{3-8}
& & \multirow{2}{*}{0.7}& \multirow{2}{*}{0.01}& \multirow{2}{*}{2\phantom{.0}} & Re& \multirow{2}{*}{-11} & \texttt{0.709~+/-~0.030}\\
& & & & & Im& & \texttt{-3.845~+/-~0.030} \ \\
\hline
\end{tabular}%
}%
\end{center}

The two-loop eight-point integral~\ref{tab:Kseries_2loopB}.{\tt 2L8P.K1} shows good convergence, but we could not obtain a result for its massive counterpart~\ref{tab:Kseries_2loopB}.{\tt 2L8P.K1$^*$} because it features a challenging maximal overlap structure (despite involving less than the 46 unique E-surfaces of integral~\ref{tab:Kseries_2loopB}.{\tt 2L8P.K1}) that we could not determine in a reasonable amount of computing time using the algorithm described in sect.~\ref{sec:source_determination}.
Beyond two loops (table~\ref{tab:Kseries_HigherLoop}), we again observe a significant improvement when considering massive internal propagators, which can partly be explained by the fact that in this case the deformation is no longer forced by the dynamic scaling of eq.~\eqref{eq:branchcut_scaling} to become zero on the focal points of existing E-surfaces.
We should mention that the four-point four-loop integrals included in the tables are at the upper end of the complexity that can currently be handled by our implementation. For massless internal propagators, the scattering kinematics {\tt Ki} does not yield a good enough convergence while the decay kinematics necessitated an adjustment of the contour deformation hyperparameters (using a value for the parameter $M$ in eq.~\ref{eq:deformation_direction} smaller than our defaults, \emph{e.g.} $M\sim 0.01$). Given that such integrals are also beyond what is of current phenomenological relevance, we present their results mostly to highlight the potential of numerical LTD.

Despite the wide range of variances obtained, we always find the central value obtained from numerical LTD integration to be within less than five sigmas away from the analytical benchmark ones (when available), as indicated by the $\Delta [\sigma]$ column of the tables. This observation is actually the most important aspect of our results, since in this work we first aim at demonstrating that our numerical implementation of LTD is robust and can therefore be \emph{predictive}. Maximising numerical efficiency and exploring the optimisations discussed in sect.~\ref{sec:optimisation} is left to future work, for which results presented in this section can serve as a comparison baseline.

\begin{landscape}
\begin{table}[tbp]
\centering
\resizebox{\columnwidth}{!}{%
\texttt{%
\begin{tabular}{@{}llrrrp{8cm}rrcllcrrr@{}}
\hline
Topology & Kin. & $\mathtt{N_{\text{C}}}$ & $\mathtt{N_{\text{E}}}$ & $\mathtt{N_{\text{S}}}$ & $\mathtt{L_{\text{max}}}$ & $\mathtt{N_{\text{p}} \ [10^9]}$ & $\mathtt{\sfrac{t}{p} \ [\mu \text{s}]}$ & Phase & Exp. & \multicolumn{1}{c}{Reference} & Numerical LTD & $\mathtt{\Delta \ [\sigma]}$ & $\mathtt{\Delta \ [\%]}$ & $\mathtt{\Delta \ [\%] |\cdot|}$ \\
\hline
\multirow{2}{*}[7pt]{%
\begin{tabular}{@{}c@{}}
   \begin{tikzpicture}
    \begin{feynman}

        \tikzfeynmanset{every vertex={empty dot,minimum size=0mm}}
    \vertex (a1);

    \vertex[right=1cm of a1] (a3);
    
    \tikzfeynmanset{every vertex={empty dot,minimum size=0mm}}
    
    \vertex[right=0.5cm of a1] (a2);
    \vertex[above=0.5cm of a2] (a4);
    \vertex[below=0.5cm of a2] (a5);

    \vertex[left=0.433cm of a2] (b1);
    \vertex[left=0.433cm of a2] (b2);
    
    \vertex[right=0.433cm of a2] (b3);
    \vertex[right=0.433cm of a2] (b4);
    
    \vertex[left=0.433cm of a2] (d1);
    \vertex[left=0.433cm of a2] (d2);
    
    \vertex[right=0.433cm of a2] (d3);
    \vertex[right=0.433cm of a2] (d4);

    \vertex[above=0.255cm of b2] (c2);    
    \vertex[above=0.255cm of b3] (c3);
    \vertex[below=0.255cm of b2] (e2);   
    \vertex[below=0.255cm of b3] (e3);

    \tikzfeynmanset{every vertex={dot,minimum size=0.8mm}}
    
    \vertex[above=0.255cm of b1] (c1);
    \vertex[above=0.255cm of b4] (c4);    
    \vertex[below=0.255cm of b1] (e1);
    \vertex[below=0.255cm of b4] (e4);
    
    \tikzfeynmanset{every vertex={empty dot,minimum size=0mm}}
    
    \vertex[left=0.15cm of c1] (q1);
    \vertex[left=0.15cm of c2] (q2);
    \vertex[left=0.15cm of e1] (q3);
    \vertex[left=0.15cm of e2] (q4);
    \vertex[left=0.15cm of a1] (q5);

   \vertex[right=0.15cm of c3] (p1);
    \vertex[right=0.15cm of c4] (p2);
    \vertex[right=0.15cm of e3] (p3);
    \vertex[right=0.15cm of e4] (p4);
    \vertex[right=0.15cm of a3] (p5);
    
        \diagram*[large]{	
        (a1)--[quarter left](a4) -- [quarter left](a3),
        (a5) --[quarter right](a3), 
        (a5)--[quarter left](a1),        
               
        (c1) -- (q1),
        (e1) -- (q3),
        (c4) -- (p2),
        (e4) -- (p4),

        }; 
    \end{feynman}
    \end{tikzpicture}
     \\
Box4E\end{tabular}}
\rule[-5pt]{0pt}{27pt}& \multirow{2}{*}{I}& \multirow{2}{*}{4}& \multirow{2}{*}{4}& \multirow{2}{*}{4}& \multirow{2}{*}{$\mathtt{[2, 2, 2, 2]}$}& \multirow{2}{*}{3\phantom{.0}}& \multirow{2}{*}{15}& Re& \multirow{2}{*}{-08}& \texttt{-6.57830}& \texttt{-6.57637~+/-~0.00122}& 1.590& 0.029& \multirow{2}{*}{0.022}\\
\rule[-5pt]{0pt}{27pt}& & & & & & & & Im& & \texttt{-7.43707}& \texttt{-7.43805~+/-~0.00121}& 0.813& 0.013& \\
\hline
\multirow{12}{*}{%
\begin{tabular}{@{}c@{}}
    \begin{tikzpicture}
    \begin{feynman}

        \tikzfeynmanset{every vertex={empty dot,minimum size=0mm}}
    \vertex (a1);
       \tikzfeynmanset{every vertex={dot,minimum size=0.8mm}} 
    \vertex[right=1cm of a1] (a3);
    
    \tikzfeynmanset{every vertex={empty dot,minimum size=0mm}}
    
    \vertex[right=0.5cm of a1] (a2);
    \vertex[above=0.5cm of a2] (a4);
    \vertex[below=0.5cm of a2] (a5);

    \vertex[left=0.433cm of a2] (b1);
    \vertex[left=0.25cm of a2] (b2);
    
    \vertex[right=0.433cm of a2] (b3);
    \vertex[right=0.25cm of a2] (b4);
    
    \vertex[left=0.433cm of a2] (d1);
    \vertex[left=0.25cm of a2] (d2);
    
    \vertex[right=0.433cm of a2] (d3);
    \vertex[right=0.25cm of a2] (d4);

    \vertex[above=0.433cm of b2] (c2);    
    \vertex[above=0.255cm of b3] (c3);
    \vertex[below=0.433cm of b2] (e2);   
    \vertex[below=0.255cm of b3] (e3);

    \tikzfeynmanset{every vertex={dot,minimum size=0.8mm}}
    
    \vertex[above=0.255cm of b1] (c1);
    \vertex[above=0.433cm of b4] (c4);    
    \vertex[below=0.255cm of b1] (e1);
    \vertex[below=0.433cm of b4] (e4);
    
    \tikzfeynmanset{every vertex={empty dot,minimum size=0mm}}
    
    \vertex[left=0.15cm of c1] (q1);
    \vertex[left=0.15cm of c2] (q2);
    \vertex[left=0.15cm of e1] (q3);
    \vertex[left=0.15cm of e2] (q4);
    \vertex[left=0.15cm of a1] (q5);

   \vertex[right=0.15cm of c3] (p1);
    \vertex[right=0.15cm of c4] (p2);
    \vertex[right=0.15cm of e3] (p3);
    \vertex[right=0.15cm of e4] (p4);
    \vertex[right=0.15cm of a3] (p5);
    
        \diagram*[large]{	
        (a1)--[quarter left](a4) -- [quarter left](a3),
        (a5) --[quarter right](a3), 
        (a5)--[quarter left](a1),        
               
        (c1) -- (q1),
        (e1) -- (q3),
        (a3) -- (p5),
        (c4) -- (p2),
        (e4) -- (p4),

        }; 
    \end{feynman}
    \end{tikzpicture} \\
1L5P\end{tabular}}
& \multirow{2}{*}{I}& \multirow{2}{*}{5}& \multirow{2}{*}{8}& \multirow{2}{*}{1}& \multirow{2}{*}{$\mathtt{[8]}$}& \multirow{2}{*}{3\phantom{.0}}& \multirow{2}{*}{15}& Re& \multirow{2}{*}{-12}& \texttt{-3.44342}& \texttt{-3.44317~+/-~0.00045}& 0.564& 0.007& \multirow{2}{*}{0.007}\\
& & & & & & & & Im& & \texttt{-2.56487}& \texttt{-2.56505~+/-~0.00046}& 0.400& 0.007& \\
\cline{2-15}
& \multirow{2}{*}{II}& \multirow{2}{*}{5}& \multirow{2}{*}{10}& \multirow{2}{*}{1}& \multirow{2}{*}{$\mathtt{[10]}$}& \multirow{2}{*}{3\phantom{.0}}& \multirow{2}{*}{15}& Re& \multirow{2}{*}{-13}& \phantom{+}\texttt{0}\phantom{.00000}& \texttt{-0.00036~+/-~0.00029}& 1.266&  & \multirow{2}{*}{0.006}\\
& & & & & & & & Im& & \texttt{ 5.97143}& \texttt{~5.97143~+/-~0.00029}& 0.003& 2e-05& \\
\cline{2-15}
& \multirow{2}{*}{III}& \multirow{2}{*}{5}& \multirow{2}{*}{8}& \multirow{2}{*}{2}& \multirow{2}{*}{$\mathtt{[7, 7]}$}& \multirow{2}{*}{3\phantom{.0}}& \multirow{2}{*}{16}& Re& \multirow{2}{*}{-12}& \texttt{-0.83905}& \texttt{-0.83888~+/-~0.00016}& 1.029& 0.020& \multirow{2}{*}{0.012}\\
& & & & & & & & Im& & \texttt{-1.71341}& \texttt{-1.71325~+/-~0.00017}& 0.937& 0.009& \\
\cline{2-15}
& \multirow{2}{*}{IV}& \multirow{2}{*}{5}& \multirow{2}{*}{8}& \multirow{2}{*}{3}& \multirow{2}{*}{$\mathtt{[7, 7, 7]}$}& \multirow{2}{*}{3\phantom{.0}}& \multirow{2}{*}{17}& Re& \multirow{2}{*}{-12}& \texttt{-3.48997}& \texttt{-3.49044~+/-~0.00054}& 0.870& 0.013& \multirow{2}{*}{0.013}\\
& & & & & & & & Im& & \texttt{-3.90013}& \texttt{-3.89965~+/-~0.00054}& 0.891& 0.012& \\
\cline{2-15}
& \multirow{2}{*}{V}& \multirow{2}{*}{5}& \multirow{2}{*}{6}& \multirow{2}{*}{4}& \multirow{2}{*}{$\mathtt{[2, 2, 3, 4]}$}& \multirow{2}{*}{3\phantom{.0}}& \multirow{2}{*}{19}& Re& \multirow{2}{*}{-10}& \texttt{ 0.89920}& \texttt{~0.90036~+/-~0.00076}& 1.519& 0.129& \multirow{2}{*}{0.027}\\
& & & & & & & & Im& & \texttt{ 4.17837}& \texttt{~4.17823~+/-~0.00080}& 0.180& 0.003& \\
\cline{2-15}
& \multirow{2}{*}{VI}& \multirow{2}{*}{5}& \multirow{2}{*}{8}& \multirow{2}{*}{5}& \multirow{2}{*}{$\mathtt{[4, 4, 5, 5, 5]}$}& \multirow{2}{*}{3\phantom{.0}}& \multirow{2}{*}{19}& Re& \multirow{2}{*}{-13}& \texttt{ 0.04119}& \texttt{~0.04227~+/-~0.00068}& 1.593& 2.634& \multirow{2}{*}{0.057}\\
& & & & & & & & Im& & \texttt{-2.18057}& \texttt{-2.18118~+/-~0.00068}& 0.891& 0.028& \\
\hline
\multirow{20}{*}{%
\begin{tabular}{@{}c@{}}
   \begin{tikzpicture}
    \begin{feynman}
    \tikzfeynmanset{every vertex={dot,minimum size=0.8mm}}
    \vertex (a1);
    
    \vertex[right=1cm of a1] (a3);
    
    \tikzfeynmanset{every vertex={empty dot,minimum size=0mm}}
    
    \vertex[right=0.5cm of a1] (a2);
    \vertex[above=0.5cm of a2] (a4);
    \vertex[below=0.5cm of a2] (a5);

    \vertex[left=0.433cm of a2] (b1);
    \vertex[left=0.25cm of a2] (b2);
    
    \vertex[right=0.433cm of a2] (b3);
    \vertex[right=0.25cm of a2] (b4);
    
    \vertex[left=0.433cm of a2] (d1);
    \vertex[left=0.25cm of a2] (d2);
    
    \vertex[right=0.433cm of a2] (d3);
    \vertex[right=0.25cm of a2] (d4);

    \vertex[above=0.433cm of b2] (c2);    
    \vertex[above=0.255cm of b3] (c3);
    \vertex[below=0.433cm of b2] (e2);   
    \vertex[below=0.255cm of b3] (e3);
    
    \tikzfeynmanset{every vertex={dot,minimum size=0.8mm}}

    \vertex[above=0.433cm of b2] (c2);

    \vertex[above=0.433cm of b4] (c4);

    \vertex[below=0.433cm of b2] (e2);

    \vertex[below=0.433cm of b4] (e4);
    
    \tikzfeynmanset{every vertex={empty dot,minimum size=0mm}}
    
    \vertex[left=0.15cm of c1] (q1);
    \vertex[left=0.15cm of c2] (q2);
    \vertex[left=0.15cm of e1] (q3);
    \vertex[left=0.15cm of e2] (q4);
    \vertex[left=0.15cm of a1] (q5);

    \vertex[right=0.15cm of c3] (p1);
    \vertex[right=0.15cm of c4] (p2);
    \vertex[right=0.15cm of e3] (p3);
    \vertex[right=0.15cm of e4] (p4);
    \vertex[right=0.15cm of a3] (p5);
    
        \diagram*[large]{	
        (a1)--[quarter left](a4) -- [quarter left](a3),
        (a5) --[quarter right](a3), 
        (a5)--[quarter left](a1),

        (c2) -- (q2),
        (e2) -- (q4),
        (a1) -- (q5),
             
        (c4) -- (p2),
        (e4) -- (p4),
        (a3) -- (p5),
        }; 
    \end{feynman}
    \end{tikzpicture} \\
1L6P\end{tabular}}
& \multirow{2}{*}{I}& \multirow{2}{*}{6}& \multirow{2}{*}{12}& \multirow{2}{*}{1}& \multirow{2}{*}{$\mathtt{[12]}$}& \multirow{2}{*}{3\phantom{.0}}& \multirow{2}{*}{20}& Re& \multirow{2}{*}{-13}& \texttt{ 0.03040}& \texttt{~0.03046~+/-~0.00006}& 1.067& 0.202& \multirow{2}{*}{0.009}\\
& & & & & & & & Im& & \texttt{-1.17683}& \texttt{-1.17691~+/-~0.00008}& 1.057& 0.007& \\
\cline{2-15}
& \multirow{2}{*}{II}& \multirow{2}{*}{6}& \multirow{2}{*}{6}& \multirow{2}{*}{2}& \multirow{2}{*}{$\mathtt{[1, 5]}$}& \multirow{2}{*}{3\phantom{.0}}& \multirow{2}{*}{21}& Re& \multirow{2}{*}{+01}& \texttt{-2.07014}& \texttt{-2.07392~+/-~0.00188}& 2.004& 0.182& \multirow{2}{*}{0.214}\\
& & & & & & & & Im& & \texttt{ 0.42343}& \texttt{~0.42593~+/-~0.00161}& 1.551& 0.590& \\
\cline{2-15}
& \multirow{2}{*}{III}& \multirow{2}{*}{6}& \multirow{2}{*}{12}& \multirow{2}{*}{2}& \multirow{2}{*}{$\mathtt{[11, 10]}$}& \multirow{2}{*}{3\phantom{.0}}& \multirow{2}{*}{20}& Re& \multirow{2}{*}{-15}& \texttt{ 1.36918}& \texttt{~1.36950~+/-~0.00052}& 0.628& 0.024& \multirow{2}{*}{0.024}\\
& & & & & & & & Im& & \texttt{-2.25901}& \texttt{-2.25957~+/-~0.00053}& 1.054& 0.025& \\
\cline{2-15}
& \multirow{2}{*}{IV}& \multirow{2}{*}{6}& \multirow{2}{*}{12}& \multirow{2}{*}{3}& \multirow{2}{*}{$\mathtt{[9, 10, 10]}$}& \multirow{2}{*}{3\phantom{.0}}& \multirow{2}{*}{22}& Re& \multirow{2}{*}{-15}& \texttt{ 1.29770}& \texttt{~1.29802~+/-~0.00038}& 0.847& 0.025& \multirow{2}{*}{0.019}\\
& & & & & & & & Im& & \texttt{-2.16590}& \texttt{-2.16555~+/-~0.00037}& 0.929& 0.016& \\
\cline{2-15}
& \multirow{2}{*}{V}& \multirow{2}{*}{6}& \multirow{2}{*}{6}& \multirow{2}{*}{4}& \multirow{2}{*}{$\mathtt{[2, 3, 3, 3]}$}& \multirow{2}{*}{3\phantom{.0}}& \multirow{2}{*}{22}& Re& \multirow{2}{*}{-14}& \texttt{-0.27217}& \texttt{-0.27225~+/-~0.00010}& 0.839& 0.032& \multirow{2}{*}{0.007}\\
& & & & & & & & Im& & \texttt{-1.20896}& \texttt{-1.20895~+/-~0.00011}& 0.098& 0.001& \\
\cline{2-15}
& \multirow{2}{*}{VI}& \multirow{2}{*}{6}& \multirow{2}{*}{9}& \multirow{2}{*}{4}& \multirow{2}{*}{$\mathtt{[4, 6, 6, 6]}$}& \multirow{2}{*}{3\phantom{.0}}& \multirow{2}{*}{23}& Re& \multirow{2}{*}{-17}& \texttt{ 2.83772}& \texttt{~2.83777~+/-~0.00040}& 0.118& 0.002& \multirow{2}{*}{0.002}\\
& & & & & & & & Im& & \texttt{ 0.83142}& \texttt{~0.83144~+/-~0.00040}& 0.059& 0.003& \\
\cline{2-15}
& \multirow{2}{*}{VII}& \multirow{2}{*}{6}& \multirow{2}{*}{10}& \multirow{2}{*}{4}& \multirow{2}{*}{$\mathtt{[7, 7, 8, 8]}$}& \multirow{2}{*}{3\phantom{.0}}& \multirow{2}{*}{23}& Re& \multirow{2}{*}{-17}& \texttt{-3.01939}& \texttt{-3.01976~+/-~0.00040}& 0.917& 0.012& \multirow{2}{*}{0.008}\\
& & & & & & & & Im& & \texttt{-7.73337}& \texttt{-7.73280~+/-~0.00047}& 1.199& 0.007& \\
\cline{2-15}
& \multirow{2}{*}{VIII}& \multirow{2}{*}{6}& \multirow{2}{*}{10}& \multirow{2}{*}{4}& \multirow{2}{*}{$\mathtt{[3, 5, 6, 7]}$}& \multirow{2}{*}{3\phantom{.0}}& \multirow{2}{*}{24}& Re& \multirow{2}{*}{-02}& \texttt{ 2.11928}& \texttt{~2.13487~+/-~0.03230}& 0.483& 0.736& \multirow{2}{*}{1.055}\\
& & & & & & & & Im& & \texttt{ 0.64030}& \texttt{~0.65770~+/-~0.03145}& 0.553& 2.717& \\
\cline{2-15}
& \multirow{2}{*}{IX}& \multirow{2}{*}{6}& \multirow{2}{*}{12}& \multirow{2}{*}{4}& \multirow{2}{*}{$\mathtt{[8, 9, 9, 10]}$}& \multirow{2}{*}{3\phantom{.0}}& \multirow{2}{*}{22}& Re& \multirow{2}{*}{-14}& \texttt{ 0.00794}& \texttt{~0.00804~+/-~0.00014}& 0.710& 1.253& \multirow{2}{*}{0.009}\\
& & & & & & & & Im& & \texttt{-1.15282}& \texttt{-1.15278~+/-~0.00014}& 0.290& 0.004& \\
\cline{2-15}
& \multirow{2}{*}{X}& \multirow{2}{*}{6}& \multirow{2}{*}{10}& \multirow{2}{*}{5}& \multirow{2}{*}{$\mathtt{[6, 6, 7, 7, 7]}$}& \multirow{2}{*}{3\phantom{.0}}& \multirow{2}{*}{24}& Re& \multirow{2}{*}{+00}& \texttt{-2.81475}& \texttt{-2.81583~+/-~0.00060}& 1.809& 0.038& \multirow{2}{*}{0.029}\\
& & & & & & & & Im& & \texttt{ 2.47327}& \texttt{~2.47308~+/-~0.00061}& 0.313& 0.008& \\
\end{tabular}%
}%
}%
\caption{\label{tab:explore_1loop} Results for one-loop four-point to six-point functions. \texttt{Box4E} has been used as an example topology throughout this work. See the main text for details.}%
\end{table}

\begin{table}[tbp]
\centering
\resizebox{\columnwidth}{!}{%
\texttt{%
\begin{tabular}{@{}llrrrlrrcllcrrr@{}}
\hline
Topology & Kin. & $\mathtt{N_{\text{C}}}$ & $\mathtt{N_{\text{E}}}$ & $\mathtt{N_{\text{S}}}$ & $\mathtt{L_{\text{max}}}$ & $\mathtt{N_{\text{p}} \ [10^9]}$ & $\mathtt{\sfrac{t}{p} \ [\mu \text{s}]}$ & Phase & Exp. & \multicolumn{1}{c}{Reference} & Numerical LTD & $\mathtt{\Delta \ [\sigma]}$ & $\mathtt{\Delta \ [\%]}$ & $\mathtt{\Delta \ [\%] |\cdot|}$ \\
\hline
\multirow{4}{*}{%
\begin{tabular}{@{}c@{}}
     
    
    \begin{tikzpicture}
    \begin{feynman}

        \tikzfeynmanset{every vertex={empty dot,minimum size=0mm}}
    \vertex (a1);
       \tikzfeynmanset{every vertex={dot,minimum size=0.8mm}} 
    \vertex[right=1cm of a1] (a3);
    
    \tikzfeynmanset{every vertex={empty dot,minimum size=0mm}}
    
    \vertex[right=0.5cm of a1] (a2);
    \vertex[above=0.5cm of a2] (a4);
    \vertex[below=0.5cm of a2] (a5);

    \vertex[left=0.433cm of a2] (b1);
    \vertex[left=0.25cm of a2] (b2);
    
    \vertex[right=0.433cm of a2] (b3);
    \vertex[right=0.25cm of a2] (b4);
    
    \vertex[left=0.433cm of a2] (d1);
    \vertex[left=0.25cm of a2] (d2);
    
    \vertex[right=0.433cm of a2] (d3);
    \vertex[right=0.25cm of a2] (d4);
    
    \tikzfeynmanset{every vertex={dot,minimum size=0.8mm}}
    
    \vertex[above=0.255cm of b1] (c1);
   
    \vertex[below=0.255cm of b1] (e1);
    \vertex[above=0.cm of a4] (an);
    
    \tikzfeynmanset{every vertex={empty dot,minimum size=0mm}}
    
    \vertex[left=0.15cm of c1] (q1);
    \vertex[left=0.15cm of c2] (q2);
    \vertex[left=0.15cm of e1] (q3);
    \vertex[left=0.15cm of e2] (q4);
    \vertex[left=0.15cm of a1] (q5);
    
     \vertex[above=0.15cm of a4] (ext);   
    \vertex[right=0.15cm of c3] (p1);

    \vertex[right=0.15cm of e3] (p3);

    \vertex[right=0.15cm of a3] (p5);
    
        \diagram*[large]{	
        (a1)--[quarter left, double](a4) -- [quarter left,double](a3),
        (a5) --[quarter right,double](a3), 
        (a5)--[quarter left, double](a1),        
               
        (c1) -- (q1),
        (e1) -- (q3),
        (a3) -- (p5),
        (a4)--(a5),
        (ext) -- (a4),

        }; 
        
    \tikzfeynmanset{every vertex={dot,minimum size=0.8mm}}
    
    \vertex[above=0.cm of c1] (n1);
   
    \vertex[below=0.cm of e1] (n2);
    \vertex[above=0.cm of an] (n3);     
        
    \end{feynman}
    \end{tikzpicture} \\
2L4P.a\end{tabular}}
& \multirow{2}{*}{I}& \multirow{2}{*}{11}& \multirow{2}{*}{2}& \multirow{2}{*}{6}& \multirow{2}{*}{$\mathtt{[2]}$}& \multirow{2}{*}{3\phantom{.0}}& \multirow{2}{*}{39}& Re& \multirow{2}{*}{-06}& \texttt{ 3.82891} \multirow{2}{*}{\cite{Frellesvig:2019byn}}& \texttt{~3.82875~+/-~0.00015}& 1.107& 0.004& \multirow{2}{*}{0.003}\\
& & & & & & & & Im& & \texttt{-4.66840}& \texttt{-4.66843~+/-~0.00017}& 0.188& 0.001& \\
\cline{2-15}
& \multirow{2}{*}{II}& \multirow{2}{*}{11}& \multirow{2}{*}{4}& \multirow{2}{*}{7}& \multirow{2}{*}{$\mathtt{[4]}$}& \multirow{2}{*}{3\phantom{.0}}& \multirow{2}{*}{42}& Re& \multirow{2}{*}{-10}& \texttt{ 2.83647} \multirow{2}{*}{\cite{Frellesvig:2019byn}}& \texttt{~2.83742~+/-~0.00072}& 1.312& 0.033& \multirow{2}{*}{0.032}\\
& & & & & & & & Im& & \texttt{ 3.38265}& \texttt{~3.38163~+/-~0.00066}& 1.558& 0.030& \\
\hline
\multirow{2}{*}[7pt]{%
\begin{tabular}{@{}c@{}}
 

      \begin{tikzpicture}
            \begin{feynman}
                        
            \tikzfeynmanset{every vertex={dot,minimum size=1mm}}
           
            \tikzfeynmanset{every vertex={empty dot,minimum size=0mm}}
             \vertex (a1);
            \vertex[right=0.5cm of a1] (a2);
            \vertex[right=0.5cm of a2] (a3);
            \vertex[below=1cm of a3] (a4);
            
            \vertex[left=0.5cm of a4] (a5);
            \vertex[left=0.5cm of a5] (a6);
                          
            \tikzfeynmanset{every vertex={dot,minimum size=0.8mm}}
            
            \vertex[right=0cm of a1] (e1);        
            \vertex[right=0cm of a3] (e2);        
            \vertex[right=0cm of a4] (e3);        
            \vertex[right=0cm of a6] (e4);    
            
            \tikzfeynmanset{every vertex={empty dot,minimum size=0mm}}                
  
            \vertex[left=0.15cm of e1] (d1);        
            \vertex[right=0.15cm of e2] (d2);        
            \vertex[right=0.15cm of e3] (d3);        
            \vertex[left=0.15cm of e4] (d4);    
            
                \diagram*[large]{	
                (a1)--(a2)--(a3)--(a4)--(a5)--(a6)--(a1), 
                (a2)--(a5),
             		(e1)--(d1),
		(e2)--(d2),
		(e3)--(d3),
		(e4)--(d4),
                           }; 
            \end{feynman}
    \end{tikzpicture} \\
2L4P.b\end{tabular}}
\rule[-5pt]{0pt}{27pt}& \multirow{2}{*}{I}& \multirow{2}{*}{15}& \multirow{2}{*}{13}& \multirow{2}{*}{8}& \multirow{2}{*}{$\mathtt{[13]}$}& \multirow{2}{*}{3\phantom{.0}}& \multirow{2}{*}{55}& Re& \multirow{2}{*}{-02}& \texttt{-5.89700} \multirow{2}{*}{\cite{Usyukina:1992jd}}& \texttt{-5.89794~+/-~0.00099}& 0.956& 0.016& \multirow{2}{*}{0.025}\\
\rule[-5pt]{0pt}{27pt}& & & & & & & & Im& & \phantom{+}\texttt{0}\phantom{.00000}& \texttt{~0.00112~+/-~0.00095}& 1.171&  & \\
\hline
\multirow{3}{*}[-2pt]{%
\begin{tabular}{@{}c@{}}
    \begin{tikzpicture}
    \begin{feynman}

    \tikzfeynmanset{every vertex={empty dot,minimum size=0mm}}
    \vertex (a1);
    
    \vertex[right=1cm of a1] (a3);
    \vertex[right=0.5cm of a1] (a2);
    \vertex[above=0.5cm of a2] (a4);
    \vertex[below=0.5cm of a2] (a5);
    
    \vertex[left=0.433cm of a2] (b1);
    \vertex[right=0.433cm of a2] (b2);
    
    \vertex[above=0.cm of a1] (c1);
    \vertex[above=0.25cm of b2] (c2);
    \vertex[below=0.25cm of b1] (d1);
    \vertex[below=0.cm of a3] (d2);
    \vertex[below=0.16666cm of a4] (l1);
    \vertex[below=0.33333cm of a4] (l2);
    \vertex[below=0.5cm of a4] (l3);
    \vertex[below=0.66666cm of a4] (l4);
    \vertex[below=0.83333cm of a4] (l5);
    
    \vertex[left=0.15cm of c1] (ec1);

    \vertex[right=0.15cm of d2] (ed2);
    \vertex[right=0.15cm of l1] (el1);
    \vertex[right=0.15cm of l2] (el2);
    \vertex[right=0.15cm of l3] (el3);
    \vertex[right=0.15cm of l4] (el4);
    \vertex[right=0.15cm of l5] (el5);

    \tikzfeynmanset{every vertex={dot,minimum size=0.8mm}}

    \vertex[below=0.cm of a3] (d2);
    
    \vertex[below=0.cm of l1] (l1);
    \vertex[below=0.cm of l2] (l2);
    \vertex[below=0.cm of l3] (l3);
    \vertex[below=0.cm of l4] (l4);
    \vertex[below=0.cm of l5] (l5);
    
        \diagram*[large]{	
        (a1)--[quarter left](a4) -- [quarter left](a3),
        (a5) --[quarter right](a3), 
        (a5)--[quarter left](a1),
        (a4) -- (a5),

        (d2) -- (ed2),
        (l1) -- (el1),
        (l1) -- (el1),
        (l2) -- (el2),
        (l3) -- (el3),
        (l4) -- (el4),
        (l5) -- (el5),
        }; 
    \end{feynman}
    \end{tikzpicture}
    \\
2L6P.a\end{tabular}}
\rule[-5pt]{0pt}{20pt}& \multirow{3}{*}{I}& \multirow{3}{*}{20}& \multirow{3}{*}{20}& \multirow{3}{*}{14}& \multirow{3}{*}{$\mathtt{[20]}$}& \multirow{3}{*}{3\phantom{.0}}& \multirow{3}{*}{88}& \multirow{2}{*}{Re}& \multirow{3}{*}{+01}& \texttt{-8.608\phantom{0}~+/-~0.009\phantom{0}} \cite{Borowka:2017idc}& \multirow{2}{*}{\texttt{-8.64045~+/-~0.00392}}& \multirow{2}{*}{ }& \multirow{2}{*}{0.045}& \multirow{3}{*}{0.064}\\
& & & & & & & & & & \texttt{-8.66\phantom{00}~+/-~0.08\phantom{00}} \cite{BeckerMultiLoop2012}& & & & \\
\rule[-5pt]{0pt}{20pt}& & & & & & & & Im& & \phantom{+}\texttt{0}\phantom{.00000}& \texttt{-0.00220~+/-~0.00393}&  &  & \\
\hline
\multirow{3}{*}[-2pt]{%
\begin{tabular}{@{}c@{}}
  

\begin{tikzpicture}
    \begin{feynman}

    \tikzfeynmanset{every vertex={empty dot,minimum size=0mm}}
    \vertex (a1);
    
    \vertex[right=1cm of a1] (a3);
    \vertex[right=0.5cm of a1] (a2);
    \vertex[above=0.5cm of a2] (a4);
    \vertex[below=0.5cm of a2] (a5);
    
    \vertex[left=0.433cm of a2] (b1);
    \vertex[right=0.433cm of a2] (b2);
    
    \vertex[above=0.25cm of b1] (c1);
    \vertex[above=0.25cm of b2] (c2);
    \vertex[below=0.25cm of b1] (d1);
    \vertex[below=0.25cm of b2] (d2);
    
    \vertex[below=0.2cm of a4] (l1);
    \vertex[below=0.2cm of l1] (l2);
    \vertex[below=0.2cm of l2] (l3);
    \vertex[below=0.2cm of l3] (l4);
    
    \vertex[left=0.15cm of c1] (ec1);
    \vertex[right=0.15cm of c2] (ec2);
    \vertex[left=0.15cm of d1] (ed1);
    \vertex[right=0.15cm of d2] (ed2);
    \vertex[right=0.15cm of l1] (el1);
    \vertex[right=0.15cm of l2] (el2); 
    \vertex[right=0.15cm of l3] (el3);
    \vertex[right=0.15cm of l4] (el4);

    \vertex[above=0.25cm of b1] (c1);   
    \vertex[below=0.25cm of b1] (d1);
    \tikzfeynmanset{every vertex={dot,minimum size=0.8mm}}

    \vertex[above=0.25cm of b2] (c2);

    \vertex[below=0.25cm of b2] (d2);
    
    \vertex[above=0.cm of l1] (l1);
    \vertex[below=0.cm of l2] (l2);
    \vertex[below=0.cm of l3] (l3);
    \vertex[below=0.cm of l4] (l4);
    
        \diagram*[large]{	
        (a1)--[quarter left](a4) -- [quarter left](a3),
        (a5) --[quarter right](a3), 
        (a5)--[quarter left](a1),
        (a4) -- (a5),
        
        (c2) -- (ec2),
        (d2) -- (ed2),
        (l1) -- (el1),
        (l2) -- (el2),
        (l3)--(el3),
	(l4)--(el4),
    
        }; 
    \end{feynman}
    \end{tikzpicture} \\
2L6P.b\end{tabular}}
\rule[-5pt]{0pt}{20pt}& \multirow{3}{*}{I}& \multirow{3}{*}{23}& \multirow{3}{*}{23}& \multirow{3}{*}{18}& \multirow{3}{*}{$\mathtt{[21, 22]}$}& \multirow{3}{*}{3\phantom{.0}}& \multirow{3}{*}{95}& \multirow{2}{*}{Re}& \multirow{3}{*}{+02}& \texttt{-1.1886\phantom{}~+/-~0.0005\phantom{}} \cite{Borowka:2017idc}& \multirow{2}{*}{\texttt{-1.19040~+/-~0.00092}}& \multirow{2}{*}{ }& \multirow{2}{*}{0.077}& \multirow{3}{*}{0.109}\\
& & & & & & & & & & \texttt{-1.17\phantom{00}~+/-~0.02\phantom{00}} \cite{BeckerMultiLoop2012}& & & & \\
\rule[-5pt]{0pt}{20pt}& & & & & & & & Im& & \phantom{+}\texttt{0}\phantom{.00000}& \texttt{~0.00147~+/-~0.00092}&  &  & \\
\hline
\multirow{3}{*}[-2pt]{%
\begin{tabular}{@{}c@{}}
  

    \begin{tikzpicture}
    \begin{feynman}
    \tikzfeynmanset{every vertex={dot,minimum size=0.8mm}}
    \vertex (a1);
    
    \vertex[right=1cm of a1] (a3);
    
    \tikzfeynmanset{every vertex={empty dot,minimum size=0mm}}
    
    \vertex[right=0.5cm of a1] (a2);
    \vertex[above=0.5cm of a2] (a4);
    \vertex[below=0.5cm of a2] (a5);

    \vertex[left=0.433cm of a2] (b1);
    \vertex[left=0.25cm of a2] (b2);
    
    \vertex[right=0.433cm of a2] (b3);
    \vertex[right=0.25cm of a2] (b4);
    
    \vertex[left=0.433cm of a2] (d1);
    \vertex[left=0.25cm of a2] (d2);
    
    \vertex[right=0.433cm of a2] (d3);
    \vertex[right=0.25cm of a2] (d4);
    
    \tikzfeynmanset{every vertex={dot,minimum size=0.8mm}}

    \vertex[above=0.433cm of b2] (c2);

    \vertex[above=0.433cm of b4] (c4);

    \vertex[below=0.433cm of b2] (e2);

    \vertex[below=0.433cm of b4] (e4);
    
    \tikzfeynmanset{every vertex={empty dot,minimum size=0mm}}
    
    \vertex[left=0.15cm of c1] (q1);
    \vertex[left=0.15cm of c2] (q2);
    \vertex[left=0.15cm of e1] (q3);
    \vertex[left=0.15cm of e2] (q4);
    \vertex[left=0.15cm of a1] (q5);

    \vertex[right=0.15cm of c4] (p2);
    \vertex[right=0.15cm of e4] (p4);
    \vertex[right=0.15cm of a3] (p5);
    
        \diagram*[large]{	
        (a1)--[quarter left](a4) -- [quarter left](a3),
        (a5) --[quarter right](a3), 
        (a5)--[quarter left](a1),
        (a4)--(a5),

        (c2) -- (q2),
        (e2) -- (q4),
        (a1) -- (q5),
             
        (c4) -- (p2),
        (e4) -- (p4),
        (a3) -- (p5),
        }; 
    \end{feynman}
    \end{tikzpicture} \\
2L6P.c\end{tabular}}
\rule[-5pt]{0pt}{20pt}& \multirow{3}{*}{I}& \multirow{3}{*}{24}& \multirow{3}{*}{24}& \multirow{3}{*}{20}& \multirow{3}{*}{$\mathtt{[19, 22, 22]}$}& \multirow{3}{*}{3\phantom{.0}}& \multirow{3}{*}{94}& \multirow{2}{*}{Re}& \multirow{3}{*}{+01}& \texttt{-7.607\phantom{0}~+/-~0.006\phantom{0}} \cite{Borowka:2017idc}& \multirow{2}{*}{\texttt{-7.62856~+/-~0.00716}}& \multirow{2}{*}{ }& \multirow{2}{*}{0.094}& \multirow{3}{*}{0.133}\\
& & & & & & & & & & \texttt{-7.8\phantom{000}~+/-~0.1\phantom{000}} \cite{BeckerMultiLoop2012}& & & & \\
\rule[-5pt]{0pt}{20pt}& & & & & & & & Im& & \phantom{+}\texttt{0}\phantom{.00000}& \texttt{-0.00052~+/-~0.00724}&  &  & \\
\hline
\multirow{3}{*}[-2pt]{%
\begin{tabular}{@{}c@{}}
    \begin{tikzpicture}
    \begin{feynman}

    \tikzfeynmanset{every vertex={empty dot,minimum size=0mm}}
    \vertex (a1);
    
    \vertex[right=1cm of a1] (a3);
    \vertex[right=0.5cm of a1] (a2);
    \vertex[above=0.5cm of a2] (a4);
    \vertex[below=0.5cm of a2] (a5);
    
    \vertex[left=0.433cm of a2] (b1);
    \vertex[right=0.433cm of a2] (b2);
    
    \vertex[above=0.cm of a1] (c1);
    \vertex[above=0.25cm of b2] (c2);
    \vertex[below=0.25cm of b1] (d1);
    \vertex[below=0.cm of a3] (d2);
    \vertex[below=0.2cm of a4] (l1);
    \vertex[below=0.4cm of a4] (l2);
    \vertex[below=0.6cm of a4] (l3);
    \vertex[below=0.8cm of a4] (l4);
    
    \vertex[left=0.15cm of c1] (ec1);

    \vertex[right=0.15cm of d2] (ed2);
    \vertex[right=0.15cm of l1] (el1);
    \vertex[right=0.15cm of l2] (el2);
    \vertex[right=0.15cm of l3] (el3);
    \vertex[right=0.15cm of l4] (el4);

    \tikzfeynmanset{every vertex={dot,minimum size=0.8mm}}
    
    \vertex[above=0.cm of a1] (c1);

    \vertex[below=0.cm of a3] (d2);
    
    \vertex[below=0.2cm of a4] (l1);
    \vertex[below=0.4cm of a4] (l2);
    \vertex[below=0.6cm of a4] (l3);
    \vertex[below=0.8cm of a4] (l4);
    
        \diagram*[large]{	
        (a1)--[quarter left](a4) -- [quarter left](a3),
        (a5) --[quarter right](a3), 
        (a5)--[quarter left](a1),
        (a4) -- (a5),
    
        (c1) -- (ec1),

        (d2) -- (ed2),
        (l1) -- (el1),
        (l1) -- (el1),
        (l2) -- (el2),
        (l3) -- (el3),
        (l4) -- (el4),
        }; 
    \end{feynman}
    \end{tikzpicture} \\
2L6P.d\end{tabular}}
\rule[-5pt]{0pt}{20pt}& \multirow{3}{*}{I}& \multirow{3}{*}{24}& \multirow{3}{*}{23}& \multirow{3}{*}{15}& \multirow{3}{*}{$\mathtt{[23]}$}& \multirow{3}{*}{3\phantom{.0}}& \multirow{3}{*}{91}& \multirow{2}{*}{Re}& \multirow{3}{*}{+01}& \texttt{-1.833\phantom{0}~+/-~0.002\phantom{0}} \cite{Borowka:2017idc}& \multirow{2}{*}{\texttt{-1.83639~+/-~0.00075}}& \multirow{2}{*}{ }& \multirow{2}{*}{0.041}& \multirow{3}{*}{0.058}\\
& & & & & & & & & & \texttt{-1.91\phantom{00}~+/-~0.02\phantom{00}} \cite{BeckerMultiLoop2012}& & & & \\
\rule[-5pt]{0pt}{20pt}& & & & & & & & Im& & \phantom{+}\texttt{0}\phantom{.00000}& \texttt{-0.00042~+/-~0.00075}&  &  & \\
\hline
\multirow{3}{*}[-2pt]{%
\begin{tabular}{@{}c@{}}
 \begin{tikzpicture}
    \begin{feynman}

    \tikzfeynmanset{every vertex={empty dot,minimum size=0mm}}
    \vertex (a1);
    
    \vertex[right=1cm of a1] (a3);
    \vertex[right=0.5cm of a1] (a2);
    \vertex[above=0.5cm of a2] (a4);
    \vertex[below=0.5cm of a2] (a5);
    
    \vertex[left=0.433cm of a2] (b1);
    \vertex[right=0.433cm of a2] (b2);
    
    \vertex[left=0.cm of a1] (c1);
    \vertex[above=0.25cm of b2] (c2);
    \vertex[below=0.cm of a2] (d1);
    \vertex[below=0.25cm of b2] (d2);
    \vertex[above=0.3333cm of a2] (l1);
    \vertex[below=0.3333cm of a2] (l2);
    
    \vertex[left=0.15cm of c1] (ec1);
    \vertex[right=0.15cm of c2] (ec2);
    \vertex[right=0.15cm of d1] (ed1);
    \vertex[right=0.15cm of d2] (ed2);
    \vertex[right=0.15cm of l1] (el1);
    \vertex[right=0.15cm of l2] (el2);     
    
    \tikzfeynmanset{every vertex={dot,minimum size=0.8mm}}
    
    \vertex[above=0.cm of a1] (c1);
    \vertex[above=0.25cm of b2] (c2);
    
    \vertex[below=0.cm of a2] (d1);
    \vertex[below=0.25cm of b2] (d2);
    
    \vertex[above=0.3333cm of a2] (l1);
    \vertex[below=0.3333cm of a2] (l2);
    
        \diagram*[large]{	
        (a1)--[quarter left](a4) -- [quarter left](a3),
        (a5) --[quarter right](a3), 
        (a5)--[quarter left](a1),
        (a4) -- (a5),
        
        (c1) -- (ec1),
        (c2) -- (ec2),
	(d1)--(ed1),
        (d2) -- (ed2),
        (l1) -- (el1),
        (l2) -- (el2),

        }; 
    \end{feynman}
    \end{tikzpicture} \\
2L6P.e\end{tabular}}
\rule[-5pt]{0pt}{20pt}& \multirow{3}{*}{I}& \multirow{3}{*}{26}& \multirow{3}{*}{26}& \multirow{3}{*}{19}& \multirow{3}{*}{$\mathtt{[25, 25]}$}& \multirow{3}{*}{3\phantom{.0}}& \multirow{3}{*}{101}& \multirow{2}{*}{Re}& \multirow{3}{*}{+01}& \texttt{-4.597\phantom{0}~+/-~0.004\phantom{0}} \cite{Borowka:2017idc}& \multirow{2}{*}{\texttt{-4.61094~+/-~0.00423}}& \multirow{2}{*}{ }& \multirow{2}{*}{0.092}& \multirow{3}{*}{0.131}\\
& & & & & & & & & & \texttt{-4.64\phantom{00}~+/-~0.08\phantom{00}} \cite{BeckerMultiLoop2012}& & & & \\
\rule[-5pt]{0pt}{20pt}& & & & & & & & Im& & \phantom{+}\texttt{0}\phantom{.00000}& \texttt{~0.00404~+/-~0.00430}&  &  & \\
\hline
\multirow{3}{*}[-2pt]{%
\begin{tabular}{@{}c@{}}
   
 
 \begin{tikzpicture}
    \begin{feynman}

    \tikzfeynmanset{every vertex={empty dot,minimum size=0mm}}
    \vertex (a1);
    
    \vertex[right=1cm of a1] (a3);
    \vertex[right=0.5cm of a1] (a2);
    \vertex[above=0.5cm of a2] (a4);
    \vertex[below=0.5cm of a2] (a5);
    
    \vertex[left=0.433cm of a2] (b1);
    \vertex[right=0.433cm of a2] (b2);
    
    \vertex[above=0.25cm of b1] (c1);
    \vertex[above=0.25cm of b2] (c2);
    \vertex[below=0.25cm of b1] (d1);
    \vertex[below=0.25cm of b2] (d2);
    \vertex[above=0.3333cm of a2] (l1);
    \vertex[below=0.3333cm of a2] (l2);
    
    \vertex[left=0.15cm of c1] (ec1);
    \vertex[right=0.15cm of c2] (ec2);
    \vertex[left=0.15cm of d1] (ed1);
    \vertex[right=0.15cm of d2] (ed2);
    \vertex[right=0.15cm of l1] (el1);
    \vertex[right=0.15cm of l2] (el2);     
    
    \tikzfeynmanset{every vertex={dot,minimum size=0.8mm}}
    
    \vertex[above=0.25cm of b1] (c1);
    \vertex[above=0.25cm of b2] (c2);
    
    \vertex[below=0.25cm of b1] (d1);
    \vertex[below=0.25cm of b2] (d2);
    
    \vertex[above=0.3333cm of a2] (l1);
    \vertex[below=0.3333cm of a2] (l2);
    
        \diagram*[large]{	
        (a1)--[quarter left](a4) -- [quarter left](a3),
        (a5) --[quarter right](a3), 
        (a5)--[quarter left](a1),
        (a4) -- (a5),
        
        (c1) -- (ec1),
        (c2) -- (ec2),
        (d1) -- (ed1),
        (d2) -- (ed2),
        (l1) -- (el1),
        (l2) -- (el2),

        }; 
    \end{feynman}
    \end{tikzpicture} \\
2L6P.f\end{tabular}}
\rule[-5pt]{0pt}{20pt}& \multirow{3}{*}{I}& \multirow{3}{*}{27}& \multirow{3}{*}{33}& \multirow{3}{*}{20}& \multirow{3}{*}{$\mathtt{[29, 32]}$}& \multirow{3}{*}{3\phantom{.0}}& \multirow{3}{*}{119}& \multirow{2}{*}{Re}& \multirow{3}{*}{+02}& \texttt{-1.0271\phantom{}~+/-~0.0003\phantom{}} \cite{Borowka:2017idc}& \multirow{2}{*}{\texttt{-1.02723~+/-~0.00111}}& \multirow{2}{*}{ }& \multirow{2}{*}{0.108}& \multirow{3}{*}{0.153}\\
& & & & & & & & & & \texttt{-1.03\phantom{00}~+/-~0.02\phantom{00}} \cite{BeckerMultiLoop2012}& & & & \\
\rule[-5pt]{0pt}{20pt}& & & & & & & & Im& & \phantom{+}\texttt{0}\phantom{.00000}& \texttt{~0.00165~+/-~0.00112}&  &  & \\
\end{tabular}%
}%
}%
\caption{\label{tab:explore_higherloopA} Results for two-loop topologies with benchmark kinematics from the literature. See the main text for details.}%
\end{table}

\begin{table}[tbp]
\centering
\resizebox{\columnwidth}{!}{%
\texttt{%
\begin{tabular}{@{}llrrrlrrcllcrrr@{}}
\hline
Topology & Kin. & $\mathtt{N_{\text{C}}}$ & $\mathtt{N_{\text{E}}}$ & $\mathtt{N_{\text{S}}}$ & $\mathtt{L_{\text{max}}}$ & $\mathtt{N_{\text{p}} \ [10^9]}$ & $\mathtt{\sfrac{t}{p} \ [\mu \text{s}]}$ & Phase & Exp. & \multicolumn{1}{c}{Reference} & Numerical LTD & $\mathtt{\Delta \ [\sigma]}$ & $\mathtt{\Delta \ [\%]}$ & $\mathtt{\Delta \ [\%] |\cdot|}$ \\
\hline
\multirow{2}{*}[7pt]{%
\begin{tabular}{@{}c@{}}
   

        \begin{tikzpicture}
            \begin{feynman}

            \tikzfeynmanset{every vertex={dot,minimum size=1mm}}

            \tikzfeynmanset{every vertex={empty dot,minimum size=0mm}}
             \vertex (a1);
            \vertex[right=0.33cm of a1] (a2);
            \vertex[right=0.33cm of a2] (a3);
             \vertex[right=0.33cm of a3] (a8);
            \vertex[below=1cm of a8] (a4);
            
            \vertex[left=0.33cm of a4] (a5);
            \vertex[left=0.33cm of a5] (a6);
            \vertex[left=0.33cm of a6] (a7);
                          
            \tikzfeynmanset{every vertex={dot,minimum size=0.8mm}}
            
            \vertex[right=0cm of a1] (e1);        
            \vertex[right=0cm of a8] (e2);        
            \vertex[right=0cm of a4] (e3);        
            \vertex[right=0cm of a7] (e4);    
            
            \tikzfeynmanset{every vertex={empty dot,minimum size=0mm}}                
  
            \vertex[left=0.15cm of e1] (d1);        
            \vertex[right=0.15cm of e2] (d2);        
            \vertex[right=0.15cm of e3] (d3);        
            \vertex[left=0.15cm of e4] (d4);    
            
                \diagram*[large]{	
                (a1)--(a2)--(a3)--(a8)--(a4)--(a5)--(a6)--(a7)--(a1), 
                (a2)--(a6),
                (a3)--(a5),
		(e1)--(d1),
		(e2)--(d2),
		(e3)--(d3),
		(e4)--(d4),
                           }; 
            \end{feynman}
    \end{tikzpicture} \\
3L4P\end{tabular}}
\rule[-5pt]{0pt}{27pt}& \multirow{2}{*}{I}& \multirow{2}{*}{56}& \multirow{2}{*}{22}& \multirow{2}{*}{49}& \multirow{2}{*}{$\mathtt{[22]}$}& \multirow{2}{*}{1\phantom{.0}}& \multirow{2}{*}{346}& Re& \multirow{2}{*}{-03}& \phantom{+}\texttt{0}\phantom{.00000} \multirow{2}{*}{\cite{Usyukina:1992jd}}& \texttt{~0.00796~+/-~0.00877}& 0.907&  & \multirow{2}{*}{0.149}\\
\rule[-5pt]{0pt}{27pt}& & & & & & & & Im& & \texttt{-6.74400}& \texttt{-6.73786~+/-~0.00856}& 0.717& 0.091& \\
\hline
\multirow{2}{*}[7pt]{%
\begin{tabular}{@{}c@{}}
 

        \begin{tikzpicture}
            \begin{feynman}

            \tikzfeynmanset{every vertex={dot,minimum size=1mm}}

            \tikzfeynmanset{every vertex={empty dot,minimum size=0mm}}
             \vertex (a1);
            \vertex[right=0.5cm of a1] (a2);
            \vertex[right=0.5cm of a2] (a3);
            \vertex[below=1cm of a3] (a4);
            
            \vertex[below=0.5cm of a1] (t1);
            \vertex[below=0.5cm of a3] (t2);
            
            \vertex[left=0.5cm of a4] (a5);
            \vertex[left=0.5cm of a5] (a6);
                          
            \tikzfeynmanset{every vertex={dot,minimum size=0.8mm}}
            
            \vertex[right=0cm of a1] (e1);        
            \vertex[right=0cm of a3] (e2);        
            \vertex[right=0cm of a4] (e3);        
            \vertex[right=0cm of a6] (e4);    
            
            \tikzfeynmanset{every vertex={empty dot,minimum size=0mm}}                
  
            \vertex[left=0.15cm of e1] (d1);        
            \vertex[right=0.15cm of e2] (d2);        
            \vertex[right=0.15cm of e3] (d3);        
            \vertex[left=0.15cm of e4] (d4);    
            
                \diagram*[large]{	
                (a1)--(a2)--(a3)--(a4)--(a5)--(a6)--(a1), 
                (a2)--(a5),
             		(e1)--(d1),
		(e2)--(d2),
		(e3)--(d3),
		(e4)--(d4),
		(t1)--(t2),
                           }; 
            \end{feynman}
    \end{tikzpicture} \\
4L4P.a\end{tabular}}
\rule[-5pt]{0pt}{27pt}& \multirow{2}{*}{I}& \multirow{2}{*}{192}& \multirow{2}{*}{44}& \multirow{2}{*}{280}& \multirow{2}{*}{$\mathtt{[44]}$}& \multirow{2}{*}{0.7}& \multirow{2}{*}{0}& Re& \multirow{2}{*}{-05}& \texttt{ 8.41610} \multirow{2}{*}{\cite{Basso:2017jwq}}& \texttt{~8.38828~+/-~0.07772}& 0.358& 0.331& \multirow{2}{*}{0.352}\\
\rule[-5pt]{0pt}{27pt}& & & & & & & & Im& & \phantom{+}\texttt{0}\phantom{.00000}& \texttt{-0.01028~+/-~0.07754}& 0.133&  & \\
\hline
\multirow{2}{*}[7pt]{%
\begin{tabular}{@{}c@{}}
         
       
       \begin{tikzpicture}
            \begin{feynman}

            \tikzfeynmanset{every vertex={dot,minimum size=1mm}}

            \tikzfeynmanset{every vertex={empty dot,minimum size=0mm}}
             \vertex (a1);
            \vertex[right=0.25cm of a1] (a2);
            \vertex[above=0.5cm of a2] (a4);
            \vertex[below=0.5cm of a2] (a5);

            \vertex[right=1cm of a1] (d1);

            \vertex[right=0.25cm of a4] (b1);
            \vertex[right=0.25cm of a5] (b2);
            \vertex[right=0.25cm of b1] (c1);
            \vertex[right=0.25cm of b2] (c2);
            
            \vertex[above=0.5cm of a1] (a7);
            \vertex[below=0.5cm of a1] (a8);
            \vertex[above=0.5cm of d1] (a9);
            \vertex[below=0.5cm of d1] (a10);
            
            \vertex[left=0.15cm of a7] (a11);
            \vertex[left=0.15cm of a8] (a12);
            \vertex[right=0.15cm of a9] (a13);
            \vertex[right=0.15cm of a10] (a14);
            
            \tikzfeynmanset{every vertex={dot,minimum size=0.8mm}}

            \vertex[above=0.5cm of a1] (e1);
            \vertex[below=0.5cm of a1] (e2);    
            
            \vertex[above=0.5cm of d1] (f1);
            \vertex[below=0.5cm of d1] (f2);

                \diagram*[large]{	
                (e1)--(a4), 
                (a5)--(e2),
		(e1)--(e2),

                (a4) -- (a5),
                (a4) -- (b1),
                (a5) -- (b2),
                (b1)--(b2),
                (b1)--(c1),
                (b2)--(c2),
                (c1)--(c2),

                (c1) --  (f1),
                (f2) -- (c2),
		        (f1)--(f2),
		        
		        (a7) -- (a11),
		        (a8) -- (a12),
		        (a9) -- (a13),
		        (a10) -- (a14),
                }; 
            \end{feynman}
    \end{tikzpicture} \\
4L4P.b\end{tabular}}
\rule[-5pt]{0pt}{27pt}& \multirow{2}{*}{I}& \multirow{2}{*}{209}& \multirow{2}{*}{33}& \multirow{2}{*}{270}& \multirow{2}{*}{$\mathtt{[33]}$}& \multirow{2}{*}{0.5}& \multirow{2}{*}{2712}& Re& \multirow{2}{*}{-04}& \texttt{ 7.41128} \multirow{2}{*}{\cite{Usyukina:1992jd}}& \texttt{~7.96654~+/-~0.11281}& 4.922& 7.492& \multirow{2}{*}{7.562}\\
\rule[-5pt]{0pt}{27pt}& & & & & & & & Im& & \phantom{+}\texttt{0}\phantom{.00000}& \texttt{~0.07617~+/-~0.11858}& 0.642&  & \\
\hline
\multirow{2}{*}[7pt]{%
\begin{tabular}{@{}c@{}}
  

      \begin{tikzpicture}
            \begin{feynman}

            \tikzfeynmanset{every vertex={dot,minimum size=1mm}}

            \tikzfeynmanset{every vertex={empty dot,minimum size=0mm}}
             \vertex (a1);
            \vertex[right=0.2cm of a1] (a2);
            \vertex[right=0.2cm of a2] (a3);
            \vertex[right=0.2cm of a3] (a4);
            \vertex[right=0.2cm of a4] (a5);
            \vertex[right=0.2cm of a5] (a6);
            
            \vertex[below=1cm of a6] (a7);
            
            \vertex[left=0.2cm of a7] (a8);
	    \vertex[left=0.2cm of a8] (a9);
	    \vertex[left=0.2cm of a9] (a10);
	    \vertex[left=0.2cm of a10] (a11);
	    \vertex[left=0.2cm of a11] (a12);
	    
	    \vertex[right=0.15cm of a6] (d1);
	    \vertex[right=0.15cm of a7] (d2);
	    \vertex[left=0.15cm of a1] (e3);
	    \vertex[left=0.15cm of a12] (e4);

            \tikzfeynmanset{every vertex={dot,minimum size=0.8mm}}

            \vertex[above=0cm of a1] (e1);
            \vertex[below=0cm of a6] (e2);    
            
            \vertex[above=0cm of a7] (f1);
            \vertex[below=0cm of a12] (f2);

                \diagram*[large]{	
                (a1)--(a6)--(a7)--(a12)--(a1),
                (a2)--(a11),
                (a3)--(a10),
                (a4)--(a9),
                (a5)--(a8),
                (d1)--(a6),
                (d2)--(a7),
                (e3)--(a1),
                (e4)--(a12),
                }; 
            \end{feynman}
    \end{tikzpicture} \\
5L4P\end{tabular}}
\rule[-5pt]{0pt}{27pt}& \multirow{2}{*}{I}& \multirow{2}{*}{780}& \multirow{2}{*}{0}& \multirow{2}{*}{0}& \multirow{2}{*}{}& \multirow{2}{*}{1.8}& \multirow{2}{*}{255}& Re& \multirow{2}{*}{-16}& \phantom{+}\texttt{0}\phantom{.00000} \multirow{2}{*}{\cite{Basso:2017jwq}}& &  &  & \multirow{2}{*}{0.843}\\
\rule[-5pt]{0pt}{27pt}& & & & & & & & Im& & \texttt{ 3.31697}& \texttt{~3.28900~+/-~0.01964}& 1.424& 0.843& \\
\hline
\multirow{2}{*}[7pt]{%
\begin{tabular}{@{}c@{}}
 
 
        \begin{tikzpicture}
            \begin{feynman}
            
            \tikzfeynmanset{every vertex={empty dot,minimum size=0mm}}
             \vertex (a1);
            \vertex[right=0.33cm of a1] (a2);
            \vertex[right=0.33cm of a2] (a3);
             \vertex[right=0.33cm of a3] (a8);
            \vertex[below=1cm of a8] (a4);
             \vertex[below=0.5cm of a8] (c1);
             
             \vertex[below=0.5cm of a1] (t1);
             \vertex[below=0.5cm of a8] (t2);             
            
            \vertex[left=0.33cm of a4] (a5);
            \vertex[left=0.33cm of a5] (a6);
            \vertex[left=0.33cm of a6] (a7);
                          
            \tikzfeynmanset{every vertex={dot,minimum size=0.8mm}}
            
            \vertex[right=0cm of a1] (e1);        
            \vertex[right=0cm of a8] (e2);        
            \vertex[right=0cm of a4] (e3);        
            \vertex[right=0cm of a7] (e4);

            \tikzfeynmanset{every vertex={empty dot,minimum size=0mm}}                
  
            \vertex[left=0.15cm of e1] (d1);        
            \vertex[right=0.15cm of e2] (d2);        
            \vertex[right=0.15cm of e3] (d3);        
            \vertex[left=0.15cm of e4] (d4);
            \vertex[right=0.15cm of c1] (d5);    
            
                \diagram*[large]{	
                (a1)--(a2)--(a3)--(a8)--(a4)--(a5)--(a6)--(a7)--(a1), 
                (a2)--(a6),
                (a3)--(a5),
		(e1)--(d1),
		(e2)--(d2),
		(e3)--(d3),
		(e4)--(d4),

		(t1)--(t2),
                           }; 
            \end{feynman}
    \end{tikzpicture} \\
6L4P.a\end{tabular}}
\rule[-5pt]{0pt}{27pt}& \multirow{2}{*}{I}& \multirow{2}{*}{2415}& \multirow{2}{*}{0}& \multirow{2}{*}{0}& \multirow{2}{*}{}& \multirow{2}{*}{14.5}& \multirow{2}{*}{1196}& Re& \multirow{2}{*}{-19}& \texttt{ 8.40449} \multirow{2}{*}{\cite{Basso:2017jwq}}& \texttt{~8.36493~+/-~0.02167}& 1.825& 0.471& \multirow{2}{*}{0.471}\\
\rule[-5pt]{0pt}{27pt}& & & & & & & & Im& & \phantom{+}\texttt{0}\phantom{.00000}& &  &  & \\
\hline
\multirow{2}{*}[7pt]{%
\begin{tabular}{@{}c@{}}
        \begin{tikzpicture}
            \begin{feynman}

            \tikzfeynmanset{every vertex={dot,minimum size=1mm}}

            \tikzfeynmanset{every vertex={empty dot,minimum size=0mm}}
             \vertex (a1);
            \vertex[right=0.16666cm of a1] (a2);
            \vertex[right=0.16666cm of a2] (a3);
            \vertex[right=0.16666cm of a3] (a4);
            \vertex[right=0.16666cm of a4] (a5);
            \vertex[right=0.16666cm of a5] (a51);            
            \vertex[right=0.16666cm of a51] (a6);
            
            \vertex[below=1cm of a6] (a7);
            
            \vertex[left=0.16666cm of a7] (a81);	    
            \vertex[left=0.16666cm of a81] (a8);
	    \vertex[left=0.16666cm of a8] (a9);
	    \vertex[left=0.16666cm of a9] (a10);
	    \vertex[left=0.16666cm of a10] (a11);

	    \vertex[left=0.16666cm of a11] (a12);
	    
	    \vertex[right=0.15cm of a6] (d1);
	    \vertex[right=0.15cm of a7] (d2);
	    \vertex[left=0.15cm of a1] (e3);
	    \vertex[left=0.15cm of a12] (e4);

            \tikzfeynmanset{every vertex={dot,minimum size=0.8mm}}

            \vertex[above=0cm of a1] (e1);
            \vertex[below=0cm of a6] (e2);    
            
            \vertex[above=0cm of a7] (f1);
            \vertex[below=0cm of a12] (f2);

                \diagram*[large]{	
                (a1)--(a6)--(a7)--(a12)--(a1),
                (a2)--(a11),
                (a3)--(a10),
                (a4)--(a9),
                (a5)--(a8),
                (d1)--(a6),
                (d2)--(a7),
                (e3)--(a1),
                (e4)--(a12),
                (a81)--(a51),
                }; 
            \end{feynman}
    \end{tikzpicture} \\
6L4P.b\end{tabular}}
\rule[-5pt]{0pt}{27pt}& \multirow{2}{*}{I}& \multirow{2}{*}{2911}& \multirow{2}{*}{0}& \multirow{2}{*}{0}& \multirow{2}{*}{}& \multirow{2}{*}{1\phantom{.0}}& \multirow{2}{*}{1200}& Re& \multirow{2}{*}{-18}& \texttt{ 0.90600} \multirow{2}{*}{\cite{Basso:2017jwq}}& \texttt{~1.09968~+/-~0.41729}& 0.464& 21.38& \multirow{2}{*}{21.38}\\
\rule[-5pt]{0pt}{27pt}& & & & & & & & Im& & \phantom{+}\texttt{0}\phantom{.00000}& &  &  & \\
\end{tabular}%
}%
}%
\caption{\label{tab:explore_higherloopB} Results for three- to six-loop ladder and fishnet integrals. The five- and six-loop configurations do not have any singular E-surfaces. See the main text for details.}%
\end{table}

\begin{table}[tbp]
\centering
\resizebox{\columnwidth}{!}{%
\texttt{%
\begin{tabular}{@{}llrrrlrrcllcrrr@{}}
\hline
Topology & Kin. & $\mathtt{N_{\text{C}}}$ & $\mathtt{N_{\text{E}}}$ & $\mathtt{N_{\text{S}}}$ & $\mathtt{L_{\text{max}}}$ & $\mathtt{N_{\text{p}} \ [10^9]}$ & $\mathtt{\sfrac{t}{p} \ [\mu \text{s}]}$ & Phase & Exp. & \multicolumn{1}{c}{Reference} & Numerical LTD & $\mathtt{\Delta \ [\sigma]}$ & $\mathtt{\Delta \ [\%]}$ & $\mathtt{\Delta \ [\%] |\cdot|}$ \\
\hline
\multirow{12}{*}{%
\begin{tabular}{@{}c@{}}
   \begin{tikzpicture}
    \begin{feynman}

        \tikzfeynmanset{every vertex={empty dot,minimum size=0mm}}
    \vertex (a1);

    \vertex[right=1cm of a1] (a3);
    
    \tikzfeynmanset{every vertex={empty dot,minimum size=0mm}}
    
    \vertex[right=0.5cm of a1] (a2);
    \vertex[above=0.5cm of a2] (a4);
    \vertex[below=0.5cm of a2] (a5);

    \vertex[left=0.433cm of a2] (b1);
    \vertex[left=0.433cm of a2] (b2);
    
    \vertex[right=0.433cm of a2] (b3);
    \vertex[right=0.433cm of a2] (b4);
    
    \vertex[left=0.433cm of a2] (d1);
    \vertex[left=0.433cm of a2] (d2);
    
    \vertex[right=0.433cm of a2] (d3);
    \vertex[right=0.433cm of a2] (d4);

    \vertex[above=0.255cm of b2] (c2);    
    \vertex[above=0.255cm of b3] (c3);
    \vertex[below=0.255cm of b2] (e2);   
    \vertex[below=0.255cm of b3] (e3);

    \tikzfeynmanset{every vertex={dot,minimum size=0.8mm}}
    
    \vertex[above=0.255cm of b1] (c1);
    \vertex[above=0.255cm of b4] (c4);    
    \vertex[below=0.255cm of b1] (e1);
    \vertex[below=0.255cm of b4] (e4);
    
    \tikzfeynmanset{every vertex={empty dot,minimum size=0mm}}
    
    \vertex[left=0.15cm of c1] (q1);
    \vertex[left=0.15cm of c2] (q2);
    \vertex[left=0.15cm of e1] (q3);
    \vertex[left=0.15cm of e2] (q4);
    \vertex[left=0.15cm of a1] (q5);

   \vertex[right=0.15cm of c3] (p1);
    \vertex[right=0.15cm of c4] (p2);
    \vertex[right=0.15cm of e3] (p3);
    \vertex[right=0.15cm of e4] (p4);
    \vertex[right=0.15cm of a3] (p5);
    
        \diagram*[large]{	
        (a1)--[quarter left](a4) -- [quarter left](a3),
        (a5) --[quarter right](a3), 
        (a5)--[quarter left](a1),        
               
        (c1) -- (q1),
        (e1) -- (q3),
        (c4) -- (p2),
        (e4) -- (p4),

        }; 
    \end{feynman}
    \end{tikzpicture}
     \\
1L4P\end{tabular}}
& \multirow{2}{*}{K1}& \multirow{2}{*}{4}& \multirow{2}{*}{5}& \multirow{2}{*}{1}& \multirow{2}{*}{$\mathtt{[5]}$}& \multirow{2}{*}{3\phantom{.0}}& \multirow{2}{*}{14}& Re& \multirow{2}{*}{-03}& \texttt{ 1.13116}& \texttt{~1.13123~+/-~0.00006}& 1.126& 0.006& \multirow{2}{*}{0.005}\\
& & & & & & & & Im& & \texttt{-0.55487}& \texttt{-0.55486~+/-~0.00005}& 0.163& 0.002& \\
\cline{2-15}
& \multirow{2}{*}{K2}& \multirow{2}{*}{4}& \multirow{2}{*}{5}& \multirow{2}{*}{1}& \multirow{2}{*}{$\mathtt{[5]}$}& \multirow{2}{*}{3\phantom{.0}}& \multirow{2}{*}{12}& Re& \multirow{2}{*}{-05}& \texttt{ 5.71928}& \texttt{~5.71929~+/-~0.00055}& 0.003& 3e-05& \multirow{2}{*}{0.005}\\
& & & & & & & & Im& & \texttt{-7.24005}& \texttt{-7.24055~+/-~0.00053}& 0.952& 0.007& \\
\cline{2-15}
& \multirow{2}{*}{K3}& \multirow{2}{*}{4}& \multirow{2}{*}{5}& \multirow{2}{*}{1}& \multirow{2}{*}{$\mathtt{[5]}$}& \multirow{2}{*}{3\phantom{.0}}& \multirow{2}{*}{12}& Re& \multirow{2}{*}{-06}& \texttt{ 1.55382}& \texttt{~1.55376~+/-~0.00012}& 0.545& 0.004& \multirow{2}{*}{0.005}\\
& & & & & & & & Im& & \texttt{-2.06994}& \texttt{-2.07005~+/-~0.00012}& 0.930& 0.005& \\
\cline{2-15}
& \multirow{2}{*}{K1$^*$}& \multirow{2}{*}{4}& \multirow{2}{*}{5}& \multirow{2}{*}{3}& \multirow{2}{*}{$\mathtt{[3, 3, 3]}$}& \multirow{2}{*}{3\phantom{.0}}& \multirow{2}{*}{16}& Re& \multirow{2}{*}{-03}& \texttt{ 1.85226}& \texttt{~1.85214~+/-~0.00012}& 1.069& 0.007& \multirow{2}{*}{0.004}\\
& & & & & & & & Im& & \texttt{-2.18400}& \texttt{-2.18397~+/-~0.00012}& 0.285& 0.002& \\
\cline{2-15}
& \multirow{2}{*}{K2$^*$}& \multirow{2}{*}{4}& \multirow{2}{*}{5}& \multirow{2}{*}{2}& \multirow{2}{*}{$\mathtt{[3, 3]}$}& \multirow{2}{*}{3\phantom{.0}}& \multirow{2}{*}{14}& Re& \multirow{2}{*}{-04}& \texttt{ 0.30270}& \texttt{~0.30272~+/-~0.00004}& 0.527& 0.007& \multirow{2}{*}{0.005}\\
& & & & & & & & Im& & \texttt{-1.08125}& \texttt{-1.08130~+/-~0.00004}& 1.313& 0.005& \\
\cline{2-15}
& \multirow{2}{*}{K3$^*$}& \multirow{2}{*}{4}& \multirow{2}{*}{3}& \multirow{2}{*}{1}& \multirow{2}{*}{$\mathtt{[3]}$}& \multirow{2}{*}{3\phantom{.0}}& \multirow{2}{*}{12}& Re& \multirow{2}{*}{-06}& \texttt{-0.17986}& \texttt{-0.17991~+/-~0.00005}& 1.054& 0.028& \multirow{2}{*}{0.007}\\
& & & & & & & & Im& & \texttt{-2.27578}& \texttt{-2.27593~+/-~0.00008}& 1.970& 0.007& \\
\hline
\multirow{12}{*}{%
\begin{tabular}{@{}c@{}}
    \begin{tikzpicture}
    \begin{feynman}

        \tikzfeynmanset{every vertex={empty dot,minimum size=0mm}}
    \vertex (a1);
       \tikzfeynmanset{every vertex={dot,minimum size=0.8mm}} 
    \vertex[right=1cm of a1] (a3);
    
    \tikzfeynmanset{every vertex={empty dot,minimum size=0mm}}
    
    \vertex[right=0.5cm of a1] (a2);
    \vertex[above=0.5cm of a2] (a4);
    \vertex[below=0.5cm of a2] (a5);

    \vertex[left=0.433cm of a2] (b1);
    \vertex[left=0.25cm of a2] (b2);
    
    \vertex[right=0.433cm of a2] (b3);
    \vertex[right=0.25cm of a2] (b4);
    
    \vertex[left=0.433cm of a2] (d1);
    \vertex[left=0.25cm of a2] (d2);
    
    \vertex[right=0.433cm of a2] (d3);
    \vertex[right=0.25cm of a2] (d4);

    \vertex[above=0.433cm of b2] (c2);    
    \vertex[above=0.255cm of b3] (c3);
    \vertex[below=0.433cm of b2] (e2);   
    \vertex[below=0.255cm of b3] (e3);

    \tikzfeynmanset{every vertex={dot,minimum size=0.8mm}}
    
    \vertex[above=0.255cm of b1] (c1);
    \vertex[above=0.433cm of b4] (c4);    
    \vertex[below=0.255cm of b1] (e1);
    \vertex[below=0.433cm of b4] (e4);
    
    \tikzfeynmanset{every vertex={empty dot,minimum size=0mm}}
    
    \vertex[left=0.15cm of c1] (q1);
    \vertex[left=0.15cm of c2] (q2);
    \vertex[left=0.15cm of e1] (q3);
    \vertex[left=0.15cm of e2] (q4);
    \vertex[left=0.15cm of a1] (q5);

   \vertex[right=0.15cm of c3] (p1);
    \vertex[right=0.15cm of c4] (p2);
    \vertex[right=0.15cm of e3] (p3);
    \vertex[right=0.15cm of e4] (p4);
    \vertex[right=0.15cm of a3] (p5);
    
        \diagram*[large]{	
        (a1)--[quarter left](a4) -- [quarter left](a3),
        (a5) --[quarter right](a3), 
        (a5)--[quarter left](a1),        
               
        (c1) -- (q1),
        (e1) -- (q3),
        (a3) -- (p5),
        (c4) -- (p2),
        (e4) -- (p4),

        }; 
    \end{feynman}
    \end{tikzpicture} \\
1L5P\end{tabular}}
& \multirow{2}{*}{K1}& \multirow{2}{*}{5}& \multirow{2}{*}{8}& \multirow{2}{*}{2}& \multirow{2}{*}{$\mathtt{[7, 7]}$}& \multirow{2}{*}{3\phantom{.0}}& \multirow{2}{*}{18}& Re& \multirow{2}{*}{-05}& \texttt{-1.90847}& \texttt{-1.90856~+/-~0.00074}& 0.120& 0.005& \multirow{2}{*}{0.006}\\
& & & & & & & & Im& & \texttt{-6.45346}& \texttt{-6.45306~+/-~0.00077}& 0.515& 0.006& \\
\cline{2-15}
& \multirow{2}{*}{K2}& \multirow{2}{*}{5}& \multirow{2}{*}{8}& \multirow{2}{*}{3}& \multirow{2}{*}{$\mathtt{[5, 5, 7]}$}& \multirow{2}{*}{3\phantom{.0}}& \multirow{2}{*}{18}& Re& \multirow{2}{*}{-06}& \texttt{-0.15108}& \texttt{-0.15137~+/-~0.00032}& 0.937& 0.197& \multirow{2}{*}{0.017}\\
& & & & & & & & Im& & \texttt{-1.80679}& \texttt{-1.80672~+/-~0.00033}& 0.210& 0.004& \\
\cline{2-15}
& \multirow{2}{*}{K3}& \multirow{2}{*}{5}& \multirow{2}{*}{8}& \multirow{2}{*}{3}& \multirow{2}{*}{$\mathtt{[5, 5, 7]}$}& \multirow{2}{*}{3\phantom{.0}}& \multirow{2}{*}{20}& Re& \multirow{2}{*}{-09}& \texttt{-0.66240}& \texttt{-0.66271~+/-~0.00032}& 0.957& 0.046& \multirow{2}{*}{0.034}\\
& & & & & & & & Im& & \texttt{-1.23531}& \texttt{-1.23567~+/-~0.00032}& 1.102& 0.029& \\
\cline{2-15}
& \multirow{2}{*}{K1$^*$}& \multirow{2}{*}{5}& \multirow{2}{*}{8}& \multirow{2}{*}{2}& \multirow{2}{*}{$\mathtt{[6, 6]}$}& \multirow{2}{*}{3\phantom{.0}}& \multirow{2}{*}{19}& Re& \multirow{2}{*}{-05}& \texttt{ 2.60399}& \texttt{~2.60394~+/-~0.00072}& 0.060& 0.002& \multirow{2}{*}{0.012}\\
& & & & & & & & Im& & \texttt{-7.94917}& \texttt{-7.95017~+/-~0.00076}& 1.320& 0.013& \\
\cline{2-15}
& \multirow{2}{*}{K2$^*$}& \multirow{2}{*}{5}& \multirow{2}{*}{8}& \multirow{2}{*}{3}& \multirow{2}{*}{$\mathtt{[4, 5, 5]}$}& \multirow{2}{*}{3\phantom{.0}}& \multirow{2}{*}{20}& Re& \multirow{2}{*}{-06}& \texttt{-0.48303}& \texttt{-0.48305~+/-~0.00059}& 0.034& 0.004& \multirow{2}{*}{0.009}\\
& & & & & & & & Im& & \texttt{-3.27695}& \texttt{-3.27664~+/-~0.00061}& 0.509& 0.009& \\
\cline{2-15}
& \multirow{2}{*}{K3$^*$}& \multirow{2}{*}{5}& \multirow{2}{*}{6}& \multirow{2}{*}{2}& \multirow{2}{*}{$\mathtt{[5, 5]}$}& \multirow{2}{*}{3\phantom{.0}}& \multirow{2}{*}{16}& Re& \multirow{2}{*}{-09}& \texttt{-1.21497}& \texttt{-1.21508~+/-~0.00020}& 0.560& 0.009& \multirow{2}{*}{0.006}\\
& & & & & & & & Im& & \texttt{-1.53129}& \texttt{-1.53126~+/-~0.00020}& 0.188& 0.002& \\
\end{tabular}%
}%
}%
\caption{\label{tab:Kseries_1loopA} Results for one-loop four-point and five-point topologies for scattering kinematics ($2 \rightarrow N$) for massless and massive propagators (indicated by a \texttt{*}). See the main text for details.}%
\end{table}

\begin{table}[tbp]
\centering
\resizebox{\columnwidth}{!}{%
\texttt{%
\begin{tabular}{@{}llrrrlrrcllcrrr@{}}
\hline
Topology & Kin. & $\mathtt{N_{\text{C}}}$ & $\mathtt{N_{\text{E}}}$ & $\mathtt{N_{\text{S}}}$ & $\mathtt{L_{\text{max}}}$ & $\mathtt{N_{\text{p}} \ [10^9]}$ & $\mathtt{\sfrac{t}{p} \ [\mu \text{s}]}$ & Phase & Exp. & \multicolumn{1}{c}{Reference} & Numerical LTD & $\mathtt{\Delta \ [\sigma]}$ & $\mathtt{\Delta \ [\%]}$ & $\mathtt{\Delta \ [\%] |\cdot|}$ \\
\hline
\multirow{12}{*}{%
\begin{tabular}{@{}c@{}}
   \begin{tikzpicture}
    \begin{feynman}
    \tikzfeynmanset{every vertex={dot,minimum size=0.8mm}}
    \vertex (a1);
    
    \vertex[right=1cm of a1] (a3);
    
    \tikzfeynmanset{every vertex={empty dot,minimum size=0mm}}
    
    \vertex[right=0.5cm of a1] (a2);
    \vertex[above=0.5cm of a2] (a4);
    \vertex[below=0.5cm of a2] (a5);

    \vertex[left=0.433cm of a2] (b1);
    \vertex[left=0.25cm of a2] (b2);
    
    \vertex[right=0.433cm of a2] (b3);
    \vertex[right=0.25cm of a2] (b4);
    
    \vertex[left=0.433cm of a2] (d1);
    \vertex[left=0.25cm of a2] (d2);
    
    \vertex[right=0.433cm of a2] (d3);
    \vertex[right=0.25cm of a2] (d4);

    \vertex[above=0.433cm of b2] (c2);    
    \vertex[above=0.255cm of b3] (c3);
    \vertex[below=0.433cm of b2] (e2);   
    \vertex[below=0.255cm of b3] (e3);
    
    \tikzfeynmanset{every vertex={dot,minimum size=0.8mm}}

    \vertex[above=0.433cm of b2] (c2);

    \vertex[above=0.433cm of b4] (c4);

    \vertex[below=0.433cm of b2] (e2);

    \vertex[below=0.433cm of b4] (e4);
    
    \tikzfeynmanset{every vertex={empty dot,minimum size=0mm}}
    
    \vertex[left=0.15cm of c1] (q1);
    \vertex[left=0.15cm of c2] (q2);
    \vertex[left=0.15cm of e1] (q3);
    \vertex[left=0.15cm of e2] (q4);
    \vertex[left=0.15cm of a1] (q5);

    \vertex[right=0.15cm of c3] (p1);
    \vertex[right=0.15cm of c4] (p2);
    \vertex[right=0.15cm of e3] (p3);
    \vertex[right=0.15cm of e4] (p4);
    \vertex[right=0.15cm of a3] (p5);
    
        \diagram*[large]{	
        (a1)--[quarter left](a4) -- [quarter left](a3),
        (a5) --[quarter right](a3), 
        (a5)--[quarter left](a1),

        (c2) -- (q2),
        (e2) -- (q4),
        (a1) -- (q5),
             
        (c4) -- (p2),
        (e4) -- (p4),
        (a3) -- (p5),
        }; 
    \end{feynman}
    \end{tikzpicture} \\
1L6P\end{tabular}}
& \multirow{2}{*}{K1}& \multirow{2}{*}{6}& \multirow{2}{*}{12}& \multirow{2}{*}{2}& \multirow{2}{*}{$\mathtt{[11, 9]}$}& \multirow{2}{*}{3\phantom{.0}}& \multirow{2}{*}{24}& Re& \multirow{2}{*}{-06}& \texttt{ 0.51025}& \texttt{~0.51018~+/-~0.00031}& 0.224& 0.014& \multirow{2}{*}{0.009}\\
& & & & & & & & Im& & \texttt{-1.54756}& \texttt{-1.54768~+/-~0.00032}& 0.380& 0.008& \\
\cline{2-15}
& \multirow{2}{*}{K2}& \multirow{2}{*}{6}& \multirow{2}{*}{12}& \multirow{2}{*}{5}& \multirow{2}{*}{$\mathtt{[8, 8, 8, 9, 10]}$}& \multirow{2}{*}{3\phantom{.0}}& \multirow{2}{*}{27}& Re& \multirow{2}{*}{-08}& \texttt{ 0.60440}& \texttt{~0.60407~+/-~0.00216}& 0.154& 0.055& \multirow{2}{*}{0.015}\\
& & & & & & & & Im& & \texttt{-6.96339}& \texttt{-6.96436~+/-~0.00213}& 0.457& 0.014& \\
\cline{2-15}
& \multirow{2}{*}{K3}& \multirow{2}{*}{6}& \multirow{2}{*}{12}& \multirow{2}{*}{3}& \multirow{2}{*}{$\mathtt{[8, 9, 10]}$}& \multirow{2}{*}{3\phantom{.0}}& \multirow{2}{*}{25}& Re& \multirow{2}{*}{-12}& \texttt{ 0.40660}& \texttt{~0.40655~+/-~0.00152}& 0.028& 0.010& \multirow{2}{*}{0.144}\\
& & & & & & & & Im& & \texttt{-2.51956}& \texttt{-2.51588~+/-~0.00157}& 2.343& 0.146& \\
\cline{2-15}
& \multirow{2}{*}{K1$^*$}& \multirow{2}{*}{6}& \multirow{2}{*}{12}& \multirow{2}{*}{4}& \multirow{2}{*}{$\mathtt{[8, 9, 9, 9]}$}& \multirow{2}{*}{3\phantom{.0}}& \multirow{2}{*}{27}& Re& \multirow{2}{*}{-06}& \texttt{ 1.30210}& \texttt{~1.30529~+/-~0.00289}& 1.107& 0.245& \multirow{2}{*}{0.192}\\
& & & & & & & & Im& & \texttt{-2.27354}& \texttt{-2.27744~+/-~0.00284}& 1.374& 0.171& \\
\cline{2-15}
& \multirow{2}{*}{K2$^*$}& \multirow{2}{*}{6}& \multirow{2}{*}{12}& \multirow{2}{*}{4}& \multirow{2}{*}{$\mathtt{[8, 8, 8, 9]}$}& \multirow{2}{*}{3\phantom{.0}}& \multirow{2}{*}{27}& Re& \multirow{2}{*}{-08}& \texttt{-2.19936}& \texttt{-2.20131~+/-~0.00241}& 0.809& 0.089& \multirow{2}{*}{0.032}\\
& & & & & & & & Im& & \texttt{-6.37931}& \texttt{-6.37841~+/-~0.00254}& 0.354& 0.014& \\
\cline{2-15}
& \multirow{2}{*}{K3$^*$}& \multirow{2}{*}{6}& \multirow{2}{*}{10}& \multirow{2}{*}{3}& \multirow{2}{*}{$\mathtt{[7, 8, 8]}$}& \multirow{2}{*}{3\phantom{.0}}& \multirow{2}{*}{22}& Re& \multirow{2}{*}{-12}& \texttt{-1.27979}& \texttt{-1.28057~+/-~0.00088}& 0.884& 0.061& \multirow{2}{*}{0.486}\\
& & & & & & & & Im& & \texttt{-2.22849}& \texttt{-2.21602~+/-~0.00088}& 14.09& 0.559& \\
\hline
\multirow{12}{*}{%
\begin{tabular}{@{}c@{}}
 

    \begin{tikzpicture}
    \begin{feynman}

        \tikzfeynmanset{every vertex={empty dot,minimum size=0mm}}
    \vertex (a1);
    \vertex[right=0.5cm of a1] (a2);
    \vertex[right=1cm of a1] (a3);
    \vertex[above=0.5cm of a2] (a4);
    \vertex[below=0.5cm of a2] (a5);
    
    \vertex[below=0.2cm of a4] (b1);    
    \vertex[below=0.2cm of b1] (b2);    
    \vertex[below=0.2cm of b2] (b3);    
    \vertex[below=0.2cm of b3] (b4);    

    \tikzfeynmanset{every vertex={dot,minimum size=0.8mm}}
    
    \vertex[right=0.4cm of b1] (v1);
    \vertex[right=0.49cm of b2] (v2);
    \vertex[right=0.4cm of b4] (v3);
    \vertex[right=0.49cm of b3] (v4);
    
    \vertex[left=0.4cm of b1] (vv1);
    \vertex[left=0.49cm of b2] (vv2);    
    \vertex[left=0.4cm of b4] (vv3);
    \vertex[left=0.49cm of b3] (vv4);
        
    \tikzfeynmanset{every vertex={empty dot,minimum size=0mm}}

    \vertex[right=0.15cm of v1] (e1);
    \vertex[right=0.15cm of v2] (e2);
    \vertex[right=0.15cm of v3] (e3);
    \vertex[right=0.15cm of v4] (e4);
    
    \vertex[left=0.15cm of vv1] (ee1);
    \vertex[left=0.15cm of vv2] (ee2);    
    \vertex[left=0.15cm of vv3] (ee3);
    \vertex[left=0.15cm of vv4] (ee4);
     
        \diagram*[large]{	
        (a1)--[quarter left](a4) -- [quarter left](a3),
        (a5) --[quarter right](a3), 
        (a5)--[quarter left](a1),        
               
	(v1)--(e1),
	(v2)--(e2),
	(v3)--(e3),
	(v4)--(e4),
				
	(vv1)--(ee1),
	(vv2)--(ee2),
	(vv3)--(ee3),
	(vv4)--(ee4),

        }; 
    \end{feynman}
    \end{tikzpicture}
     \\
1L8P\end{tabular}}
& \multirow{2}{*}{K1}& \multirow{2}{*}{8}& \multirow{2}{*}{23}& \multirow{2}{*}{2}& \multirow{2}{*}{$\mathtt{[22, 18]}$}& \multirow{2}{*}{3\phantom{.0}}& \multirow{2}{*}{37}& Re& \multirow{2}{*}{-10}& \texttt{ 5.09917}& \texttt{~5.10300~+/-~0.00400}& 0.958& 0.075& \multirow{2}{*}{0.086}\\
& & & & & & & & Im& & \texttt{-1.62799}& \texttt{-1.62544~+/-~0.00373}& 0.685& 0.157& \\
\cline{2-15}
& \multirow{2}{*}{K2}& \multirow{2}{*}{8}& \multirow{2}{*}{23}& \multirow{2}{*}{9}& \multirow{2}{*}{$\mathtt{[14, 15, 16, 15, 14, 16, 19, 19, 18]}$}& \multirow{2}{*}{3\phantom{.0}}& \multirow{2}{*}{47}& Re& \multirow{2}{*}{-12}& \texttt{ 4.20915}& \texttt{~4.21309~+/-~0.00421}& 0.934& 0.093& \multirow{2}{*}{0.134}\\
& & & & & & & & Im& & \texttt{-1.95289}& \texttt{-1.95771~+/-~0.00394}& 1.223& 0.247& \\
\cline{2-15}
& \multirow{2}{*}{K3}& \multirow{2}{*}{8}& \multirow{2}{*}{23}& \multirow{2}{*}{12}& \multirow{2}{*}{$\mathtt{[15, 15, 15, 14, 16, 18, 14, 18, 18, 16, 18, 18]}$}& \multirow{2}{*}{3\phantom{.0}}& \multirow{2}{*}{52}& Re& \multirow{2}{*}{-19}& \texttt{ 1.27379}& \texttt{~1.26931~+/-~0.00486}& 0.923& 0.352& \multirow{2}{*}{1.004}\\
& & & & & & & & Im& & \texttt{-0.82567}& \texttt{-0.84023~+/-~0.00503}& 2.898& 1.764& \\
\cline{2-15}
& \multirow{2}{*}{K1$^*$}& \multirow{2}{*}{8}& \multirow{2}{*}{23}& \multirow{2}{*}{4}& \multirow{2}{*}{$\mathtt{[20, 19, 19, 18]}$}& \multirow{2}{*}{3\phantom{.0}}& \multirow{2}{*}{37}& Re& \multirow{2}{*}{-09}& \texttt{-0.35693}& \texttt{-0.35626~+/-~0.00057}& 1.168& 0.187& \multirow{2}{*}{0.082}\\
& & & & & & & & Im& & \texttt{-1.46806}& \texttt{-1.46911~+/-~0.00058}& 1.822& 0.072& \\
\cline{2-15}
& \multirow{2}{*}{K2$^*$}& \multirow{2}{*}{8}& \multirow{2}{*}{23}& \multirow{2}{*}{7}& \multirow{2}{*}{$\mathtt{[14, 14, 16, 17, 18, 17, 18]}$}& \multirow{2}{*}{3\phantom{.0}}& \multirow{2}{*}{45}& Re& \multirow{2}{*}{-12}& \texttt{-1.14718}& \texttt{-1.16905~+/-~0.00794}& 2.754& 1.906& \multirow{2}{*}{1.004}\\
& & & & & & & & Im& & \texttt{-2.70587}& \texttt{-2.72569~+/-~0.00967}& 2.050& 0.732& \\
\cline{2-15}
& \multirow{2}{*}{K3$^*$}& \multirow{2}{*}{8}& \multirow{2}{*}{21}& \multirow{2}{*}{6}& \multirow{2}{*}{$\mathtt{[17, 17, 16, 15, 14, 14]}$}& \multirow{2}{*}{3\phantom{.0}}& \multirow{2}{*}{37}& Re& \multirow{2}{*}{-08}& \texttt{-0.57515}& \texttt{-0.57605~+/-~0.00196}& 0.459& 0.156& \multirow{2}{*}{0.048}\\
& & & & & & & & Im& & \texttt{-4.04221}& \texttt{-4.04047~+/-~0.00202}& 0.858& 0.043& \\
\end{tabular}%
}%
}%
\caption{\label{tab:Kseries_1loopB} Results for one-loop six-point and eight-point topologies for scattering kinematics ($2 \rightarrow N$) for massless and massive propagators (indicated by a \texttt{*}). See the main text for details.}%
\end{table}

\begin{table}[tbp]
\centering
\resizebox{\columnwidth}{!}{%
\texttt{%
\begin{tabular}{@{}llrrrlrrcllcrrr@{}}
\hline
Topology & Kin. & $\mathtt{N_{\text{C}}}$ & $\mathtt{N_{\text{E}}}$ & $\mathtt{N_{\text{S}}}$ & $\mathtt{L_{\text{max}}}$ & $\mathtt{N_{\text{p}} \ [10^9]}$ & $\mathtt{\sfrac{t}{p} \ [\mu \text{s}]}$ & Phase & Exp. & \multicolumn{1}{c}{Reference} & Numerical LTD & $\mathtt{\Delta \ [\sigma]}$ & $\mathtt{\Delta \ [\%]}$ & $\mathtt{\Delta \ [\%] |\cdot|}$ \\
\hline
\multirow{12}{*}{%
\begin{tabular}{@{}c@{}}
 

      \begin{tikzpicture}
            \begin{feynman}
                        
            \tikzfeynmanset{every vertex={dot,minimum size=1mm}}
           
            \tikzfeynmanset{every vertex={empty dot,minimum size=0mm}}
             \vertex (a1);
            \vertex[right=0.5cm of a1] (a2);
            \vertex[right=0.5cm of a2] (a3);
            \vertex[below=1cm of a3] (a4);
            
            \vertex[left=0.5cm of a4] (a5);
            \vertex[left=0.5cm of a5] (a6);
                          
            \tikzfeynmanset{every vertex={dot,minimum size=0.8mm}}
            
            \vertex[right=0cm of a1] (e1);        
            \vertex[right=0cm of a3] (e2);        
            \vertex[right=0cm of a4] (e3);        
            \vertex[right=0cm of a6] (e4);    
            
            \tikzfeynmanset{every vertex={empty dot,minimum size=0mm}}                
  
            \vertex[left=0.15cm of e1] (d1);        
            \vertex[right=0.15cm of e2] (d2);        
            \vertex[right=0.15cm of e3] (d3);        
            \vertex[left=0.15cm of e4] (d4);    
            
                \diagram*[large]{	
                (a1)--(a2)--(a3)--(a4)--(a5)--(a6)--(a1), 
                (a2)--(a5),
             		(e1)--(d1),
		(e2)--(d2),
		(e3)--(d3),
		(e4)--(d4),
                           }; 
            \end{feynman}
    \end{tikzpicture} \\
2L4P.b\end{tabular}}
& \multirow{2}{*}{K1}& \multirow{2}{*}{15}& \multirow{2}{*}{12}& \multirow{2}{*}{8}& \multirow{2}{*}{$\mathtt{[12]}$}& \multirow{2}{*}{3\phantom{.0}}& \multirow{2}{*}{53}& Re& \multirow{2}{*}{-06}& \texttt{-1.08406} \multirow{2}{*}{\cite{Usyukina:1992jd}}& \texttt{-1.08656~+/-~0.00127}& 1.971& 0.230& \multirow{2}{*}{0.090}\\
& & & & & & & & Im& & \texttt{ 2.86821}& \texttt{~2.86702~+/-~0.00125}& 0.951& 0.041& \\
\cline{2-15}
& \multirow{2}{*}{K2}& \multirow{2}{*}{15}& \multirow{2}{*}{10}& \multirow{2}{*}{8}& \multirow{2}{*}{$\mathtt{[10]}$}& \multirow{2}{*}{3\phantom{.0}}& \multirow{2}{*}{55}& Re& \multirow{2}{*}{-08}& \texttt{ 3.11053} \multirow{2}{*}{\cite{Usyukina:1992jd}}& \texttt{~3.09646~+/-~0.00696}& 2.021& 0.452& \multirow{2}{*}{0.140}\\
& & & & & & & & Im& & \texttt{ 9.53885}& \texttt{~9.53952~+/-~0.00706}& 0.094& 0.007& \\
\cline{2-15}
& \multirow{2}{*}{K3}& \multirow{2}{*}{15}& \multirow{2}{*}{10}& \multirow{2}{*}{8}& \multirow{2}{*}{$\mathtt{[10]}$}& \multirow{2}{*}{3\phantom{.0}}& \multirow{2}{*}{56}& Re& \multirow{2}{*}{-10}& \texttt{ 1.70372} \multirow{2}{*}{\cite{Usyukina:1992jd}}& \texttt{~1.70253~+/-~0.00285}& 0.419& 0.070& \multirow{2}{*}{0.025}\\
& & & & & & & & Im& & \texttt{ 4.56497}& \texttt{~4.56488~+/-~0.00291}& 0.031& 0.002& \\
\cline{2-15}
& \multirow{2}{*}{K1$^*$}& \multirow{2}{*}{15}& \multirow{2}{*}{9}& \multirow{2}{*}{11}& \multirow{2}{*}{$\mathtt{[7, 8]}$}& \multirow{2}{*}{3\phantom{.0}}& \multirow{2}{*}{62}& Re& \multirow{2}{*}{-06}& \texttt{ 2.802\phantom{0}~+/-~0.008\phantom{}} \multirow{2}{*}{\cite{Borowka:2017idc}}& \texttt{~2.80094~+/-~0.00023}&  & 0.008& \multirow{2}{*}{0.008}\\
& & & & & & & & Im& & \texttt{ 3.345\phantom{0}~+/-~0.008\phantom{0}}& \texttt{~3.34866~+/-~0.00025}&  & 0.007& \\
\cline{2-15}
& \multirow{2}{*}{K2$^*$}& \multirow{2}{*}{15}& \multirow{2}{*}{6}& \multirow{2}{*}{13}& \multirow{2}{*}{$\mathtt{[4, 4]}$}& \multirow{2}{*}{3\phantom{.0}}& \multirow{2}{*}{77}& Re& \multirow{2}{*}{-08}& \texttt{ 7.9\phantom{000}~+/-~0.7\phantom{00}} \multirow{2}{*}{\cite{Borowka:2017idc}}& \texttt{~8.15559~+/-~0.00123}&  & 0.015& \multirow{2}{*}{0.017}\\
& & & & & & & & Im& & \texttt{ 6.9\phantom{000}~+/-~0.7\phantom{000}}& \texttt{~6.10277~+/-~0.00124}&  & 0.020& \\
\cline{2-15}
& \multirow{2}{*}{K3$^*$}& \multirow{2}{*}{15}& \multirow{2}{*}{7}& \multirow{2}{*}{8}& \multirow{2}{*}{$\mathtt{[7]}$}& \multirow{2}{*}{3\phantom{.0}}& \multirow{2}{*}{55}& Re& \multirow{2}{*}{-10}& \texttt{ 3.1\phantom{000}~+/-~0.1\phantom{00}} \multirow{2}{*}{\cite{Borowka:2017idc}}& \texttt{~3.10306~+/-~0.00021}&  & 0.007& \multirow{2}{*}{0.009}\\
& & & & & & & & Im& & \texttt{ 0.1\phantom{000}~+/-~0.1\phantom{000}}& \texttt{~0.09376~+/-~0.00020}&  & 0.212& \\
\hline
\multirow{12}{*}{%
\begin{tabular}{@{}c@{}}
  
 
    \begin{tikzpicture}
    \begin{feynman}

        \tikzfeynmanset{every vertex={empty dot,minimum size=0mm}}
    \vertex (a1);
       \tikzfeynmanset{every vertex={dot,minimum size=0.8mm}} 
    \vertex[right=1cm of a1] (a3);
    
    \tikzfeynmanset{every vertex={empty dot,minimum size=0mm}}
    
    \vertex[right=0.5cm of a1] (a2);
    \vertex[above=0.5cm of a2] (a4);
    \vertex[below=0.5cm of a2] (a5);

    \vertex[left=0.433cm of a2] (b1);
    \vertex[left=0.25cm of a2] (b2);
    
    \vertex[right=0.433cm of a2] (b3);
    \vertex[right=0.25cm of a2] (b4);
    
    \vertex[left=0.433cm of a2] (d1);
    \vertex[left=0.25cm of a2] (d2);
    
    \vertex[right=0.433cm of a2] (d3);
    \vertex[right=0.25cm of a2] (d4);
    
    \tikzfeynmanset{every vertex={dot,minimum size=0.8mm}}
    
    \vertex[above=0.255cm of b1] (c1);
    \vertex[above=0.433cm of b4] (c4);    
    \vertex[below=0.255cm of b1] (e1);
    \vertex[below=0.433cm of b4] (e4);
    
    \tikzfeynmanset{every vertex={empty dot,minimum size=0mm}}
    
    \vertex[left=0.15cm of c1] (q1);
    \vertex[left=0.15cm of c2] (q2);
    \vertex[left=0.15cm of e1] (q3);
    \vertex[left=0.15cm of e2] (q4);
    \vertex[left=0.15cm of a1] (q5);

    \vertex[right=0.15cm of c4] (p2);
    \vertex[right=0.15cm of e4] (p4);
    \vertex[right=0.15cm of a3] (p5);
    
        \diagram*[large]{	
        (a1)--[quarter left](a4) -- [quarter left](a3),
        (a5) --[quarter right](a3), 
        (a5)--[quarter left](a1),        
               
        (c1) -- (q1),
        (e1) -- (q3),
        (a3) -- (p5),
        (c4) -- (p2),
        (e4) -- (p4),
        (a4)--(a5),

        }; 
    \end{feynman}
    \end{tikzpicture} \\
2L5P\end{tabular}}
& \multirow{2}{*}{K1}& \multirow{2}{*}{19}& \multirow{2}{*}{17}& \multirow{2}{*}{14}& \multirow{2}{*}{$\mathtt{[16, 16]}$}& \multirow{2}{*}{3\phantom{.0}}& \multirow{2}{*}{80}& Re& \multirow{2}{*}{-07}& \multicolumn{1}{c}{n/a}& \texttt{~0.27368~+/-~0.00131}&  & 0.479& \multirow{2}{*}{0.125}\\
& & & & & & & & Im& & \multicolumn{1}{c}{n/a}& \texttt{~1.44760~+/-~0.00129}&  & 0.089& \\
\cline{2-15}
& \multirow{2}{*}{K2}& \multirow{2}{*}{19}& \multirow{2}{*}{13}& \multirow{2}{*}{19}& \multirow{2}{*}{$\mathtt{[8, 8, 12]}$}& \multirow{2}{*}{3\phantom{.0}}& \multirow{2}{*}{86}& Re& \multirow{2}{*}{-09}& \multicolumn{1}{c}{n/a}& \texttt{~1.08568~+/-~0.00342}&  & 0.315& \multirow{2}{*}{0.230}\\
& & & & & & & & Im& & \multicolumn{1}{c}{n/a}& \texttt{~1.78725~+/-~0.00339}&  & 0.190& \\
\cline{2-15}
& \multirow{2}{*}{K3}& \multirow{2}{*}{19}& \multirow{2}{*}{13}& \multirow{2}{*}{19}& \multirow{2}{*}{$\mathtt{[8, 8, 12]}$}& \multirow{2}{*}{3\phantom{.0}}& \multirow{2}{*}{86}& Re& \multirow{2}{*}{-13}& \multicolumn{1}{c}{n/a}& \texttt{~2.09848~+/-~0.00648}&  & 0.309& \multirow{2}{*}{0.313}\\
& & & & & & & & Im& & \multicolumn{1}{c}{n/a}& \texttt{~2.04022~+/-~0.00648}&  & 0.318& \\
\cline{2-15}
& \multirow{2}{*}{K1$^*$}& \multirow{2}{*}{19}& \multirow{2}{*}{14}& \multirow{2}{*}{16}& \multirow{2}{*}{$\mathtt{[13, 12]}$}& \multirow{2}{*}{3\phantom{.0}}& \multirow{2}{*}{80}& Re& \multirow{2}{*}{-07}& \multicolumn{1}{c}{n/a}& \texttt{~1.51586~+/-~0.00027}&  & 0.018& \multirow{2}{*}{0.019}\\
& & & & & & & & Im& & \multicolumn{1}{c}{n/a}& \texttt{~1.31451~+/-~0.00027}&  & 0.021& \\
\cline{2-15}
& \multirow{2}{*}{K2$^*$}& \multirow{2}{*}{19}& \multirow{2}{*}{10}& \multirow{2}{*}{20}& \multirow{2}{*}{$\mathtt{[8, 9]}$}& \multirow{2}{*}{3\phantom{.0}}& \multirow{2}{*}{97}& Re& \multirow{2}{*}{-09}& \multicolumn{1}{c}{n/a}& \texttt{~1.97798~+/-~0.01394}&  & 0.705& \multirow{2}{*}{0.799}\\
& & & & & & & & Im& & \multicolumn{1}{c}{n/a}& \texttt{~1.13209~+/-~0.01173}&  & 1.036& \\
\cline{2-15}
& \multirow{2}{*}{K3$^*$}& \multirow{2}{*}{19}& \multirow{2}{*}{12}& \multirow{2}{*}{18}& \multirow{2}{*}{$\mathtt{[10, 10, 10]}$}& \multirow{2}{*}{3\phantom{.0}}& \multirow{2}{*}{84}& Re& \multirow{2}{*}{-13}& \multicolumn{1}{c}{n/a}& \texttt{~2.00638~+/-~0.00061}&  & 0.030& \multirow{2}{*}{0.043}\\
& & & & & & & & Im& & \multicolumn{1}{c}{n/a}& \texttt{-0.08277~+/-~0.00060}&  & 0.730& \\
\end{tabular}%
}%
}%
\caption{\label{tab:Kseries_2loopA} Results for two-loop topologies for scattering kinematics ($2 \rightarrow N$) for massless and massive propagators (indicated by a \texttt{*}). When there is no reference result, $\Delta [\%]$ and $\Delta [\%] |\cdot|$ refer to the Monte-Carlo accuracy relative to the central value. See the main text for details.}%
\end{table}

\begin{table}[tbp]
\centering
\resizebox{\columnwidth}{!}{%
\texttt{%
\begin{tabular}{@{}llrrrlrrcllcrrr@{}}
\hline
Topology & Kin. & $\mathtt{N_{\text{C}}}$ & $\mathtt{N_{\text{E}}}$ & $\mathtt{N_{\text{S}}}$ & $\mathtt{L_{\text{max}}}$ & $\mathtt{N_{\text{p}} \ [10^9]}$ & $\mathtt{\sfrac{t}{p} \ [\mu \text{s}]}$ & Phase & Exp. & \multicolumn{1}{c}{Reference} & Numerical LTD & $\mathtt{\Delta \ [\sigma]}$ & $\mathtt{\Delta \ [\%]}$ & $\mathtt{\Delta \ [\%] |\cdot|}$ \\
\hline
\multirow{4}{*}{%
\begin{tabular}{@{}c@{}}
    \begin{tikzpicture}
    \begin{feynman}

    \tikzfeynmanset{every vertex={empty dot,minimum size=0mm}}
    \vertex (a1);
    
    \vertex[right=1cm of a1] (a3);
    \vertex[right=0.5cm of a1] (a2);
    \vertex[above=0.5cm of a2] (a4);
    \vertex[below=0.5cm of a2] (a5);
    
    \vertex[left=0.433cm of a2] (b1);
    \vertex[right=0.433cm of a2] (b2);
    
    \vertex[above=0.cm of a1] (c1);
    \vertex[above=0.25cm of b2] (c2);
    \vertex[below=0.25cm of b1] (d1);
    \vertex[below=0.cm of a3] (d2);
    \vertex[below=0.16666cm of a4] (l1);
    \vertex[below=0.33333cm of a4] (l2);
    \vertex[below=0.5cm of a4] (l3);
    \vertex[below=0.66666cm of a4] (l4);
    \vertex[below=0.83333cm of a4] (l5);
    
    \vertex[left=0.15cm of c1] (ec1);

    \vertex[right=0.15cm of d2] (ed2);
    \vertex[right=0.15cm of l1] (el1);
    \vertex[right=0.15cm of l2] (el2);
    \vertex[right=0.15cm of l3] (el3);
    \vertex[right=0.15cm of l4] (el4);
    \vertex[right=0.15cm of l5] (el5);

    \tikzfeynmanset{every vertex={dot,minimum size=0.8mm}}

    \vertex[below=0.cm of a3] (d2);
    
    \vertex[below=0.cm of l1] (l1);
    \vertex[below=0.cm of l2] (l2);
    \vertex[below=0.cm of l3] (l3);
    \vertex[below=0.cm of l4] (l4);
    \vertex[below=0.cm of l5] (l5);
    
        \diagram*[large]{	
        (a1)--[quarter left](a4) -- [quarter left](a3),
        (a5) --[quarter right](a3), 
        (a5)--[quarter left](a1),
        (a4) -- (a5),

        (d2) -- (ed2),
        (l1) -- (el1),
        (l1) -- (el1),
        (l2) -- (el2),
        (l3) -- (el3),
        (l4) -- (el4),
        (l5) -- (el5),
        }; 
    \end{feynman}
    \end{tikzpicture}
    \\
2L6P.a\end{tabular}}
& \multirow{2}{*}{K1}& \multirow{2}{*}{20}& \multirow{2}{*}{20}& \multirow{2}{*}{15}& \multirow{2}{*}{$\mathtt{[19, 14]}$}& \multirow{2}{*}{3\phantom{.0}}& \multirow{2}{*}{100}& Re& \multirow{2}{*}{-09}& \multicolumn{1}{c}{n/a}& \texttt{~4.58688~+/-~0.05132}&  & 1.119& \multirow{2}{*}{1.059}\\
& & & & & & & & Im& & \multicolumn{1}{c}{n/a}& \texttt{~5.04144~+/-~0.05075}&  & 1.007& \\
\cline{2-15}
& \multirow{2}{*}{K1$^*$}& \multirow{2}{*}{20}& \multirow{2}{*}{17}& \multirow{2}{*}{24}& \multirow{2}{*}{$\mathtt{[12, 13, 13, 13, 13]}$}& \multirow{2}{*}{3\phantom{.0}}& \multirow{2}{*}{116}& Re& \multirow{2}{*}{-09}& \multicolumn{1}{c}{n/a}& \texttt{-1.04316~+/-~0.35247}&  & 33.79& \multirow{2}{*}{10.99}\\
& & & & & & & & Im& & \multicolumn{1}{c}{n/a}& \texttt{-4.42468~+/-~0.35421}&  & 8.005& \\
\hline
\multirow{4}{*}{%
\begin{tabular}{@{}c@{}}
  

\begin{tikzpicture}
    \begin{feynman}

    \tikzfeynmanset{every vertex={empty dot,minimum size=0mm}}
    \vertex (a1);
    
    \vertex[right=1cm of a1] (a3);
    \vertex[right=0.5cm of a1] (a2);
    \vertex[above=0.5cm of a2] (a4);
    \vertex[below=0.5cm of a2] (a5);
    
    \vertex[left=0.433cm of a2] (b1);
    \vertex[right=0.433cm of a2] (b2);
    
    \vertex[above=0.25cm of b1] (c1);
    \vertex[above=0.25cm of b2] (c2);
    \vertex[below=0.25cm of b1] (d1);
    \vertex[below=0.25cm of b2] (d2);
    
    \vertex[below=0.2cm of a4] (l1);
    \vertex[below=0.2cm of l1] (l2);
    \vertex[below=0.2cm of l2] (l3);
    \vertex[below=0.2cm of l3] (l4);
    
    \vertex[left=0.15cm of c1] (ec1);
    \vertex[right=0.15cm of c2] (ec2);
    \vertex[left=0.15cm of d1] (ed1);
    \vertex[right=0.15cm of d2] (ed2);
    \vertex[right=0.15cm of l1] (el1);
    \vertex[right=0.15cm of l2] (el2); 
    \vertex[right=0.15cm of l3] (el3);
    \vertex[right=0.15cm of l4] (el4);

    \vertex[above=0.25cm of b1] (c1);   
    \vertex[below=0.25cm of b1] (d1);
    \tikzfeynmanset{every vertex={dot,minimum size=0.8mm}}

    \vertex[above=0.25cm of b2] (c2);

    \vertex[below=0.25cm of b2] (d2);
    
    \vertex[above=0.cm of l1] (l1);
    \vertex[below=0.cm of l2] (l2);
    \vertex[below=0.cm of l3] (l3);
    \vertex[below=0.cm of l4] (l4);
    
        \diagram*[large]{	
        (a1)--[quarter left](a4) -- [quarter left](a3),
        (a5) --[quarter right](a3), 
        (a5)--[quarter left](a1),
        (a4) -- (a5),
        
        (c2) -- (ec2),
        (d2) -- (ed2),
        (l1) -- (el1),
        (l2) -- (el2),
        (l3)--(el3),
	(l4)--(el4),
    
        }; 
    \end{feynman}
    \end{tikzpicture} \\
2L6P.b\end{tabular}}
& \multirow{2}{*}{K1}& \multirow{2}{*}{23}& \multirow{2}{*}{23}& \multirow{2}{*}{15}& \multirow{2}{*}{$\mathtt{[22, 19]}$}& \multirow{2}{*}{3\phantom{.0}}& \multirow{2}{*}{91}& Re& \multirow{2}{*}{-09}& \multicolumn{1}{c}{n/a}& \texttt{~1.17336~+/-~0.00888}&  & 0.757& \multirow{2}{*}{0.303}\\
& & & & & & & & Im& & \multicolumn{1}{c}{n/a}& \texttt{~3.99809~+/-~0.00896}&  & 0.224& \\
\cline{2-15}
& \multirow{2}{*}{K1$^*$}& \multirow{2}{*}{23}& \multirow{2}{*}{20}& \multirow{2}{*}{20}& \multirow{2}{*}{$\mathtt{[18, 17, 18]}$}& \multirow{2}{*}{3\phantom{.0}}& \multirow{2}{*}{103}& Re& \multirow{2}{*}{-09}& \multicolumn{1}{c}{n/a}& \texttt{~5.35217~+/-~0.00153}&  & 0.029& \multirow{2}{*}{0.033}\\
& & & & & & & & Im& & \multicolumn{1}{c}{n/a}& \texttt{~3.81579~+/-~0.00150}&  & 0.039& \\
\hline
\multirow{4}{*}{%
\begin{tabular}{@{}c@{}}
  

    \begin{tikzpicture}
    \begin{feynman}
    \tikzfeynmanset{every vertex={dot,minimum size=0.8mm}}
    \vertex (a1);
    
    \vertex[right=1cm of a1] (a3);
    
    \tikzfeynmanset{every vertex={empty dot,minimum size=0mm}}
    
    \vertex[right=0.5cm of a1] (a2);
    \vertex[above=0.5cm of a2] (a4);
    \vertex[below=0.5cm of a2] (a5);

    \vertex[left=0.433cm of a2] (b1);
    \vertex[left=0.25cm of a2] (b2);
    
    \vertex[right=0.433cm of a2] (b3);
    \vertex[right=0.25cm of a2] (b4);
    
    \vertex[left=0.433cm of a2] (d1);
    \vertex[left=0.25cm of a2] (d2);
    
    \vertex[right=0.433cm of a2] (d3);
    \vertex[right=0.25cm of a2] (d4);
    
    \tikzfeynmanset{every vertex={dot,minimum size=0.8mm}}

    \vertex[above=0.433cm of b2] (c2);

    \vertex[above=0.433cm of b4] (c4);

    \vertex[below=0.433cm of b2] (e2);

    \vertex[below=0.433cm of b4] (e4);
    
    \tikzfeynmanset{every vertex={empty dot,minimum size=0mm}}
    
    \vertex[left=0.15cm of c1] (q1);
    \vertex[left=0.15cm of c2] (q2);
    \vertex[left=0.15cm of e1] (q3);
    \vertex[left=0.15cm of e2] (q4);
    \vertex[left=0.15cm of a1] (q5);

    \vertex[right=0.15cm of c4] (p2);
    \vertex[right=0.15cm of e4] (p4);
    \vertex[right=0.15cm of a3] (p5);
    
        \diagram*[large]{	
        (a1)--[quarter left](a4) -- [quarter left](a3),
        (a5) --[quarter right](a3), 
        (a5)--[quarter left](a1),
        (a4)--(a5),

        (c2) -- (q2),
        (e2) -- (q4),
        (a1) -- (q5),
             
        (c4) -- (p2),
        (e4) -- (p4),
        (a3) -- (p5),
        }; 
    \end{feynman}
    \end{tikzpicture} \\
2L6P.c\end{tabular}}
& \multirow{2}{*}{K1}& \multirow{2}{*}{24}& \multirow{2}{*}{22}& \multirow{2}{*}{16}& \multirow{2}{*}{$\mathtt{[20, 21]}$}& \multirow{2}{*}{3\phantom{.0}}& \multirow{2}{*}{89}& Re& \multirow{2}{*}{-09}& \multicolumn{1}{c}{n/a}& \texttt{~4.90974~+/-~0.01407}&  & 0.286& \multirow{2}{*}{0.375}\\
& & & & & & & & Im& & \multicolumn{1}{c}{n/a}& \texttt{-2.13974~+/-~0.01434}&  & 0.670& \\
\cline{2-15}
& \multirow{2}{*}{K1$^*$}& \multirow{2}{*}{24}& \multirow{2}{*}{20}& \multirow{2}{*}{22}& \multirow{2}{*}{$\mathtt{[17, 17, 17, 17]}$}& \multirow{2}{*}{3\phantom{.0}}& \multirow{2}{*}{108}& Re& \multirow{2}{*}{-08}& \multicolumn{1}{c}{n/a}& \texttt{~1.05934~+/-~0.15850}&  & 14.96& \multirow{2}{*}{14.87}\\
& & & & & & & & Im& & \multicolumn{1}{c}{n/a}& \texttt{~1.03698~+/-~0.15312}&  & 14.77& \\
\hline
\multirow{4}{*}{%
\begin{tabular}{@{}c@{}}
    \begin{tikzpicture}
    \begin{feynman}

    \tikzfeynmanset{every vertex={empty dot,minimum size=0mm}}
    \vertex (a1);
    
    \vertex[right=1cm of a1] (a3);
    \vertex[right=0.5cm of a1] (a2);
    \vertex[above=0.5cm of a2] (a4);
    \vertex[below=0.5cm of a2] (a5);
    
    \vertex[left=0.433cm of a2] (b1);
    \vertex[right=0.433cm of a2] (b2);
    
    \vertex[above=0.cm of a1] (c1);
    \vertex[above=0.25cm of b2] (c2);
    \vertex[below=0.25cm of b1] (d1);
    \vertex[below=0.cm of a3] (d2);
    \vertex[below=0.2cm of a4] (l1);
    \vertex[below=0.4cm of a4] (l2);
    \vertex[below=0.6cm of a4] (l3);
    \vertex[below=0.8cm of a4] (l4);
    
    \vertex[left=0.15cm of c1] (ec1);

    \vertex[right=0.15cm of d2] (ed2);
    \vertex[right=0.15cm of l1] (el1);
    \vertex[right=0.15cm of l2] (el2);
    \vertex[right=0.15cm of l3] (el3);
    \vertex[right=0.15cm of l4] (el4);

    \tikzfeynmanset{every vertex={dot,minimum size=0.8mm}}
    
    \vertex[above=0.cm of a1] (c1);

    \vertex[below=0.cm of a3] (d2);
    
    \vertex[below=0.2cm of a4] (l1);
    \vertex[below=0.4cm of a4] (l2);
    \vertex[below=0.6cm of a4] (l3);
    \vertex[below=0.8cm of a4] (l4);
    
        \diagram*[large]{	
        (a1)--[quarter left](a4) -- [quarter left](a3),
        (a5) --[quarter right](a3), 
        (a5)--[quarter left](a1),
        (a4) -- (a5),
    
        (c1) -- (ec1),

        (d2) -- (ed2),
        (l1) -- (el1),
        (l1) -- (el1),
        (l2) -- (el2),
        (l3) -- (el3),
        (l4) -- (el4),
        }; 
    \end{feynman}
    \end{tikzpicture} \\
2L6P.d\end{tabular}}
& \multirow{2}{*}{K1}& \multirow{2}{*}{24}& \multirow{2}{*}{20}& \multirow{2}{*}{26}& \multirow{2}{*}{$\mathtt{[16, 7, 14, 14, 4]}$}& \multirow{2}{*}{3\phantom{.0}}& \multirow{2}{*}{136}& Re& \multirow{2}{*}{-08}& \multicolumn{1}{c}{n/a}& \texttt{~1.90487~+/-~0.05753}&  & 3.020& \multirow{2}{*}{2.017}\\
& & & & & & & & Im& & \multicolumn{1}{c}{n/a}& \texttt{-3.55267~+/-~0.05746}&  & 1.617& \\
\cline{2-15}
& \multirow{2}{*}{K1$^*$}& \multirow{2}{*}{24}& \multirow{2}{*}{17}& \multirow{2}{*}{30}& \multirow{2}{*}{$\mathtt{[13, 12, 12, 12, 2]}$}& \multirow{2}{*}{3\phantom{.0}}& \multirow{2}{*}{144}& Re& \multirow{2}{*}{-08}& \multicolumn{1}{c}{n/a}& \texttt{-2.97419~+/-~0.00961}&  & 0.323& \multirow{2}{*}{0.367}\\
& & & & & & & & Im& & \multicolumn{1}{c}{n/a}& \texttt{-2.18847~+/-~0.00957}&  & 0.437& \\
\hline
\multirow{4}{*}{%
\begin{tabular}{@{}c@{}}
 \begin{tikzpicture}
    \begin{feynman}

    \tikzfeynmanset{every vertex={empty dot,minimum size=0mm}}
    \vertex (a1);
    
    \vertex[right=1cm of a1] (a3);
    \vertex[right=0.5cm of a1] (a2);
    \vertex[above=0.5cm of a2] (a4);
    \vertex[below=0.5cm of a2] (a5);
    
    \vertex[left=0.433cm of a2] (b1);
    \vertex[right=0.433cm of a2] (b2);
    
    \vertex[left=0.cm of a1] (c1);
    \vertex[above=0.25cm of b2] (c2);
    \vertex[below=0.cm of a2] (d1);
    \vertex[below=0.25cm of b2] (d2);
    \vertex[above=0.3333cm of a2] (l1);
    \vertex[below=0.3333cm of a2] (l2);
    
    \vertex[left=0.15cm of c1] (ec1);
    \vertex[right=0.15cm of c2] (ec2);
    \vertex[right=0.15cm of d1] (ed1);
    \vertex[right=0.15cm of d2] (ed2);
    \vertex[right=0.15cm of l1] (el1);
    \vertex[right=0.15cm of l2] (el2);     
    
    \tikzfeynmanset{every vertex={dot,minimum size=0.8mm}}
    
    \vertex[above=0.cm of a1] (c1);
    \vertex[above=0.25cm of b2] (c2);
    
    \vertex[below=0.cm of a2] (d1);
    \vertex[below=0.25cm of b2] (d2);
    
    \vertex[above=0.3333cm of a2] (l1);
    \vertex[below=0.3333cm of a2] (l2);
    
        \diagram*[large]{	
        (a1)--[quarter left](a4) -- [quarter left](a3),
        (a5) --[quarter right](a3), 
        (a5)--[quarter left](a1),
        (a4) -- (a5),
        
        (c1) -- (ec1),
        (c2) -- (ec2),
	(d1)--(ed1),
        (d2) -- (ed2),
        (l1) -- (el1),
        (l2) -- (el2),

        }; 
    \end{feynman}
    \end{tikzpicture} \\
2L6P.e\end{tabular}}
& \multirow{2}{*}{K1}& \multirow{2}{*}{26}& \multirow{2}{*}{21}& \multirow{2}{*}{34}& \multirow{2}{*}{$\mathtt{[16, 9, 9, 14, 15, 9, 7]}$}& \multirow{2}{*}{3\phantom{.0}}& \multirow{2}{*}{163}& Re& \multirow{2}{*}{-07}& \multicolumn{1}{c}{n/a}& \texttt{~2.87833~+/-~0.00951}&  & 0.330& \multirow{2}{*}{0.386}\\
& & & & & & & & Im& & \multicolumn{1}{c}{n/a}& \texttt{~1.99937~+/-~0.00961}&  & 0.481& \\
\cline{2-15}
& \multirow{2}{*}{K1$^*$}& \multirow{2}{*}{26}& \multirow{2}{*}{18}& \multirow{2}{*}{43}& \multirow{2}{*}{$\mathtt{[13, 12, 7, 7, 12, 12, 12, 12, 7, 5]}$}& \multirow{2}{*}{3\phantom{.0}}& \multirow{2}{*}{172}& Re& \multirow{2}{*}{-07}& \multicolumn{1}{c}{n/a}& \texttt{~1.67332~+/-~0.00578}&  & 0.346& \multirow{2}{*}{0.482}\\
& & & & & & & & Im& & \multicolumn{1}{c}{n/a}& \texttt{-0.21788~+/-~0.00571}&  & 2.620& \\
\hline
\multirow{4}{*}{%
\begin{tabular}{@{}c@{}}
   
 
 \begin{tikzpicture}
    \begin{feynman}

    \tikzfeynmanset{every vertex={empty dot,minimum size=0mm}}
    \vertex (a1);
    
    \vertex[right=1cm of a1] (a3);
    \vertex[right=0.5cm of a1] (a2);
    \vertex[above=0.5cm of a2] (a4);
    \vertex[below=0.5cm of a2] (a5);
    
    \vertex[left=0.433cm of a2] (b1);
    \vertex[right=0.433cm of a2] (b2);
    
    \vertex[above=0.25cm of b1] (c1);
    \vertex[above=0.25cm of b2] (c2);
    \vertex[below=0.25cm of b1] (d1);
    \vertex[below=0.25cm of b2] (d2);
    \vertex[above=0.3333cm of a2] (l1);
    \vertex[below=0.3333cm of a2] (l2);
    
    \vertex[left=0.15cm of c1] (ec1);
    \vertex[right=0.15cm of c2] (ec2);
    \vertex[left=0.15cm of d1] (ed1);
    \vertex[right=0.15cm of d2] (ed2);
    \vertex[right=0.15cm of l1] (el1);
    \vertex[right=0.15cm of l2] (el2);     
    
    \tikzfeynmanset{every vertex={dot,minimum size=0.8mm}}
    
    \vertex[above=0.25cm of b1] (c1);
    \vertex[above=0.25cm of b2] (c2);
    
    \vertex[below=0.25cm of b1] (d1);
    \vertex[below=0.25cm of b2] (d2);
    
    \vertex[above=0.3333cm of a2] (l1);
    \vertex[below=0.3333cm of a2] (l2);
    
        \diagram*[large]{	
        (a1)--[quarter left](a4) -- [quarter left](a3),
        (a5) --[quarter right](a3), 
        (a5)--[quarter left](a1),
        (a4) -- (a5),
        
        (c1) -- (ec1),
        (c2) -- (ec2),
        (d1) -- (ed1),
        (d2) -- (ed2),
        (l1) -- (el1),
        (l2) -- (el2),

        }; 
    \end{feynman}
    \end{tikzpicture} \\
2L6P.f\end{tabular}}
& \multirow{2}{*}{K1}& \multirow{2}{*}{27}& \multirow{2}{*}{27}& \multirow{2}{*}{22}& \multirow{2}{*}{$\mathtt{[24, 21, 24]}$}& \multirow{2}{*}{3\phantom{.0}}& \multirow{2}{*}{121}& Re& \multirow{2}{*}{-08}& \multicolumn{1}{c}{n/a}& \texttt{-0.95486~+/-~0.00890}&  & 0.932& \multirow{2}{*}{0.368}\\
& & & & & & & & Im& & \multicolumn{1}{c}{n/a}& \texttt{~3.28530~+/-~0.00889}&  & 0.271& \\
\cline{2-15}
& \multirow{2}{*}{K1$^*$}& \multirow{2}{*}{27}& \multirow{2}{*}{24}& \multirow{2}{*}{34}& \multirow{2}{*}{$\mathtt{[19, 20, 20, 20, 20]}$}& \multirow{2}{*}{3\phantom{.0}}& \multirow{2}{*}{152}& Re& \multirow{2}{*}{-08}& \multicolumn{1}{c}{n/a}& \texttt{~2.55104~+/-~0.00208}&  & 0.082& \multirow{2}{*}{0.097}\\
& & & & & & & & Im& & \multicolumn{1}{c}{n/a}& \texttt{-1.63019~+/-~0.00205}&  & 0.126& \\
\hline
\multirow{2}{*}[7pt]{%
\begin{tabular}{@{}c@{}}
   

    \begin{tikzpicture}
    \begin{feynman}

    \tikzfeynmanset{every vertex={empty dot,minimum size=0mm}}
    \vertex (a1);
    
    \vertex[right=1cm of a1] (a3);
    \vertex[right=0.5cm of a1] (a2);
    \vertex[above=0.5cm of a2] (a4);
    \vertex[below=0.5cm of a2] (a5);
    
    \vertex[left=0.433cm of a2] (b1);
    \vertex[right=0.433cm of a2] (b2);
    
    \vertex[above=0.25cm of b1] (c1);
    \vertex[above=0.25cm of b2] (c2);
    \vertex[below=0.25cm of b1] (d1);
    \vertex[below=0.25cm of b2] (d2);
    \vertex[below=0.2cm of a4] (l1);
    \vertex[below=0.4cm of a4] (l2);
    \vertex[below=0.6cm of a4] (l3);
    \vertex[below=0.8cm of a4] (l4);
    
    \vertex[left=0.15cm of c1] (ec1);
    \vertex[right=0.15cm of c2] (ec2);
    \vertex[left=0.15cm of d1] (ed1);
    \vertex[right=0.15cm of d2] (ed2);
    \vertex[right=0.15cm of l1] (el1);
    \vertex[right=0.15cm of l2] (el2);
    \vertex[right=0.15cm of l3] (el3);
    \vertex[right=0.15cm of l4] (el4);

    \tikzfeynmanset{every vertex={dot,minimum size=0.8mm}}
    
    \vertex[above=0.25cm of b1] (c1);
    \vertex[above=0.25cm of b2] (c2);
    
    \vertex[below=0.25cm of b1] (d1);
    \vertex[below=0.25cm of b2] (d2);
    
    \vertex[below=0.2cm of a4] (l1);
    \vertex[below=0.4cm of a4] (l2);
    \vertex[below=0.6cm of a4] (l3);
    \vertex[below=0.8cm of a4] (l4);
    
        \diagram*[large]{	
        (a1)--[quarter left](a4) -- [quarter left](a3),
        (a5) --[quarter right](a3), 
        (a5)--[quarter left](a1),
        (a4) -- (a5),
    
        (c1) -- (ec1),
        (c2) -- (ec2),
        (d1) -- (ed1),
        (d2) -- (ed2),
        (l1) -- (el1),
        (l1) -- (el1),
        (l2) -- (el2),
        (l3) -- (el3),
        (l4) -- (el4),
        }; 
    \end{feynman}
    \end{tikzpicture} \\
2L8P\end{tabular}}
\rule[-5pt]{0pt}{27pt}& \multirow{2}{*}{K1}& \multirow{2}{*}{39}& \multirow{2}{*}{46}& \multirow{2}{*}{40}& \multirow{2}{*}{$\mathtt{[37, 42, 41, 40]}$}& \multirow{2}{*}{3\phantom{.0}}& \multirow{2}{*}{237}& Re& \multirow{2}{*}{-12}& \multicolumn{1}{c}{n/a}& \texttt{-5.15438~+/-~0.03310}&  & 0.642& \multirow{2}{*}{0.544}\\
\rule[-5pt]{0pt}{27pt}& & & & & & & & Im& & \multicolumn{1}{c}{n/a}& \texttt{~6.78546~+/-~0.03243}&  & 0.478& \\
\end{tabular}%
}%
}%
\caption{\label{tab:Kseries_2loopB} Results for two-loop topologies for scattering kinematics ($2 \rightarrow N$) for massless and massive propagators (indicated by a \texttt{*}). When there is no reference result, $\Delta [\%]$ and $\Delta [\%] |\cdot|$ refer to the Monte-Carlo accuracy relative to the central value. See the main text for details.}%
\end{table}

\begin{table}[tbp]
\centering
\resizebox{\columnwidth}{!}{%
\texttt{%
\begin{tabular}{@{}llrrrp{5cm}rrcllcrrr@{}}
\hline
Topology & Kin. & $\mathtt{N_{\text{C}}}$ & $\mathtt{N_{\text{E}}}$ & $\mathtt{N_{\text{S}}}$ & $\mathtt{L_{\text{max}}}$ & $\mathtt{N_{\text{p}} \ [10^9]}$ & $\mathtt{\sfrac{t}{p} \ [\mu \text{s}]}$ & Phase & Exp. & \multicolumn{1}{c}{Reference} & Numerical LTD & $\mathtt{\Delta \ [\sigma]}$ & $\mathtt{\Delta \ [\%]}$ & $\mathtt{\Delta \ [\%] |\cdot|}$ \\
\hline
\multirow{12}{*}{%
\begin{tabular}{@{}c@{}}
   

        \begin{tikzpicture}
            \begin{feynman}

            \tikzfeynmanset{every vertex={dot,minimum size=1mm}}

            \tikzfeynmanset{every vertex={empty dot,minimum size=0mm}}
             \vertex (a1);
            \vertex[right=0.33cm of a1] (a2);
            \vertex[right=0.33cm of a2] (a3);
             \vertex[right=0.33cm of a3] (a8);
            \vertex[below=1cm of a8] (a4);
            
            \vertex[left=0.33cm of a4] (a5);
            \vertex[left=0.33cm of a5] (a6);
            \vertex[left=0.33cm of a6] (a7);
                          
            \tikzfeynmanset{every vertex={dot,minimum size=0.8mm}}
            
            \vertex[right=0cm of a1] (e1);        
            \vertex[right=0cm of a8] (e2);        
            \vertex[right=0cm of a4] (e3);        
            \vertex[right=0cm of a7] (e4);    
            
            \tikzfeynmanset{every vertex={empty dot,minimum size=0mm}}                
  
            \vertex[left=0.15cm of e1] (d1);        
            \vertex[right=0.15cm of e2] (d2);        
            \vertex[right=0.15cm of e3] (d3);        
            \vertex[left=0.15cm of e4] (d4);    
            
                \diagram*[large]{	
                (a1)--(a2)--(a3)--(a8)--(a4)--(a5)--(a6)--(a7)--(a1), 
                (a2)--(a6),
                (a3)--(a5),
		(e1)--(d1),
		(e2)--(d2),
		(e3)--(d3),
		(e4)--(d4),
                           }; 
            \end{feynman}
    \end{tikzpicture} \\
3L4P\end{tabular}}
& \multirow{2}{*}{K1}& \multirow{2}{*}{56}& \multirow{2}{*}{17}& \multirow{2}{*}{49}& \multirow{2}{*}{$\mathtt{[17]}$}& \multirow{2}{*}{1\phantom{.0}}& \multirow{2}{*}{357}& Re& \multirow{2}{*}{-09}& \texttt{-2.42423} \multirow{2}{*}{\cite{Usyukina:1992jd}}& \texttt{-2.43299~+/-~0.03927}& 0.223& 0.361& \multirow{2}{*}{0.471}\\
& & & & & & & & Im& & \texttt{-3.40035}& \texttt{-3.41797~+/-~0.03956}& 0.445& 0.518& \\
\cline{2-15}
& \multirow{2}{*}{K2}& \multirow{2}{*}{56}& \multirow{2}{*}{17}& \multirow{2}{*}{49}& \multirow{2}{*}{$\mathtt{[17]}$}& \multirow{2}{*}{1\phantom{.0}}& \multirow{2}{*}{366}& Re& \multirow{2}{*}{-11}& \texttt{-5.30309} \multirow{2}{*}{\cite{Usyukina:1992jd}}& \texttt{-5.36759~+/-~0.14110}& 0.457& 1.216& \multirow{2}{*}{1.246}\\
& & & & & & & & Im& & \texttt{-1.07803}& \texttt{-1.05826~+/-~0.13399}& 0.148& 1.834& \\
\cline{2-15}
& \multirow{2}{*}{K3}& \multirow{2}{*}{56}& \multirow{2}{*}{17}& \multirow{2}{*}{49}& \multirow{2}{*}{$\mathtt{[17]}$}& \multirow{2}{*}{1\phantom{.0}}& \multirow{2}{*}{378}& Re& \multirow{2}{*}{-14}& \texttt{-4.47047} \multirow{2}{*}{\cite{Usyukina:1992jd}}& \texttt{-4.46226~+/-~0.10022}& 0.082& 0.184& \multirow{2}{*}{1.462}\\
& & & & & & & & Im& & \texttt{-0.66383}& \texttt{-0.72941~+/-~0.09918}& 0.661& 9.879& \\
\cline{2-15}
& \multirow{2}{*}{K1$^*$}& \multirow{2}{*}{56}& \multirow{2}{*}{7}& \multirow{2}{*}{55}& \multirow{2}{*}{$\mathtt{[7]}$}& \multirow{2}{*}{1\phantom{.0}}& \multirow{2}{*}{379}& Re& \multirow{2}{*}{-09}& \multicolumn{1}{c}{n/a}& \texttt{-3.89588~+/-~0.00173}&  & 0.044& \multirow{2}{*}{0.043}\\
& & & & & & & & Im& & \multicolumn{1}{c}{n/a}& \texttt{~3.89127~+/-~0.00165}&  & 0.043& \\
\cline{2-15}
& \multirow{2}{*}{K2$^*$}& \multirow{2}{*}{56}& \multirow{2}{*}{7}& \multirow{2}{*}{61}& \multirow{2}{*}{$\mathtt{[5, 5]}$}& \multirow{2}{*}{1\phantom{.0}}& \multirow{2}{*}{454}& Re& \multirow{2}{*}{-11}& \multicolumn{1}{c}{n/a}& \texttt{-3.15581~+/-~0.00639}&  & 0.203& \multirow{2}{*}{0.208}\\
& & & & & & & & Im& & \multicolumn{1}{c}{n/a}& \texttt{~2.97368~+/-~0.00633}&  & 0.213& \\
\cline{2-15}
& \multirow{2}{*}{K3$^*$}& \multirow{2}{*}{56}& \multirow{2}{*}{12}& \multirow{2}{*}{49}& \multirow{2}{*}{$\mathtt{[12]}$}& \multirow{2}{*}{1\phantom{.0}}& \multirow{2}{*}{364}& Re& \multirow{2}{*}{-14}& \multicolumn{1}{c}{n/a}& \texttt{-0.10876~+/-~0.00096}&  & 0.883& \multirow{2}{*}{0.072}\\
& & & & & & & & Im& & \multicolumn{1}{c}{n/a}& \texttt{~1.86939~+/-~0.00095}&  & 0.051& \\
\hline
\multirow{12}{*}{%
\begin{tabular}{@{}c@{}}
         
      
      \begin{tikzpicture}
            \begin{feynman}

            \tikzfeynmanset{every vertex={dot,minimum size=1mm}}

            \tikzfeynmanset{every vertex={empty dot,minimum size=0mm}}
             \vertex (a1);
            \vertex[right=0.33cm of a1] (a2);
            \vertex[right=0.33cm of a2] (a3);
             \vertex[right=0.33cm of a3] (a8);
            \vertex[below=1cm of a8] (a4);
             \vertex[below=0.5cm of a8] (c1);
            
            \vertex[left=0.33cm of a4] (a5);
            \vertex[left=0.33cm of a5] (a6);
            \vertex[left=0.33cm of a6] (a7);
                          
            \tikzfeynmanset{every vertex={dot,minimum size=0.8mm}}
            
            \vertex[right=0cm of a1] (e1);        
            \vertex[right=0cm of a8] (e2);        
            \vertex[right=0cm of a4] (e3);        
            \vertex[right=0cm of a7] (e4);    
            \vertex[below=0.cm of c1] (c1);
            
            \tikzfeynmanset{every vertex={empty dot,minimum size=0mm}}                
  
            \vertex[left=0.15cm of e1] (d1);        
            \vertex[right=0.15cm of e2] (d2);        
            \vertex[right=0.15cm of e3] (d3);        
            \vertex[left=0.15cm of e4] (d4);
            \vertex[right=0.15cm of c1] (d5);    
            
                \diagram*[large]{	
                (a1)--(a2)--(a3)--(a8)--(a4)--(a5)--(a6)--(a7)--(a1), 
                (a2)--(a6),
                (a3)--(a5),
		(e1)--(d1),
		(e2)--(d2),
		(e3)--(d3),
		(e4)--(d4),
		(c1)--(d5),
                           }; 
            \end{feynman}
    \end{tikzpicture}
     \\
3L5P\end{tabular}}
& \multirow{2}{*}{K1}& \multirow{2}{*}{71}& \multirow{2}{*}{24}& \multirow{2}{*}{80}& \multirow{2}{*}{$\mathtt{[23, 23]}$}& \multirow{2}{*}{1\phantom{.0}}& \multirow{2}{*}{490}& Re& \multirow{2}{*}{-10}& \multicolumn{1}{c}{n/a}& \texttt{-1.06298~+/-~0.02843}&  & 2.675& \multirow{2}{*}{2.922}\\
& & & & & & & & Im& & \multicolumn{1}{c}{n/a}& \texttt{-0.88557~+/-~0.02875}&  & 3.246& \\
\cline{2-15}
& \multirow{2}{*}{K2}& \multirow{2}{*}{71}& \multirow{2}{*}{20}& \multirow{2}{*}{80}& \multirow{2}{*}{$\mathtt{[14, 19]}$}& \multirow{2}{*}{1\phantom{.0}}& \multirow{2}{*}{503}& Re& \multirow{2}{*}{-06}& \multicolumn{1}{c}{n/a}& \texttt{-3.28794~+/-~0.07308}&  & 2.223& \multirow{2}{*}{3.202}\\
& & & & & & & & Im& & \multicolumn{1}{c}{n/a}& \texttt{-0.29022~+/-~0.07635}&  & 26.31& \\
\cline{2-15}
& \multirow{2}{*}{K3}& \multirow{2}{*}{71}& \multirow{2}{*}{20}& \multirow{2}{*}{103}& \multirow{2}{*}{$\mathtt{[13, 13, 19]}$}& \multirow{2}{*}{1\phantom{.0}}& \multirow{2}{*}{589}& Re& \multirow{2}{*}{-17}& \multicolumn{1}{c}{n/a}& \texttt{-1.61475~+/-~0.14277}&  & 8.841& \multirow{2}{*}{12.07}\\
& & & & & & & & Im& & \multicolumn{1}{c}{n/a}& \texttt{~0.25654~+/-~0.13621}&  & 53.10& \\
\cline{2-15}
& \multirow{2}{*}{K1$^*$}& \multirow{2}{*}{71}& \multirow{2}{*}{17}& \multirow{2}{*}{99}& \multirow{2}{*}{$\mathtt{[15, 14]}$}& \multirow{2}{*}{1\phantom{.0}}& \multirow{2}{*}{563}& Re& \multirow{2}{*}{-10}& \multicolumn{1}{c}{n/a}& \texttt{-1.26220~+/-~0.00124}&  & 0.098& \multirow{2}{*}{0.106}\\
& & & & & & & & Im& & \multicolumn{1}{c}{n/a}& \texttt{~1.06124~+/-~0.00123}&  & 0.116& \\
\cline{2-15}
& \multirow{2}{*}{K2$^*$}& \multirow{2}{*}{71}& \multirow{2}{*}{3}& \multirow{2}{*}{57}& \multirow{2}{*}{$\mathtt{[3]}$}& \multirow{2}{*}{1\phantom{.0}}& \multirow{2}{*}{427}& Re& \multirow{2}{*}{-07}& \multicolumn{1}{c}{n/a}& \texttt{~4.58640~+/-~0.00609}&  & 0.133& \multirow{2}{*}{0.180}\\
& & & & & & & & Im& & \multicolumn{1}{c}{n/a}& \texttt{~1.80523~+/-~0.00645}&  & 0.357& \\
\cline{2-15}
& \multirow{2}{*}{K3$^*$}& \multirow{2}{*}{71}& \multirow{2}{*}{20}& \multirow{2}{*}{102}& \multirow{2}{*}{$\mathtt{[17, 17, 18]}$}& \multirow{2}{*}{1\phantom{.0}}& \multirow{2}{*}{572}& Re& \multirow{2}{*}{-18}& \multicolumn{1}{c}{n/a}& \texttt{-1.05359~+/-~0.01706}&  & 1.619& \multirow{2}{*}{0.396}\\
& & & & & & & & Im& & \multicolumn{1}{c}{n/a}& \texttt{~5.92117~+/-~0.01660}&  & 0.280& \\
\hline
\multirow{2}{*}[7pt]{%
\begin{tabular}{@{}c@{}}
 

        \begin{tikzpicture}
            \begin{feynman}

            \tikzfeynmanset{every vertex={dot,minimum size=1mm}}

            \tikzfeynmanset{every vertex={empty dot,minimum size=0mm}}
             \vertex (a1);
            \vertex[right=0.5cm of a1] (a2);
            \vertex[right=0.5cm of a2] (a3);
            \vertex[below=1cm of a3] (a4);
            
            \vertex[below=0.5cm of a1] (t1);
            \vertex[below=0.5cm of a3] (t2);
            
            \vertex[left=0.5cm of a4] (a5);
            \vertex[left=0.5cm of a5] (a6);
                          
            \tikzfeynmanset{every vertex={dot,minimum size=0.8mm}}
            
            \vertex[right=0cm of a1] (e1);        
            \vertex[right=0cm of a3] (e2);        
            \vertex[right=0cm of a4] (e3);        
            \vertex[right=0cm of a6] (e4);    
            
            \tikzfeynmanset{every vertex={empty dot,minimum size=0mm}}                
  
            \vertex[left=0.15cm of e1] (d1);        
            \vertex[right=0.15cm of e2] (d2);        
            \vertex[right=0.15cm of e3] (d3);        
            \vertex[left=0.15cm of e4] (d4);    
            
                \diagram*[large]{	
                (a1)--(a2)--(a3)--(a4)--(a5)--(a6)--(a1), 
                (a2)--(a5),
             		(e1)--(d1),
		(e2)--(d2),
		(e3)--(d3),
		(e4)--(d4),
		(t1)--(t2),
                           }; 
            \end{feynman}
    \end{tikzpicture} \\
4L4P.a\end{tabular}}
\rule[-5pt]{0pt}{27pt}& \multirow{2}{*}{K1$^*$}& \multirow{2}{*}{192}& \multirow{2}{*}{14}& \multirow{2}{*}{408}& \multirow{2}{*}{$\mathtt{[13, 13]}$}& \multirow{2}{*}{0.5}& \multirow{2}{*}{3602}& Re& \multirow{2}{*}{-09}& \multicolumn{1}{c}{n/a}& \texttt{~1.28725~+/-~0.00637}&  & 0.495& \multirow{2}{*}{0.281}\\
\rule[-5pt]{0pt}{27pt}& & & & & & & & Im& & \multicolumn{1}{c}{n/a}& \texttt{~2.95568~+/-~0.00642}&  & 0.217& \\
\hline
\multirow{2}{*}[7pt]{%
\begin{tabular}{@{}c@{}}
         
       
       \begin{tikzpicture}
            \begin{feynman}

            \tikzfeynmanset{every vertex={dot,minimum size=1mm}}

            \tikzfeynmanset{every vertex={empty dot,minimum size=0mm}}
             \vertex (a1);
            \vertex[right=0.25cm of a1] (a2);
            \vertex[above=0.5cm of a2] (a4);
            \vertex[below=0.5cm of a2] (a5);

            \vertex[right=1cm of a1] (d1);

            \vertex[right=0.25cm of a4] (b1);
            \vertex[right=0.25cm of a5] (b2);
            \vertex[right=0.25cm of b1] (c1);
            \vertex[right=0.25cm of b2] (c2);
            
            \vertex[above=0.5cm of a1] (a7);
            \vertex[below=0.5cm of a1] (a8);
            \vertex[above=0.5cm of d1] (a9);
            \vertex[below=0.5cm of d1] (a10);
            
            \vertex[left=0.15cm of a7] (a11);
            \vertex[left=0.15cm of a8] (a12);
            \vertex[right=0.15cm of a9] (a13);
            \vertex[right=0.15cm of a10] (a14);
            
            \tikzfeynmanset{every vertex={dot,minimum size=0.8mm}}

            \vertex[above=0.5cm of a1] (e1);
            \vertex[below=0.5cm of a1] (e2);    
            
            \vertex[above=0.5cm of d1] (f1);
            \vertex[below=0.5cm of d1] (f2);

                \diagram*[large]{	
                (e1)--(a4), 
                (a5)--(e2),
		(e1)--(e2),

                (a4) -- (a5),
                (a4) -- (b1),
                (a5) -- (b2),
                (b1)--(b2),
                (b1)--(c1),
                (b2)--(c2),
                (c1)--(c2),

                (c1) --  (f1),
                (f2) -- (c2),
		        (f1)--(f2),
		        
		        (a7) -- (a11),
		        (a8) -- (a12),
		        (a9) -- (a13),
		        (a10) -- (a14),
                }; 
            \end{feynman}
    \end{tikzpicture} \\
4L4P.b\end{tabular}}
\rule[-5pt]{0pt}{27pt}& \multirow{2}{*}{K1$^*$}& \multirow{2}{*}{209}& \multirow{2}{*}{13}& \multirow{2}{*}{292}& \multirow{2}{*}{$\mathtt{[9, 10, 10]}$}& \multirow{2}{*}{0.5}& \multirow{2}{*}{3140}& Re& \multirow{2}{*}{-12}& \multicolumn{1}{c}{n/a}& \texttt{-4.34119~+/-~0.01166}&  & 0.269& \multirow{2}{*}{0.319}\\
\rule[-5pt]{0pt}{27pt}& & & & & & & & Im& & \multicolumn{1}{c}{n/a}& \texttt{-2.77244~+/-~0.01160}&  & 0.419& \\
\end{tabular}%
}%
}%
\caption{\label{tab:Kseries_HigherLoop} Results for three- and four-loop topologies for scattering kinematics ($2 \rightarrow N$) for massless and massive propagators (indicated by a \texttt{*}). When there is no reference result, $\Delta [\%]$ and $\Delta [\%] |\cdot|$ refer to the Monte-Carlo accuracy relative to the central value. See the main text for details.}%
\end{table}
\end{landscape}

\subsection{Divergent one-loop four- and five-point scalar integrals}
\label{sec:scalar_results}
\begin{figure}[t!]%
 \centering
 \subfloat[Divergent box topology.]{\includegraphics[width=0.8\textwidth]{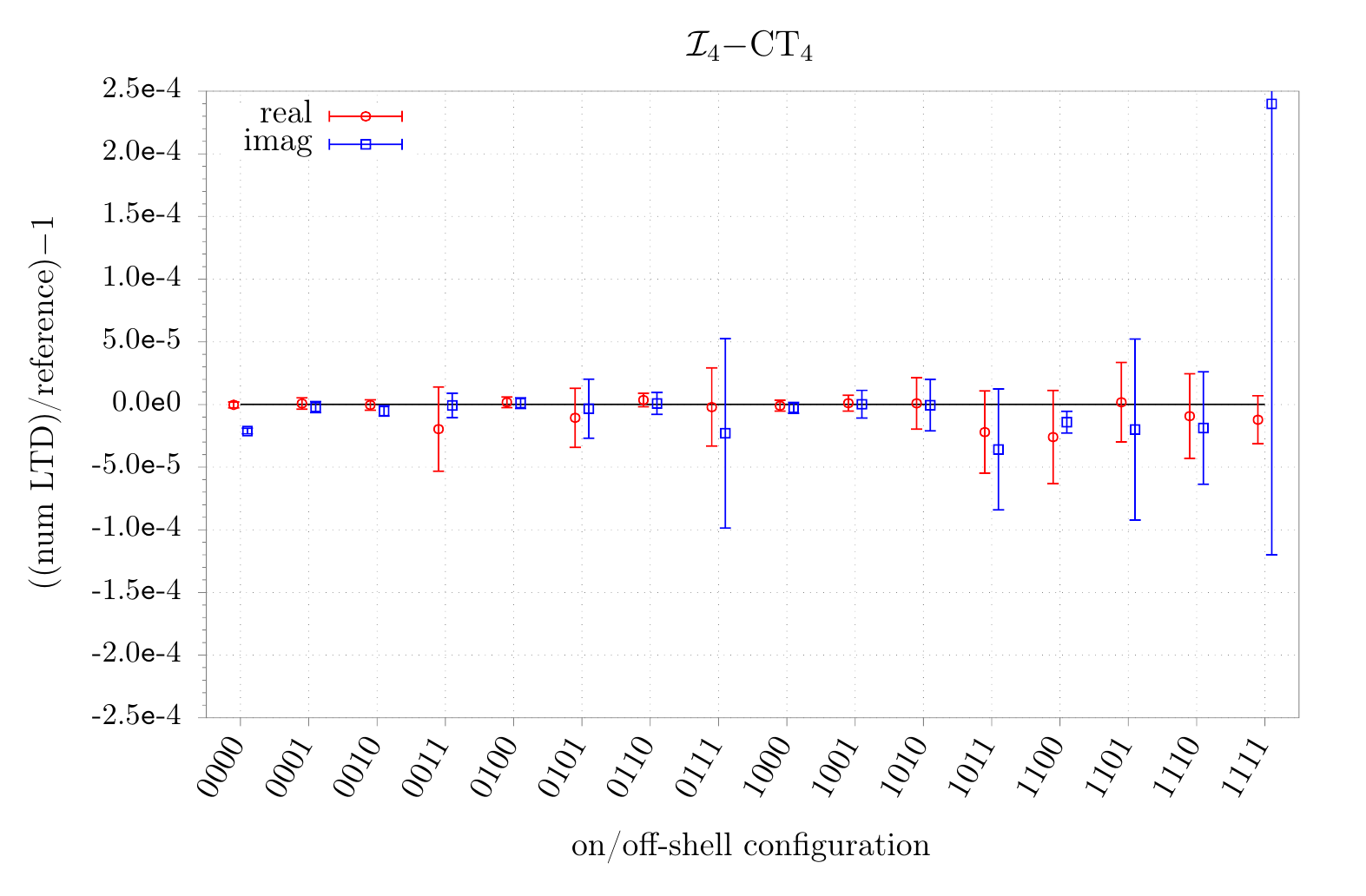}\label{plot:boxscan}}\\
 \subfloat[Divergent pentagon topology.]{\includegraphics[width=0.8\textwidth]{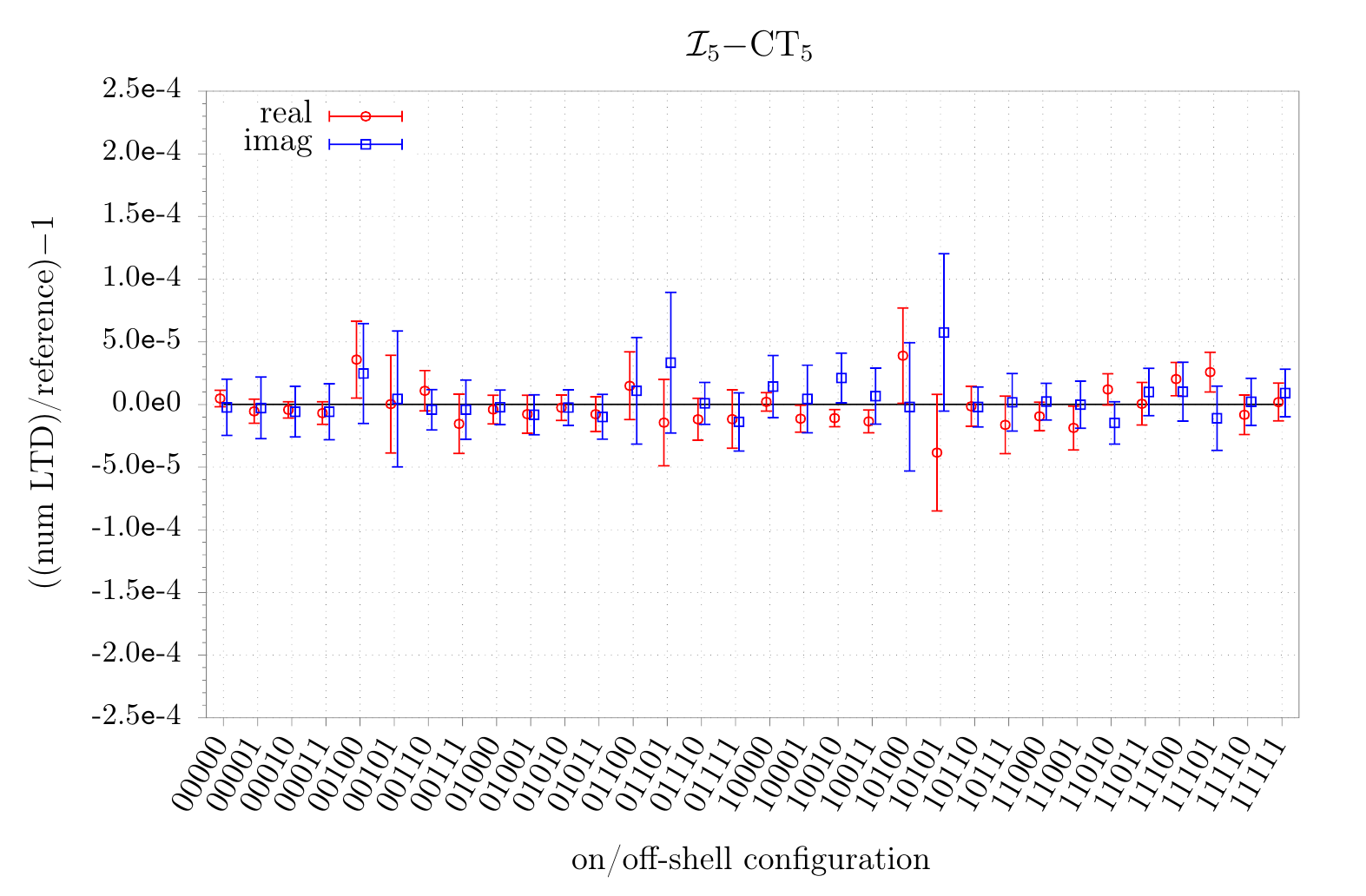}\label{plot:pentascan}} \caption{Results for the computation of divergent four- and five-point scalar one-loop integrals. We show the real and imaginary part of the expression integrated with LTD, compared with the analytic expression computed with {\sc\small MadLoop5}~\cite{Hirschi:2011pa,Alwall:2014hca}(\texttt{ML5}) and \texttt{qcdloop} \cite{Carrazza:2016gav}. The (nominal) horizontal axis shows different phase-space configurations using a binary notation, where a 1 (resp. 0) in the $i$th position signifies that the $i$th external momentum is on-shell with $p_i^2=0$ (resp. off-shell, that is with $p_i^2\ne0$). All but one of the central values are within 0.005\% of the analytical result. The outlier with configuration 1111 lies within 0.024\% of the analytical result and has a relative standard error of 0.036\%.}
 \label{fig:scalar_scans}
\end{figure}
We apply the subtraction scheme presented in sect.~\ref{sec:amplitudes} to one-loop four- and five-point functions with massless propagators.
For a randomly selected phase-space configuration, we go through all combinations of setting external momenta on-shell.
For both the box and pentagon kinematics, we set $s_{12}=1$. 
For the box topology, when one of the external momenta is massive, we set $m_1^2=\frac{1}{4}$, $m_2^2=\frac{1}{8}$, $m_3^2=\frac{2}{9}$, $m_4^2=\frac{1}{9}$, respectively.
For the pentagon topology, the masses are set to $m_1=0.10$, $m_2=0.11$, $m_3=0.12$, $m_4=0.13$, $m_5=0.14$.
The results for these different configurations are shown in fig.~\ref{fig:scalar_scans}, where the particular combination of masses for the external momenta is labelled by a binary number with the convention that a 1 in the $i$th position means that the $i$th external momentum is massless.
We use the \texttt{Cuhre} integrator from \texttt{Cuba} package~\cite{Hahn:2004fe} with 200 million sample points. 
The time for each evaluation is independent of the mass configuration and is similar with the one presented in tab.~\ref{tab:explore_1loop}.

Both the four-point (``box'') and five-point (``pentagon'') function can be integrated with high accuracy and precision: all but one of the central values are within a $0.005\%$ of the analytical result. Only the imaginary part of the box topology with all the external momenta on-shell has a large uncertainty. The reason is that the central value of this integral is ten times smaller than for the other box configurations. However, even this point lies within 0.024\% of the analytical result and has a relative standard error of 0.036\%.

The analytic expression of the box integral and the triangle integrals required to construct the analytical expression for the counterterms have been computed using {\sc\small qcdloop}~\cite{Carrazza:2016gav}. The pentagon integral has been obtained using {\sc\small MadLoop5}~\cite{Hirschi:2011pa,Alwall:2014hca} ({\sc\small ML5} henceforth).

\subsection{One-loop amplitude for $q \bar{q} \rightarrow \gamma_1 \gamma_2$ and $q \bar{q} \rightarrow \gamma_1 \gamma_2 \gamma_3$}
\label{sec:amplitude_results}

\begin{figure}[t!]%
 \centering
 \subfloat[Integration of the one-loop $d \bar{d} \rightarrow \gamma_1 \gamma_2$ amplitude.]{\includegraphics[trim={0.cm 8.5cm 3cm 7.5cm},clip, width=1.0\textwidth]{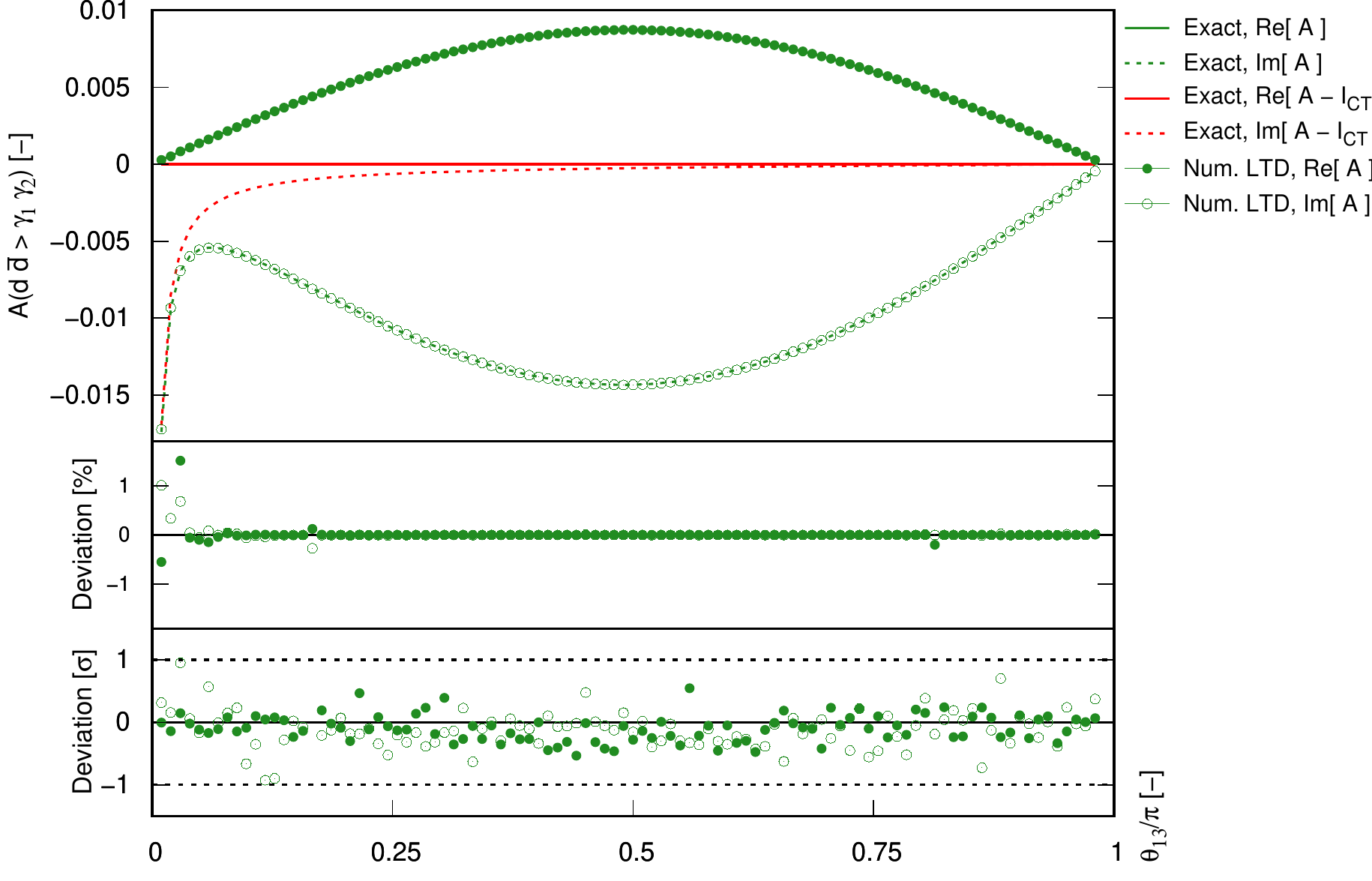}\label{plot:dd2a_re}}\\
 \subfloat[Regulated one-loop $d \bar{d} \rightarrow \gamma_1 \gamma_2$ amplitude.]{\includegraphics[width=0.9\textwidth]{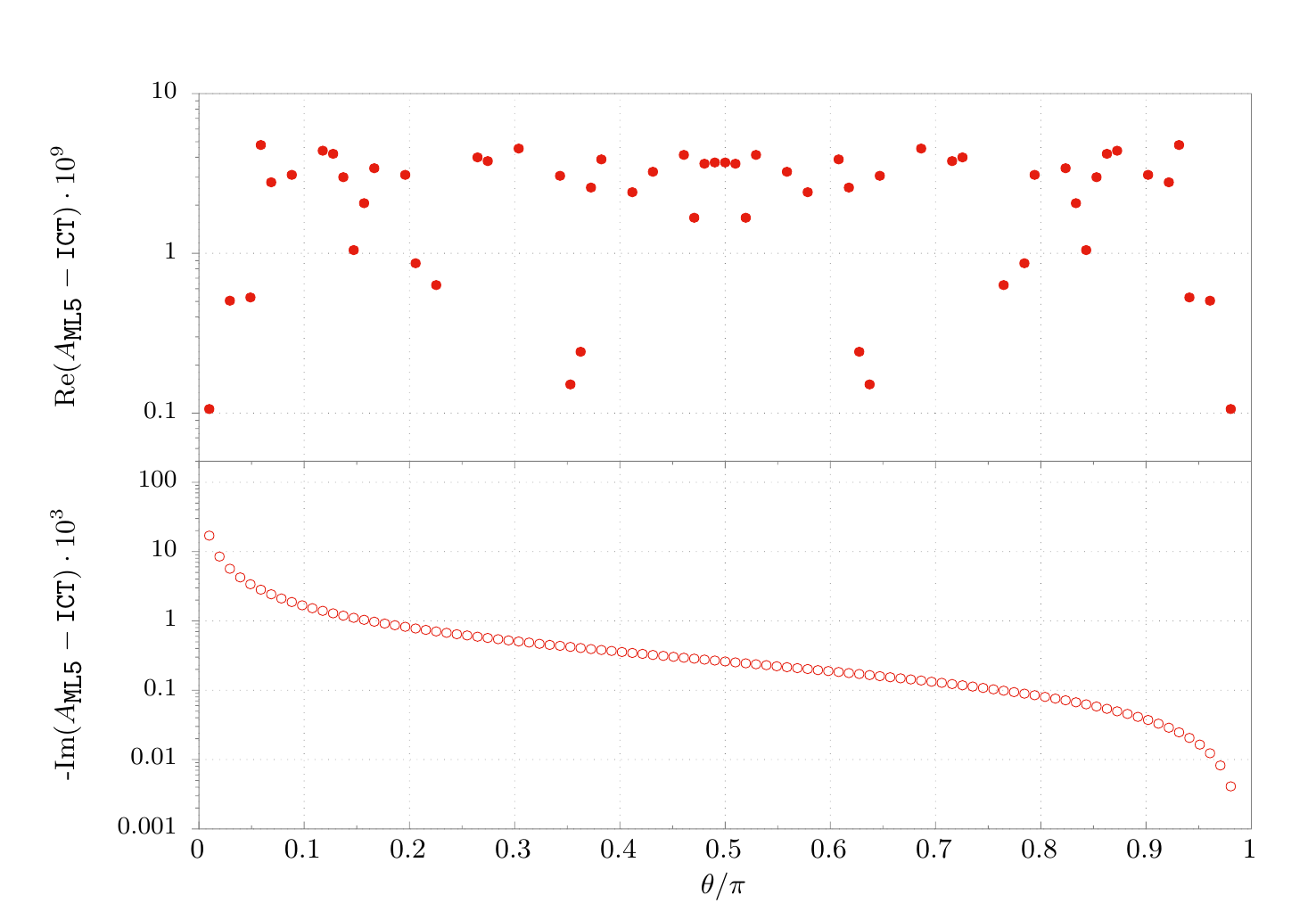}\label{plot:dd2a_ctsub}}\\
 \caption{A scan of our results using numerical LTD for the $d \bar{d} \rightarrow \gamma\gamma$ amplitude for various scattering angle $\theta_{13}$. In figure (a-b) we can see the results computed with LTD compared with the analytic expression obtained with {\sc\small MadLoop5}~\cite{Hirschi:2011pa,Alwall:2014hca} (\texttt{ML5}). In the last plot we show the result of the integral of the the finite regulated integrand that we actually integrate numerically. This corresponds to subtracting the integrated counterterms to the exact analytic result for the amplitude.}%
 \label{fig:dd2A}%
\end{figure}
In this section we present the results from the integration of the amplitudes $d\bar{d}$ to two and three photons.
For simplicity, we kept the order of the final photons fixed; the actual result for the amplitude can then be recovered by permuting through the final-state photon momenta.
The helicities are defined following the HELAS convention \cite{Murayama:1992gi}, and are taken positive for all the external particles.
The evaluation of the numerator, involving contractions of Lorentz and spinor indices, is performed numerically at run-time. This is not an efficient way to perform the numerator algebra, but the aim of this work is to highlight how LTD can be used to obtain results for physical and divergent expressions.

The analytic expressions have been compared with {\sc\small ML5} with $g_s=1.21771$, $g =0.30795$ and $\mu_r = 91.1880$ as couplings.
We also remind the reader that the results from {\sc\small ML5} are rescaled by an overall factor $(4\pi)^\epsilon / \Gamma(1-\epsilon)$.

For the $d\bar{d}\rightarrow\gamma_1\gamma_2$ process, we consider the process in its centre-of-mass rest frame, with the quarks aligned along the $z$-axis.
The result will only depend on the scattering energy and angle. 
The former is kept fixed and corresponds to a simple rescaling of the integral and the latter is varied in a scan and plotted in fig.~\ref{fig:dd2A}.
We used the \texttt{Cuhre} integrator from \texttt{Cuba} package~\cite{Hahn:2004fe} with two million evaluations.
In the last plot of fig.~\ref{fig:dd2A} we notice that the result is almost completely determined by the integrated counterterms. 
This is especially true for the real part, where one can see that resulting regulated integral is six orders of magnitude smaller than the finite part of the analytic expression.

As for the case of scalar divergent integrals, we use the \texttt{Cuhre} integrator with however only $2$ million sample points in this case. 
Despite this relatively low statistics, a large fraction of the results already have relative error below $0.05\%$. In the upper plot of fig.~\ref{fig:dd2A} we show the relative deviation with a large scale in order to highlight the few points that are not within this small error. One important observation however is that the Monte-Carlo error reported is reliable, as highlighted by the fact that all discrepancies are smaller than one (in modulus) when expressed in unit of the Monte-Calo standard deviation $\sigma$.

\begin{figure}[t!]%
 \centering
 \subfloat[Real part of the amplitude]{\includegraphics[width=0.49\textwidth]{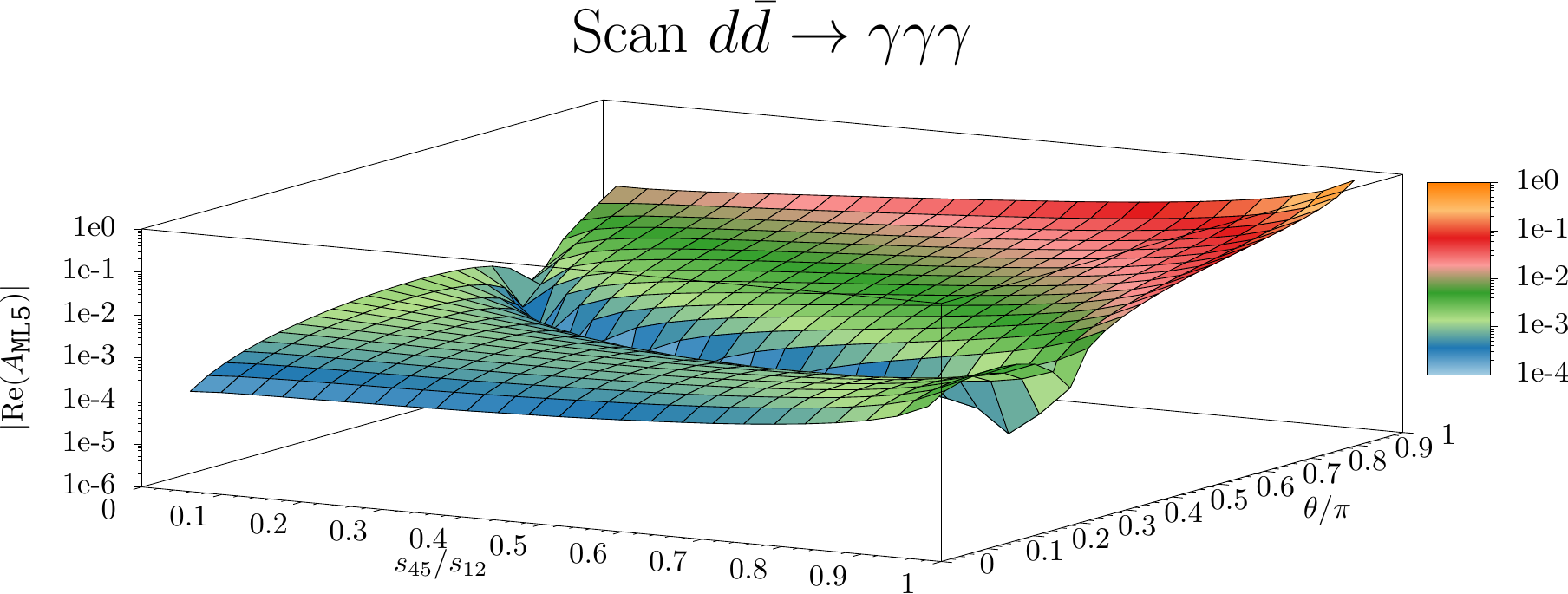}\label{plot:dd3a_amp_re}}
 \subfloat[Imaginary part of the amplitude.]{\includegraphics[width=0.49\textwidth]{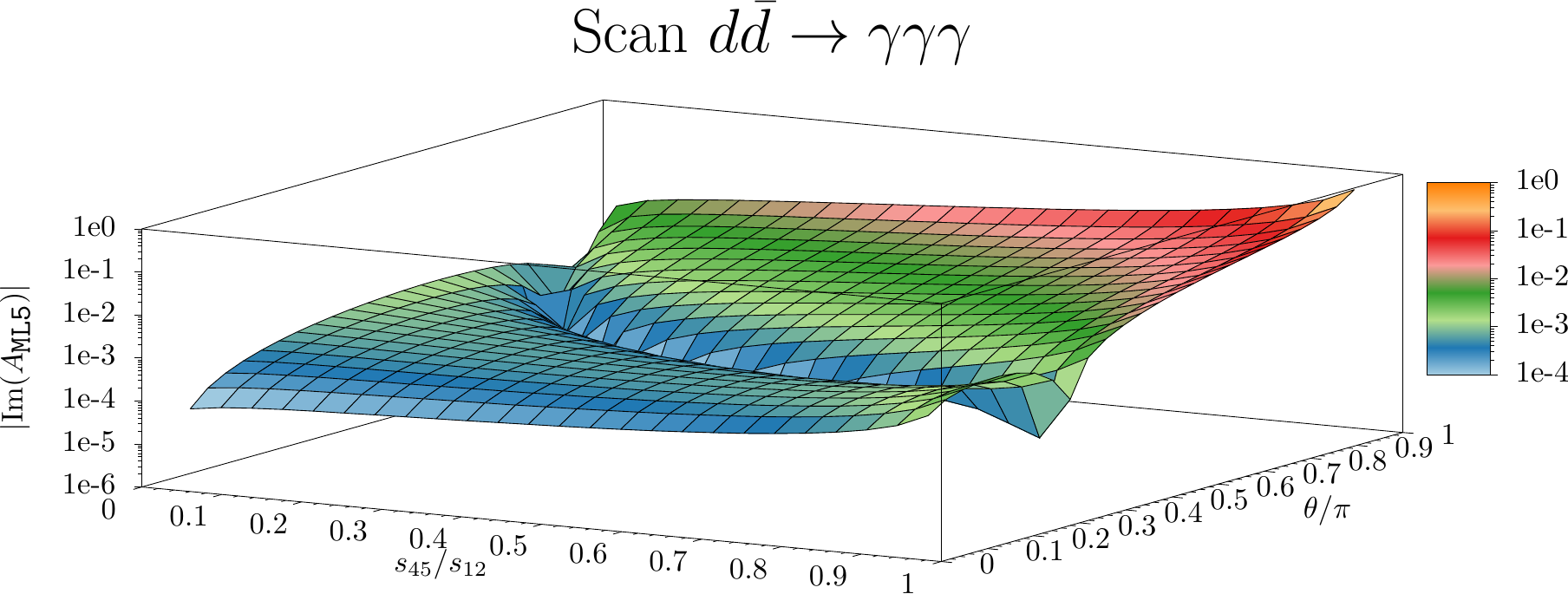}\label{plot:dd3a_amp_im}}\\
 \subfloat[Real part of the regulated amplitude.]{\includegraphics[width=0.49\textwidth]{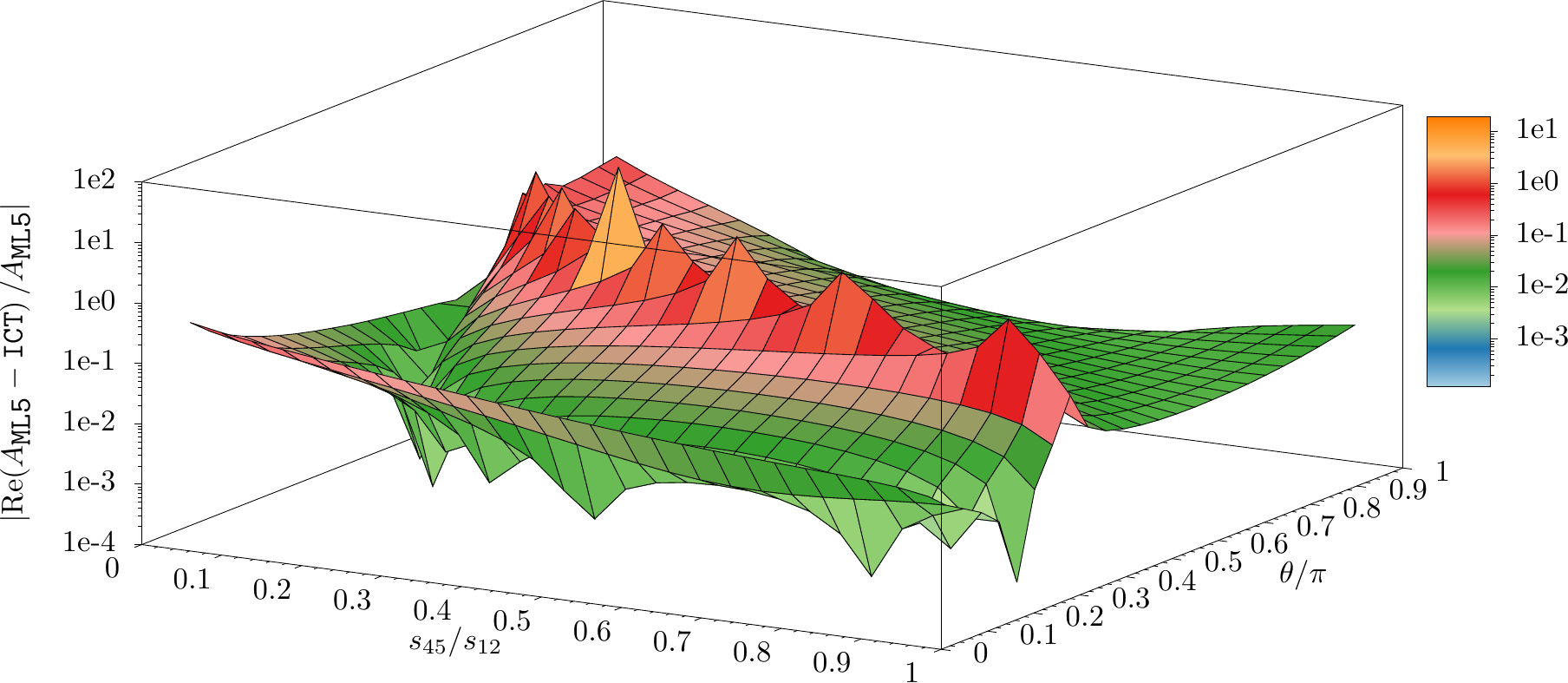}\label{plot:dd3a_ctsub_re}}
 \subfloat[Imaginary part of the regulated amplitude.]{\includegraphics[width=0.49\textwidth]{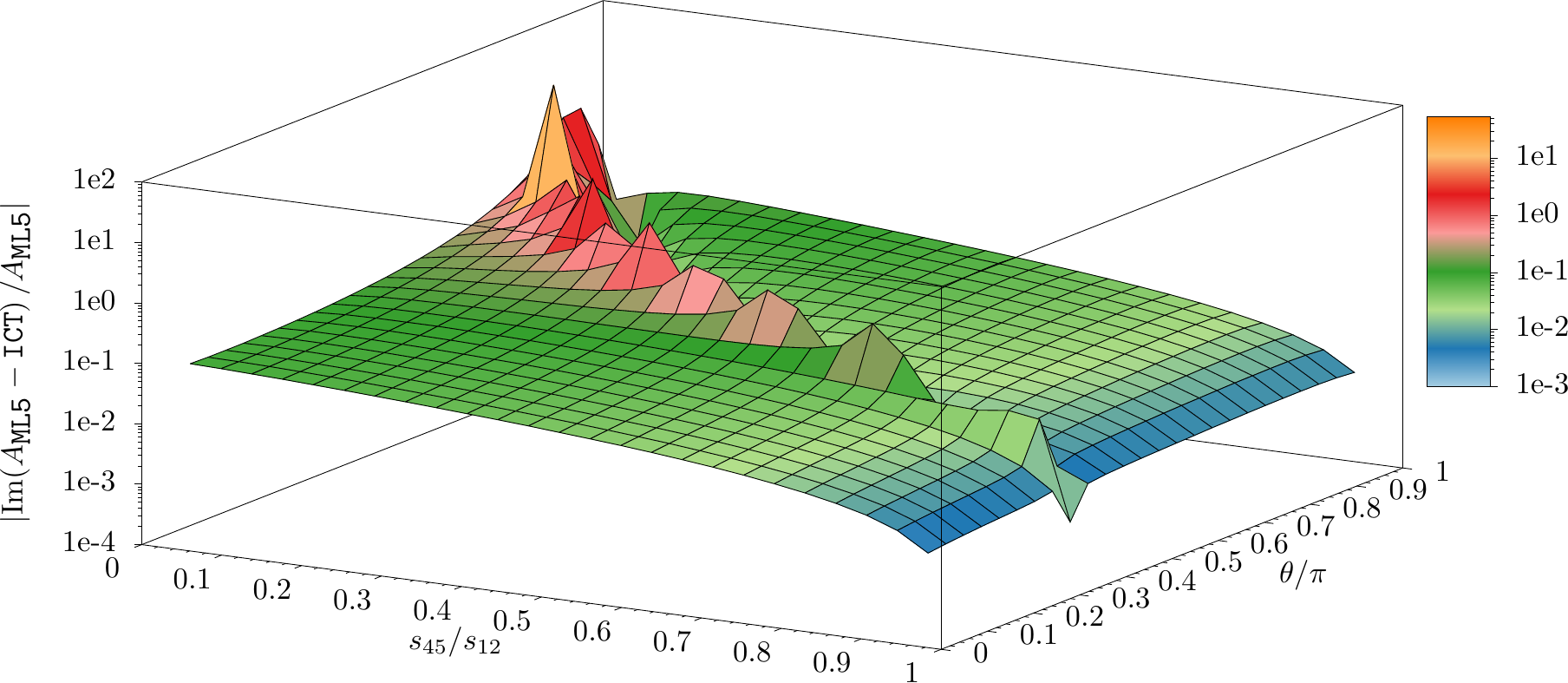}\label{plot:dd3a_ctsub_im}}\\
 \subfloat[Accuracy and precision of the real part of the LTD integration.]{\includegraphics[width=0.49\textwidth]{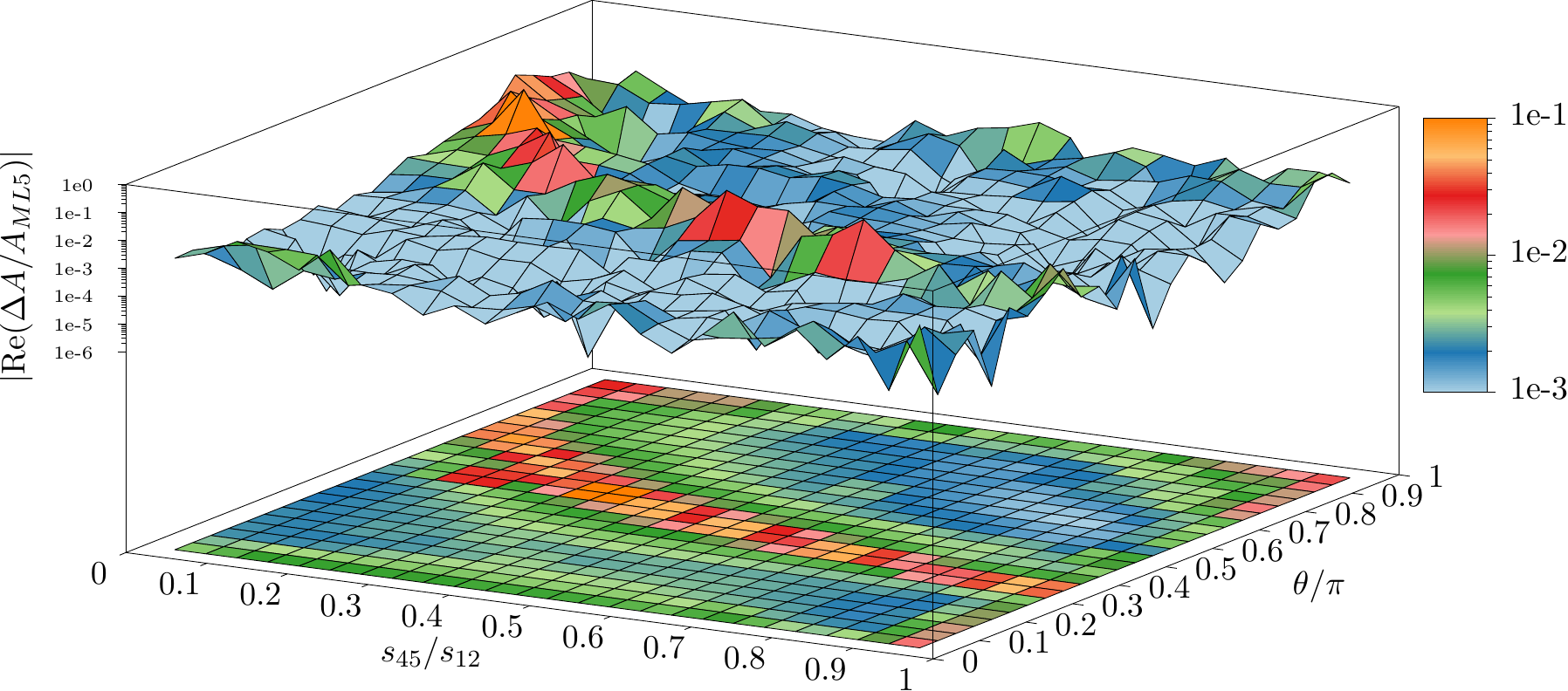}\label{plot:dd3a_rel_re}}\hfill%
 \subfloat[Accuracy and precision of the imaginary part of the LTD integration.]{\includegraphics[width=0.49\textwidth]{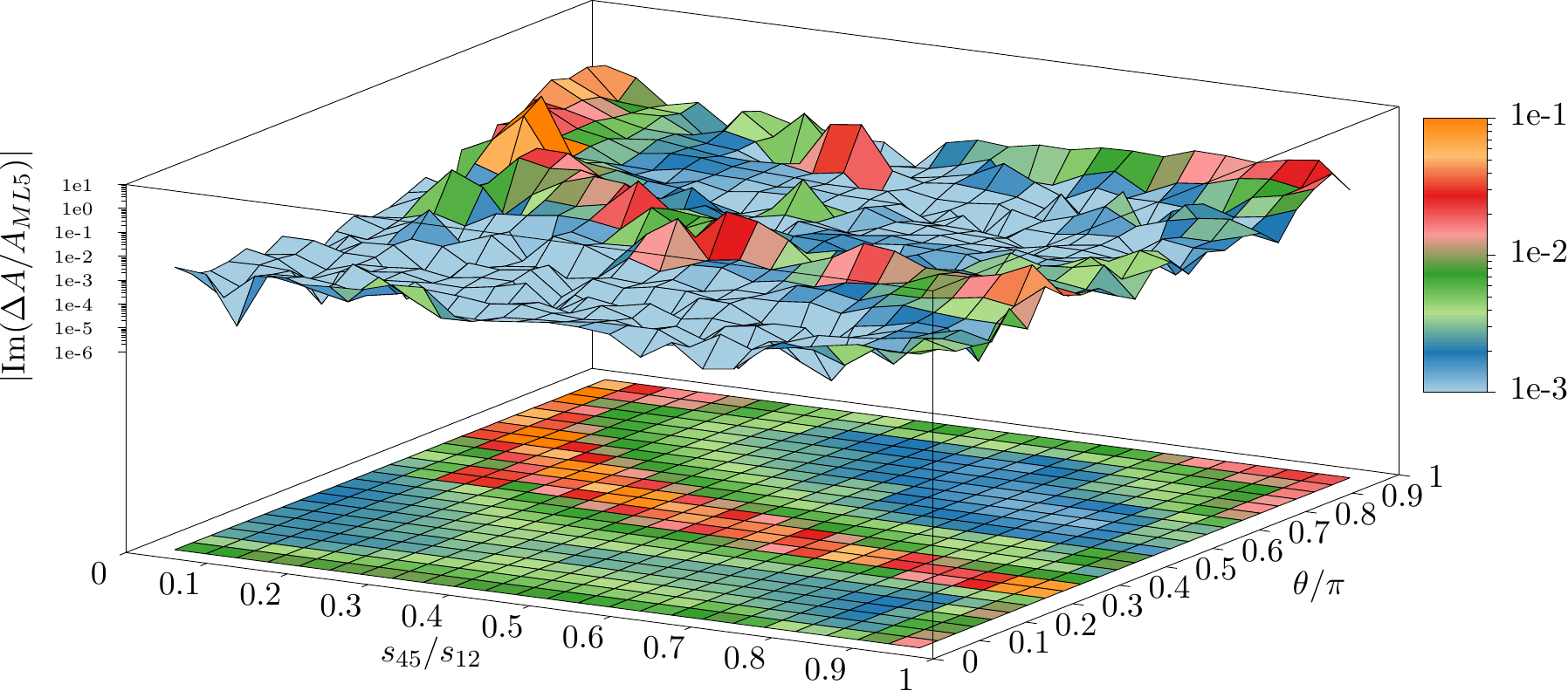}\label{plot:dd3a_rel_im}}\\
 \caption{A scan for $d \bar{d} \rightarrow \gamma_1\gamma_2\gamma_3$. The results are absolute values plotted on a log scale.
 The first row (a -- b) shows the real and the imaginary part of the amplitude computed with \texttt{ML5}. 
 The second row (c -- d) shows the relative difference between the analytic expression and the integrated counterterms.
 The last row (e -- f) shows the LTD integration. They are a combination of two plots: the surface above shows the relative error of the central value compared with the analytic expression, the flat surface below shows the Monte Carlo error for the point right above.}%
 \label{fig:dd3A}%
\end{figure}

In fig.~\ref{fig:dd3A} we show a scan of $d\bar{d}\rightarrow\gamma_1\gamma_2\gamma_3$.
In the same way as for the two-photon production case, we consider the scattering in the centre-of-mass rest frame. This time however, the number of unspecified and non-trivial degrees of freedom is four so that keeping a fixed energy $s_{12}=1$ leaves us with three parameters.
For the kinematic configuration  $d(p_1)\bar{d}(p_2)\rightarrow \gamma_1(p_3)\gamma_2(p_4)\gamma_3(p_5)$, we choose to scan in the angle $\theta_{13}=\angle(p_1,p_3)$, and $s_{45}$ which gives an indication of how collinear the momenta $p_4$ and $p_5$ are.
We fix the remaining degree of freedom by forcing the process on a plane, which allows for the configuration where $p_4$ is collinear to $p_1$, thus resulting in the valley shown in plots (a -- b) of \reffig{fig:dd3A}.
For $d \bar{d} \rightarrow \gamma_1\gamma_2\gamma_3$, we observe that the relative contribution from the integrated counterterms is not as large as for $d\bar{d}\rightarrow\gamma_1\gamma_2$, because this five-point amplitude has more contributions that are IR-finite (specifically D4-6 from fig.~\ref{fig:1loop_ddAAA}) and therefore not captured by the counterterms.

We can see that the relative error is $<1\%$ for most of the points in the scan as shown in the upper plane of plot (e--f) from fig.~\ref{fig:1loop_ddAAA}).
In the lower part of the same plots, the precision of the result with an error that is also $<1\%$ for most of the points. 
Along the valley, the \emph{relative} accuracy is not as good as in the other regions, which is to be expected when the central value of the integrated expression becomes smaller that the values around it.
Similarly as to elsewhere in this subsection, the results were obtained using the \texttt{Cuhre} integrator and $2$ million sample points.
The low number of samples is due to two mainly two reasons: first, we used a naive implementation of the numerator containing spinor chains that are recomputed numerically for each evaluation and second, despite the measure taken for improving the UV behaviour of the integrand, probing that region still requires many evaluations in quadruple precision thus increasing the overall evaluation time by roughly one order of magnitude when compared to the corresponding scalar topologies.
\\
In the present work, we put no effort in optimising the numerator expression which we leave to future work. The main objective of these results is to demonstrate  the viability of computing physical amplitudes with numerical LTD by combining the contour deformation together with the necessary infrared and ultraviolet counterterms.
Optimising the implementation of the numerator will allow us to handle more complicated amplitudes and to consider higher integration statistics.

%

\FloatBarrier

\section{Conclusion}
\label{sec:conclusion}

The ongoing and future research programme of the LHC calls for improving on the theoretical accuracy of the simulation of many scattering processes. A formidable effort from the high energy physics community over the last decades lead to the computation of many higher-order corrections of key relevance.
However, computing QCD amplitudes beyond two loops and/or four scales remains extremely challenging, even with modern analytical techniques.
We identify this problem as being one of the main bottlenecks whose resolution demands a radically new approach.
This observation is what motivates our work on numerical Loop-Tree Duality, as its strength and limitations are orthogonal, and thus complementary, to those of the canonical paradigms for predicting collider observables.
The potential of numerical LTD is reinforced by the promising perspective it entails regarding its eventual combination with real-emission contributions.
In our recent work of ref.~\cite{Capatti:2019ypt}, we presented our first developments and generalisation of LTD and, encouraged by our findings, we proceeded in this work to extend its range of applicability.

First, we established a contour deformation for regulating the threshold singularities exhibited by loop integrals when considering physical scattering kinematics.
In accordance with our long-term goals, we built a solution that is prone to automation and made no compromise regarding the generality of numerical LTD: availability of computational resources should remain the only limiting factor.
Moreover, we insisted that the validity of the contour deformation should be independent of the particular values of its hyperparameters, thus guaranteeing the predictive power of numerical LTD.
We demonstrated that our construction and implementation achieves these objectives by applying it to over 100 different representative configurations, ranging from one-loop boxes to four-loop 2x2 fishnets.

Second, we presented our first step towards computing divergent integrals and physical amplitudes.
This requires combining the LTD expression with \emph{local} integrand-level counterterms regularising divergences occurring for ultraviolet, soft and/or collinear loop momenta configurations.
We described this subtraction procedure at one loop and showcase explicit examples for divergent scalar four- and five-point integrals, as well as for the one-loop amplitude of the production of two and three photons.
This paves the way for a first application of numerical LTD to the numerical computation of two-loop divergent scalar integrals and of complete two-loop amplitudes, using the local counterterms introduced in refs.~\cite{Anastasiou:2018rib,Anastasiou:2019xxx}.

In this work, we focused on further developing numerical LTD in a way that is provably correct, general and that demonstrates predictive power. Therefore, we did not tune our hyperparameters for the hundreds of cases we studied and, although already satisfactory, the numerical convergence and run-time speed showcased by our results are by no means final.
We leave their improvement to future work.

The ability to locally regulate ultraviolet and infrared singularities at higher loops and the performance of the numerical convergence are two key difficulties whose resolution will determine the eventual viability of numerical LTD. Our work shows a clear path for this novel approach to significantly contribute to the effort of meeting the theoretical accuracy goal set by the needs of current collider experiments.

\section{Acknowledgements}
We would like to thank Francesco Moriello for providing us with the opportunity of applying our method to a case of practical interest, Stephen Jones for helping us produce comparison results with {\sc\small pySecDec}, Mao Zeng and Babis Anastasiou for providing insights on the amplitude subtraction and Armin Schweitzer and Rayan Haindl for fruitful discussions.
We also thank Lance Dixon for interesting discussions and suggesting the application of our work to the class of fishnet loop integrals, which are now known analytically~\cite{Basso:2017jwq}.
This project has received funding from the European Research Council (ERC) under grant agreement No 694712 (PertQCD) and SNSF grant No 179016. Numerical results presented in this work used computational resources from the Piz Daint cluster, administered by the Swiss National Supercomputing Centre (CSCS).

\newpage
\appendix
\section{Loop-Tree Duality example at two loops}
\label{sec:ltd_two_loops}

In this section we demonstrate explicitly how the LTD formula can be obtained for a two-loop two-point topology, the double-triangle, by iteratively applying residue theorem for each loop momentum's energy integration.
This explicit computation will highlight the cancellation of residues involving Heaviside functions and will explicitly derive the two-loop cut structure. \par

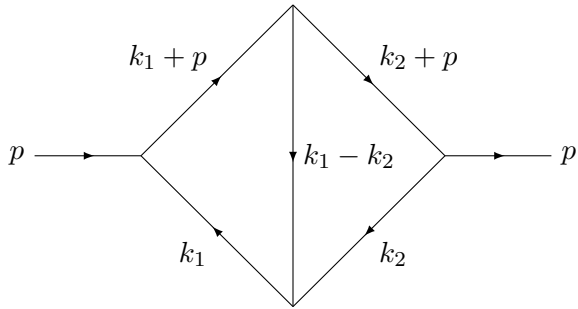
\begin{figure}[hbt]
    \centering
    \begin{tikzpicture}
        \pgfmathsetmacro{\r}{2}
        \coordinate (A) at (-1.7*\r,0);
        \coordinate (B) at (-1*\r,0);
        \coordinate (C) at (1*\r,0);
        \coordinate (D) at (1.7*\r,0);
        \coordinate (E) at ({\r*cos(90)}, {\r*sin(90)});
        \coordinate (F) at ({\r*cos(270)}, {\r*sin(270)});
        \draw[->-] (B) --node[auto]{$k_1+p$} (E);
        \draw[->-] (E) --node[auto]{$k_2+p$} (C);
        \draw[->-] (C) --node[auto]{$k_2$} (F);
        \draw[->-] (F) --node[auto]{$k_1$} (B);
        \draw[->-] (A) node[left]{$p$} -- (B);
        \draw[->-] (C) -- (D) node[right]{$p$};
        \draw[->-] (E) -- node[auto]{$k_1-k_2$} (F);
    \end{tikzpicture}
    \caption{The double-triangle diagram in terms of a particular choice of loop momentum basis and momentum routing.}
    \label{fig:ltd_example_double_triangle}
\end{figure}

We start with the double-triangle integrand
\begin{align}
    f
    &=
    \frac{1}{k_1^2 - m_1^2 + \mathrm{i}\delta}
    \frac{1}{(k_1+p)^2 - m_2^2 + \mathrm{i}\delta}
    \frac{1}{(k_1-k_2)^2 - m_3^2 + \mathrm{i}\delta}
    \frac{1}{(k_2+p)^2 - m_4^2 + \mathrm{i}\delta}
    \frac{1}{k_2^2 - m_5^2 + \mathrm{i}\delta} \\
    &=
    \frac{1}{(k_1^0)^2-E_1^2}
    \frac{1}{(k_1^0+p^0)^2-E_2^2}
    \frac{1}{(k_1^0-k_2^0)^2-E_3^2}
    \frac{1}{(k_2^0+p^0)^2-E_4^2}
    \frac{1}{(k_2^0)^2-E_5^2},
\end{align}
with the on-shell energies $E_i = \sqrt{\vec{q}_i^{\,2} + m_i^2 -\mathrm{i}\delta}$ and the real external four-momentum $p$.
The loop integral we consider is
\begin{align}
    I = \int \frac{\mathrm{d}^4 k_1}{(2\pi)^4}\frac{\mathrm{d}^4 k_2}{(2\pi)^4} f.
\end{align} \par
where $f$ can be seen as a meromorphic function in $k_1^0$ on $\mathbb{C}$.
It has three poles located in the lower half-plane:
\begin{align}
    k_1^{(1)} = E_1, \quad
    k_1^{(2)} = E_2 - p^0, \quad
    k_1^{(3)} = E_3 + k_2^0.
\end{align}
We then integrate $f$ along a contour $[-R,R]$ closing on an arc $C_R$ in the lower half-plane in the limit of $R\rightarrow \infty$.
With residue theorem and using that the integral along $C_R$ vanishes (from the requirement of UV-convergence of the integrand), we find that
\begin{align}
    \int_{-\infty}^\infty \frac{\mathrm{d}k_1^0}{2\pi} f
    = -i \sum_{i=1}^3 \Res_{k_1^{(i)}}[f],
\end{align}
where the three residues are
\begin{align}
    \Res_{k_1^{(1)}}[f] &=
        \frac{1}{2E_1}
        \frac{1}{(E_1+p^0)^2-E_2^2}
        \frac{1}{(E_1-k_2^0)^2-E_3^2}
        \frac{1}{(k_2^0+p^0)^2-E_4^2}
        \frac{1}{(k_2^0)^2-E_5^2}, \\
    \Res_{k_1^{(2)}}[f] &= 
        \frac{1}{(E_2-p^0)^2-E_1^2}
        \frac{1}{2E_2}
        \frac{1}{(E_2-p^0-k_2^0)^2-E_3^2}
        \frac{1}{(k_2^0+p^0)^2-E_4^2}
        \frac{1}{(k_2^0)^2-E_5^2}, \\
    \Res_{k_1^{(3)}}[f] &=
        \frac{1}{(E_3+k_2^0)^2-E_1^2}
        \frac{1}{(E_3+k_2^0+p^0)^2-E_2^2}
        \frac{1}{2E_3}
        \frac{1}{(k_2^0+p^0)^2-E_4^2}
        \frac{1}{(k_2^0)^2-E_5^2}.
\end{align}
Now we consider each residue $\Res_{k_1^{(i)}}[f]$ as a meromorphic function in $k_2^0$ on $\mathbb{C}$.
For the first residue, the poles located in the lower half-plane are:
\begin{align}
    k_2^{(1,1)} &= E_4 - p^0, \quad
    k_2^{(1,2)} = E_5, \\
    k_2^{(1,3)} &= E_1 + E_3, \quad
    k_2^{(1,4)} = E_1 - E_3 \ \ \text{if} \ \Im[k_2^{(1,4)}] < 0,
\end{align}
for the second at
\begin{align}
    k_2^{(2,1)} &= k_2^{(1,1)}, \quad
    k_2^{(2,2)} = k_2^{(1,2)}, \\
    k_2^{(2,3)} &= E_2 - p^0 + E_3, \quad
    k_2^{(2,4)} = E_2 - p^0 - E_3 \ \ \text{if} \ \Im[k_2^{(2,4)}] < 0,
\end{align}
and for the third at
\begin{align}
    k_2^{(3,1)} &= k_2^{(1,1)}, \quad
    k_2^{(3,2)} = k_2^{(1,2)}, \\
    k_2^{(3,3)} &= k_2^{(1,4)} \ \ \text{if} \ \Im[k_2^{(1,4)}]  < 0, \quad
    k_2^{(3,4)} = k_2^{(2,4)} \ \ \text{if} \ \Im[k_2^{(2,4)}] < 0.
\end{align}
We see that each residue has poles at $k_2^{(1,1)}$ and $k_2^{(1,2)}$. Note that there are two poles at $k_2^{(1,4)}$ and $k_2^{(2,4)}$, which are located in either the lower or the upper complex half-plane depending on the values of $\vec{k}_1$ and $\vec{k}_2$. \par

As before, we now integrate the sum of the three residues along a contour $[-R,R]$ closing on an arc $C_R$ in the lower half-plane in the limit of $R\rightarrow \infty$.
With residue theorem and using that the integral along $C_R$ vanishes, we find that
\begin{align}
\label{eq:explicit_double_triangle}
    \int_{-\infty}^\infty \frac{\mathrm{d}k_2^0}{2\pi} &\int_{-\infty}^\infty \frac{\mathrm{d}k_1^0}{2\pi} f 
    =  -\mathrm{i} \int_{-\infty}^\infty \frac{\mathrm{d}k_2^0}{2\pi} \sum_{i=1}^3 \Res_{k_1^{(i)}}[f] \\
    &=  (-\mathrm{i})^2
    \bigg(
        \sum_{i=1}^3 \Res_{k_2^{(1,i)}k_1^{(1)}}[f]
        +\sum_{i=1}^3 \Res_{k_2^{(2,i)}k_1^{(2)}}[f]
        +\sum_{i=1}^2 \Res_{k_2^{(3,i)}k_1^{(3)}}[f] \\
        &\phantom{=(-\mathrm{i})^2\bigg(}+ \left(
            \Res_{k_2^{(1,4)}k_1^{(1)}}[f] + \Res_{k_2^{(3,3)}k_1^{(3)}}[f]
        \right) \Theta(-\Im[k_2^{(1,4)}]) \\
        &\phantom{=(-\mathrm{i})^2\bigg(}+ \left(
            \Res_{k_2^{(2,4)}k_1^{(2)}}[f] + \Res_{k_2^{(3,4)}k_1^{(3)}}[f]
        \right) \Theta(-\Im[k_2^{(2,4)}])
    \bigg),
\end{align}
where we used the short form $\Res_{ab}[f] \equiv \Res_{a}[\Res_{b}[f]]$ and the Heaviside function $\Theta$. \par

The twelve residues are
\begin{align}
    \Res_{k_2^{(1,1)}k_1^{(1)}}[f] &=
        \frac{1}{2E_1}
        \frac{1}{(E_1+p^0)^2-E_2^2}
        \frac{1}{(E_1-E_4+p^0)^2-E_3^2}
        \frac{1}{2E_4}
        \frac{1}{(E_4-p^0)^2-E_5^2}, \\
    \Res_{k_2^{(1,2)}k_1^{(1)}}[f] &=
        \frac{1}{2E_1}
        \frac{1}{(E_1+p^0)^2-E_2^2}
        \frac{1}{(E_1-E_5)^2-E_3^2}
        \frac{1}{(E_5+p^0)^2-E_4^2}
        \frac{1}{2E_5}, \\
    \Res_{k_2^{(1,3)}k_1^{(1)}}[f] &=
        \frac{1}{2E_1}
        \frac{1}{(E_1+p^0)^2-E_2^2}
         \frac{1}{2E_3}
        \frac{1}{(E_1+E_3+p^0)^2-E_4^2}
        \frac{1}{(E_1+E_3)^2-E_5^2}, \\
    \Res_{k_2^{(1,4)}k_1^{(1)}}[f] &= -
        \frac{1}{2E_1}
        \frac{1}{(E_1+p^0)^2-E_2^2}
         \frac{1}{2E_3}
        \frac{1}{(E_1-E_3+p^0)^2-E_4^2}
        \frac{1}{(E_1-E_3)^2-E_5^2}, \\
    \Res_{k_2^{(2,1)}k_1^{(2)}}[f] &= 
        \frac{1}{(E_2-p^0)^2-E_1^2}
        \frac{1}{2E_2}
        \frac{1}{(E_2-E_4)^2-E_3^2}
        \frac{1}{2E_4}
        \frac{1}{(E_4-p^0)^2-E_5^2}, \\
    \Res_{k_2^{(2,2)}k_1^{(2)}}[f] &= 
        \frac{1}{(E_2-p^0)^2-E_1^2}
        \frac{1}{2E_2}
        \frac{1}{(E_2-E_5-p^0)^2-E_3^2}
        \frac{1}{(E_5+p^0)^2-E_4^2}
        \frac{1}{2E_5}, \\
    \Res_{k_2^{(2,3)}k_1^{(2)}}[f] &= 
        \frac{1}{(E_2-p^0)^2-E_1^2}
        \frac{1}{2E_2}
        \frac{1}{2E_3}
        \frac{1}{(E_1+E_3+p^0)^2-E_4^2}
        \frac{1}{(E_1+E_3)^2-E_5^2}, \\
    \Res_{k_2^{(2,4)}k_1^{(2)}}[f] &= -
        \frac{1}{(E_2-p^0)^2-E_1^2}
        \frac{1}{2E_2}
        \frac{1}{2E_3}
        \frac{1}{(E_1-E_3+p^0)^2-E_4^2}
        \frac{1}{(E_1-E_3)^2-E_5^2}, \\
    \Res_{k_2^{(3,1)}k_1^{(3)}}[f] &=
        \frac{1}{(E_3+E_4-p^0)^2-E_1^2}
        \frac{1}{(E_3+E_4)^2-E_2^2}
        \frac{1}{2E_3}
        \frac{1}{2E_4}
        \frac{1}{(E_4-p^0)^2-E_5^2}, \\
    \Res_{k_2^{(3,2)}k_1^{(3)}}[f] &=
        \frac{1}{(E_3+E_5)^2-E_1^2}
        \frac{1}{(E_3+E_5+p^0)^2-E_2^2}
        \frac{1}{2E_3}
        \frac{1}{(E_5+p^0)^2-E_4^2}
        \frac{1}{2E_5}, \\
    \Res_{k_2^{(3,3)}k_1^{(3)}}[f] &= -\Res_{k_2^{(1,4)}k_1^{(1)}}[f], \\
    \Res_{k_2^{(3,4)}k_1^{(3)}}[f] &= -\Res_{k_2^{(2,4)}k_1^{(2)}}[f].
\end{align}
It follows that the four residues coming together with a Heaviside function cancel pairwise and eight residues remain. 
The pairwise cancellation of the Heaviside functions is directly related to dual cancellations between H-surfaces.
\par
We observe that we can write eq.~\eqref{eq:explicit_double_triangle} more compactly and generally as
\begin{align}
    \int_{-\infty}^\infty \frac{\mathrm{d}k_2^0}{2\pi} \int_{-\infty}^\infty \frac{\mathrm{d}k_1^0}{2\pi} f
    &=
    (-\mathrm{i})^2 \sum_{\mathbf{b} \in \mathcal{B}}
    \left.
    \frac{1}{\prod_{i \in \mathbf{b}} 2 E_i}
    \frac{1}{\prod_{i \in \mathbf{e}\setminus\mathbf{b}} D_i}
    \right\vert_{\{q_j^0 = \sigma_j^\mathbf{b}E_j\}_{j \in \mathbf{b}}}
\end{align}
where, in the present double-triangle example, we have that $\mathcal{B}=\{\{1,3\},\{1,4\},\{1,5\},\{2,3\},$ $\{2,4\},\{2,5\},\{3,4\},\{3,5\}\}$ is the set of all loop momentum bases and $\mathbf{e}= \{1,2,3,4,5\}$ the set of all edges.
The energy $q_i^0$ flowing in the Feynman propagator $D_i = (q_i^0)^2 - E_i^2$ can then be expressed as a linear combination of loop momentum energy basis elements $\{q_j^0| j\in \mathbf{b}\}$ for any $\mathbf{b} \in \mathcal{B}$ and the energy $p^0$ of the external momentum.
The cut structure signs, i.e. the signs of the energy cuts that put propagators $j\in\mathbf{b}$ on-shell, are denoted as $\sigma^\mathbf{b}_j$ for $j \in \mathbf{b}$.
By comparison with the eight residues computed above, we find
\begin{align}
    (\sigma_1^{\{1,3\}},\sigma_3^{\{1,3\}})
    &= (\sigma_2^{\{2,3\}},\sigma_3^{\{2,3\}})
    = (+1,-1) \label{eq:cs1}\\
    (\sigma_1^{\{1,4\}},\sigma_4^{\{1,4\}})
    &= (\sigma_1^{\{1,5\}},\sigma_5^{\{1,5\}})
    = (\sigma_2^{\{2,4\}},\sigma_4^{\{2,4\}}
    = (\sigma_2^{\{2,5\}},\sigma_5^{\{2,5\}})
    = (+1,+1) \label{eq:cs2}\\
    (\sigma_3^{\{3,4\}},\sigma_4^{\{3,4\}})
    &= (\sigma_3^{\{3,5\}},\sigma_5^{\{3,5\}})
    = (+1,+1). \label{eq:cs3}
\end{align}
The cut structure is a result of the propagator's signatures (i.e. the initial choice of momentum routing in the loop graph), the choice of integration order and of the contour closure (in either the upper or lower complex half-plane) of each energy integration.
We stress that since the signature is independent of the contribution to the internal momentum flow coming from external legs, the cut structure is independent of which particular propagator of a given loop line is being cut, as already suggested by the cut structure signs above. When accounting for this degeneracy, one can limit oneself to only reporting the cut structure for a given combination of loop lines (as opposed to propagators) being cut. In that case, \emph{any two-loop} integral will \emph{always} feature exactly three cut-structures (as opposed to twelve in the listing of eqs.~\eqref{eq:cs1} -- \eqref{eq:cs3}).
\par
Equipped with the above, our general LTD identity applied to the double-triangle integral reads:
\begin{align}
    I
    &=
    (-\mathrm{i})^2
    \int
    \frac{\mathrm{d}^3 \vec{k}_1}{(2\pi)^3}
    \frac{\mathrm{d}^3 \vec{k}_2}{(2\pi)^3}
    \sum_{\mathbf{b} \in \mathcal{B}}
    \left.
    \frac{1}{\prod_{i \in \mathbf{b}} 2 E_i}
    \frac{1}{\prod_{i \in \mathbf{e}\setminus\mathbf{b}} D_i}
    \right\vert_{\{q_j^0 = \sigma_j^\mathbf{b}E_j\}_{j \in \mathbf{b}}}.
\end{align}

\section{Expression for the $q\bar{q}\rightarrow \gamma_1 \gamma_2 \gamma_3$ amplitude and its counterterms}
\label{sec:diagram_expressions}
In order to provide an explicit parametrisation of all the integrals that appear in the computation of the $q\bar{q}\rightarrow\gamma_1\gamma_2\gamma_3$, we give the expression for the diagrams and the counterterms.
The individual diagrams can be written as explicit integrals using dimensional regularisation, since in general they contain singularities.

The integrals appearing in \reffig{fig:1loop_ddAAA} are given by:
\begin{align}
I_1
&=C_1\,\mu^{2\epsilon} (4\pi)^2\int \frac{\dd^d k}{(2\pi)^d}
	\frac{\bar v_2 \pol{1} (-\slashed p_{23})\gamma^\mu\momsl{3} \pol{2} \momsl{4} \pol{3} \momsl{5}\gamma_\mu u_1}
		{s_{23}\prop{1}\prop{3}\prop{4}\prop{5}}
\\
I_2
&=C_1\,\mu^{2\epsilon} (4\pi)^2\int \frac{\dd^d k}{(2\pi)^d}
	\frac{\bar v_2 \gamma^\mu \momsl{2}\pol{1} \momsl{3} \pol{2} \momsl{4} \gamma_\mu(\slashed p_{15}) \pol{3} u_1}
		{s_{15}\prop{1}\prop{2}\prop{3}\prop{4}}
\\
I_3
&=C_1\,\mu^{2\epsilon} (4\pi)^2\int \frac{\dd^d k}{(2\pi)^d}
	\frac{\bar v_2 \gamma^\mu\momsl{2}\pol{1} \momsl{3} \pol{2} \momsl{4} \pol{3} \momsl{5}\gamma_\mu u_1}
		{\prop{1}\prop{2}\prop{3}\prop{4}\prop{5}}
\\
I_4
&=C_1\,\mu^{2\epsilon} (4\pi)^2\int \frac{\dd^d k}{(2\pi)^d}
	\frac{\bar v_2 \pol{1} (-\slashed p_{23})\pol{2} (\slashed p_{15})\gamma^\mu\momsl{4}\gamma_\mu(\slashed p_{15}) \pol{3} u_1}
		{s_{23}s_{15}^2\prop{1}\prop{4}}
\\
I_5
&=C_1\,\mu^{2\epsilon} (4\pi)^2\int \frac{\dd^d k}{(2\pi)^d}
	\frac{\bar v_2 \pol{1} (-\slashed p_{23}) \gamma^\mu\momsl{3} \pol{2} \momsl{4} \gamma_\mu(\slashed p_{15}) \pol{3} u_1}
		{s_{23}s_{15}\prop{1}\prop{3}\prop{4}}
\\
I_6
&=C_1\,\mu^{2\epsilon} (4\pi)^2\int \frac{\dd^d k}{(2\pi)^d}
	\frac{\bar v_2 \pol{1} (-\slashed p_{23})\gamma^\mu\momsl{3} \gamma_\mu(-\slashed p_{23}) \pol{2} (\slashed p_{15}) \pol{3} u_1}
		{s_{23}^2s_{15}\prop{1}\prop{3}}
\\
I_7
&=C_1\,\mu^{2\epsilon} (4\pi)^2\int \frac{\dd^d k}{(2\pi)^d}
	\frac{\bar v_2 \pol{1} (-\slashed p_{23}) \pol{2} (\slashed p_{15})\gamma^\mu \momsl{4} \pol{3} \momsl{5} \gamma_\mu u_1}
		{s_{23}s_{15}\prop{1}\prop{4}\prop{5}}
\\
I_8
&=C_1\,\mu^{2\epsilon} (4\pi)^2\int \frac{\dd^d k}{(2\pi)^d}
	\frac{\bar v_2 \gamma^\mu\momsl{2}\pol{1} \momsl{3}\gamma_\mu (-\slashed p_{23})\pol{2} (\slashed p_{15}) \pol{3}  u_1}
		{s_{23}s_{15}\prop{1}\prop{2}\prop{3}}
\end{align}
The IR counterterm reads:
\begin{equation}
I_{\text{IR}}=C_1\,\mu^{2\epsilon} (4\pi)^2\int \frac{\dd^d k}{(2\pi)^d}
	\frac{\bar v_2 \gamma^\mu\momsl{2}\pol{1} \momsl{s23} \pol{2} \momsl{s15} \pol{3} \momsl{5} \gamma_\mu u_1}
		{s_{23}s_{15}\prop{1}\prop{2}\prop{5}}
\end{equation}
The UV counterterms read:
\begin{align}
\label{A:bubble_UVCT}
I_{\text{UV}_4}
=&\quad C_1\,\mu^{2\epsilon} (4\pi)^2\int \frac{\dd^d k}{(2\pi)^d}
	\frac{\bar v_2 \pol{1} (-\slashed p_{23})\pol{2} (\slashed p_{15})\gamma^\mu\left(\momsl{uv}-\frac{\momsl{uv}\momsl{s15}\momsl{uv}}{\prop{uv}}\right)\gamma_\mu(\slashed p_{15}) \pol{3} u_1}
		{s_{23}s_{15}^2[\prop{uv}]^2}
\\
\label{A:triangle_UVCT}
I_{\text{UV}_5}
=&\quad C_1\,\mu^{2\epsilon} (4\pi)^2\int \frac{\dd^d k}{(2\pi)^d}
	\frac{\bar v_2 \pol{1} (-\slashed p_{23}) \gamma^\mu\momsl{uv} \pol{2} \momsl{uv} \gamma_\mu(\slashed p_{15}) \pol{3} u_1}
		{s_{23}s_{15}[\prop{uv}]^3}
\\
I_{\text{UV}_6}
=&\quad C_1\,\mu^{2\epsilon} (4\pi)^2\int \frac{\dd^d k}{(2\pi)^d}
	\frac{\bar v_2 \pol{1} (-\slashed p_{23})\gamma_\mu\left(\momsl{uv}-\frac{\momsl{uv}\momsl{s23}\momsl{uv}}{\prop{uv}}\right) \gamma_\mu(-\slashed p_{23}) \pol{2} (\slashed p_{15}) \pol{3} u_1}
		{s_{23}^2s_{15}[\prop{uv}]^2}
\\
I_{\text{UV}_7}
=&\quad C_1\,\mu^{2\epsilon} (4\pi)^2\int \frac{\dd^d k}{(2\pi)^d}
	\frac{\bar v_2 \pol{1} (-\slashed p_{23}) \pol{2} (\slashed p_{15})\gamma^\mu \momsl{uv} \pol{3} \momsl{uv} \gamma_\mu u_1}
		{s_{23}s_{15}[\prop{uv}]^3}
\\
I_{UV_8}
=&\quad C_1\,\mu^{2\epsilon} (4\pi)^2\int \frac{\dd^d k}{(2\pi)^d}
	\frac{\bar v_2 \gamma^\mu\momsl{uv}\pol{1} \momsl{uv}\gamma_\mu (-\slashed p_{23})\pol{2} (\slashed p_{15}) \pol{3}  u_1}
		{s_{23}s_{15}[\prop{uv}]^3}
\\
I_{\text{UV}_{\text{IR}}}
=&\quad C_1\,\mu^{2\epsilon} (4\pi)^2\int \frac{\dd^d k}{(2\pi)^d}
	\frac{\bar v_2 \gamma^\mu\momsl{uv}\pol{1} \momsl{s23} \pol{2} \momsl{s15} \pol{3} \momsl{uv} \gamma_\mu u_1}
		{s_{23}s_{15}[\prop{uv}]^3}
\end{align}

\section{Loop-Tree Duality with raised propagators}
\label{sec:raised_propagators}
When a diagram contains raised propagators, the Minkowski representation of the integrand features complex poles in the energy with order higher than one. Thus, in order to generalise the integration of the energy component of loop momenta carried out in sect.~\ref{sec:ltd_intro}, it is necessary to use the definition of higher-order residues~\cite{Bierenbaum:2012th}.

Raised propagators generally appear at higher loops when a diagram has a propagator insertion on a propagator. They also appear as a result of using Integration by Parts identities. The UV counterterm we constructed also features a raised propagator, since in the UV limit every propagator scales as $1/\left(k^2-\mu_{\text{UV}}^2\right)$. 

Applying residue theorem to a general integral with raised propagators we obtain:
\begin{align}
\begin{split}
\oint_{x_+} d k_0& \frac{F(k_0,\vec{k})}{(k_0-x_+)^{1+n}(k_0-x_-)^{1+n}}
= \left. \frac{1}{n!}\pdv{^n}{k_0^n}\frac{F(k_0,\vec{k})}{(k_0-x_-)^{1+n}}\right|_{k_0=x_+}
\\
 &=\frac{1}{n!}\sum_{m=0}^n (-1)^{n-m}\frac{(2n-m)!}{(n-m)!m!}\frac{\left. \partial^m_{k_0}F(k_0,\vec{k})\right|_{k_0=x_+}}{(x_+-x_-)^{1+2n-m}} \,.
\end{split}
\end{align}
For the processes considered in this paper that needs UV regulation, namely the one-loop QCD corrections to the $d\bar d$ to photons, the numerator function $F$ will consist of a spinor contraction containing a product of order $n$ in the loop momentum $k$ and the other propagator excluded from this particular residue.

\bibliographystyle{JHEP}
\bibliography{biblio.bib}

\end{document}